\documentclass[11pt, a4paper, twoside]{Thesis} 
 \pdfoutput=1
\usepackage{dcolumn}   
\usepackage{bm}        
\usepackage{amssymb}   
\usepackage{amsmath}
\usepackage{sidecap}
\usepackage{graphicx,psfrag,subfigure}
\usepackage{adjustbox}
\usepackage{color}
\usepackage{slashed}
\usepackage{xcolor}
\usepackage{wrapfig}
\usepackage{lscape}
\usepackage{rotating}
\usepackage{tikz}
\usetikzlibrary{shapes.geometric, arrows}
\tikzstyle{startstop} = [rectangle, rounded corners, minimum width=2cm, minimum height=1cm,text centered, draw=black, fill=red!30,xshift=4cm]
\tikzstyle{io} = [trapezium, trapezium left angle=70, trapezium right angle=110, minimum width=1cm, minimum height=1cm, text centered, draw=black, fill=blue!30]
\tikzstyle{process} = [rectangle, minimum width=1cm, minimum height=1cm, text centered, draw=black, fill=orange!30]
\tikzstyle{decision} = [diamond, minimum width=3cm, minimum height=1cm, text centered, draw=black, fill=green!30]
\tikzstyle{arrow} = [thick,->,>=stealth]
\usepackage{multirow}

\newcommand{\bea}{\begin{eqnarray}}
\newcommand{\eea}{\end{eqnarray}}
\newcommand{\bee}{\begin{eqnarray*}}
\newcommand{\eee}{\end{eqnarray*}}
\newcommand{\be}{\begin{equation}}
\newcommand{\ee}{\end{equation}}

\usepackage{appendix}
\usepackage{lineno}
\modulolinenumbers[5]
\usepackage{array}
\usepackage{float}
\usepackage{placeins}
\usepackage{stackengine}
\usepackage{url}
\usepackage{numprint}
\usepackage{caption}
\usepackage{booktabs}  
\nprounddigits{3}
\newcolumntype{P}[1]{>{\centering\arraybackslash}p{#1}}
\newcolumntype{M}[1]{>{\centering\arraybackslash}m{#1}}
\setstackEOL{\#}
\setstackgap{L}{12pt}
\usepackage[utf8]{inputenc}
\usepackage{palatino}
\usepackage{mathtools}
\usepackage{layout}
\usepackage{afterpage}
\usepackage{lipsum}
\usepackage[Glenn]{fncychap}

\usepackage[colorlinks]{hyperref}
\hypersetup{
    colorlinks=true,
    linkcolor=blue,
    filecolor=magenta,      
    urlcolor=cyan,
}
\def\lsim{\mathrel{\raise.3ex\hbox{$<$\kern-.75em\lower1ex\hbox{$\sim$}}}}
\def\gsim{\mathrel{\raise.3ex\hbox{$>$\kern-.75em\lower1ex\hbox{$\sim$}}}}

\usepackage{fancyhdr}
\usepackage[avantgarde]{quotchap}
\usepackage{epsfig,epsf}
\usepackage{amsthm}
\usepackage{amsfonts}
\usepackage[all]{xy}
\usepackage{enumitem}
\usepackage{cite}
\usepackage{doi}
\usepackage{lipsum}
\usepackage{setspace}
\AtBeginEnvironment{subappendices}{%
\chapter*{Appendix}
\addcontentsline{toc}{chapter}{Appendices}
\counterwithin{figure}{section}
\counterwithin{table}{section}
}
\graphicspath{{Pictures/}} 
\hypersetup{urlcolor=black, colorlinks=false} 
\thesistitle{\sc  Aspects of Heavy Supersymmetry}
\supervisor{Prof. Sudhir K. Vempati}
\degree{Doctor of Philosophy}
\degreemajor{The Faculty of Science}
\authors{Priyanka Lamba}
\university{Indian Institute of Science}
\department{Centre for High Energy Physics}
\placelng{Bengaluru-560012, India}
\datesub{October, 2020}
\datesig{October, 2020}
\title{\ttitle} 

\begin{document}

\frontmatter 
\setstretch{1.6} 
\fancyhead{} 
\rhead{\thepage} 
\lhead{} 
\newcommand{\HRule}{\rule{\linewidth}{0.5mm}} 
\maketitle


\clearpage 

\clearpage
\newpage\null\thispagestyle{empty}\newpage

\Declaration

\setstretch{1.3} 
\dedication{\addtocontents{toc}{}
\begin{flushright}
 {\huge\calligra My Parents and Brother}  
\end{flushright}
	}
	\clearpage

\setstretch{1.3} 
\newpage\null\thispagestyle{empty}\newpage
\acknowledgements{\addtocontents{toc}{}
Foremost, I express my sincere gratitude to my supervisor Prof. Sudhir Vempati for giving me his time and invaluable guidance, for his patience with my temper and English, and for pushing me to work hard. I am thankful to him for giving me full freedom to make mistakes while also motivating me to find a way out of it. Thanks for helping me as a guide and a friend in handling difficult situations, personally and professionally. And most of all I thank you for keeping us in mind even in your tough times and for reminding me to always keep the big picture in mind.\\
I thank all the professors of CHEP for course work and discussions about Physics and socially and for giving a friendly and peaceful environment. I want to thank Prof. Chethan Krishnan for being patient with me in class with a lot of questions in multiple course works and for his straightforward attitude and advice. I have been motivated by his presence and by hearing his energy in Physics discussions. I want to thank Prof. Apoorva Patel for pedagogical lectures that taught us how to learn physics from beautiful math equations. I thank Prof. Sachin Vaidya for helping us learn group theory in a neat fashion. Thanks to Prof. Justin David for teaching us how to do long calculations with patience. Thanks to Prof. Aninda Sinha for clarifying doubts by explicit calculations and introducing various topics for further perusal. Thanks to Prof. B. Ananthanarayan for teaching us SM and forcing us to learn very useful tools in Mathematica that have been very important in research. Thanks to Prof. Rohini Godbole for teaching us supersymmetric phenomenology , with great energy and passion, that helped in my research work and for always being patient with my questions. I also want to thank Prof. Godbole for her discussions about and continuous motivation to women in STEM. Thanks to Prof. Biplob Bhattacharya and Prof. Diptiman Sen for discussions on different issues and for being approachable to students for any type of discussion.\\
I want to thank Prof. Xerxes Tata, Prof. Giancarlo D'Ambrosio, Prof. Sunil Mukhi, Prof. Dileep Jatkar, Prof. Namit Mahajan, Prof. Suvrat Raju for teaching us various useful topics in IISc and SERC school.
I want to specially thank my teachers and professors for their motivation and faith in my abilities that helped me reach the research field; my school teacher Yudhister Sharma, my college Prof. Saini, Prof. Anjali Krishnamurthy; my MSc dissertation guide Prof. Debajyoti Choudhury for his vital advice in directing me when I had little knowledge of PhD and its workings, for his continued advice, Physics discussions and motivation, especially in my difficult times, and for always believing in me.\\  
I want to sincerely thank my collaborators Prof. Emilian Dudas and Prof. Stephane Lavignac for Physics discussions that have been instructional and great learning experiences, and for being patient with and helping in my research work.\\
I want to thank Prof. Yann Mambrini, Dr. Maira Dutra, Dr. Nicolás Bernal for helping me learn dark matter during my visit in Orsay and for making my visit memorable.\\
I want to thank the SERC computation facility, especially for giving me access to the supercomputer "Cray-XC40 system SahasraT", without it, it is challenging to do numerics on time.\\
I also take this opportunity to thank Mrs. Mallika for keeping our offices and department always very neat and clean, to Mr. Keshava for taking care of all administrative regulations, and for ensuring that we do not have to deal with official bureaucracies. I am thankful to Sarvana for his help in various departmental work.\\
Now it is time to deeply thanks Family and friends whom we always take for granted, without them their is no color in life. I want to deeply thank my parents and brother and sister-in law and grandmother for giving me a lot of love and care and simultaneously helping me become independent and courageous. I thank them for believing in my decisions and for always understanding me and my philosophy of life. Thanks to my lovely brother for becoming all time best friend. Thanks to my cute and lovely nephews for making me laugh and be happy even in stressful times. I want to thank my partner, Sachin, and my in-laws for their all time encouragement and faith and extending me their unfailing support. I want to thank my whole maternal and paternal family for their support, especially my cousins Arti di, Priti di, Ujala di, Avinash bhaiya, Surjit, Sunita di and my Bauji for continuous support. \\
I want to thanks my friends without them it would be very difficult to live outside home for 13 years. I want to thank my friend Bhawana and Shiuli for being available any time for me during my PhD. I want to thank Bhawana for bearing me with my temper and for always being by my side in both good and bad times and bearing my talkative nature and always coming with smile and a lot of positivity. I want to thank Shiuli for many long physics discussions and life philosophical discussions and help me come out with my rigid and tough rules of life and help in my personal growth. Thanks for giving a lot of time for random things and being with me on all beautiful trips. I also want to thank Shiuli for proofreading my thesis and help me correct grammatical mistakes. I want to thank both of you for making my life easier in PhD. I thank Vikas, my curious acquaintance for increasing my patience and perseverance, for his passion in cooking and sharing with us and for keeping the batchmates together; Mahesh, Riya, Moid for philosophical discussions and comfortable companionship, Lata for not letting me forget my people's food. I also thank my off-campus friends Megha, Lata Yadav, Neha, Sunanda, Swati, Priya, Sunil, Divya di, Shifa, Simran.  \\
Thanks to my friendly postdocs Charanjit and Mathew for a lot of useful Physics discussion and for supporting me emotionally and advising me on different aspects of PhD.\\
I want to thank Pawan bhaiya for discussing Physics and always listening to my worries and for giving life lessons and being a well-wisher. I learned how to engage with all kinds of conversations with a smile. I feel the loss of your presence.\\
I want to thank my seniors Gaurav da, Rahool da, Ashwani, Anjani, Sayan, Mahul, Surabhi, Ranjani, Parveen bhaiya, Avinash, Abhiram, Adhip bhaiya, Jayita, Ayon da, Akhilesh, Satish bhaiya, my juniors Alam, Prabhat, Rhitaja, Deepak, Sipaz, Brijesh for making the department lively and making PhD life less stressful via get-togethers and for making me more social. Thanks to my batchmates Ratan, Amit, Dipankar, Aditya, Jagannath, Sudeb, Sruthi, Sheena, Sreeshma, Srishti, Sanhita, Arnab, Tushar, Debi, Shoubhik, Sanyukta, Khushboo.\\
I want to express my sincere gratitude towards University Grants Commission (UGC) for granting a full time fellowship while pursuing my PhD at Indian Institute of Science. I also thank Mr. Manohar at administrative department for making our life easy by handling extensive paperwork for fellowship and grants for visits.\\
I thank the CNRS LIA (Laboratoire International Associ) THEP (Theoretical
High Energy Physics), the INFRE-HEPNET (IndoFrench Network on High Energy Physics) of
CEFIPRA/IFCPAR (Indo-French Centre for the Promotion of Advanced Research) and Prof. Fawzi Boudzema for providing me with funding for visits to CPhT, Ecole Polytechnique, LPhT, CEA Saclay and LPT Orsay for my research work. I also thank
CEFIPRA for the individual project grant ”Glimpses of New Physics''.\\
I thank the IISc administration and community for the beautiful campus that has played a great role in harboring positivity and promoting good mental health. 

}

\clearpage 
\newpage\null\thispagestyle{empty}\newpage
\setstretch{1.3} 
\Synopsis{{}
The recent discovery of a Higgs boson with a mass around 125 GeV, taken together with  experimental results from  flavor factories, dark matter direct  detection, and searches for SUSY particles in the LHC suggest that supersymmetric particles could be heavy in the range of multi-TeV beyond the reach of LHC. In typical supersymmetry breaking models, supersymmetry breaking in the hidden sector  is parametrised  by a single spurion field mediated at a specific scale, which we call single scale breaking. Although in this case, heavy-scale SUSY breaking models have very large fine-tuning, they are simultaneously economic from the phenomenological perspective. In a typical heavy-scale SUSY model, the gauginos and higgsinos are still around the TeV scale since they can be protected by chiral symmetry, and contain a dark matter candidate. The heavy sfermions relax bounds coming from flavor changing processes and CP violation, and also  increase the radiative corrections to the  Higgs mass.\\
The aim of the thesis is to study the implications of the heavy supersymmetry while relaxing the assumption of single scale mediation for supersymmetry breaking. In the first study of this thesis, we consider MSSM with $N_{HS}$ sequestered hidden sectors at a high scale, contributing to supersymmetry breaking.  Each hidden sector communicates supersymmetry breaking to the visible (MSSM) sector through effective interactions. Considering a random distribution for the spurion parameters leads a normal distribution for the soft parameter  with mean values and standard deviations that are analytically computable. We study the probability of getting Higgs mass in the correct range while having successful electroweak symmetry breaking. We show that the probability 
distribution is peaked when the quanta of supersymmetric breaking is around $\tilde{m} = 220 $ GeV for the parametrisation of the spurion fields we have considered. For these regions we study the supersymmetric spectrum. \\
In the next work, we study fine-tuning, where each of the spurion's contribution is  para-metrised as $\tilde{m} M_{Pl} c_\alpha$. We treat each spurion field as an independent source of supersymmetry breaking. Requiring minimal fine tuning from each sector,  gives $C^{2N_{HS}-1}_r$ solutions. In fact we find there is only one solution independent of all the RG coefficients, where all the sectors contribute coherently. The fine-tuning becomes almost negligible even with a small number of hidden sectors,  $N_{HS}=20$. The coherent SUSY framework also has a well-tempered dark matter region due to high cancellation in gaugino soft terms. It also has regions of coannihilation  with charginos.\\ 
A concrete realisation with a large number of hidden sectors is presented in the next chapter, where we consider a Stringy landscape like scenario inspired by Bousso-Polchinski's solution to the cosmological constant problem. The spurion fields are given in terms of quantized four form fluxes whose vacuum expectation values set the supersymmetry breaking scales. Coupled to supergravity, we compute the soft spectrum in this framework assuming a uniform  distribution for the quantized flux charges. We have shown that this framework  naturally leads to a suppression of the flavor violating entries as $1/\sqrt{N_{HS}}$. There is further suppression due to the renormalisation group running at the weak scale,\ especially for the hadronic mass insertions. \\
In the last part of the thesis, we consider single scale supersymmetry breaking and study the implications of the heavy spectrum in the context of SUSY GUT. We consider SUSY SU(5) with a novel decoupling scenario named flavored 'split-generation', where $\mathcal{O}(1)$ flavor violation can be present in the model.  In this scenario, first and second generation of sfermions are assumed to be heavy (order of 10s of TeV) and the remaining SUSY spectrum lies around a few TeV. Two codes have been developed for this work. 1) A modified version of SuSeFLAV has been developed where 
we calculate the full two-loop $\beta-$coefficients and one-loop threshold at the two different  scales.  2) A code for full SUSY SU(5) proton decay analysis. In this work, we study gauge coupling unification and the two dominant proton decay channels ($p \rightarrow e^+ \pi^0$ and $p \rightarrow K^+ \bar\nu$) both with and without flavor mixing in a heavy and light (third) generation. The flavored ‘split-generation' scenario leads to peculiar cancellations in the amplitudes. The rate of $p \rightarrow \pi^0 e^+$ is highly sensitive to amount of flavor violation present as it opens the gluino contribution to the amplitudes. On the other hand, we find that sensitivity of the rates of  $p \rightarrow K^+ \bar\nu$ is related to the flavor of the neutrino emitted. The implications for future proton decay experiments like Hyper-K, DUNE and JUNO are reported. 
 
}
\clearpage 

\setstretch{1.3} 
\listofpublications{\addtocontents{toc}{}
\begin{itemize}
\item  “Diluting SUSY flavour problem on the Landscape”, Emilian Dudas, Priyanka
Lamba, and Sudhir K. Vempati (Published in Phys.Lett.B 804 (2020) 135404)
\item “A fresh look at proton decay in SUSY SU(5)”, Charanjit Kaur, Steph\'ane Lavignac,
Priyanka Lamba and Sudhir K. Vempati (In preparation, soon to be online)
\item “Coherent SUSY”, Priyanka
Lamba and Sudhir K. Vempati (In preparation, soon to be online)
\item “MSSM with Multiple Hidden Sectors", Priyanka Lamba and Sudhir K. Vempati (In preparation, soon to be online)
\end{itemize}

\hspace{.5cm} Codes developed for the work presented in the thesis:  

\begin{itemize}
\item “Sproton\_decay” , Priyanka Lamba

({\color{blue}https://github.com/Priyanka-Lamba/Sproton\_decay})
\item "SuSeFLAV\_H", Priyanka Lamba \textit{et. al}

({\color{blue}https://github.com/Priyanka-Lamba/SuSeFLAV\_H})
\end{itemize}
}

\clearpage 

\pagestyle{fancy} 

\lhead{\bfseries\texttt{Contents}} 
\tableofcontents 

\lhead{\bfseries\texttt{List of Figures}} 
\listoffigures 

\lhead{\bfseries\texttt{List of Tables}} 
\listoftables 


\clearpage 

\setstretch{1.5} 

\lhead{\bfseries\texttt{Abbreviations}} 
\listofsymbols{ll} 
{
\textbf{ATLAS} & A Toroidal LHC ApparatuS\\
\textbf{BBN} & Big Bang Nucleosynthesis\\
\textbf{BG} & Barbieri-Giudice\\
\textbf{BP} & Bousso-Polchinski\\
\textbf{BSM} & Beyond Standard Model\\
\textbf{CC} & Charged Current\\
\textbf{CKM} & Cabibbo-Kobayashi-Maskawa\\
\textbf{C.L.} & Confidence Level\\
\textbf{CLT} & Central Limit Theorem\\
\textbf{CMB} & Cosmic Microwave Background\\
\textbf{CMS} &  Compact Muon Solenoid \\
\textbf{CMSSM} & Constrained Minimal Supersymmetric Standard Model\\
\textbf{CP} & Charge Conjugation Parity Symmetry\\
\textbf{DM} & Dark Matter\\
\textbf{EW} & Electro-Weak\\
\textbf{EWSB} &  Electro-Weak Symmetry Breaking\\
\textbf{FC} & Flavor-Changing\\
\textbf{FCNC} & Flavor-Changing Neutral Currents\\
\textbf{FP} & Focus-Point\\
\textbf{GIM} & Glashow–Iliopoulos–Maiani \\
\textbf{GMSB} & Gauge Mediated Supersymmetry Breaking\\
\textbf{GUT} & Grand Unified Theory\\
\textbf{HB} &  Hyperbolic Branch\\
\textbf{LEP} & Large Electron–Positron Collider\\
\textbf{LHC} & Large Hadron Collider\\
\textbf{LSP} & Lightest Supersymmetric Particle \\
\textbf{mGMSB} & Minimal Gauge Mediated Supersymmetry Breaking\\
\textbf{MSSM} & Minimal Supersymmetric Standard Model\\
\textbf{mSUGRA} & Minimal Supergravity\\
\textbf{NK} & Nomura-Kitano\\
\textbf{NNLO} & Next-to-Next-Leading Order\\
\textbf{QCD} &  Quantum Chromodynamics\\
\textbf{QFT} & Quantum Field Theory\\
\textbf{RGE} & Renormalisation Group Equation\\
\textbf{RG} & Renormalisation Group\\
\textbf{SM} & Standard Model\\
\textbf{SSM} & Supersymmetric Standard Model\\
\textbf{SUGRA} & Supergravity\\
\textbf{SUSY} & Supersymmetry \\
\textbf{vev} & Vacuum Expectation Value\\
\textbf{WIMP} & Weakly Interacting Massive Particle \\
\textbf{Y} & Hypercharge 
}

%
%
%


\clearpage 




%
%
%
%


\mainmatter 

\pagestyle{fancy} 


\chapter{Introduction} 

\label{Chapter 1} 

\lhead{Chapter 1. \emph{Introduction}} 

 \onehalfspacing
\section{Motivation}
Supersymmetry is one of the most popular models describing physics beyond the Standard Model. Supersymmetric field theories have many exciting properties, such as non-renormalizable theorems, exact renormalization groups, non-perturbative solutions. Supersymmetry is a non-trivial extension of the Poincare group in QFT,  and it incorporates gravity when it is made local. Soft supersymmetry breaking enables us to keep most of these properties intact. Supersymmetry also has excellent phenomenological properties such as a solution of the hierarchy problem, lightest supersymmetric particle (LSP) as Dark matter, exact unification of gauge couplings, and stabilization of the EW vacuum. In recent years, there has been a significant change in the perception of supersymmetric theories.
Traditionally, supersymmetric particles have been expected to lie around the scale of 1 TeV so as to be discoverable at the LHC. The last few years have brought a paradigm shift in the understanding of physics beyond Standard Model, particularly in supersymmetric theories. The discovery of the Higgs particle at the LHC in 2012 and the subsequent non-discovery of supersymmetric partners of the Standard Model particles have perhaps dashed the expectations. The current thesis, however, takes the viewpoint that indications for a more massive supersymmetric spectrum were probably already present, taking into consideration the heaviness of the discovered Higgs mass and constraints from indirect searches such as flavor, CP violation, and dark matter. The null results from the LHC have just reinforced these expectations. In the following, we summarise the various direct and indirect constraints on the MSSM parameter space before discussing the probable heaviness of the supersymmetric spectrum.
\section{Higgs Discovery}
\begin{figure} [h!]
\centering    
\subfigure{\includegraphics[width=7cm,height=6cm]{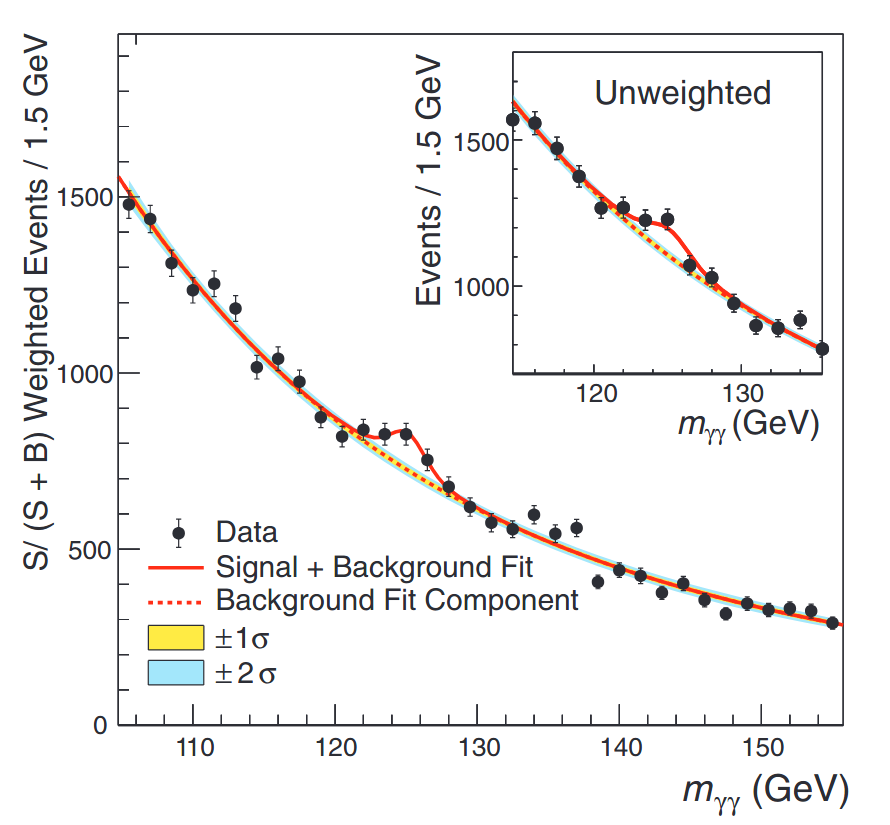}}
\subfigure{\includegraphics[width=7cm,height=6cm]{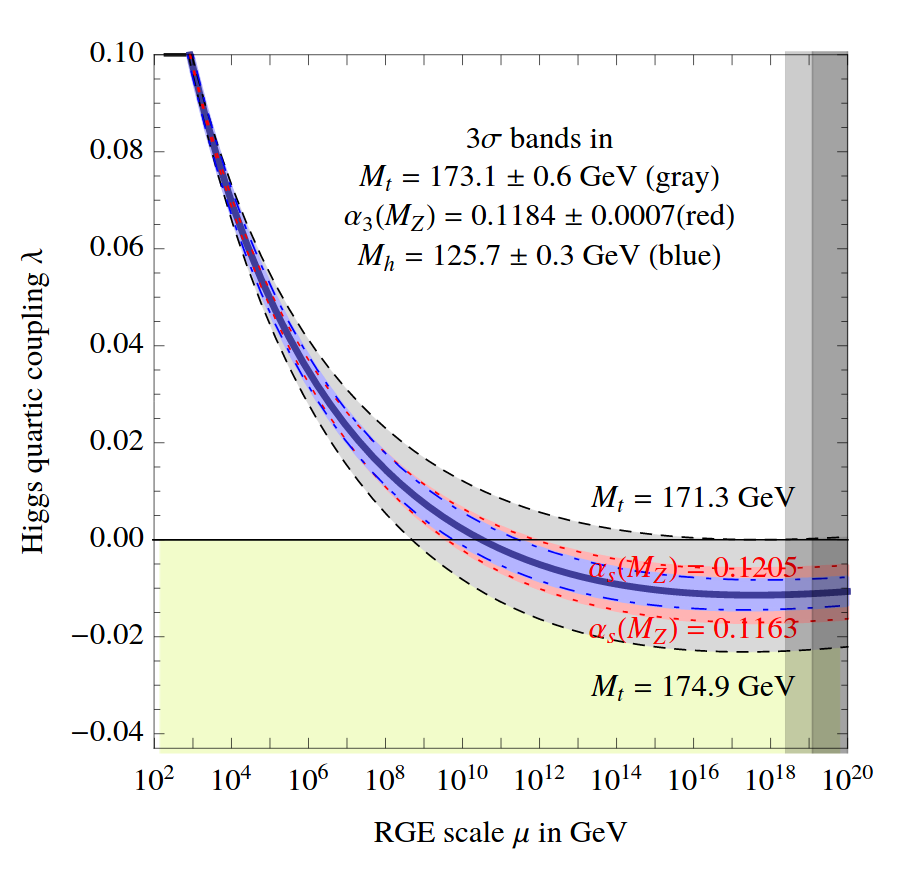}}
\caption{\textbf{left:} The invariant mass distribution of di-photon pair for the 7 and 8 TeV data colleted by CMS in 2011 and 2012, respectively from Higgs boson search in the $h \rightarrow \gamma \gamma$ channel\cite{CMS:2012nga}. \textbf{right:} RG evolution of the quartic coupling $\lambda$ in the SM \cite{Degrassi:2012ry}.
}
\label{fig:CMShiggs}
\end{figure}

\begin{figure} [h!]
\centering    
\subfigure{\includegraphics[width=7cm,height=6cm]{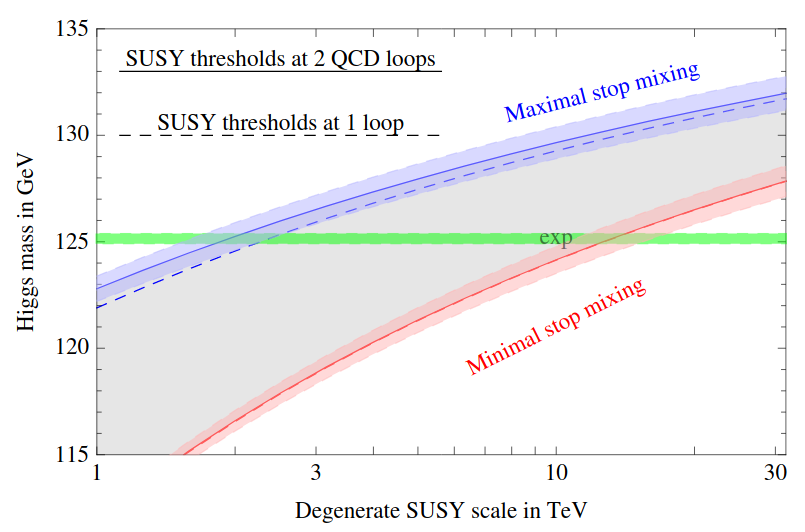}}
\subfigure{\includegraphics[width=7cm,height=6cm]{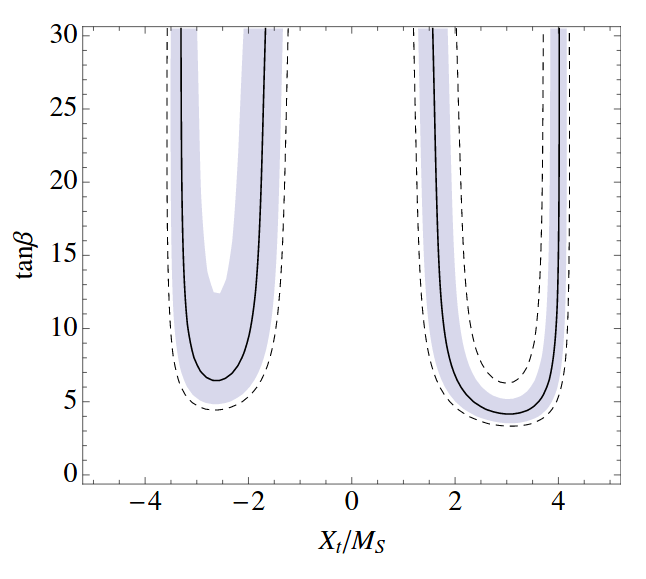}}
\caption{\textbf{left:} The Higgs mass as a function of the SUSY sclae, with degenerate spectrum of superpartners and $\tan{\beta}$ \cite{Bagnaschi:2014rsa}. \textbf{right:} The stops masses are set at 2-3 TeV and the solid curve is $m_h=125$ GeV with $m_t=173.2$ GeV. The band around curve and the dashed lines correspond to varying of Higgs mass and top mass by 2-3 GeV, respectively\cite{Draper:2011aa}.
}
\label{fig:higgsconstriant}
\end{figure}

In the sixties, theoretical physicists proposed a field (the Higgs Field) that permeates the universe and gives energy to the vacuum. The Higgs field gives masses to all SM particles, which was the last missing piece of puzzle of the SM. The discovery of the Higgs boson in 2012 at the ATLAS and CMS\cite{Chatrchyan:2012ufa,Aad:2012tfa,CMS:2012nga,ATLAS:2012oga} with mass of 125.09 GeV completed the already impressive triumph of the Standard Model as the correct low-energy description of the three gauge interactions. The Higgs boson's main experimentally accessible decay channels are $h\rightarrow \gamma\gamma$, $ZZ^*$ $\rightarrow 4l$,  $WW^*\rightarrow 2l2\nu$, $b\bar{b}$, $\tau\bar{\tau}$. The dominant discovery channel at both ATLAS and CMS was $h\rightarrow \gamma\gamma$, it being the cleanest channel. All other decays channels have also been analyzed. The peak positions in  $h\rightarrow \gamma\gamma$  channels with the best mass resolution are consistent with the observation of a single new particle (the Higgs boson), see fig. \ref{fig:CMShiggs}.
 Different channels help us to understand different properties of the Higgs boson, like  $h\rightarrow \gamma\gamma$ indicates that its spin could be 0 or 2\cite{yang:1950rg}, $h\rightarrow ZZ^*$ $\rightarrow 4l$, $h\rightarrow  WW^*\rightarrow 2l\;2\nu$ determine CP properties of the Higgs\cite{Choi:2002jk,Ellis:2012wg} and $h\rightarrow b\bar{b}, h\rightarrow \tau\bar{\tau}$ help differentiate between normal spin 0 particle and a remnant spin 0 particle of spontaneous breaking which gives mass to gauge bosons, fermions\cite{Guralnik:1964eu,Higgs:1964pj,Englert:1964et} (i.e. heavier fermion has larger branching ratio than a lighter one in the decay of the Higgs boson). The measurement from Tevatron\cite{Group:2012zca} and ATLAS/CMS experiments tells us, it is a remnant spin 0 particle of spontaneous breaking with even CP. Now we can ask which model this Higgs boson lies in, i.e., the SM, the supersymmetric SM, a composite Higgs model, a string model, or some other model. The effective way to look for new physics is to examine the Higgs boson coupling's deviation from that of the SM. The inclusion of vacuum stability gives more constraints on the SM type Higgs boson. The analysis of the vacuum stability is sensitive to quantum corrections. An NNLO analysis in the SM demanding the vacuum to be absolutely stable up to the Planck scale requires\cite{Degrassi:2012ry}:
 \be
 m_h > 129.4 + 1.4\left(\frac{m_t-173.1}{0.7}\right)-0.5\left(\frac{\alpha_s(m_Z)-0.1184}{0.0007}\right) \pm 1.0
 \ee
 where masses are in GeV. '$\pm1.0$' is the theoretical error in the evaluation of $m_h$. The experimental errors on the top mass and the analysis of ref. \cite{Degrassi:2012ry} gives 
 \be
 m_h > 129.4 \pm 1.8\; \mbox{GeV}
\ee
This excludes the vacuum stability for $m_h < 126$ GeV in SM at the 2$\sigma$ level, see fig. (\ref{fig:CMShiggs}. The observed Higgs mass of $125.01$ GeV would give vacuum stability up to $10^{10}$ GeV scale. The stability of the electroweak vacuum up to Planck scale would then require new physics.
		\begin{figure}[h!]
			\centering
			\includegraphics[width=10.5cm,height=7cm]{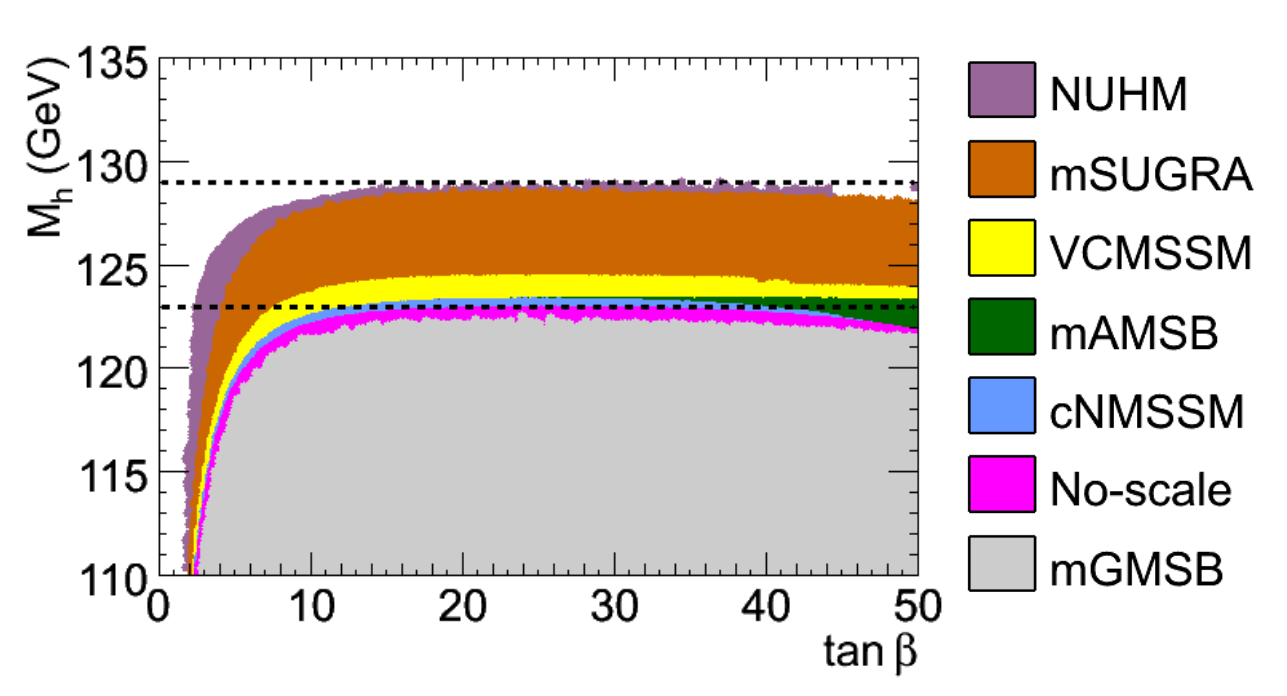}
			\caption{The value of $M_S$ is taken to be less than 3 TeV. This analysis for different models has been done in ref. \cite{Arbey:2012dq,Arbey:2011ab}.}
			\label{fig:higgsattack}
		\end{figure}\\
 Suppose this Higgs boson is a supersymmetric SM Higgs boson, what information can the Higgs discovery then give us about the type of SUSY spectrum or its breaking mechanism? If supersymmetry is exact, there is no radiative correction to Higgs mass due to cancellation between the SM particles and its superpartners. But as we know, SUSY is broken softly (see \ref{Appendix B}) and that gives finite logarithmic correction to Higgs mass. Due to large top Yukawa, the dominant one-loop correction comes from the top-stop sector using  effective potential at one loop,  given
 by\cite{Nath:2012nh,Carena:2002es}
 \be
 m_h^2\lsim m_Z^2\cos^2(2\beta) + \frac{3g^2m_t^4}{8\pi^2m_W^2}\left[\ln{\frac{M_S^2}{m_t^2}}+\frac{X_t^2}{M_S^2}\left(1-\frac{X_t^2}{12M_S^2}\right)\right]
 \ee
where $M_S$ is an average stop mass, $X_t$ is the trilinear coupling (top-squark mixing parameter), and $m_Z^2\cos^2(2\beta)$ is tree -level contribution in MSSM. The measured value of Higgs would require either a large mass of stop of around 10 TeV \cite{Kitano:2006gv,Bagnaschi:2018igf} or large trilinear coupling for lighter stop mass (of less than 2-3 TeV) which might conflict with charge and color breaking minima\cite{Chowdhury:2013dka}, see fig. (\ref{fig:higgsconstriant}). In terms of supersymmetry breaking, the discovery of the Higgs has strong implications for many models. As we can see from fig. (\ref{fig:higgsattack}), some of them like minimal gauge mediation (mGMSB) and its variations are disfavoured in the light of the Higgs discovery\cite{Draper:2011aa} or require a rather heavy spectrum in the range of multi-TeV\cite{Bagnaschi:2018igf}. While supergravity grand unified models like mSUGRA, CMSSM, etc. are more likely.

\section{LHC Results}

\begin{figure}[h!]
\centering
\subfigure[]{%
\includegraphics[width=7cm,height=6cm]{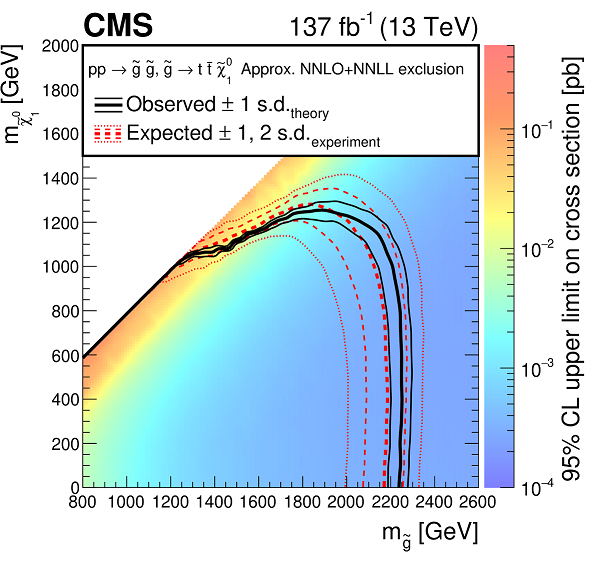}}%
\hfill
\subfigure[]{%
\includegraphics[width=7cm,height=6cm]{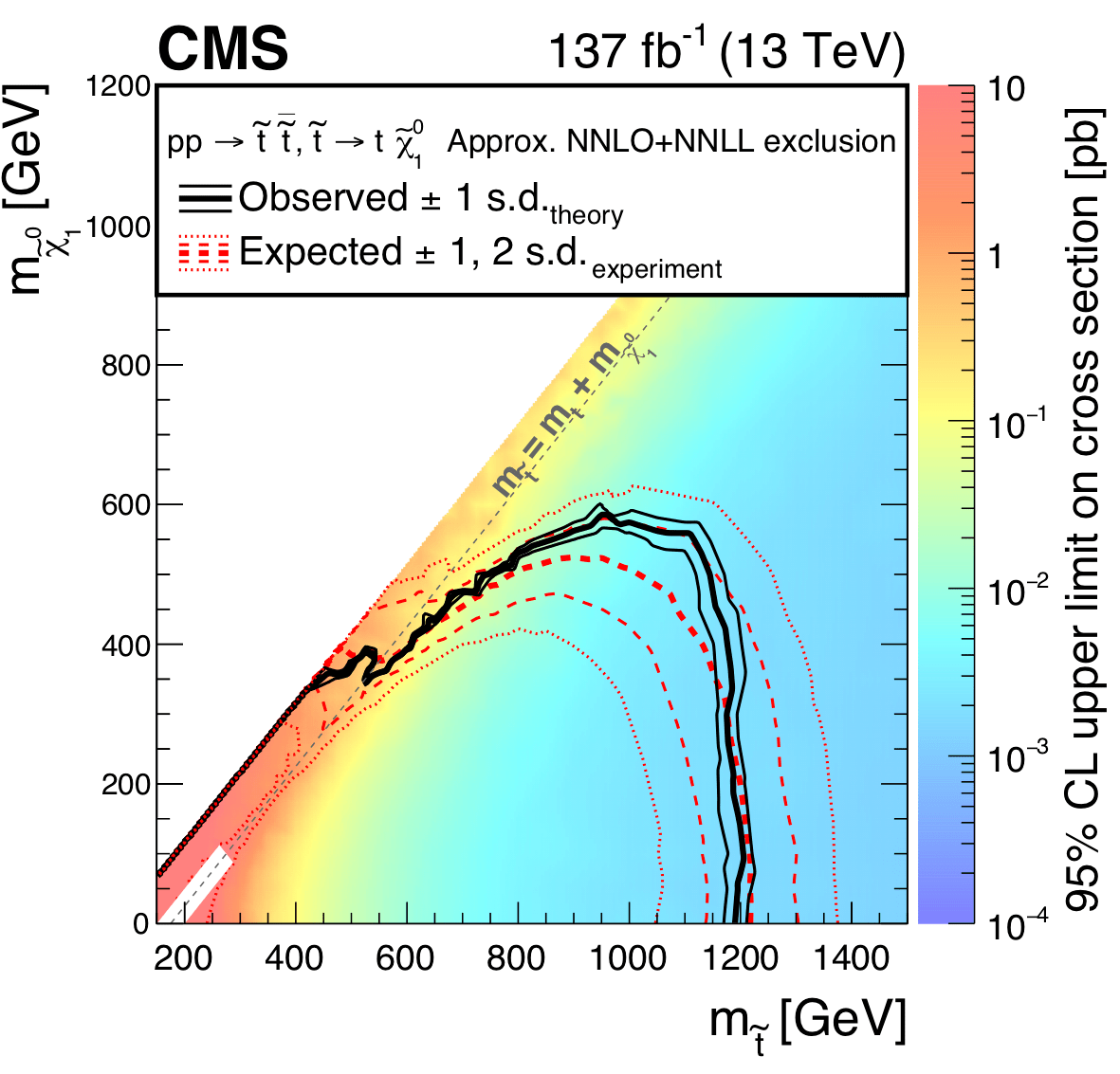}}%
\vskip\baselineskip
\subfigure[]{%
\includegraphics[width=7cm,height=6cm]{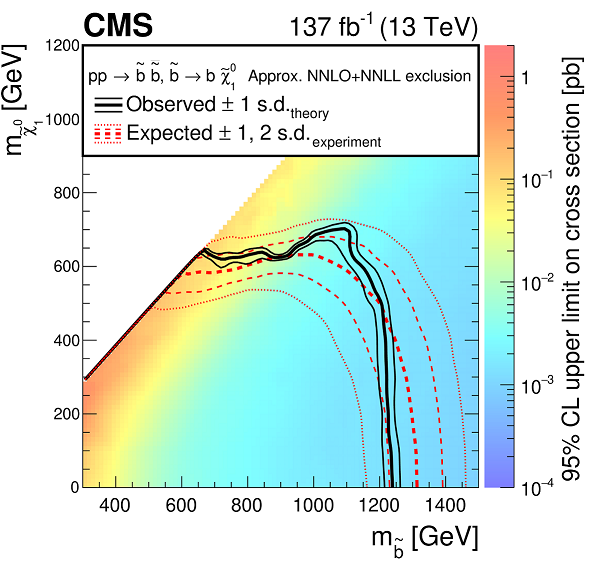}}%
\hfill
\subfigure[]{%
\includegraphics[width=7cm,height=6cm]{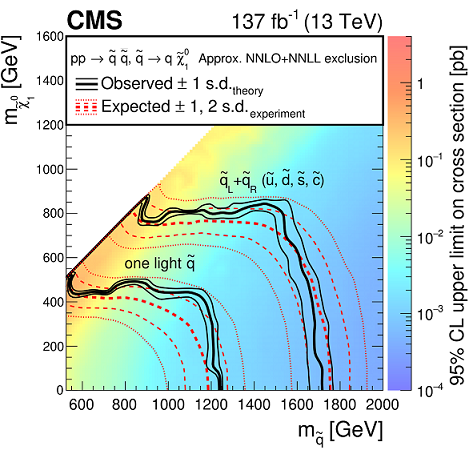}}%
\caption{Exclusion limits at 95$\%$ C.L. fig. (a) for direct gluino pair production where the gluinos decay to top quarks, fig. (b) for top squark pair production, fig. (c) for bottom squark pair  production and fig. (d) for light-flavor squark pair production. In all these figures, unity branching fraction for indicated decay is assumed\cite{Sirunyan:2019xwh}.}
\label{fig:CMSresult}
\end{figure}

Apart from the Higgs boson, the LHC has neither seen any new particles nor any indications of new physics. Supersymmetry has been considered the most attractive theory and therefore a lot of dedicated searches have been carried out in both ATLAS and CMS, to discover or to understand the nature of the SUSY spectrum. They consider “simplified models’’ to perform searches \cite{Alves:2011wf,Canepa:2019hph}. The different simplified search results can be imposed on different SUSY models, and we can consider these results as lower bounds on the spectrum. They study both R-parity conserving and R-parity violating decays and production. The LHC excludes SUSY below 1-2 TeV, see table \ref{tab:LHCbound}. These limits strongly challenge weak scale SUSY and almost completely rule it out, directing us towards a heavy scale supersymmetry. These limits are typically obtained when the sparticle decay mode is considered with a 100$\%$ branching fraction or with some simplified spectrum, and therefore these bounds can also be relaxed by considering a degenerate spectrum, compressed spectrum, or with a detailed study of branching ratios in particular models.
\begin{table}[h!]
    \centering
    \begin{tabular}{||c|c||}
    \hline
        Particle &  Mass (TeV)\\
        \hline
       Gluino  & $> 2.2$ \\
       \hline
       Stop    & $> 1.2$ \\
       \hline
       Sbottom & $> 1.2$ \\
       \hline
       Squark (1st two generation) &$ > 1.7$\\
       \hline
    \end{tabular}
    \caption{LHC lower bounds on SUSY spectrum using simplified models, see fig. \ref{fig:CMSresult}}
    \label{tab:LHCbound}
\end{table}
\section{WIMP Dark Matter Status}
\label{sectiondark}
		\begin{figure}[h!]
			\centering
			\includegraphics[width=11.5cm,height=8cm]{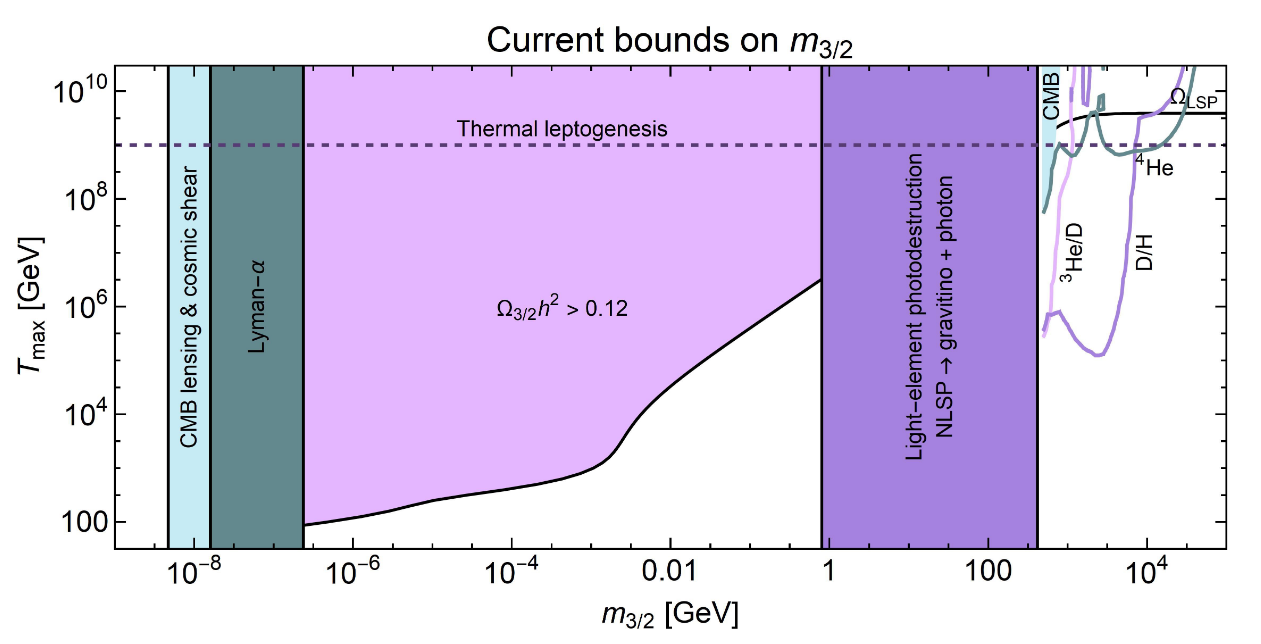}
			\caption{A combination of bounds coming from different experiments in the $m_{3/2}-T_{max}$ plane, where $T_{max}$ is the maximum temperature from which radiation dominated universe starts and reheating temperature in the inflationary models\cite{Hook:2018sai}.}
			\label{fig:gravitinomass}
		\end{figure} 
In 1933, Zwicky\cite{Zwicky:1937zza,Zwicky:1933gu} analyzed the motion of galaxies in the Coma cluster. Clusters are bound systems of galaxies, and their masses can be calculated using the Virial theorem. He showed that the mass necessary to explain the unity of the Coma cluster was much higher than the luminous mass of the cluster. It was the first evidence of non-luminous matter. Later, V. Rubin measured the rotation curves of galaxies in 1969\cite{Rubin:1970zza}, and the flatness of rotation curves at a larger distance could not be explained only by luminous matter. They needed a large amount of non-luminous matter to explain it. Later this non-luminous matter was named “Dark matter” and since then, various experimental observations have confirmed the existence of dark matter, like the kinematics of virial systems and rotating spiral galaxies, gravitational lensing of background objects, gravitational lensing and X-ray observations from the Bullet Cluster, and studies of CMB. The Planck collaboration has precisely measured the dark matter relic density, $\Omega h^2=0.120\pm0.001$ at $68\%$ C.L. \cite{Aghanim:2018eyx}, using CMB anisotropies. The most popular candidate for dark matter is Weakly Interacting Massive Particle (WIMP), which satisfies the relic abundance of DM.  There are majorly three ways to detect dark matter
\begin{itemize}
\item Indirect detection: Indirect searches from DM annihilation and decay are conducted using various satellites and telescopes (Fermi-LAT, Planck, PAMELA, AMS, H.E.S.S.). We have to deal with a lot of astrophysical backgrounds in these experiments.
\item Direct detection: In these experiments, we try to detect the scattering of dark matter from the halo on target nuclei on Earth to calculate mass and coupling of dark matter with SM. Heavily shielded underground detectors are used to detect dark matter. These detectors measure the amount of energy deposited by DM during scattering with target nuclei (DAMA, LUX, CDMS, CoGeNT, CRESST, PandaX, XENON1T, etc.).
\item Searches in colliders: Limits on dark matter are obtained from searches of dileptons, dijet resonances, monojet, single photons with missing transverse energy, at LHC and LEP.
\end{itemize}
\begin{figure}[h!] 
\centering    
\subfigure{\includegraphics[width=7cm,height=6cm]{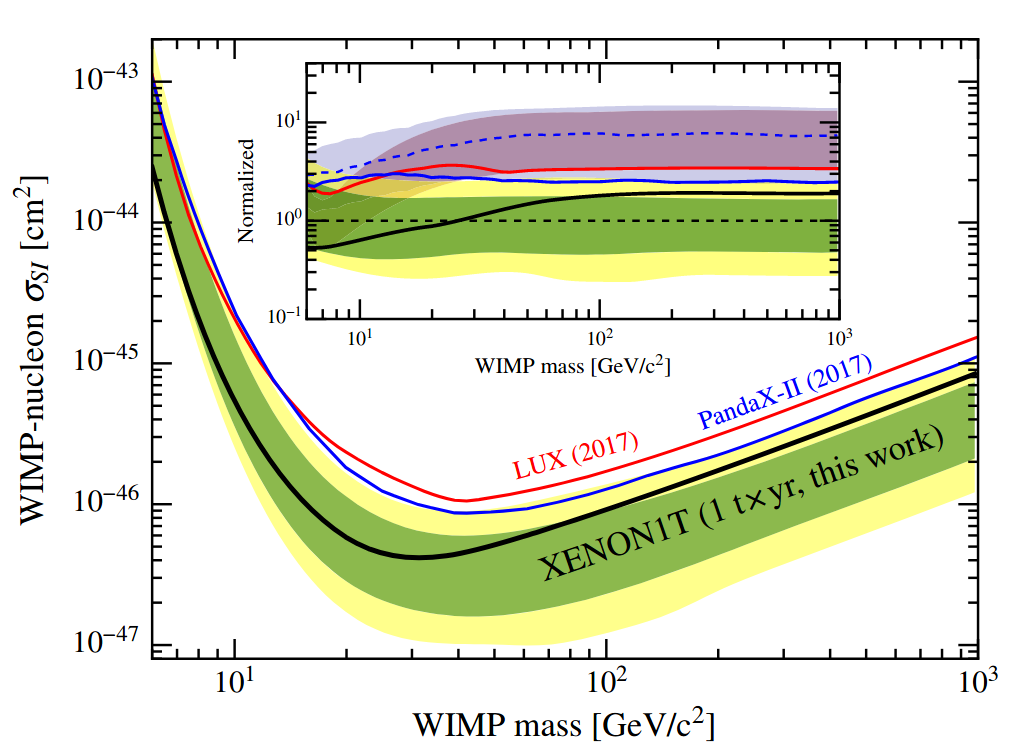}}
\subfigure{\includegraphics[width=7cm,height=6cm]{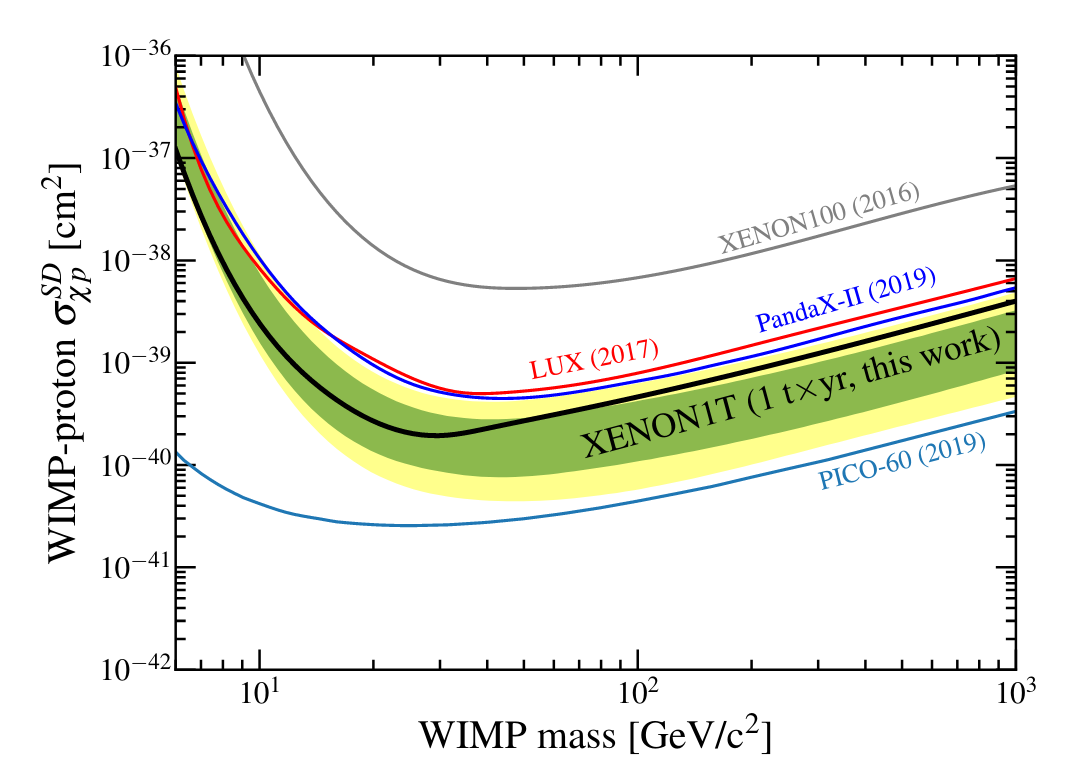}}
\caption{\textbf{left:} Spin-independent WIMP nucleon cross-section vs WIMP mass\cite{Aprile:2018dbl}. Xenon1T gives the strongest bound. \textbf{right:} Spin-dependent WIMP nucleon cross-section vs WIMP mass\cite{Aprile:2019dbj}.PICO-60 $C_3F_8$\cite{Amole:2019fdf} gives the strongest bound.
}
\label{fig:SISDWIMP}
\end{figure}
The exciting phenomenological feature of R-parity conserving SUSY is that the LSP is stable and is a good candidate for DM. The nature of LSP depends on the SUSY breaking mechanism. In some SUSY models like SUGRA, the LSP is the neutralino (pure gaugino, pure higgsino, or mixed state of gaugino and higgsino) whereas in other SUSY models like GMSB, the LSP is gravitino. Various experimental bounds like the LHC searches for missing energies, cosmological data, and direct detection of dark matter are hunting for SUSY dark matter from all directions. For example, gravitino as a DM has many bounds coming from cosmological data, see fig. \ref{fig:gravitinomass}, neutralino dark matter has strong constraints from direct detection and LHC searches. The dark matter density is also a big constraint on LSP masses. Consider the lightest neutralino which has the following form
\be
\chi^0_1 = N_{\tilde{B} 1} \tilde{B}^0 + N_{\tilde{W} 1} \tilde{W}^0 + N_{\tilde{H_u} 1} \tilde{H}_u^0 +
N_{\tilde{H_d} 1} \tilde{H}_d^0
\ee
where N's are mixing parameters, and look at the various possible compositions of the LSP to satisfy the relic density.\\
\noindent
(a) \textit{Pure bino:} The annihilation cross-section is given by \cite{ArkaniHamed:2006mb}
\be
\langle \sigma_{\chi} v \rangle = \frac{3 g^4 \tan\theta_W^4 r(1+r^2)}{2 \pi m_{\tilde{e}_R}^2 x (1 +r)^4}
\ee
where $x= M_1/T$ the mass of the bino over the temperature and  $r \equiv M_1^2/m_{\tilde{e}_R}^2$, $\theta_W$ is the weak mixing angle, or the Weinberg angle. The relic density in this case is given by 
\be
\Omega_{\tilde{B}}h^2 = 1.3 \times 10^{-2}  \left( \frac{  m_{\tilde{e}_R}}{100~ \text{GeV}}\right)^2 \frac{ (1+r)^4}{r^2 (1+r)^2} \left(1+ 0.07 \log\frac{\sqrt{r} ~~100 ~\text{GeV}}{m_{\tilde{e}_R}} \right)
\ee
The above relic density is typically large for reasonable range of parameters. One thus typically invokes co-annihilating partners which have a mass that is very close to that of the bino, there-by increasing the cross-section and bringing down the relic density to observed values. \\
\noindent
(b) \textit{Pure wino:} The annihilation cross-section of the dark matter particle is proportional to $g^4$, the weak coupling:
\be
\langle \sigma_{\chi} v \rangle  =  \left( \frac{3 g^4}{16 \pi M_2^2} \right)
\ee
where $M_2$ stands for the wino mass. The relic density is approximately given by
\be
\Omega_{\tilde{W}}h^2 \sim 0.13   \left( \frac{M_2}{2.5~ \text{TeV}}\right)^2
\ee
The observed relic density requires a heavy neutralino of the order of 2.3 TeV. \\
\noindent
(c) \textit{Pure higgsino:} The annihilation cross-section of the dark matter particle  is given by
 \be
\langle \sigma_{\chi} v \rangle = \frac{3 g^4}{512 \pi \mu^2} \left( 21 + 3 \tan\theta_W^2  + 11 \tan\theta_W^4 \right)
\ee
The relic density, in this case, is given by 
\be
\Omega_{\tilde{H}}h^2 \sim 0.10   \left( \frac{\mu}{1~ \text{TeV}}\right)^2
\ee
A neutralino of 1 TeV is required to satisfy the relic density.\\ In summary, a pure bino neutralino can be light but it requires co-annihilating partners (or some other mechanism) to give the correct relic density, whereas both a pure higgsino and a pure wino would have to be close to a TeV or larger. Admixtures of various components (known as well-tempered dark matter\cite{ArkaniHamed:2006mb}) can, however, give the right relic density. 
		\begin{figure}[h!]
			\centering
			\includegraphics[width=10.5cm,height=7cm]{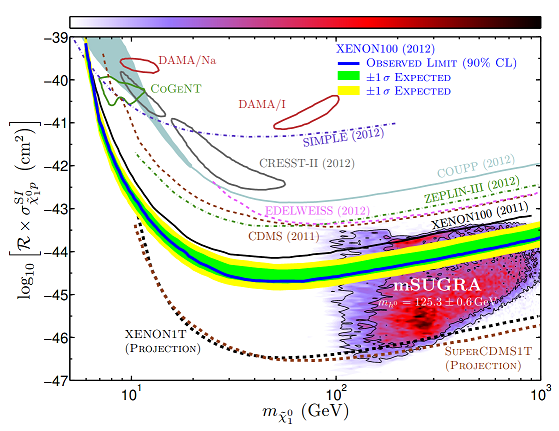}
			\caption{The spin-independent neutralino-proton cross section as a function of the neutralino mass for the model mSUGRA. It is seen that most of the parameter space is ruled out by current XENON-1T data \cite{Nath:2012nh}.}
			\label{fig:sugradrak}
		\end{figure}
There has been tremendous progress in the direct detection experiments of  WIMP dark matter. The spin-independent WIMP-nucleon cross-section is constrained to be smaller than a few $\times ~10^{-47}~\text{cm}^{-2}$ for DM masses of $\mathcal{O}(20-100)$ GeV \cite{Aprile:2018dbl,Aprile:2019dbj,Amole:2019fdf}, shown here for DM masses up to 1 TeV (fig. \ref{fig:SISDWIMP}). Many supersymmetric models with TeV scale supersymmetry predict a WIMP of this mass and cross-sections of this order. Taken together with relic density requirements, they strongly constrain most regions of supersymmetric parameter space. It leaves only some special regions like coannihilation regions, funnel regions (for example, see \cite{Bagnaschi:2017tru, Costa:2017gup}), and those corresponding to well-tempered dark matter. In fig. (\ref{fig:sugradrak}), we see direct detection measurements rule out almost all parameter space of DM in mSUGRA models (usually LSP is bino in this model). Similarly, the cosmological bound on gravitino mass, together with Higgs mass and direct searches at LHC, excludes minimal gauge mediation with high reheating temperatures, see fig. (\ref{fig:gravitinomass}). LHC searches also have not found any signal of dark matter and constrains neutralino mass to $>600$ GeV, see fig. (\ref{fig:CMSresult}). \\
Therefore, dark matter studies of supersymmetric theories need either heavy supersymmetry or a SUSY breaking mechanism that gives a well-tempered dark matter or more coannihilation and funnel region.
		
\section{Flavor Constraints}
 \begin{table}[h!]
\centering
\begin{tabular}{|c|c|c|}
\hline
Observable/Reaction & Measurement/Upper limits & Reference\\
\hline
	$\Delta M_{B^0}$ & $3.354\times 10^{-13}\mathtt{GeV}$ & \cite{Tanabashi:2018oca}\\
	$\Delta M_K$  & $3.484\times 10^{-15}\mathtt{GeV}$ & \cite{Tanabashi:2018oca} \\
	$\epsilon_K$ &  $1.596\times 10^{-3}$ & \cite{Tanabashi:2018oca} \\
	$a_{\mu}$ & $11659208.0\times10^{-10}$ & \cite{Bennett:2006fi} \\
	BR($\bar{B}\rightarrow X_s\,\gamma$) &  $3.29 \times10^{-4}$ & \cite{Lees:2012wg}\\
	Br($B\rightarrow X\,\mu\,\nu$) & 0.1086 &  \cite{Tanabashi:2018oca} \\
	Br($B\rightarrow X\, e\,\nu$)  & 0.1086&  \cite{Tanabashi:2018oca} \\
	Br($B_s\rightarrow \mu^+\,\mu-$) & $2.4\times 10^{-9}$&  \cite{Tanabashi:2018oca} \\
	Br($B^+\rightarrow \bar{D}^0\,l^+\,\nu$) & $2.27\times10^{-2}$ &  \cite{Tanabashi:2018oca} \\
	Br($B^-\rightarrow \bar{\pi}^0\,l^-\,\bar{\nu}$) & $7.80\times10^{-5}$ &  \cite{Tanabashi:2018oca} \\
	Br($K_L\rightarrow \mu^+\,\mu^-$) & $6.84\times 10^{-9}$ &  \cite{Tanabashi:2018oca} \\
	Br($K^+\rightarrow \mu^+\,\nu$) &  0.6356&  \cite{Tanabashi:2018oca} \\	
	\hline
	BR($\mu\rightarrow e\,\gamma$) & $ < 4.2\times10^{-13}$ & \cite{TheMEG:2016wtm}   \\
	BR($\mu\rightarrow$ e\, e\, e) & $< 1.0\times 10^{-12}$ & \cite{Bellgardt:1987du} \\
	BR($\tau\rightarrow \mu\,\gamma$)  & $<  4.4\times 10^{-8}$ & \cite{Amhis:2016xyh} \\
	BR($\tau\rightarrow e\,\gamma$)   & $< 3.3\times 10^{-8}$ & \cite{Amhis:2016xyh}  \\
	BR($\tau\rightarrow \mu\, \mu\, \mu$) & $< 2.1\times 10^{-8}$ & \cite{Amhis:2016xyh}   \\
	BR($\tau\rightarrow$ e\, e\, e) & $< 2.7\times 10^{-8}$ & \cite{Amhis:2016xyh}  \\
	BR($\tau\rightarrow \pi^0\,$ e) & $< 8.0 \times 10^{-8}$ & \cite{Miyazaki:2007jp}\\
	BR($\tau\rightarrow \pi^0\, \mu$) & $< 1.1 \times 10^{-7}$ & \cite{Aubert:2006cz}\\
	BR($\tau\rightarrow \rho^0\,$ e) & $< 1.8 \times 10^{-8}$ & \cite{Miyazaki:2011xe} \\
	BR($\tau\rightarrow \rho^0\, \mu$) & $ < 1.2 \times 10^{-8}$ & \cite{Miyazaki:2011xe}\\
	\hline
\end{tabular}
\caption{The experimental bounds on different flavor changing processes.}
\label{tab:flavorbound}
\end{table}

In the SM, flavor can change through either a neutral current or charged current. The flavor changing neutral current (FCNC) is absent in the SM at tree-level by GIM mechanism. In the charged current (CC) sector, flavor is violated in the SM by the well known $V_{CKM}$ which is entirely fixed by experiments. A few flavor changing transitions have been observed experimentally while many flavor changing processes have upper bounds from different experiments. From table \ref{tab:flavorbound}, many upper bounds are several orders of magnitude above the SM expectations and these give indirect bounds on the scale of new physics.
		\begin{figure}[h!]
			\centering
			\includegraphics[width=13.5cm,height=12.5cm]{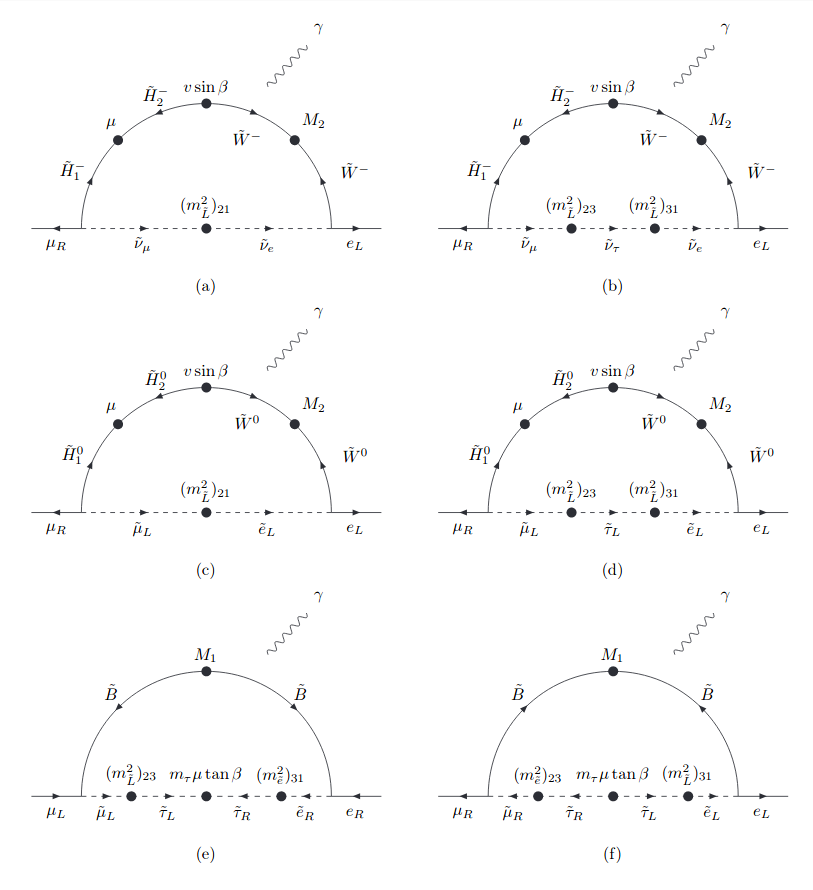}
			\caption{ One loop Feynman diagrams for $\mu \rightarrow e\, \gamma$ when the off-diagonal terms of slepton mass matrices are non-negligible, ref. \cite{Hisano:1998fj}}
			\label{fig:mueg}
		\end{figure} 
Let us assume a simple extension of the SM Lagrangian using an effective field theory approach for a new contribution in the flavor changing process:
\be
\mathcal{L}=\mathcal{L}_{SM} + \mathcal{L}_{eff}, \quad
\mathcal{L}_{eff} =\frac{1}{\Lambda^{d-4}}\sum_{i}^{}c_i(\mu)Q^i(\mu)
\ee
where $Q^i$'s are a set of higher dimension operators which will contribute to flavor changing processes; all these operators respect the SM gauge symmetry. Experiments put bounds on $c_i(\mu)/\lambda^{d-4}$, where $\Lambda$ is the scale of new physics. Both hadronic and leptonic FC processes tell us that new physics should be above 100 TeV for $\mathcal{O}(1)$ flavor violation.
In MSSM (details are given in Appendix A), soft-SUSY breaking terms for scalar masses are:
\bea
 \mathcal{L}_{soft} &= &m_{Q_{ii}}^2  \tilde{Q}_i^\dagger \tilde{Q}_i +
m_{u_{ii}}^2  \tilde{u^c}_i^\star \tilde{u^c}_i  + 
m_{d_{ii}}^2  \tilde{d^c}_i^\star \tilde{d^c}_i + 
m_{L_{ii}}^2  \tilde{L}_i^\dagger \tilde{L}_i + 
m_{e_{ii}}^2  \tilde{e^c}_i^\star \tilde{e^c}_i  \\ \nonumber 
&+& \left(\Delta_{i \neq j}^{u,d}\right)_{LL}  \tilde{Q}_i^\dagger \tilde{Q}_j + 
\left(\Delta_{i \neq j}^{u}\right)_{RR}  \tilde{u^c}_i^\star \tilde{u}^c_j  +
\left(\Delta_{i \neq j}^{d}\right)_{RR}  \tilde{d^c}_i^\star \tilde{d}^c_j  \\ \nonumber
&+& \left(\Delta_{i \neq j}^{l}\right)_{LL}  \tilde{L}_i^\dagger \tilde{L}_j + 
 \left(\Delta_{i \neq j}^{l}\right)_{RR}  \tilde{e}_i^\star \tilde{e^c}_j  + \ldots 
\eea 
Trilinear scalar couplings:
\be
\mathcal{L}_{soft} = A^u_{ij}\tilde{Q}_i\tilde{u}^c_jH_2 + A^d_{ij}\tilde{Q}_i\tilde{d}^c_jH_1 + A^e_{ij}\tilde{L}_i\tilde{e}^c_jH_1
\ee
The soft mass term $\Delta_{ij}(i\neq j,LL/RR)$ and the trilinear scalar terms $\Delta_{ij}\propto A_{ij}(\Delta_{ij}\propto A_{ij} \mbox{for}, i\neq j,LR)$ can violate flavor in both squark and slepton sector. Furthermore, all these couplings can be complex and will give CP violation in addition to the CKM phase. These terms can contribute dominantly compared to the SM amplitudes to various flavor violating processes at the weak scale, such as flavor oscillations ($K^0\leftrightarrow \bar{K}^0$), flavor violating decays like $b\rightarrow s \gamma$ and even in flavor violating decays which do not have any SM contributions like $\mu \rightarrow e \gamma$, see fig. (\ref{fig:mueg}).
 
To analyze the phenomenological impact of flavor changing processes on flavor violating terms, a useful tool called Mass Insertion approximation can be used. In this approximation, the flavor changing is encoded in non-diagonal sfermion propagators with flavor diagonal gaugino vertices. These propagators are then expanded, assuming that the flavor changing parts are much smaller than the flavor diagonal ones. Then we define flavor violating parameters, $\delta_{ij}^f\equiv \Delta_{ij}^f/m_{\tilde{f}}^2$, known as mass insertions; where $\Delta_{ij}^f$ are the flavor-violating off-diagonal entries appearing in the f=(u,d,l) sfermion mass matrices and $m_{\tilde{f}}^2$ is the average sfermion mass. The mass insertions are further sub-divided into LL/RR/LR/RL types, labeled by the corresponding SM fermions' chirality.

\begin{table}[h!]
    \centering
    \begin{tabular}{||c|c|c|c||}
    \hline
        Type of $\delta^l$ & $\mu \rightarrow e\,\gamma$ ($\delta_{12}$) & $\tau \rightarrow \mu\,\gamma$ ($\delta_{23}$) & $\tau \rightarrow e\,\gamma$ ($\delta_{13}$) \\
        \hline
         LL & $6\times 10^{-4}$ & 0.12 & 0.15\\
         LR/RL & $1 \times 10^{-5}$ & 0.03 & 0.04\\
         \hline
    \end{tabular}
    \caption{The bounds are obtained by scanning over $m_0 < 380$ GeV, $M_{1/2} < 160$ GeV, $|A_0|\leq 3m_0$ and $5<\tan{\beta}<15$ \cite{Ciuchini:2007ha}.}
    \label{tab:leptondelta}
\end{table}

\begin{table}[h!]
    \centering
    \begin{tabular}{||c|c|c|c|c||}
    \hline
      $\delta_{ij}^d$   & LL & LR & RL & RR \\
      \hline
      12     &  $1.4\times 10^{-2}$ & $9.0\times 10^{-5}$ & $9.0\times 10^{-5}$ & $9.0\times 10^{-3}$\\
      13 & $ 9.0\times 10^{-2}$ & $1.7\times 10^{-2}$ & $1.7\times 10^{-2}$ & $7.0\times 10^{-2}$\\
      23 & $0.16$ & $4.5\times 10^{-3}$ & $6.0\times 10^{-3}$ & 0.22\\
      \hline
    \end{tabular}
    \caption{The bounds are obtained by scanning over $m_0 < 380$ GeV, $M_{1/2} < 160$ GeV, $|A_0|\leq 3m_0$ and $5<\tan{\beta}<15$\cite{Ciuchini:2007ha}. The constraints on $(\delta^d_{12})_{AB}$ (AB=LL,RR,LR,RL) are taken from the measurements of $\Delta M_K, \epsilon$ and $\epsilon'/\epsilon$. The $(\delta^d_{13})_{AB}$ constraints are taken from the measurements of $\Delta M_{B_d}$ and $2\beta$ and the $(\delta^d_{23})_{AB}$ constraints are taken from the measurements of $\Delta M_{B_s}$, $b \rightarrow s\, \gamma$ and $b \rightarrow s\, l^+\,L^-$. }
    \label{tab:quarkdelta}
\end{table}
Tables \ref{tab:quarkdelta} and \ref{tab:leptondelta} are lists of constraints on $\delta$'s from different hadronic and leptonic flavor changing processes. In both sectors, $\delta$'s are suppressed to $10^{-5}$ for SUSY scale near the EW scale. Let's take an example to understand these bounds. Assume that flavor violating term is only $(\delta^l_{12})_{LL}$ (all others are zero), and all SUSY-particles have same mass, then branching fraction of $\mu \rightarrow e\, \gamma$ is\cite{Hisano:2001qz}:
\be
Br(\mu \rightarrow e \gamma) \simeq 1.18 \times 10^{-6}\left(\frac{\tan{\beta}}{15}\right)^2 \left(\frac{1\,\mbox{TeV} }{m_{SUSY}}\right)^4(\delta^l_{12})_{LL}^2
\ee
This requires either for SUSY particle to have masses around 50-60 TeV with flavor violation $\mathcal{O}(1)$ or a 1 TeV scale SUSY spectrum with suppressed flavor violating terms $(\delta^l_{12})_{LL}<6\times 10^{-4}$. LR bounds further push SUSY masses to be greater than 100 TeV for $\mathcal{O}(1)$ flavor violation. The flavor constraints are 50-60 times stronger than LHC bounds with flavor violation in supersymmetry. These limits show that the flavor violating terms should be typically at least a couple of orders of magnitude suppressed compared to the flavor conserving soft terms. The flavor problem requires heavy scale SUSY of around 100 TeV, which decouples soft masses from weak scale physics, or some alignment mechanisms. As we can see from tables \ref{tab:quarkdelta} and \ref{tab:leptondelta}, flavor violation in the first two generations of sfermion masses is more constrained from experiments than the third generation. Then the decoupling of the first two generations can avoid bounds from various experiments. The other way to avoid these bounds is to suppress flavor violating entries to zero through some symmetries or by some flavor suppressing SUSY breaking mechanism.
\section{Proton Decay Constraints on Minimal SUSY GUT SU(5)}
In the SU(5) GUT model, proton decay arises from four fermion operators (D=6 operators) while in supersymmetric SU(5) GUT model, proton decay can arise from both D=4 and D=5 operators. In an R-parity conserving scenario, D=4 operators will be zero and D=5 operators give the leading-order contribution due to the exchange of color triplet Higgsinos. They are suppressed by Higgs-triplet mass. Details of these operators are given in Chapter \ref{Chapter 5}. The correct gauge coupling unification in minimal SUSY GUT SU(5) model puts a bound on triplet-Higgs mass of $M_T\leq 2.5\times 10^{16}$ GeV with SUSY spectrum around 1 TeV. In this model, the dominant decay mode of the proton is $p\rightarrow K^+ \bar{\nu}$
with an estimated lifetime of $\lsim 10^{30}$ year\cite{Goto:1998qg}. The current limits coming from Super-K, $\tau(p\rightarrow K^+ \bar{\nu})\geq 5.9\times 10^{33}$ years, completely rules out the weak-scale supersymmetry. Decoupling of sfermions can save minimal SUSY SU(5) models\cite{Hisano:2013exa}.
\section{Heavy scale Supersymmetry}
As we discussed above,  data from various experiments (see table\ref{tab:bounds}) give us hints of heavy scale supersymmetry. In heavy scale supersymmetry, the most unpleasant aspects of low scale supersymmetry like flavor and CP violation constraints, fast proton decay constraints, Higgs mass constraints, and LHC constraints are eliminated. It does not affect SUSY's nice features like gauge couplings unification.
\begin{table}[h!]
    \centering
    \begin{tabular}{||c|c||}
    \hline
      Constraints   & Bounds on SUSY Spectrum  \\
      \hline
      Flavor and CP violation   & $1^{st}$ two generations $>100$ TeV\\
      Proton decay in minimal SUSY-SU(5) & sfermions $>50$ TeV\\
      LHC & $> 2.2$ TeV\\
      Higgs & $> 2$ TeV or $> 8$ TeV with $A=0$\\
      \hline
    \end{tabular}
    \caption{Summary of data coming from various direct and indirect experiments.}
    \label{tab:bounds}
\end{table}

\begin{figure}[h!] 
\centering    
\includegraphics[width=15cm,height=7cm]{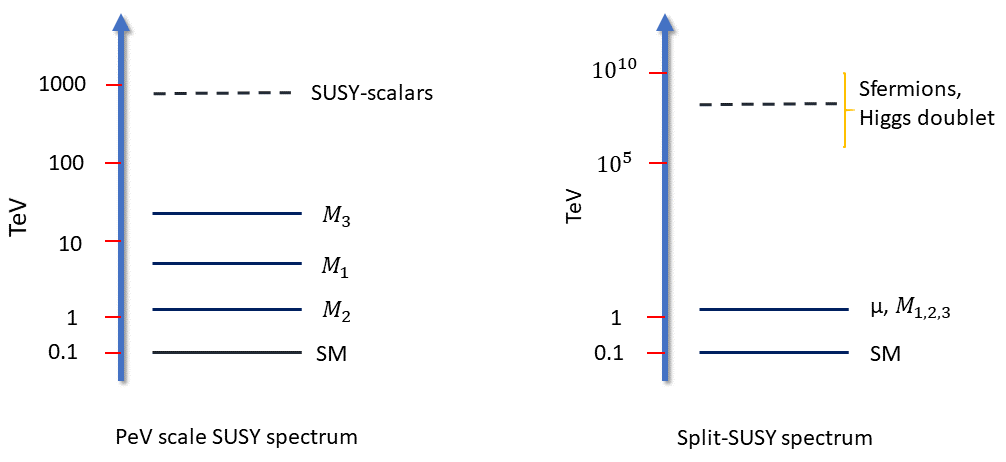}
\caption{Examples of heavy scale SUSY spectrum. \textbf{left:} Spectrum of PeV scale SUSY proposed by \cite{Wells:1997af}. \textbf{right:} Spectrum of Split-SUSY proposed by\cite{ArkaniHamed:2004fb,Giudice:2004tc}}
\label{fig:heavysusy}
\end{figure}

It is possible for SUSY to be heavy, all the way upto GUT scale, and to be completely decoupled from low scale physics. In this case, it does not have a phenomenological interest. Other possibilities are, see fig. (\ref{fig:heavysusy}), scalars of supersymmetry are at heavy scale, but gauginos are at the TeV scale (chiral symmetries can protect them). They will be seen in future experiments (33 TeV, 100 TeV, and high luminosity collider).
The Higgs mass is correlated to scalar masses, and for Higgs mass 125.01 GeV, the scalar masses should be in the range of 10 TeV to $10^5$ TeV\cite{Arvanitaki:2012ps}. For $\tan{\beta}\gsim 3$, scalar masses should be less than 100 TeV or at least the third generation of scalar masses should be light (few TeV). There are models like mini-split\cite{Arvanitaki:2012ps} in which all scalars are around 100 TeV, and gauginos are around the TeV scale, extensively studied in the literature. Similarly, there are models where generation dependent SUSY spectrum is considered\cite{Randall:2012dm,Cohen:1996vb,Dimopoulos:1995mi}, in which 1$^{st}$ two-generations are heavy but third generation and gauginos are around a few TeV. These models have very nice phenomenology and are accessible at future experiments. 
\begin{figure}[h!] 
\centering    
\includegraphics[width=15cm,height=7cm]{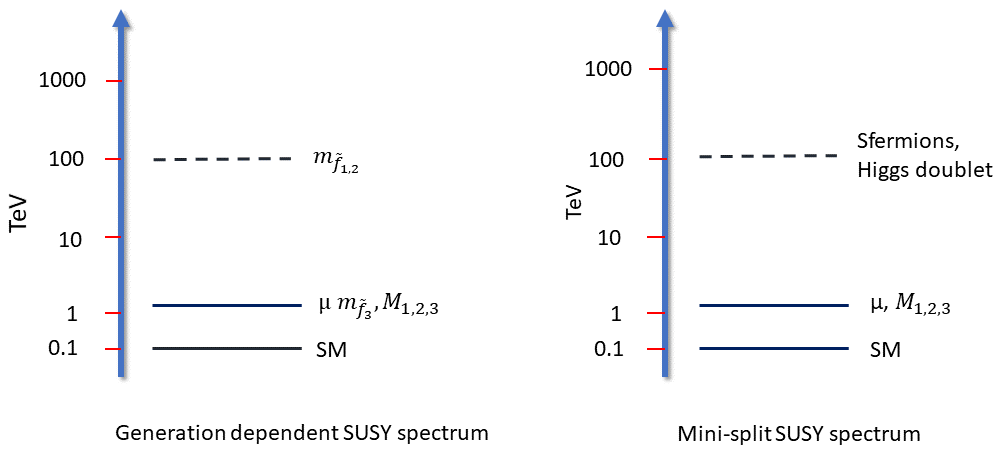}
\caption{Spectra of SUSY models which have scalar masses near 100 TeV.}
\label{fig:miniheavy}
\end{figure}

\section{Fine-tuning}
When we start developing some theory, we want it to be natural, i.e., a natural theory should not require fine-tuning of its parameters to reproduce experimental data. If a theory has fine-tuning, but there is a physical explanation for the fine-tuning,  then there is no problem with that theory. Otherwise, we should take naturalness as a principle to make predictions and to guide the theoretical study. We know that the SM is an effective theory, and the scalar receives a loop correction that is quadratically dependent on the cutoff. If we do not have any theory in between M$_{Pl}$ and EW scale, then SM Higgs will get correction proportional to $M_{Pl}^2$, and to get 125 GeV mass of Higgs, we need $\mathcal{O}(10^{-30})$ cancellation within tree level theory parameters. But we do not have any physical reason for that to happen. This problem is resolved in supersymmetric models due to a cancellation between the fermion and sfermion loops, which results in the quadratic dependence on the cutoff being replaced by a logarithmic dependence.
However, in some of the supersymmetric models, EWSB take place radiatively. The minimization of Higgs potential requires:
\begin{equation}
\frac{1}{2}m_Z^2=\frac{m_{H_d}^2-m_{H_u}^2\tan^2\beta}{\tan^2\beta-1}-\mu^2, \quad\quad\quad
2B\mu=\sin2\beta(m_{H_d}^2+m_{H_u}^2+2\mu^2)
\end{equation}
For moderate and large values of $\tan\beta$, having $m_{H_d}^2>>m_Z^2$ does not significantly affect the following equation:
\be
\frac{M_Z^2}{2} \approx -\mu^2 - m_{Hu}^2
\ee
where $\mu$ is the Higgs mixing parameter and $m_{Hu}$ is the mass of the Higgs boson that couples to the top quark. To get correct EWSB, we need  $\mu$ and $m_{Hu}$ near the EWSB scale. Otherwise, a large cancellation is needed between $\mu$ and $m_{Hu}$ to get a correct EWSB, that reintroduce the fine-tuning problem. LHC results constrain stop masses to be greater than 1.2 TeV and these have direct one loop log contribution in Higgs mass. Although Higgs mass can be satisfied with a few TeV scale stops, $m_{Hu}$ directly gets radiative correction proportional to stop mass square that will increase $m_{Hu}$ value, and then we need large cancellation between parameters for correct EWSB. \\
Now the question is, how do we define fine-tuning?
Much literature exists on this, all of which have different physical intuitions or reasoning. Few of these definitions are BG measure \cite{Barbieri:1987fn}, Nomura-Kitano measure \cite{Kitano:2006gv}, EW measure \cite{Baer:2013gva}, Anderson and Castano measure \cite{Anderson:1994dz}, Athron and Miller measure \cite{ Athron:2007ry}, Akula measure \cite{Akula:2011jx} etc. To understand fine-tuning, 
We will start with two different measures, BG measure of fine-tuning (BG) and Nomura-Kitano measure of fine-tuning (NK), which are often used in literature. The full details of these two will be discussed in Chapter \ref{Chapter 3}.
\begin{figure}[h!]%
	\centering
	\subfigure[]{%
		\label{fig:NKFT}%
		\includegraphics[width=7.5cm,height=6cm]{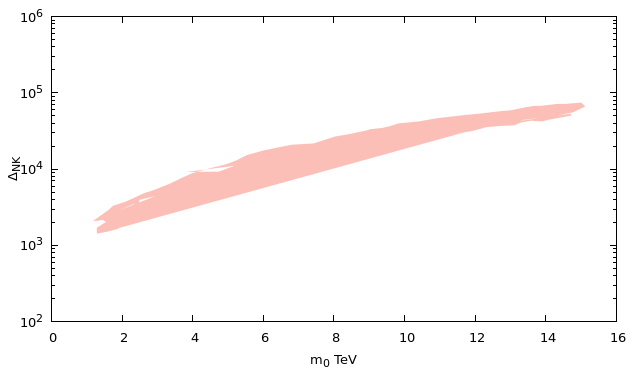}}%
	\hfill
	\subfigure[]{%
		\label{fig:BGFT}%
		\includegraphics[width=7.5cm,height=6cm]{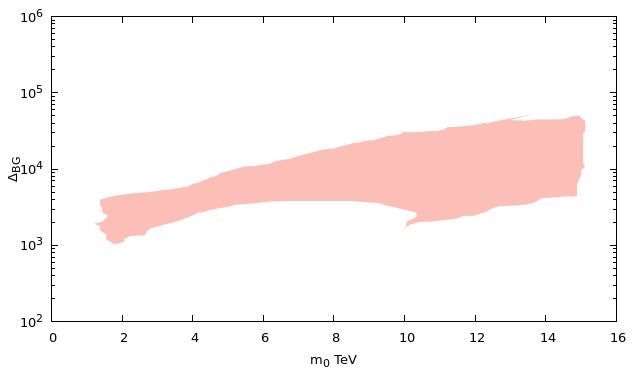}}%
	
	\caption{Fig. (a) and fig. (b) are plots of $\Delta_{NK}$ and $\Delta_{BG}$ vs $m_0$ from a scan over mSUGRA model parameter space for $\tan(\beta)=10$. Dip around $m_0=9$ TeV in $\Delta_{BG}$ is the location of the HB/FP regions\cite{Baer:2013gva}. }
	\label{fig:tataFT}
\end{figure} 
In the MSSM:
\be
m_h^2= \mu^2+m_{H_u}^2\vert_{tree}+m_{H_u}^2\vert_{rad}
\ee	
	where $\mu$ is the supersymmetric mass for the Higgs doublets,   $m_{H_u}^2\vert_{tree}$ is the tree level and $m_{H_u}^2\vert_{rad}$ is the radiative contribution to the soft supersymmetry-breaking mass squared for
	$H_u$. 
The amount of fine-tuning defined by Nomura and Kitano (NK) from $m_{H_u}\vert_{rad}$ term is given by
	\be
	\Delta^{-1}=\frac{m_{Higgs}^2}{2m_{H_u}^2\vert_{rad}}
	\ee
where the dominant contribution to $m_{H_u}^2\vert_{rad}$ arises from top-stop loop:
	\be	
	m_{H_u}^2\vert_{rad}\simeq-\frac{3y_t^2}{8\pi^2}(m_{Q_3}^2+m_{U_3}^2+|A_t|^2)\log\left(\frac{M_{mess}}{m_{\tilde{t}}}\right)
	\ee
	where $y_t$ is the top Yukawa coupling , $m_{Q_3}^2$ and $m_{U_3}^2$ are soft supersymmetry breaking masses for third-generation doublet squark  $Q_3$, singlet up-type squark $U_3$, $A_t$ is the trilinear scalar interaction parameter for the top squarks ($A=y_tA_t$), $M_{mess}$ represents the scale at which squark and slepton masses are generated , and $m_{\tilde{t}}$ is the scale of the top squark masses determined by $m_{Q_3}^2$ , $m_{U_3}^2$ and $A_t$ .  In NK fine-tuning, absence of fine-tuning is defined by $\Delta^{-1}\geq 20 \% $. We need small $m_{\tilde{t}}$ with large $A_t$ and small $\log\left(\frac{M_{mess}}{m_{\tilde{t}}}\right)$ . In minimal gauge mediation models, we have $M_{mess}$ of few TeV scale, $A_t$ is almost zero. If we demand $\Delta^{-1}\geq 20 \% $ , then $m_{\tilde{t}}\leq450 $ GeV\cite{Kitano:2006gv}. Such low mass stops with zero trilinear coupling is already ruled out by both LHC and Higgs mass data. To satisfy Higgs mass we either need stop mass to be larger than 6-7 TeV or a large A-term (which needs gauge mediation scale larger than 100 TeV). Both these requirements will make NK fine-tuning worse in these models.\\
Barbieri-Giudice measure is the well known traditional measure of fine-tuning, defined as 
\be
\Delta_{BG}= \max(\Gamma_i)
\ee
where
\be
\Gamma_i\equiv \lvert\frac{\partial \log m_Z^2}{\partial\log a_i}\lvert\equiv |\frac{a_i\partial m_Z^2}{ m_Z^2\partial a_i}|
\ee
where the $a_i$ constitute the fundamental parameters of the
model. Thus, $\Delta_{BG}$ measures the fractional change in $m_Z^2$
due to fractional variation in high scale parameters $a_i$ .
The $\Gamma_i$ is known as sensitivity coefficient.
In high scale SUSY breaking models like mSUGRA, both BG, fig.  (\ref{fig:BGFT}), and NK fine-tuning measures, fig. (\ref{fig:NKFT}), are very large even for small value of $m_0$ parameter \cite{Baer:2013gva}. This is in the case of the conventional notion of SUSY breaking originating from the dynamics of a single hidden sector. BG fine-tuning has a dip at a high $m_0$ parameter in FP/HB region, and the minimum value is still greater than $10^3$. Note that experiments are not in favor of weak scale SUSY. Furthermore, if we will increase the SUSY scale in a single hidden sector scenario, we have to fine-tune the SUSY parameters to give the correct EWSB scale. While the notion of single sector SUSY breaking is convenient as a simplifying premise, as we explain later, fine-tuning becomes almost negligible in a multiple hidden sector scenario.\\

In Chapters \ref{Chapter 2} and \ref{Chapter 3}, we explore multiple sequestered hidden sector SUSY breaking phenomenology. In chapter \ref{Chapter 2},  we write the effective operators contributing to SUSY breaking soft terms in multiple hidden sector scenario. We study the low-scale implications of this model, such as the soft spectrum and the probability of getting electroweak symmetry breaking (EWSB)
and the Higgs mass in the right range.\\
In chapter \ref{Chapter 3}, we study fine-tuning in this scenario. In this model, minimization of fine-tuning gives one solution independent of the RG coefficients, which demands all the hidden sectors contribute in a coherent manner. We study the phenomenology of this set up in detail including dark matter.\\
In chapter \ref{Chapter 4}, we study supergravity Lagrangian containing hidden sector fields, whose auxiliary fields contain the four-form fluxes, coupled to matter fields, and compute soft terms and the supersymmetric masses. This scenario naturally leads to a suppression of the flavor violating entries as $1/\sqrt{N_{HS}}$. The result crucially depends on the form of soft terms in the supergravity potential in the presence of a large number of hidden sector fields. This scenario naturally leads to well tempered neutralino dark matter.\\
In chapter \ref{Chapter 5}, we study a novel decoupling scenario named `split-generation'. In this scenario, the first and second generation of sfermions are assumed to be heavy (order of 10 TeV) and the remaining SUSY spectrum lies around a few TeV. We calculate the full two-loop $\beta$ -coefficients and the one-loop threshold at the two scales of this scenario. We developed codes for full SUSY SU(5) proton decay analysis. We also study gauge coupling unification and the two dominant proton decay channels ($p \rightarrow e^+ \pi^0$ and $p \rightarrow K^+ \bar\nu$) both with and without flavor mixing for a heavy and light (third) generation. The flavored ‘split-generation' scenario leads to peculiar cancellations in the different dressing diagrams. We find that the $p \rightarrow K^+ \bar\nu$ rates for different flavors of neutrinos are very sensitive to flavor violation.\\
In chapter \ref{Chapter 6}, we present an outlook for future explorations.\\
In appendix A, I give a basic overview of MSSM and its mass spectrum and establish the notation used throughout the thesis. In appendix B, I give a brief overview of supersymmetry breaking and give examples of gauge and gravity mediation and discuss their pros and cons.

\chapter{MSSM with Multiple Hidden Sectors}
\label{Chapter 2}

\begin{SCfigure}%
		\centering
		\caption{Consider the low energy effective field theory describing N sequestered hidden sectors which are assumed to be completely decoupled from each other and each hidden sector communicates supersymmetry breaking to the MSSM sector by gravitational interactions.}	
			\includegraphics[width=7.5cm,height=7.5cm]{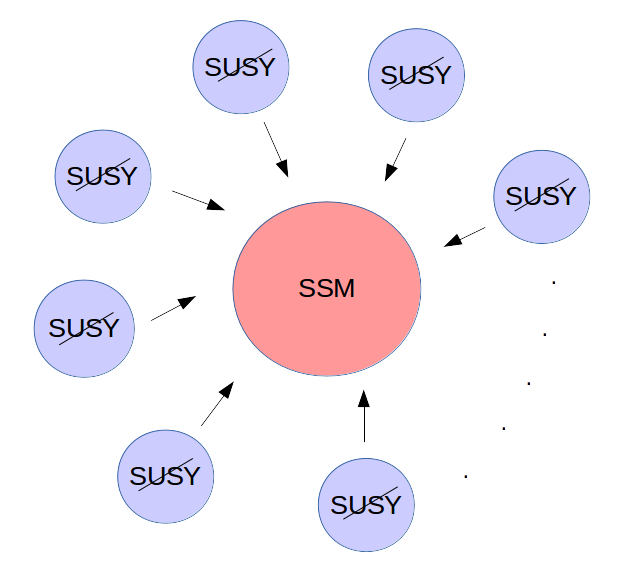}	
			\label{multihidden}
	\end{SCfigure}

\lhead{Chapter 2. \emph{MSSM with Multiple Hidden Sectors}}
 \onehalfspacing
Usually, SUSY is broken in a single hidden sector which is gauge singlet under the SM gauge group and communicates to the visible sector via gravity/gauge interactions. Single hidden sector breaking gives the spectrum at high scale, in terms of the scale of SUSY breaking $F_{\alpha}$. The simplest models of mSUGRA/CMSSM and gauge mediated SUSY breaking are strongly constrained, as has been discussed in Chapter \ref{Chapter 1}. Considering the constraints coming from the flavor, dark matter, Higgs mass, and LHC, single sector supersymmetry breaking would prefer a split or mini-split supersymmetric mass spectrum. In my thesis, I address these issues using multiple hidden sectors of SUSY breaking. 
Let us consider that these multiple hidden sectors are completely decoupled from each other (generally, there is always some interaction between hidden sectors through gravity and kinetic mixing between hidden sectors).  Each hidden sector is described by a single spurion field $T_\alpha$ with auxiliary component $F_{T_\alpha}$. SUSY is spontaneously broken in each hidden sector independently via F-term breaking at a scale of $F_{T_\alpha}$. Each hidden sector communicates supersymmetry breaking to the visible (MSSM) sector by gravitational interactions (similar set up can be envisaged with non-gravitational interactions also), see fig. (\ref{multihidden}). 
	Large number of sequestered hidden sectors have also been considered in models with multiple (pseudo) goldstini\cite{Cheung:2010mc,Cheung:2010qf,Benakli:2007zza}.
\section{Effective Operators with $N_{HS}$ Sequestered Hidden Sectors}	
Let $N_{HS}$ denote the number of spurion fields describing the sequestered hidden sectors. Higher-dimensional operators and gravitational interactions lead to interactions between the hidden sectors and the visible sector.  One writes an effective action for the visible sector fields at a high scale, treating the hidden sector fields as non-dynamical background fields. This is justified if they are very heavy compared to the observable fields.  One then writes a set of renormalization group equations for the higher-dimensional operators and evolves them into the infrared, ignoring the hidden sector dynamics. Supersymmetry breaking F and D components of the background hidden sector fields then give rise to visible sector soft supersymmetry breaking parameters (we have neglected any D-type supersymmetry breaking contributions). Integrating the gravity multiplet and the hidden sector fields, we have the following operators contributing to supersymmetry breaking soft terms and the parametrization of soft terms:
	\begin{equation}
		\label{sugraoperators1}
	M^{A}= \sum_{\alpha=1}^{N_{HS}}\frac{s_{\alpha }^A}{M_{Pl}}\int d^2\theta T_{\alpha} W^{A} W^{A}
	\quad \quad \quad
	M^{A} =  \frac{1}{M_{Pl}}\sum_{\alpha=1}^{N_{HS}}s_{\alpha }^A F_{T_{\alpha}} 
	\end{equation}
	\begin{equation}
		\label{sugraoperators2}
	m_{\tilde{f}_{ij}}^2= \sum_{\alpha=1}^{N_{HS}}\frac{d_{\alpha, i j}}{M_{Pl}^2}\int d^4\theta T^{\dagger}_{\alpha}T_{\alpha}Q^{\dagger}_iQ_j 
	\quad \quad \quad
	m_{\tilde{f}_{ij}}^2 = \frac{1}{M_{Pl}^2}\sum_{\alpha=1}^{N_{HS}}d_{\alpha, ij}F_{T_{\alpha}}^{\dagger}F_{T_{\alpha}} 
	\end{equation}
	\begin{equation}
		\label{sugraoperators3}
	A_{ijk}=\sum_{\alpha=1}^{N_{HS}}\frac{a_{\alpha, i jk}}{M_{Pl}}\int d^2\theta T_{\alpha}Q_iQ_jQ_k
	\quad \quad \quad	
	A_{ijk}=\frac{1}{M_{Pl}}\sum_{\alpha=1}^{N_{HS}}a_{\alpha, ijk}F_{T_{\alpha}}
	\end{equation}
	\begin{equation}
		\label{sugraoperators4}
	B_{H_u H_d}= \sum_{\alpha=1}^{N_{HS}}\frac{b_{\alpha }}{M_{Pl}^2}\int d^4\theta T_{\alpha}T^{\dagger}_{\alpha}H_u H_d	
		\quad \quad \quad
	B_{H_uH_d}=\frac{1}{M_{Pl}^2}\sum_{\alpha=1}^{N_{HS}}b_{\alpha}F_{T_{\alpha}}F_{T_{\alpha}}^{\dagger}
	\end{equation}
	\begin{equation}
		\label{sugraoperators5}
	\mu = \sum_{\alpha=1}^{N_{HS}}\frac{q_{\alpha }}{M_{Pl}}\int d^4\theta T^{\dagger}_{\alpha}H_u H_d
			\quad \quad \quad
				\mu=\frac{1}{M_{Pl}}\sum_{\alpha=1}^{N_{HS}}q_{\alpha}F_{T_{\alpha}}^{\dagger}	
	\end{equation}
	where $i,j$ represent the generation indices and $A$ runs over the three gauge groups
	SU(3)$_c$, SU(2)$_L$, U(1)$_Y$.  $d_\alpha, s_\alpha, a_\alpha$, $b_\alpha$ and $q_\alpha$ are complex parameters which are of
	$\mathcal{O}(1)$.  

In Chapters \ref{Chapter 2} and \ref{Chapter 3} we consider the sequestered multiple SUSY-breaking sectors communicating through gravitational couplings to the Supersymmetric Standard Model (SSM). Such models naturally appear in string theory\cite{Giddings:2001yu}, where there may be several independent sources of supersymmetry breaking.
In chapter \ref{Chapter 4} of the thesis, I study the implications of the string theory landscape for observable sector soft SUSY breaking terms where logic and setup we are using are originated from Bousso and Polchinski (BP) approach to the cosmological constant problem\cite{Bousso:2000xa}.
 	
 	 
 	
 	
 
 
In our case, we consider instead MSSM interacting with $N_{HS}$ hidden sectors at the Planck scale. We consider that auxiliary fields of the hidden sector fields contain discrete charges contributing to supersymmetry breaking. 
The main features of our framework are:
		\begin{itemize}
			\item Integer quanta parametrizing the soft supersymmetry breaking contribution from each hidden sector 
			\item Minimal number of parameters representing coupling between hidden sector fields and visible sector fields (MSSM). Since these couplings depend on moduli vev's and interactions, we parametrize them by continuous random parameters taking values inside a compact interval around zero.  
			\item Assume gravity mediation for simplicity.  A similar scan can be done for gauge mediation, although the details will be quantitatively different. 
			\item We consider a uniform distribution of charges in each hidden sector. Since each charge is a random variable, the central limit theorem gives Normal-type distributions for the soft terms.
		\end{itemize}
The low energy implications of our framework are studied across three chapters as:\\

\vspace{0.4cm}
    \begin{tikzpicture}[node distance=2cm]
  \node (start) [startstop] {Multi-SUSY breaking};
  \node (pro2a) [process, left of=start, yshift=-2cm,xshift=-2cm] {Chap. 1: Spectrum $\&$ Probabilities};
\node (pro2b) [process, below of=start, yshift=-2cm, xshift=2cm] {Chap. 2: Coherent SUSY};
\node (pro2c) [process, right of=start,yshift=-2cm, xshift=4cm] {Chap. 3: Supergravity};
\draw [arrow] (start)-- (pro2a);
\draw [arrow] (start)-- (pro2b);
\draw [arrow] (start)-- (pro2c);
\end{tikzpicture}

In this chapter, I discuss the probability of getting correct Higgs mass and EWSB, input distribution of soft parameters, and the spectrum at low energy. In chapter \ref{Chapter 3}, I discuss the fine-tuning of the electroweak scale and propose the "Coherent SUSY" idea with its phenomenology. In chapter \ref{Chapter 4}, I study this scenario using supergravity and its subsequent phenomenology.
\section{High scale parameters in this scenario}
The spurion fields are determined by discrete charges. We express them as 
	\be
	\label{Fterm}
	F_{\alpha} =  c_{\alpha} \tilde{m}M_{Pl}
	\ee
	where $c_{\alpha}$ is an integer, while $\tilde{m}$ sets the typical supersymmetric breaking scale (quanta). We assume $F_{\alpha}$ and thus in turn, $c_{\alpha}$ to have a discrete distribution of charges in $\alpha^{th}$ hidden sector. The soft parameters in terms of charges:
	\begin{eqnarray}
&& (m_0^2)_{ij} = {\tilde m}^2 \sum_{\alpha}  d_{\alpha, ij}  c_\alpha^2 \ , \ \nonumber \\
&& M_{1/2}^A = {\tilde m} \sum_{\alpha}  s_{\alpha}^A  c_\alpha \ , \ \nonumber \\
 && A_{i jk} = {\tilde m}  \sum_{\alpha}  a_{\alpha, i jk}  c_\alpha \ , \  \label{softpara}
 \end{eqnarray}
	As we know the Central Limit Theorem establishes that when independent random variables are added, their properly normalized sum tends toward a normal distribution, even if the original variables themselves are not normally distributed. The mean and variance of the normal distribution is
\[\mu_{nor}=\sum_{\alpha}\mu_{\alpha}\]
\[\sigma^2_{nor}=\sum_{\alpha}\sigma_{\alpha}^2\]
where $\mu_{\alpha}$, $\sigma_{\alpha}$ are the mean and variance of each independent random variable distribution, respectively. We consider each $c_{\alpha}$ is the discrete uniform distribution ${\cal U}\lbrace a,b\rbrace$ of random numbers for $\alpha^{th}$ hidden sector with mean $\mu_{uni}=\frac{a+b}{2}$ and variance $\sigma^2_{uni}=\frac{(b-a+1)^2-1}{12}$. If $c_1,....,c_N$ are independent and identical discrete uniform distributions of random numbers then the mean and variance of normal distribution is:
\[\mu_{nor}=N\mu_{uni}\] 
\[\sigma^2_{nor}=N\sigma_{uni}^2\]
assuming all complex coupling parameters are one. The soft parameter has normal distribution ${\cal N}(\mu_{nor},\sigma^2_{nor})$.
\section{EWSB and Higgs Probability}
This section discusses how the probability of getting the correct electroweak vacuum goes with number of hidden sectors $N_{HS}$. In principle, we can take $N_{HS}$ as large as we want, but we have to simultaneously check whether a large number of hidden sectors are favorable to electroweak symmetry breaking and correct Higgs mass.
We define two probabilities:
\begin{itemize}
	\item EWSB Probability: Probability of finding correct EWSB in distribution of charges. EWSB Probability is the ratio of the number of AOK points to all points of distribution. AOK points are points which are satisfying EWSB and do not have any flags listed below in table. And each point of distribution is sum of $\sum_{\alpha=1}^{N_{HS}}F_{\alpha}^2/M_{Pl}^2$ and $\sum_{\alpha=1}^{N_{HS}}F_\alpha/M_{Pl}$ for $m_0$ and $M_{1/2}$, respectively. 
	\item Higgs Probability: Probability of finding correct Higgs mass (122 GeV$<M_h<$128 GeV) from the points which satisfy EWSB conditions.
\end{itemize}
We have three free parameters $N_{HS},\tilde{m}$ and distribution of $c_{\alpha}$ with different range of charges. 
\begin{figure}[h!]
	\centering
	\subfigure[]{%
		\label{fig:probEWSB20}%
		\includegraphics[width=8cm,height=5cm]{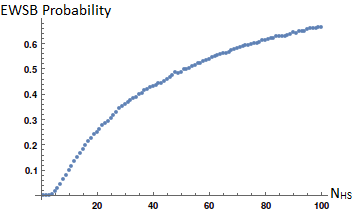}}%
	\hfill
	\subfigure[]{%
		\label{fig:probHiggs20}%
		\includegraphics[width=7cm,height=5cm]{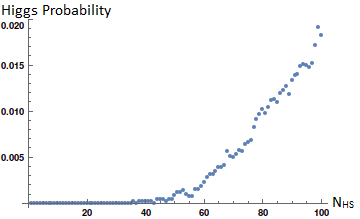}}%
	\vskip\baselineskip
	\subfigure[]{%
		\label{fig:probEWSB40}%
		\includegraphics[width=7cm,height=5cm]{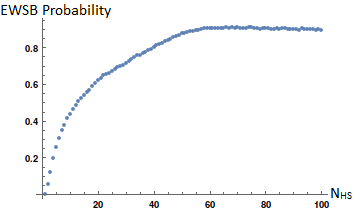}}%
	\hfill
	\subfigure[]{%
		\label{fig:probHiggs40}%
		\includegraphics[width=7cm,height=5cm]{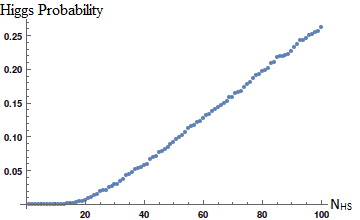}}%
	\vskip\baselineskip
	\subfigure[]{%
		\label{fig:probEWSB220}%
		\includegraphics[width=7cm,height=5cm]{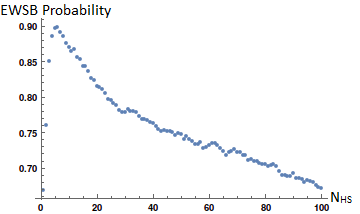}}%
	\hfill
	\subfigure[]{%
		\label{fig:probHiggs220}%
		\includegraphics[width=7cm,height=5cm]{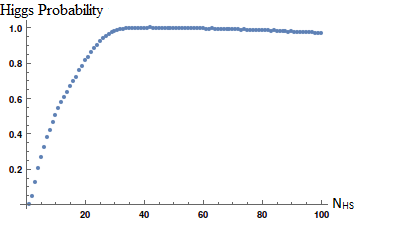}}%
	\caption{The EWSB and Higgs probability with number of hidden sectors for three different values of $\tilde{m}$. Fig. (a) and fig. (b) for $\tilde{m}=20$GeV, fig. (c) and fig. (d) for $\tilde{m}=40$GeV, fig. (e) and fig. (f) for $\tilde{m}=175$GeV.}
	\label{probabilitydiscrete}
\end{figure}
We consider soft parameters as input parameters in SuSeFLAV. For each hidden sector we have a uniform distribution of F-term vev between $-10\tilde{m}M_{Pl}$ to $10\tilde{m}M_{Pl}$. We scan $10^4$ points from each uniform distribution. Each hidden sector is contributing in total $M_0^2\propto F_{\alpha}^2$ and $M_{1/2}, A_0\propto F_{\alpha}$, where `$\alpha$' denotes $\alpha^{th}$ hidden sector. All other input parameters are given in table in appendix. \\
In figure \ref{probabilitydiscrete}, left side figures are showing EWSB probability behavior w.r.t. $N_{HS}$. Right hand side figures are showing Higgs probability behavior w.r.t. to $N_{HS}$. \\
 We consider N$_{HS}=$ 1 to 100 for three different values of $\tilde{m}$ to see region of $\tilde{m}$ and $N_{HS}$ which gives us higher probability of getting EWSB and correct Higgs mass. The fig. (\ref{fig:probEWSB20}) and fig. (\ref{fig:probHiggs20}) are corresponding to $\tilde{m}=20$ GeV. In fig (\ref{fig:probEWSB20}), we can see that the probability of finding EWSB increases as N$_{HS}$ increases. But the probability of finding correct Higgs mass with EWSB remains zero till N$_{HS}$=40 and then starts increasing, but even at N$_{HS}$=100, it is very small. The fig. (\ref{fig:probEWSB40}) and fig. (\ref{fig:probHiggs40}) are corresponding to $\tilde{m}=40$ GeV. In fig (\ref{fig:probEWSB40}), we can see that after N$_{HS}$=60 the probability of finding EWSB reaches around .9 and becomes constant with N$_{HS}$. But the probability of finding correct Higgs mass with EWSB is zero till N$_{HS}$=20 and then increases linearly but still at N$_{HS}$=100, it is 1/4. The fig. (\ref{fig:probEWSB220}) and fig. (\ref{fig:probHiggs220}) are corresponding to $\tilde{m}=175$ GeV. In fig. (\ref{fig:probEWSB220}), the probability of finding EWSB reaches a maximum of 0.9 in between N$_{HS}$=1 to 20 and then starts falling down but overall probability is still high. But the probability of finding correct Higgs mass with EWSB reaches maximum of 1 around N$_{HS}$=35 and starts decreasing gradually with N$_{HS}$\\
 As we can see that for $\tilde{m}=175$ GeV the probability for EWSB and correct Higgs mass is higher than at lower $\tilde{m}$ values, with $c_{\alpha}\epsilon [-10:10]$. We thus choose $\tilde{m}=220$ GeV with $N_{HS}=20$ as a benchmark point where we have almost $90\%$ probability of getting both EWSB and correct Higgs mass.
		
\section{Input Distribution}
Due to the large number of contributions, the observable soft terms become random variables with Normal-type distributions around an average value, where mean values and standard deviations are analytically computable. \\
In our set-up, boundary conditions at GUT scale are following (assuming universal coupling of all scalar, gaugino and trilinear coupling):
\[m_{0}^2=\frac{d_1F_1^2}{M_{Pl}^2}+\frac{d_2F_2^2}{M_{Pl}^2}+.....+\frac{d_{N_{HS}}F_{N_{HS}}^2}{M_{Pl}^2}\]
\[M_{1/2}=\frac{s_1F_1}{M_{Pl}}+\frac{s_2F_2}{M_{Pl}}+.....+\frac{s_{N_{HS}}F_{N_{HS}}}{M_{Pl}}\]
\[A_{0}=\frac{a_1F_1}{M_{Pl}}+\frac{a_2F_2}{M_{Pl}}+.....+\frac{a_{N_{HS}}F_{N_{HS}}}{M_{Pl}}\]
where $F_{\alpha}=M_{Pl}\tilde{m}c_{\alpha}$. For discrete models, $d_{\alpha},s_{\alpha},a_{\alpha}=1$ and boundary condition will reduce to the following form:
\be
m_{0}^2=\tilde{m}^2\sum_{\alpha=1}^{N_{HS}}c_{\alpha}^2\quad\quad
M_{1/2}=\tilde{m}\sum_{\alpha=1}^{N_{HS}}c_{\alpha}\quad\quad
A_{0}=\tilde{m}\sum_{\alpha=1}^{N_{HS}}c_{\alpha}
\ee
where $c_{\alpha}$ are discrete uniformly distributed charges in $\alpha^{th}$ hidden sector. Soft SUSY breaking parameters will get contribution from each hidden sector with mean \be\langle m_0^2\rangle=N\langle c_{\alpha}^2\rangle\tilde{m}^2=N_{HS}\left(\frac{(b+1)(2b+1)}{6}\right)\tilde{m}^2\quad\quad\langle M_{1/2},A_0\rangle=N\langle c_{\alpha}\rangle\tilde{m}=0\ee 
where 'b' is maximum value of uniform discrete distribution in $\alpha^{th}$ hidden sector. Standard deviation in soft SUSY parameters are:
\begin{equation*}
\Delta m_0^2=\sqrt{N_{HS}(\langle m_0^4\rangle_{\alpha}-\langle m_0^2\rangle_{\alpha}^2)}=\sqrt{N_{HS}(\langle c_{\alpha}^4\rangle\tilde{m}^4-\langle c_{\alpha}^2\rangle^2\tilde{m}^4)}
\end{equation*}
\be
\Delta m_0^2=\frac{\tilde{m}^2\sqrt{N_{HS}}}{6}\sqrt{-\frac{11}{5}-6b-b^2+6b^3+\frac{16b^4}{5}}
\ee
\begin{equation*}
\Delta M_{1/2}=\sqrt{N_{HS}(\langle M_{1/2}^2\rangle_{\alpha}-\langle M_{1/2}\rangle_{\alpha}^2)}=\sqrt{N_{HS}(\langle c_{\alpha}^2\rangle\tilde{m}^2-0)}
\end{equation*}
\be
\Delta M_{1/2}=\tilde{m}\sqrt{N_{HS}}\sqrt{\frac{(b+1)(2b+1)}{6}}
\ee

\begin{figure}[h!]
		\centering
		\subfigure[]{%
			\label{fig:m0sqhis}%
			\includegraphics[width=7.2cm,height=6.2cm]{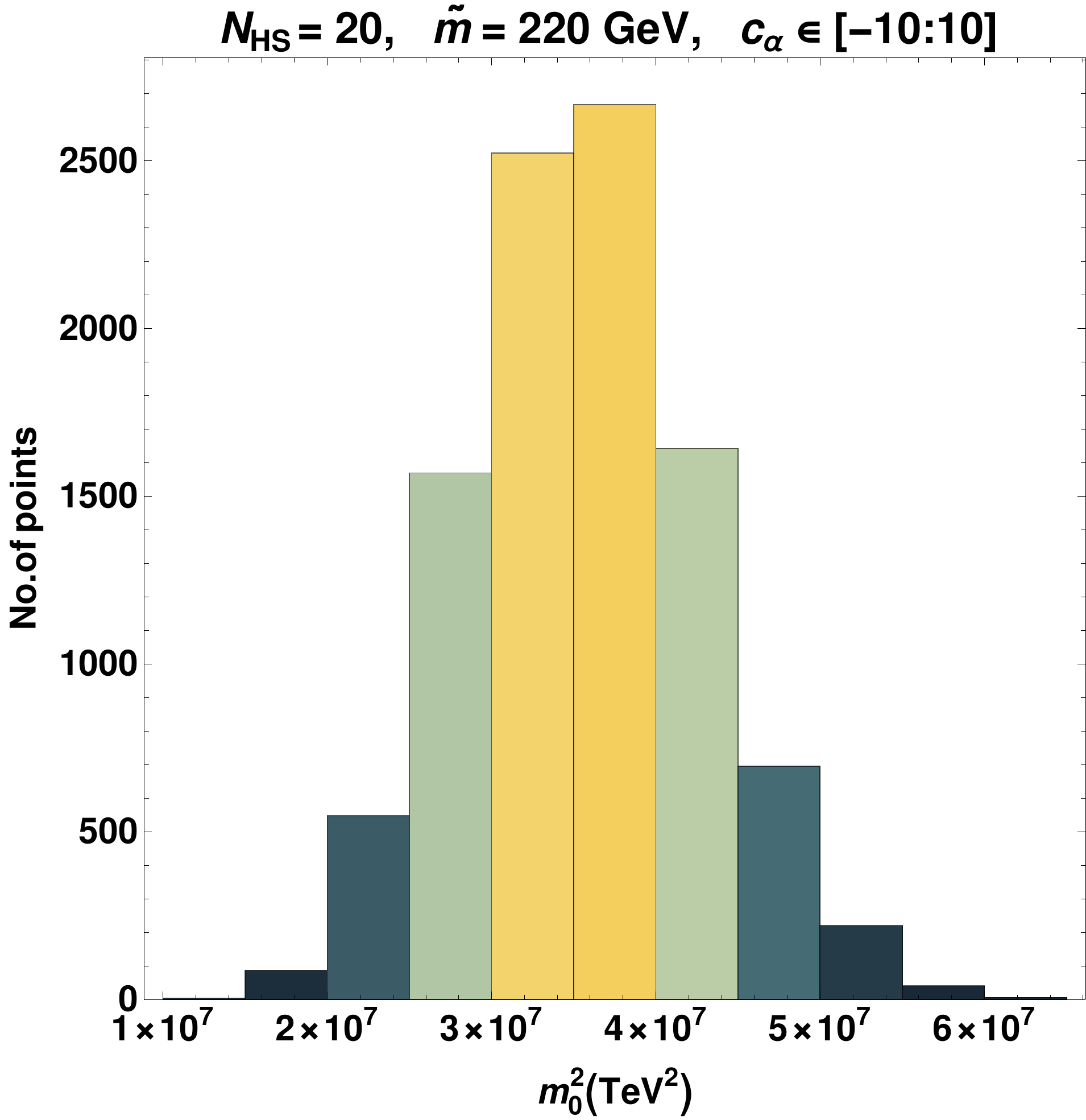}}%
		\hfill
		\subfigure[]{%
			\label{fig:m12his}%
			\includegraphics[width=7.2cm,height=6.2cm]{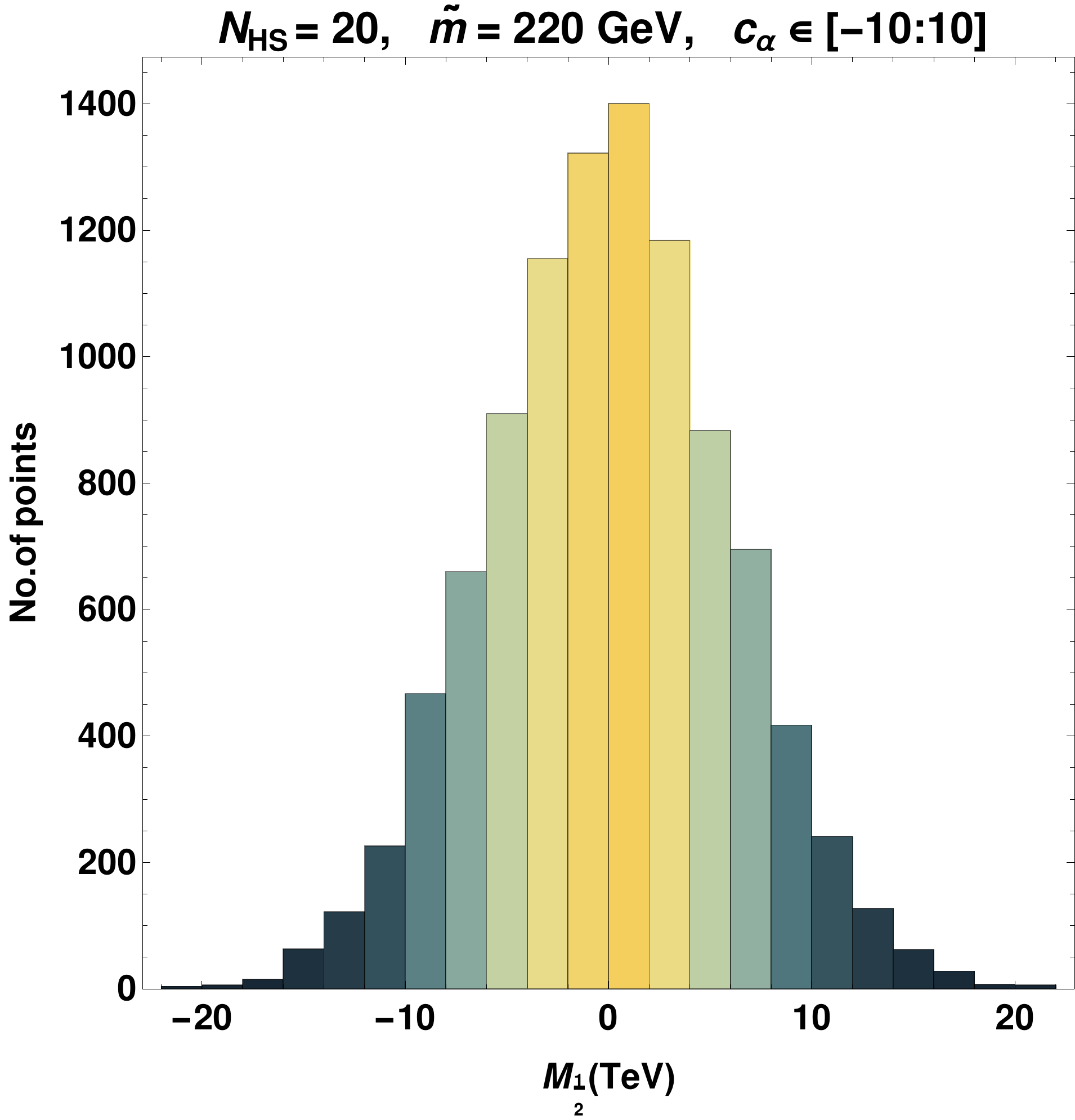}}%
		\caption{These plots are distribution of soft parameters $m_0^2$ and $M_{1/2}$, which are sum of contributions coming from $N_{HS}$ hidden sectors. These have normal distribution with mean of $\langle m_0^2\rangle=N_{HS}\langle F_1^2\rangle/M_{Pl}^2$ and $\langle M_{1/2}\rangle=N_{HS}\langle F_1\rangle/M_{Pl}$. $A_0$ have same distribution as $M_{1/2}$ for $s_{\alpha}=1$. }
		\label{inputhistogram}
	\end{figure}
\section{Spectrum}
\begin{figure}[h!]
		\centering
		\subfigure[]{%
			\includegraphics[width=7.2cm,height=6.2cm]{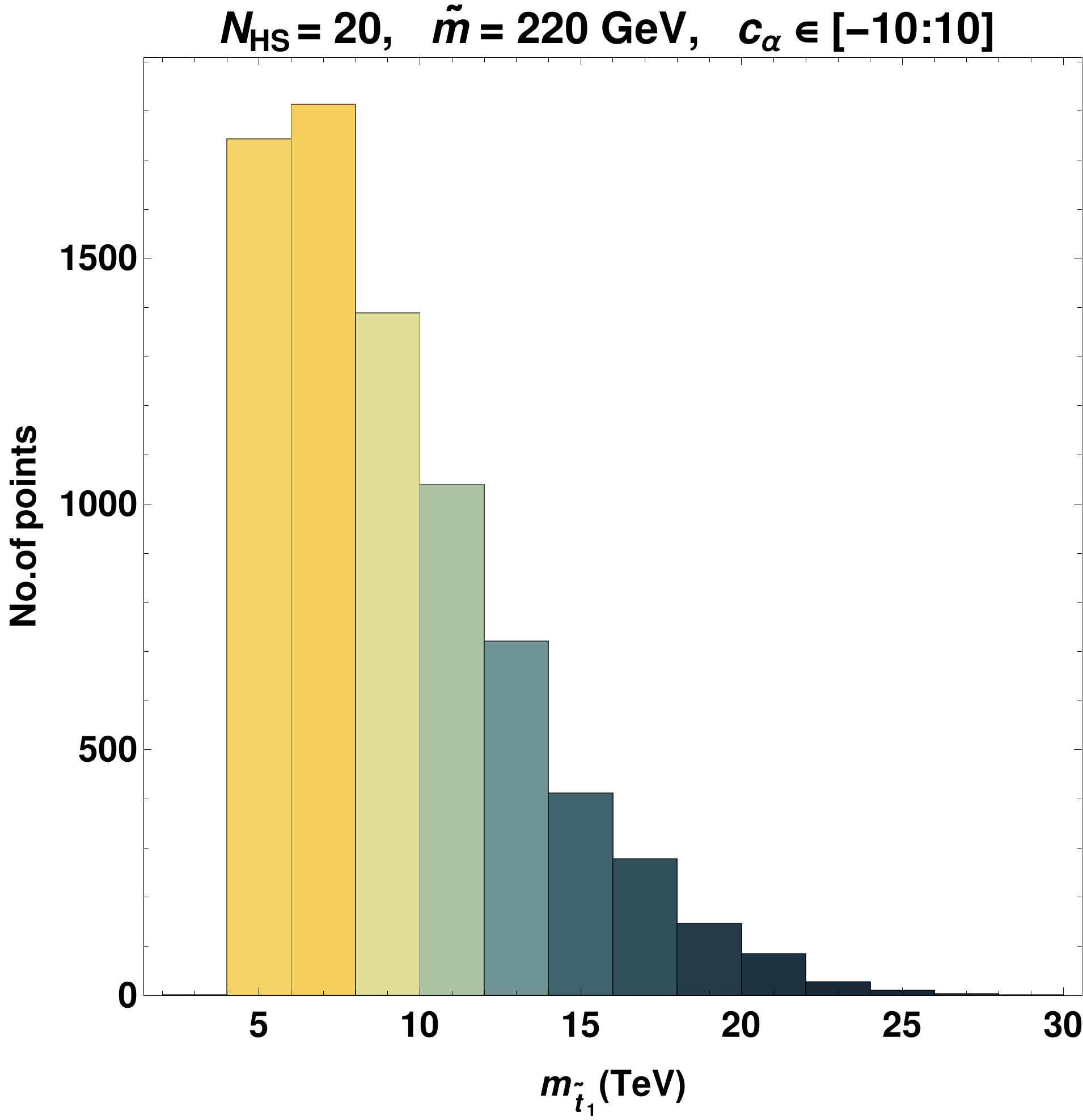}}%
		\hfill
		\subfigure[]{%
			\includegraphics[width=7.2cm,height=6.2cm]{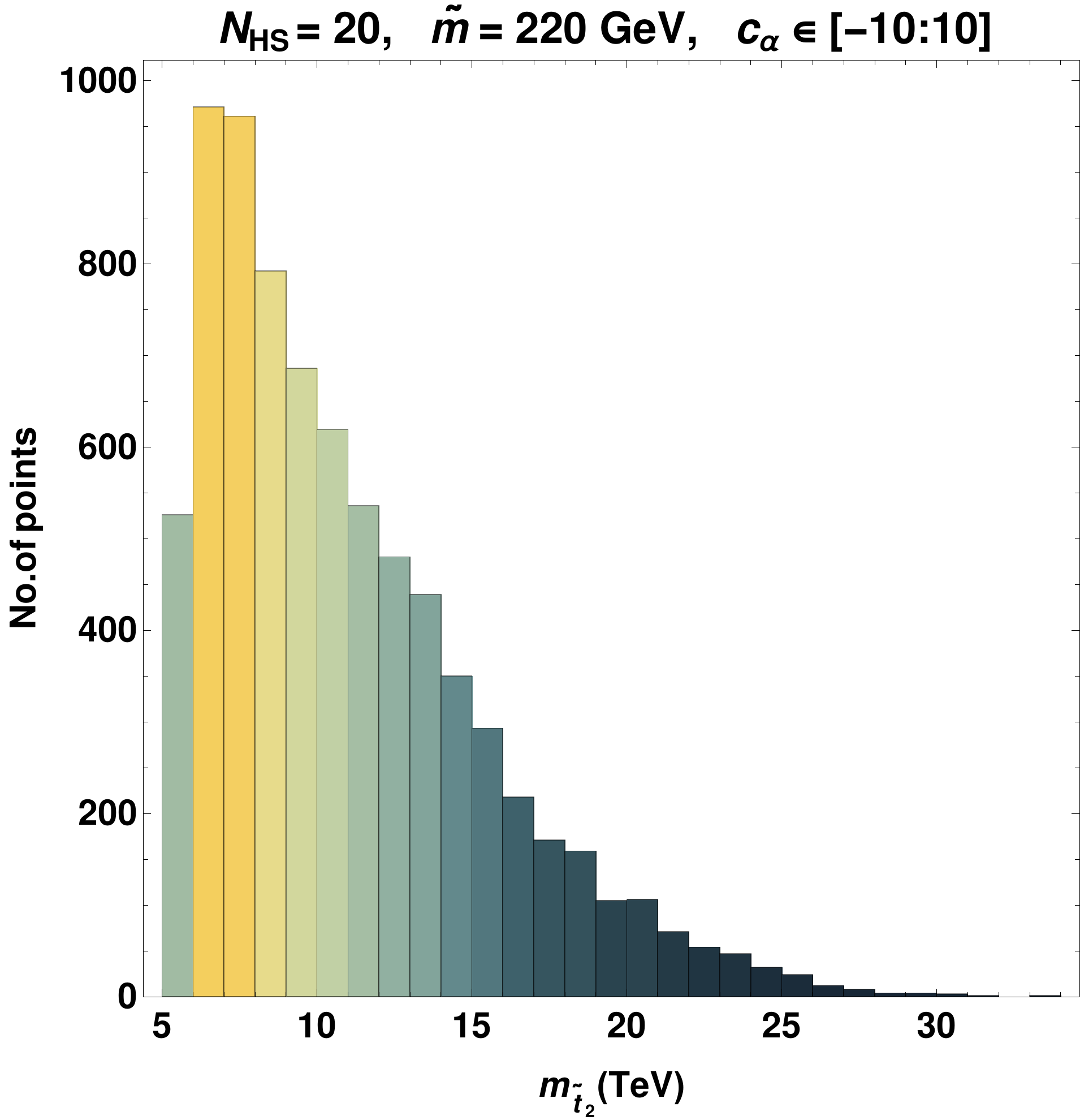}}%
	\vskip\baselineskip
		\subfigure[]{%
			\includegraphics[width=7.2cm,height=6.2cm]{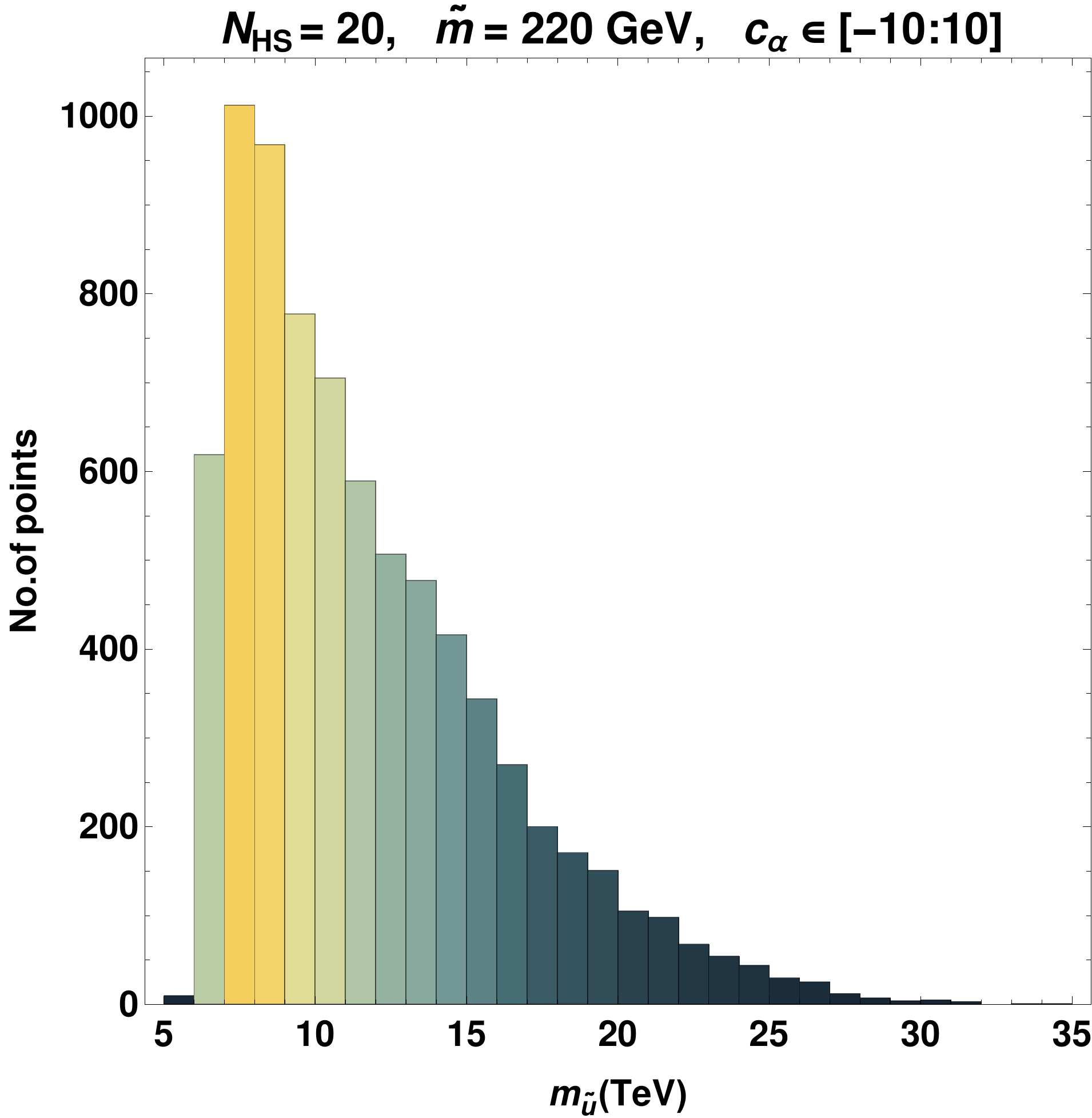}}%
		\hfill
		\subfigure[]{%
			\includegraphics[width=7.2cm,height=6.2cm]{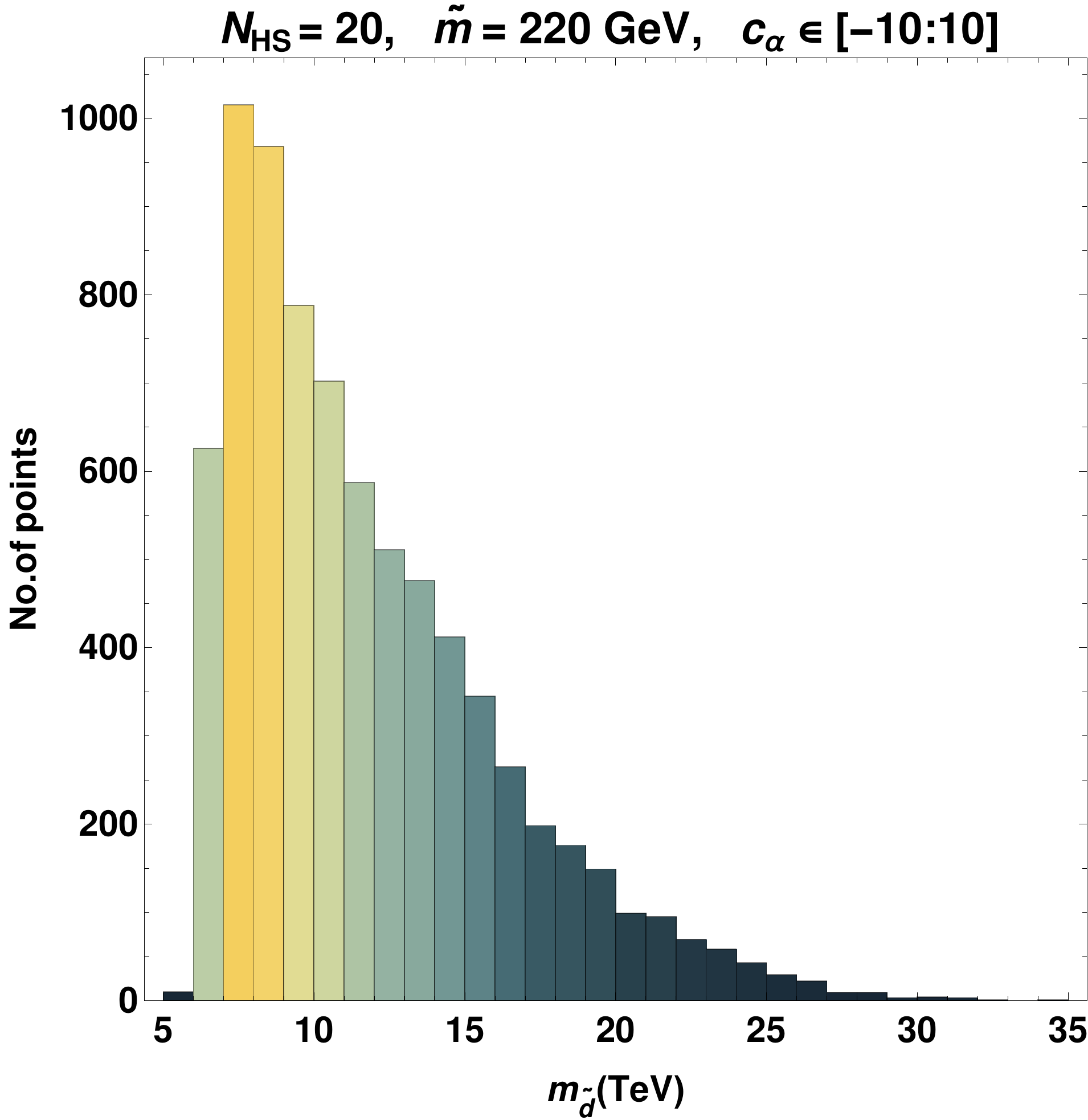}}%
	\vskip\baselineskip
		\subfigure[]{%
			\includegraphics[width=7.2cm,height=6.2cm]{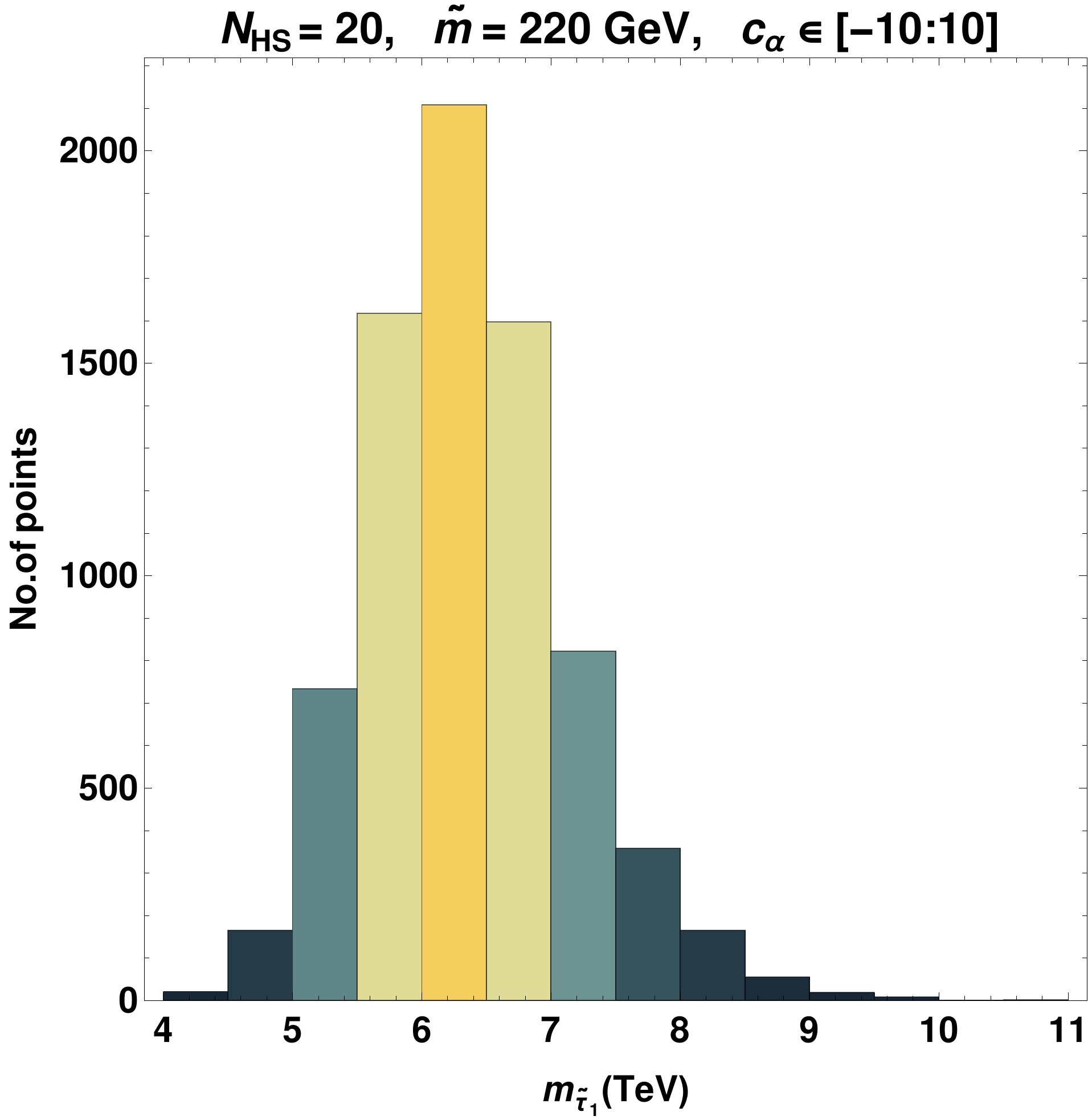}}%
		\hfill
		\subfigure[]{%
			\includegraphics[width=7.2cm,height=6.2cm]{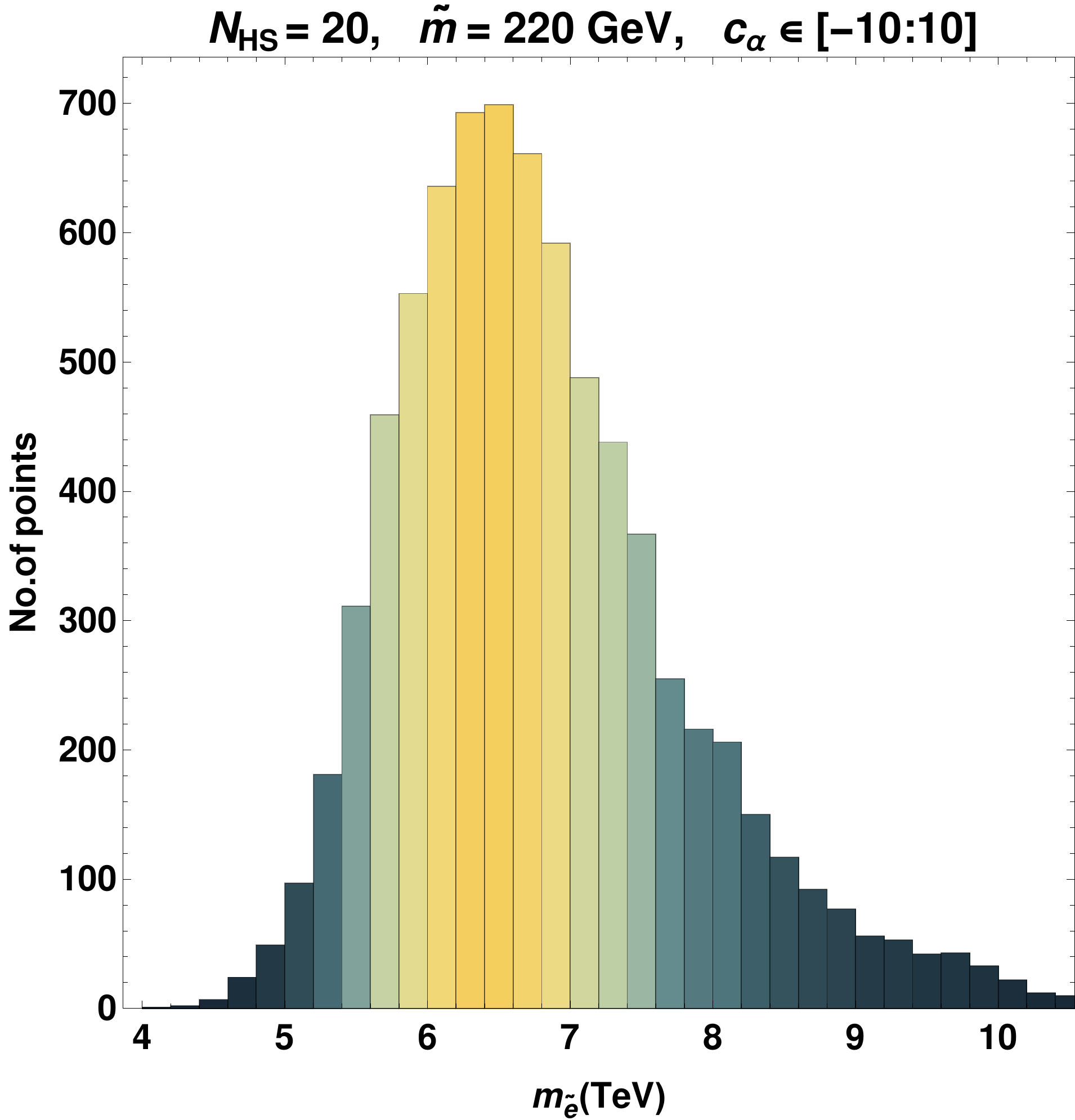}}%
		\caption{Histogram of sfermions for $N_{HS}=20$.}
		\label{sfermionspectrum}
	\end{figure}

\begin{figure}[h!]
		\centering
		\subfigure[]{%
			\label{fig:gluinomas}%
			\includegraphics[width=7.2cm,height=6.2cm]{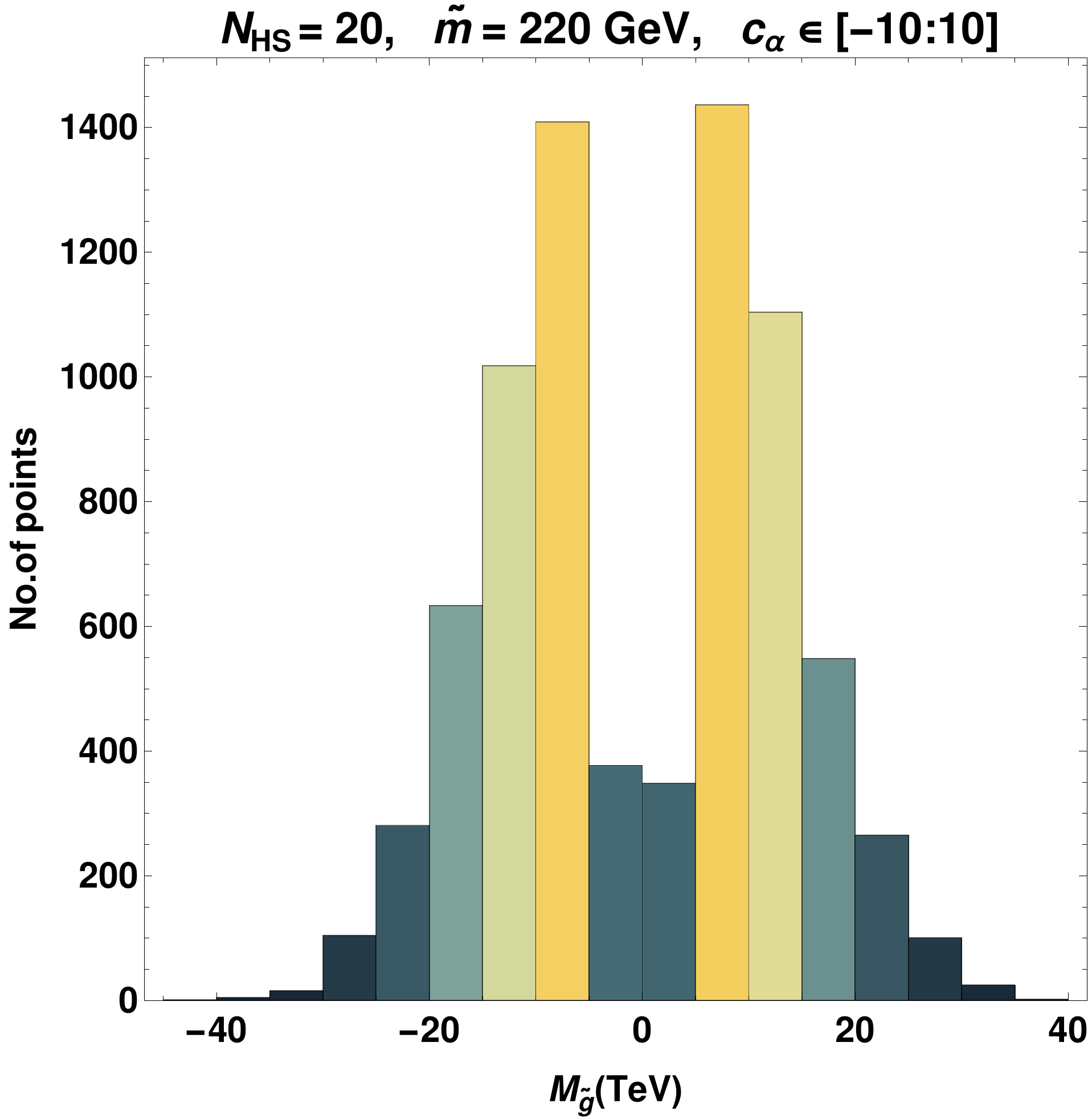}}%
		\hfill
		\subfigure[]{%
			\label{fig:murgmass}%
			\includegraphics[width=7.2cm,height=6.2cm]{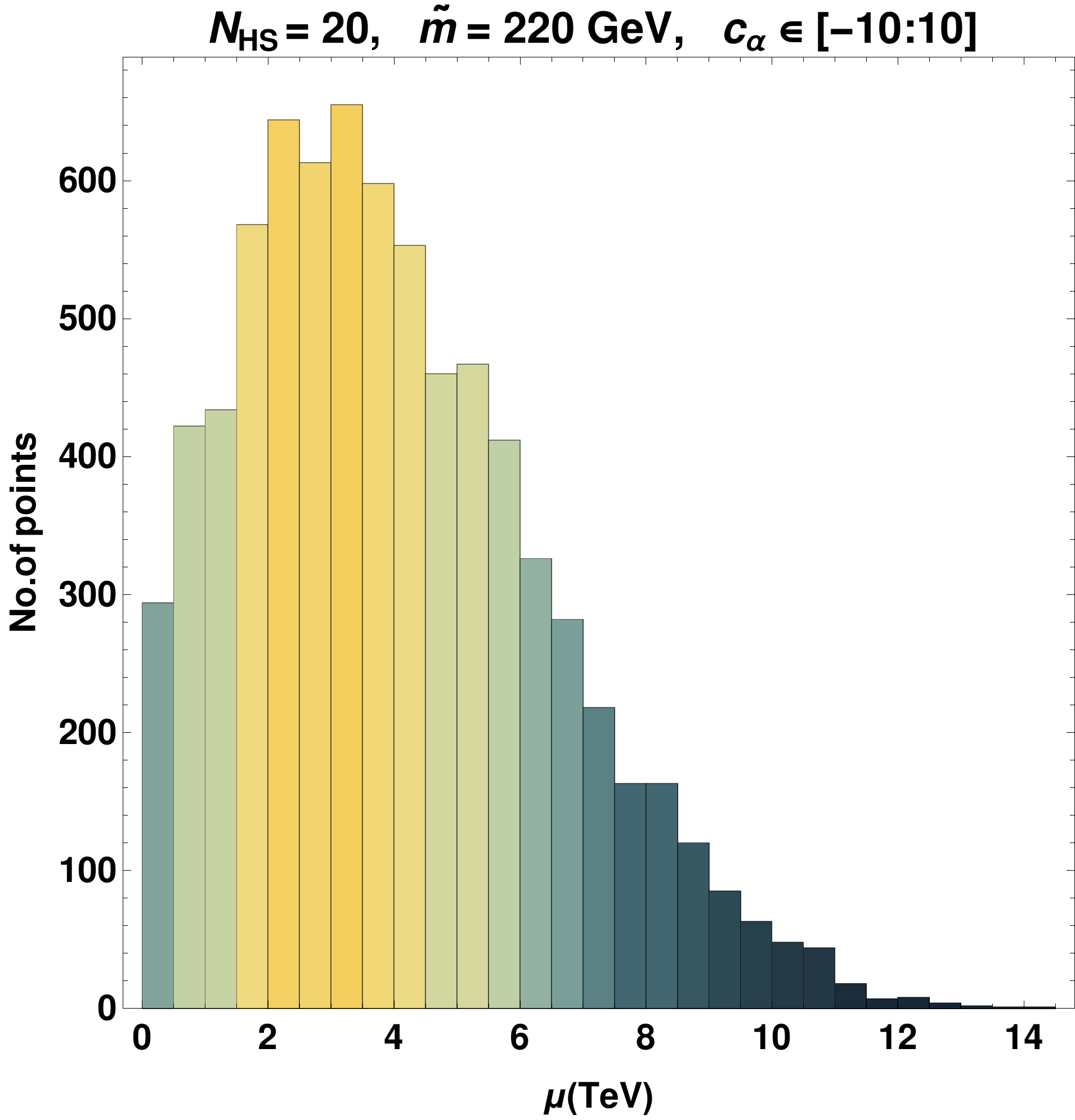}}%
	\vskip\baselineskip
		\subfigure[]{%
			\label{fig:mLSPtest}%
			\includegraphics[width=7.2cm,height=6.2cm]{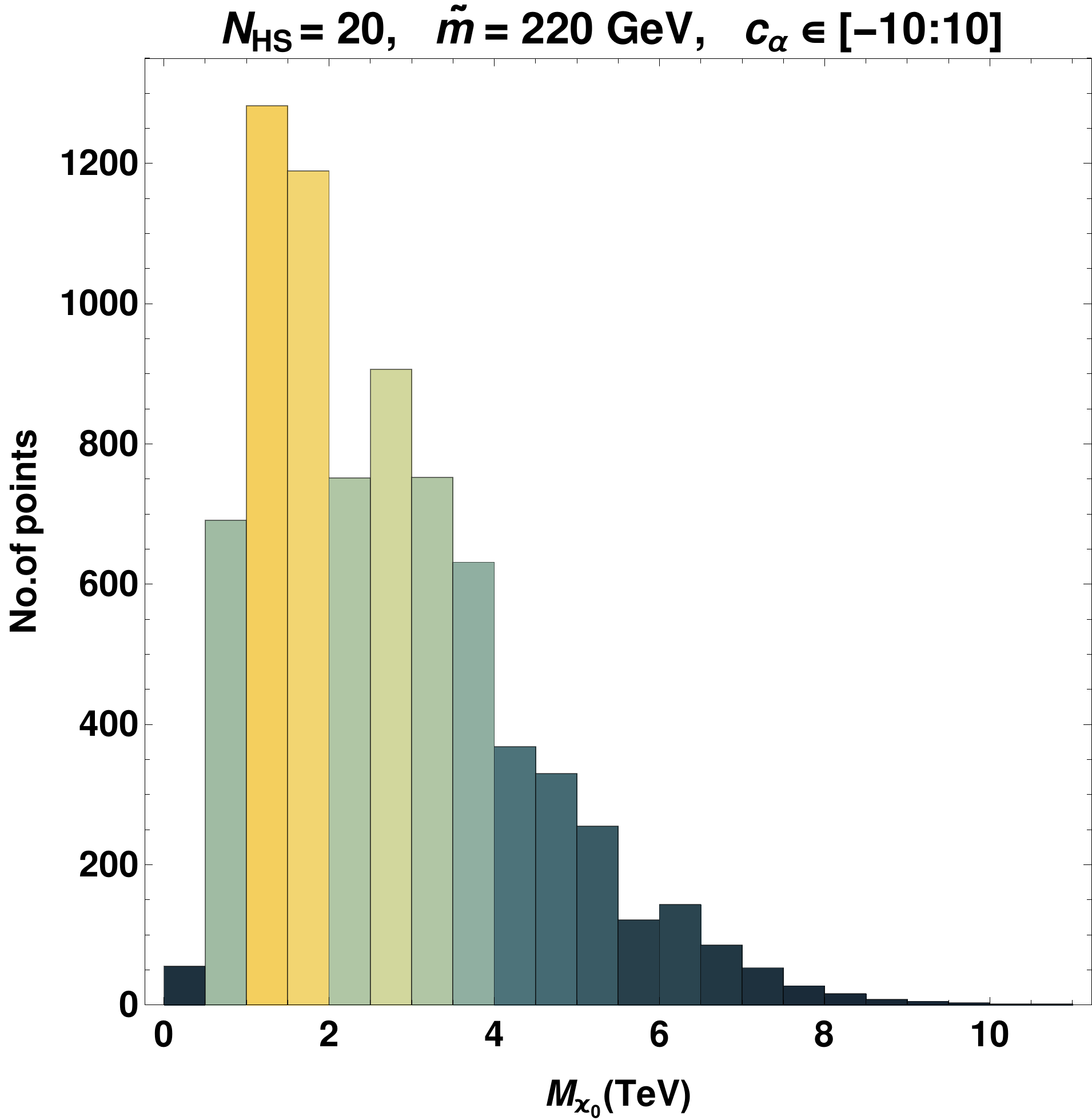}}%
		\hfill
		\subfigure[]{%
			\label{fig:mchargini}%
			\includegraphics[width=7.2cm,height=6.2cm]{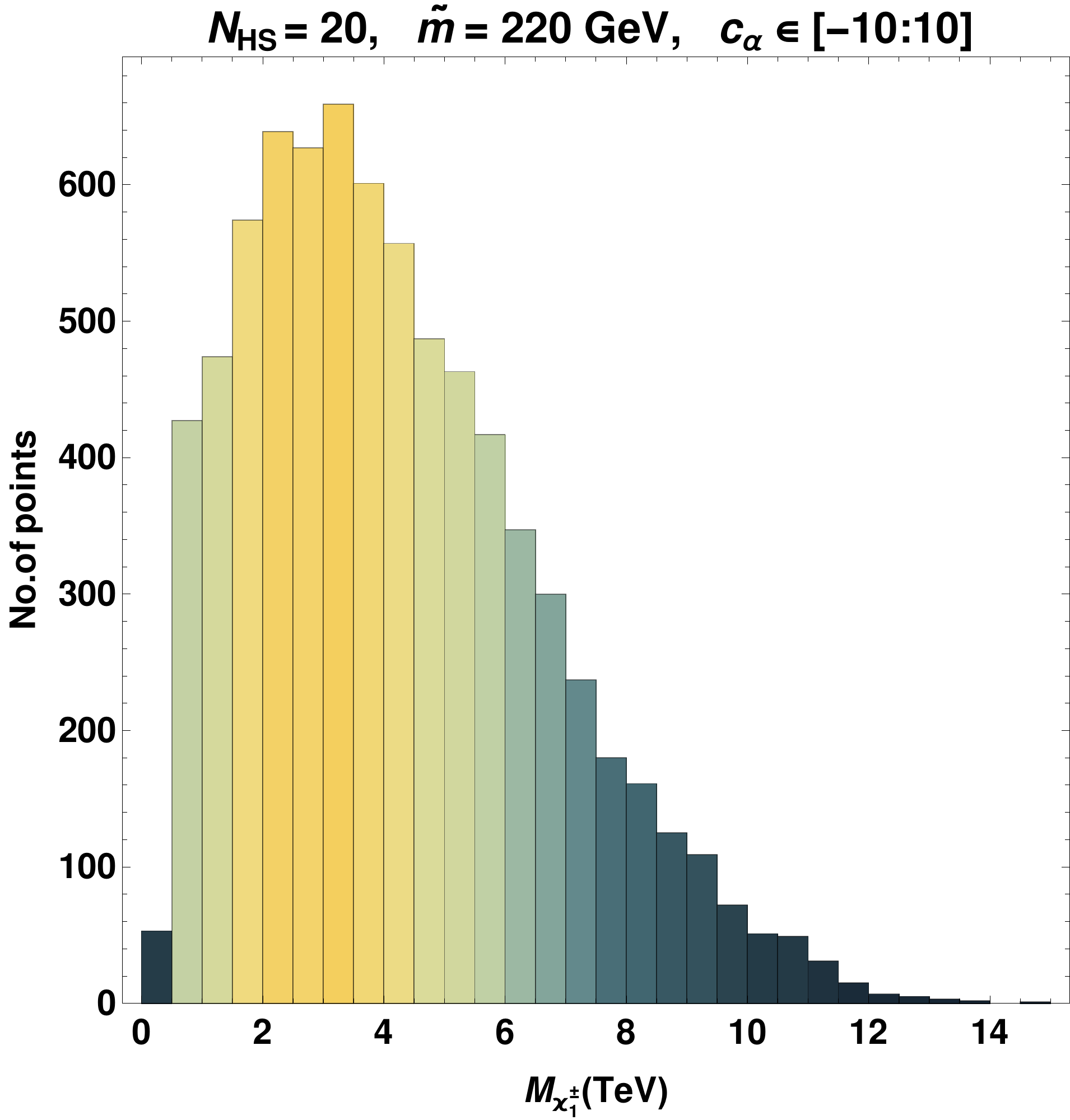}}%
		\caption{Histogram of gluino fig. (a), higgsino fig. (b), lightest neutralino fig. (c) and lightest chargino fig. (d).  }
		\label{gauginospectrum}
	\end{figure}

\begin{figure}
    \centering
    \includegraphics{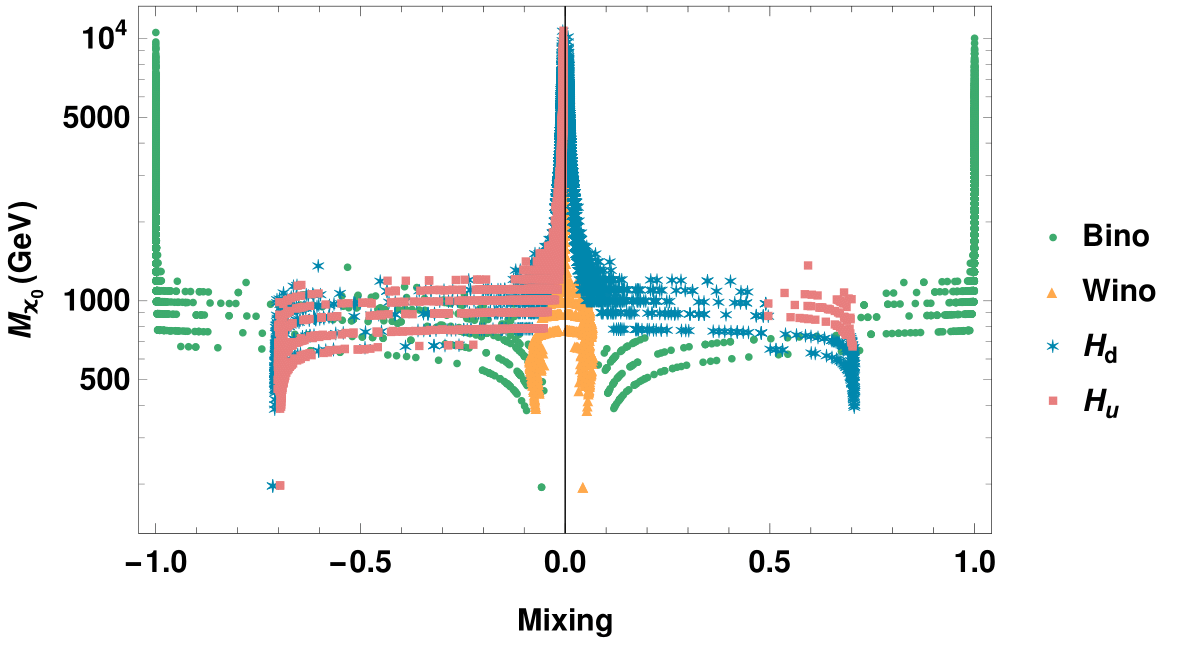}
    \caption{This is plot of mixing of higgsinos, bino, wino in lightest neutralino. y-axis is mass of lightest neutralino.}
    \label{mixingLSP}
\end{figure}

 We use full two-loop Renormalization group (RG) equations to relate high energy boundary conditions to those at low energy. And we compute spectrum for points which are satisfying EWSB, correct Higgs mass (122 GeV $<m_h<$ to 128 GeV) and do not have any flags listed in the table below. We choose $\tilde{m}=220$ GeV, $c_{\alpha}\epsilon [-10:10 ]$ and $N_{HS}=20$. All scalar masses are greater than 4 TeV in this scenario, see fig. ({\ref{sfermionspectrum}}). The sleptons have more normal-type distribution due to less dependence on gaugino masses. Squarks have large contributions at one-loop level from gauginos. Gaugino have both maximum and minimum cancellation giving a large range of gluino mass (1 to 40 TeV), see fig. (\ref{fig:gluinomas}). In minimum cancellation region gauginos have large values and that increases the tail of the distribution of squarks. In maximum cancellation regions, $\mu$ values are small, see fig. (\ref{fig:murgmass}), and we have some parameter space where LSP mass is less than a TeV, see fig. (\ref{fig:mLSPtest}). The lightest neutralino have large bino-higgsinos mixing with mass less than 2 TeV, see fig. (\ref{mixingLSP}): these regions are also due to large cancellation in gaugino sector. The contribution of gaugino masses in squarks and sleptons masses can be seen from the following semi-analytical solution of RGE's:
  \be
  m_{Q3}^2=0.6m_0^2+5.5M_{1/2}^2 \quad \quad  \quad  m_{Q12}^2=m_0^2+6.5M_{1/2}^2
  \ee
  \be
  m_{U3}^2=0.2m_0^2+4.1M_{1/2}^2 \quad \quad  \quad  m_{U12}^2=m_0^2+6.2M_{1/2}^2
  \ee
  \be
  m_{D3}^2=m_0^2+5.8M_{1/2}^2  \quad \quad  \quad  m_{D12}^2=m_0^2+6.1M_{1/2}^2
  \ee
  \be
  m_{L3}^2=m_0^2+0.5M_{1/2}^2  \quad \quad  \quad  m_{L12}^2=m_0^2+0.5M_{1/2}^2
  \ee
  \be
  m_{E3}^2=m_0^2+0.16M_{1/2}^2  \quad \quad  \quad    m_{E12}^2=m_0^2+0.15M_{1/2}^2
  \ee
  	

\clearpage	

\section{Conclusion}
We study the probability of having electroweak symmetry breaking with correct Higgs mass in multiple hidden sectors scenario. We choose $\tilde{m}=220$ GeV with $N_{HS}=20$, in which we have almost $90\%$ probability of getting both EWSB and correct Higgs mass, and study the supersymmetric spectrum. The sfermion masses are heavier than 4 TeV and gaugino masses have both high and low values depending on cancellation level in gaugino parameters. We have some parameter space with heavy scalars and light gauginos and neutralino. This spectrum manifests all the features that we discussed in chapter \ref{Chapter 1} and found to be favorable in keeping with all existing experimental constraints. This parameter space is realised due to large cancellations, and thus crucially depends on the setup of multiple hidden sectors.  \\
Given our interest in the large cancellation region, we note that increasing the charge range leads to higher probability of exact charge cancellations. But, increasing the range of charges requires a smaller value of $\tilde{m}$, to have the same probability as before of getting the correct EWSB and Higgs mass. Therefore, in the next chapter, we proceed by taking $\tilde{m}=100$ GeV with charge distribution in range [-100:100]. We study the associated spectrum, fine-tuning and dark matter phenomenology.   
  \begin{subappendices} 
\section{Details of Computation}
We use \textbf{SuSeFLAV}\cite{Chowdhury:2011zr} to calculate SUSY spectrum. We take $10^4$ random points from discrete uniform  distribution of fluxes in each hidden sector. Each hidden sector distribution has a different seed.
SuSeFLAV generates spectrum only for AOK (i.e. none of the error flags given in table below arise) point, giving flags for all other points. Details of flags and input parameters are given in following table. All points in plots of this chapter satisfy correct Higgs mass (122.0<$m_h<128.0$ GeV), LHC lower bound on gluino ($>2.2$ TeV) and stop ($>1.2$ TeV) masses. We check Higgs mass computation of SuSeFLAV with \textbf{SusyHD}\cite{Vega:2015fna} and it matches within 1 GeV. Using the points that satisfy all bounds, we calculate fine-tuning. We select points which have LSP mass less than 2 TeV and calculate dark matter density and direct-detection cross section using \textbf{micrOMEGAs}\cite{Belanger:2010pz}. We use SuSeFLAV slha.out files as input for micrOMEGAs.  \\
 
\begin{tabular}{|c|c|}
	
	\hline{\mbox{Input parameter}} & {\rm{Value}}  \\
	\hline
	$\tan\beta$ &  10 \\
	$sign(\mu)$ & +1 \\
   $M_t$ &  173.1   GeV \\
	$M_b$ & 4.78  GeV  \\
	
   $M_\tau$ &  1.7768  GeV \\
	$\alpha^{-1}$ & 127.91  \\
	$\alpha_s$ & 0.1181  \\
	$m_Z$ & 91.1876  GeV  \\

	case & CKM \\
	quark mixing & 1 \\
	spectrum tolerance & $10^{-3}$ \\
	\hline
\end{tabular}
\quad
\begin{tabular}{|c|c|}
	\hline{\mbox{Name of flags}}&{\mbox{Description of flags}}  \\
	\hline
	AOK &  Everything is fine \\
	BMUNEG & $B_{\mu}$  is negative at $ M_{SUSY}$  \\
	REWSB &  $|\mu|^2<0 $ at $ M_{SUSY}$ \\
	TACSPEC & Spectrum is tachyonic at $ M_{SUSY}$  \\
	
	FSNC &   Final spectrum non-convergent \\
	TACMh &  Lightest CP-even neutral higgs tachyonic   \\
	TACMA &  CP-odd neutral higgs tachyonic \\
	TACSUP &  up-squarks sector tachyonic \\
	LEPH &  Lightest higgs mass below LEP limit  \\
	LEPC &   Lightest chargino mass $ <103.5$ GeV \\
	LSPSTAU &  Lightest stau is LSP  \\
	\hline
	\hline
\end{tabular}
\section{Few more details}
We divide this discussion into two scenarios.
\begin{enumerate}
    \item Discrete model: This model has $d_{\alpha ij}=s_{\alpha i}=a_{\alpha ijk}=1$. To understand the behavior of probabilities with respect to $N_{HS},\tilde{m}$ and distribution of $c_{\alpha}$ with different range of flux charges, we further divide it into two cases:
    \begin{itemize}
        \item Case-I: Each $c_{\alpha}$ has discrete uniform distribution from [-10:10], three
	different values of of $\tilde{m}\approx$ 20, 40, 175 GeV with large range of $N_{HS} = 1, 100$. This is discussed above.
	\item Case-II: Fixed value of $\tilde{m}=220 $ GeV, small range of $N_{HS} = 1, 20$ with four sets of discrete uniform distributions of $c_{\alpha}$ [-10:10], [-30:30], [-50:50] and [-70:70].
    \end{itemize}
    \item Continuous model: These model have $d_{\alpha ij}, s_{\alpha i}, a_{\alpha ijk}$ as continuous parameters with values in between [-1:1]. 
\end{enumerate}
\textbf{Discrete model: Case-II}

\begin{figure}[h!]%
			\centering
			\subfigure[]{%
				\label{fig:fluxrangeEWSB}%
				\includegraphics[width=7cm,height=5cm]{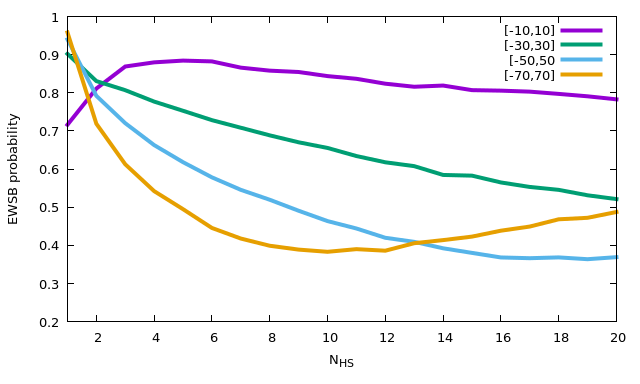}}%
			\hfill
			\subfigure[]{%
				\label{fig:fluxrangeHiggs}%
				\includegraphics[width=7cm,height=5cm]{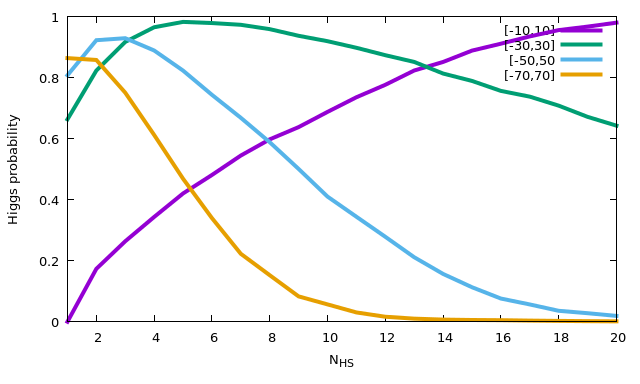}}%
			\caption{Fig. (a) and fig. (b) are plots of EWSB probability and Higgs probability, respectively with $N_{HS}$ for different range of fluxes. Purple color for flux distribution range [-10,10], green color for flux distribution range [-30,30], sky blue color for flux distribution range [-50,50] and orange color for flux distribution range [-70,70].}
			\label{fluxrange}
		\end{figure}

We choose one of the region with fixed value of $\tilde{m}=220 $ GeV and $N_{HS}$ varying between 1 to 20, where both EWSB and Higgs probability is high. In fig. (\ref{fig:probHiggs220}), we see effect of $\tilde{m}$ on on probability of getting correct EWSB and Higgs mass. In this case, we see the effect of range of flux distribution on probability of EWSB and Higgs mass. We consider four different range of discrete uniform distributions of $c_{\alpha}$ [-10:10], [-30:30], [-50:50] and [-70:70]. In fig. (\ref{fluxrange}).
		
\subsection{Continuous model}
We choose one of the region with fixed value of $\tilde{m}=220 $ GeV and $N_{HS}$ varying between 1 to 20, where both EWSB and Higgs probability is high with flux discrete distribution in range [-10:10]. In this case, we want to study how coupling parameter will effect probabilities. We choose $d_{\alpha ij}, s_{\alpha i}, a_{\alpha ijk}$ as continuous parameters with values in between [-1:1], which we leads cancellation in scalar mass sector together with cancellation in gaugino sector.
	\begin{figure}[h!]
		\centering
		\subfigure[]{%
			\label{fig:firsta}%
			\includegraphics[width=7cm,height=5cm]{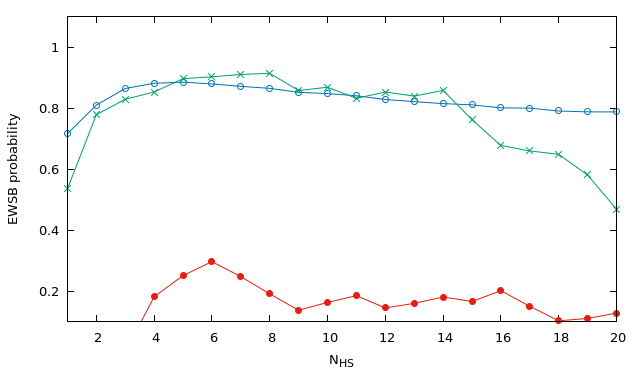}}%
		\hfill
		\subfigure[]{%
			\label{fig:seconda}%
			\includegraphics[width=7cm,height=5cm]{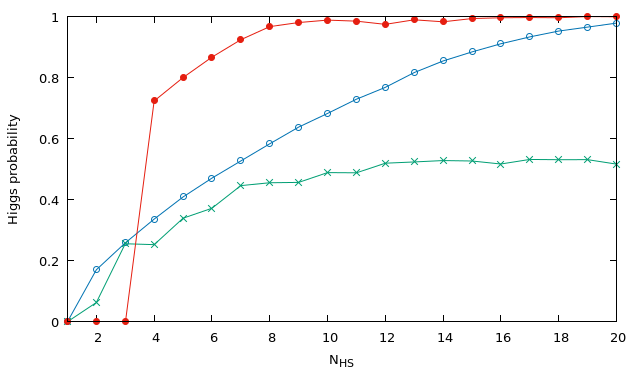}}%
		\caption{In these plots $\tilde{m}=220$ GeV.The blue color is w.r.t. $d_{\alpha},s_{\alpha},a_{\alpha}=1$, green color for $d_{\alpha},s_{\alpha},a_{\alpha}\in [-1:1]$ with different value for each $N_{HS}$ and all scalar masses, gaugino masses and trilinear couplings have same value of $d_{\alpha},s_{\alpha},a_{\alpha}$ for each $N_{HS}$ and red color for $d_{\alpha},s_{\alpha},a_{\alpha}\in [-1:1]$ with different value for each $N_{HS}$ and for each $N_{HS}$, all scalar masses, gaugino masses and trilinear couplings have different value of $d_{\alpha},s_{\alpha},a_{\alpha}$ respectively.}
	\end{figure}
  \end{subappendices}  
 \chapter{Coherent SUSY}
\label{Chapter 3}

\lhead{Chapter 3. \emph{Coherent SUSY}}
 \onehalfspacing
\section{Motivation}
	We consider soft supersymmetry breaking terms from multiple sequestered hidden sectors.
Fine tuning considerations then force the contribution from each such hidden sector to be 
roughly equal. The soft terms in this case are
 proportional to integer charges. The simplest solution of equal fine tuning from each 
sector would mean same charges for each sector.  This would naturally lead to cancellation in 
the gauge sector while  no cancellation in the soft mass squared sector, leading to natural 
realisation of focus-point/hyperbolic branch regions. Dark matter in this case is well tempered
between higgisino-wino or higgsino-bino like scenarios. The 
fine-tuning from each hidden sector is roughly smaller by a factor $N_{HS}$, the number 
of hidden sectors. 

\section{Introduction} 
Supersymmetry is a theoretically rich theory with beautiful mathematical features. Its theoretical features do not depend on the SUSY breaking scale. SUSY also has excellent phenomenological features sensitive to SUSY breaking scale like the solution to gauge hierarchy problem, dark matter, grand unification, etc. As discussed in the last chapter, we have a multiple hidden sectors scenario that gives EWSB and Higgs mass with high probability. The nice features of the spectrum merit a detailed study of the solution to gauge hierarchy problem and the viability of dark matter. We undertake these in this chapter. Here, I start by introducing the gauge hierarchy problem.

 In the Standard Model, there is a scalar particle Higgs boson with physical mass $m_h=125.09$ GeV. There is no symmetry to protect Higgs mass in the SM; that's why its bare mass receives large quantum corrections. For example, the correction to Higgs mass due to a fermion in the loop is :
\be
m_h^2\approx m_{h_0}^2-\frac{\lambda_f^2}{8\pi^2}N_c^f\int^{\Lambda}\frac{d^4p}{p^2}\approx m_{h_0}^2+\frac{\lambda_f^2}{8\pi^2}N_c^f\Lambda^2
\ee
where $m_{h_0}$ is the bare mass parameter, $\lambda_f$ is Yukawa coupling of fermion f, $N_c^f$ is the number of colors and second term is a one-loop correction coming from fermion loop. Similarly, there will be correction terms coming from the gauge bosons loop. These one-loop corrections are quadratically dependent on the cut-off scale $\Lambda$ (cut-off for the quadratically divergent integral). For large $\Lambda$, we need a large degree of cancellation between bare mass parameter and one-loop corrections. In the case of fermions, mass corrections are protected by chiral symmetries, and dimensionless couplings' quantum corrections are proportional to the logarithm of the cut-off scale. At this stage, we can simply define the measure of naturalness (fine-tuning) as:
\be
\Delta\equiv \frac{m_{h1-loop}^2}{m_h^2}
\ee
We consider $\Lambda$ as the scale of new physics. If we assume there is no new physics, we still need new physics at the scale of $M_{Pl}$ to describe quantum gravity, which implies $\Delta\sim 10^{30}$. That means a fine-tuning of 1 part in 10$^{30}$ is required in cancellation between bare mass and one-loop correction to get correct physical Higgs mass. This is known as the gauge hierarchy problem. There are two way to solve this problem:
\begin{itemize}
    \item A new physics around the Electroweak scale.
    \item Some symmetry which protects Higgs mass.
\end{itemize}
Supersymmetry is a symmetry that protects Higgs mass, and its breaking also introduces a new scale of physics. A solution to gauge hierarchy problem through weak-scale supersymmetry is well-known\cite{Veltman:1980mj,Witten:1981nf,Kaul:1981wp,Vissani:1997ys,Feng:2013pwa}. If SUSY is exact, the Higgs mass will get no perturbative corrections (all loop corrections will cancel exactly). As we know, SUSY is not exact. In the SUSY breaking scenario, Higgs mass becomes:
\be
m_h^2\approx m_{h_0}^2-\frac{\lambda_f^2}{8\pi^2}N_c^f\int^{\Lambda}\frac{d^4p}{p^2}\approx m_{h_0}^2+\frac{\lambda_f^2}{8\pi^2}N_c^f(m_{\tilde{f}}^2-m_f^2)\log\left(\frac{\Lambda^2}{m_{\tilde{f}}^2}\right)
\ee
where $m_{\tilde{f}}$ is mass of superpartner of fermion f. Even with SUSY breaking, Higgs mass's quadratic dependence reduces to a logarithmic dependence on the cut-off scale, which will be suppressed by the loop factor. Few conclusive points regarding naturalness:
\begin{itemize}
    \item The Higgs mass corrections are sensitive to SUSY partner masses, and the naturalness criteria constrain different SUSY particles differently. For example, the one-loop correction does not put constraints on the first two generations of sfermions due to Yukawa suppression, but it puts constraints on third-generation squarks masses due to large correction arising from large Yukawa couplings. The first two generations become important at two-loop Higgs mass corrections.
    \item The naturalness bounds depend on what level of fine-tuning we consider as natural. In literature, theorists demand fine-tuning to be less than $10\% $ to have a natural theory. Again, this depends on how we define fine-tuning.
\end{itemize}
In this chapter, I discuss two measures of fine-tuning, BG measure of fine-tuning\cite{Barbieri:1987fn} and Nomura-Kitano measure of fine-tuning\cite{Kitano:2006gv}, which are rigorously used in literature.

\section{Recap of Fine-tuning in Case of Conventional Single Hidden Sector Scenario}
\begin{figure}[h!]%
	\centering
	\subfigure[]{%
		\label{fig:tataN1}%
		\includegraphics[width=7.5cm,height=6cm]{coherentsusy/NKm0tata.png}}%
	\hfill
	\subfigure[]{%
		\label{fig:NKmulti}%
		\includegraphics[width=7.5cm,height=6cm]{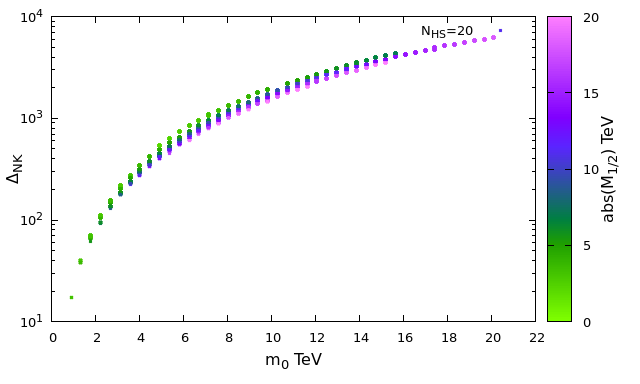}}%
	
	\caption{Plots of $\Delta_{NK}$  vs $m_0$. In fig. (a), single hidden sector is considered and a scan is done over mSUGRA model parameter space for $\tan(\beta)=10$\cite{Baer:2013gva}. In fig. (b), multiple hidden sectors are considered with $\tilde{m}=100$ GeV, charges distributed in range [-100:100] and $N_{HS}=20$. All point shown in fig. (b) are allowed from EWSB, Higgs mass and LHC constraints on stop and gluino masses. }
	\label{comptataN20}
\end{figure} 
	
\begin{itemize}
	\item \textbf{Nomura- Kitano fine-tuning}: In the SM, one may calculate the physical mass of the Higgs boson as:
	\[m_h^2=m_h^2|_{tree}+\delta m_h^2|_{rad}\]
	where $\delta m_h^2|_{rad}$ is proportional to cutoff scale square. Since $m_h^2|_{tree}$ and $\delta m_h^2|_{rad}$ are independent, naturally we would expect $m_h^2|_{tree}>\delta m_h^2|_{rad}$, otherwise if $\delta m_h^2|_{rad}>>m_h^2$, then we have to tune $m_h|_{tree}$ to be very high for obtaining correct physical Higgs mass. Using this argument we can define a fine tuning measure as:
	\[\Delta_{SM}\equiv \frac{\delta m_h^2|_{rad}}{(m_h^2/2)}\]
	which compares the radiative correction to the physical
	Higgs boson mass. A similar measure has been defined for supersymmetric models by Nomura and Kitano. In the MSSM:
	\[m_h^2= \mu^2+m_{H_u}^2\vert_{tree}+m_{H_u}^2\vert_{rad}\]	
	where $\mu$ is the supersymmetric mass for the Higgs doublets, and   $m_{H_u}^2\vert_{tree}$ and $m_{H_u}^2\vert_{rad}$ represent
	the tree-level and radiative contributions to the soft supersymmetry-breaking mass squared for
	$H_u$, respectively. The dominant contribution to $m_{H_u}^2\vert_{rad}$ arises from top-stop loop:
	\be	
	m_{H_u}^2\vert_{rad}\simeq-\frac{3y_t^2}{8\pi^2}(m_{Q_3}^2+m_{U_3}^2+|A_t|^2)\log\left(\frac{M_{mess}}{m_{\tilde{t}}}\right)
	\ee
	where $y_t$ is the top Yukawa coupling , $m_{Q_3}^2$ and $m_{U_3}^2$ are soft supersymmetry breaking masses for third-generation doublet quark $Q_3$ and singlet up-type quark $U_3$, respectively, $A_t$ is the trilinear scalar
	interaction parameter for the top squarks ($A=y_tA_t$), $M_{mess}$ represents the scale at which squark and slepton masses are generated , and $m_{\tilde{t}}$ is the scale of the top squark masses determined by $m_{Q_3}^2$ , $m_{U_3}^2$ and $A_t$ .
	For fine-tuning to be absent, each term in the right-hand-side of eq.(8) should not be much
	larger than the left-hand-side . The amount of fine tuning from $m_{H_u}^2\vert_{rad}$ term is then given by
	\[\Delta^{-1}_{NK}=\frac{m_{Higgs}^2}{2m_{H_u}^2\vert_{rad}}\]		
	This approach has two possible shortcomings, which are different from the case of the SM\cite{Baer:2013gva}:
	\begin{itemize}
		\item In this definition, finite contribution of $m_{Hu}^2$ is ignored. The fine tuning is calculated using the RG of $m_{Hu}^2$ and RG’s have only infinite contributions. In SUSY $m_{H_u}^2\vert_{tree}$ and $m_{H_u}^2\vert_{rad}$ are not independent and the value of $m_{H_u}^2\vert_{tree}$ feeds directly and indirectly into the evaluation of $m_{H_u}^2\vert_{rad}$. In fact, the larger the value of
		$m_{H_u}^2\vert_{tree}$, the larger is the cancellation in correction $m_{H_u}^2\vert_{rad}$.
		\item For high scale SUSY models, EW symmetry is broken radiatively by $m_{H_u}^2\vert_{rad}$
		being driven to large negative values. This suggests
		a regrouping of terms:
		\[m_h^2= \mu^2+(m_{H_u}^2\vert_{tree}+m_{H_u}^2\vert_{rad})\]
		where instead both $\mu^2$ and $(m_{H_u}^2\vert_{tree}+m_{H_u}^2\vert_{rad})$ should be comparable to $m_h^2$.
	\end{itemize}
\item \textbf{BG fine-tuning:}
This is well-known, traditional measure of fine-tuning. BG measure is defined as: 
\[\Delta_{BG}= \max(\Gamma_i)\]
where
\[\Gamma_i\equiv \lvert\frac{\partial \log m_Z^2}{\partial\log a_i}\lvert\equiv |\frac{a_i\partial m_Z^2}{ m_Z^2\partial a_i}|\]
where the $a_i$ constitute the fundamental parameters of the
model. Thus, $\Delta_{BG}$ measures the fractional change in $m_Z^2$
due to fractional variation in high scale parameters $a_i$ .
The $\Gamma_i$ are known as sensitivity coefficients.
An advantage of BG fine-tuning over NK fine-tuning is that it maintains
the correlation between $m_{H_u}^2\vert_{tree}$ and $m_{H_u}^2\vert_{rad}$ by replacing
$m_{H_u}^2(M_{SUSY})=m_{H_u}^2\vert_{tree}$ +$m_{H_u}^2\vert_{rad}$ by its expression in terms
of high scale parameters.\\

\end{itemize}
The conventional notion of SUSY breaking originates from the dynamics of a single hidden sector, in which NK fine-tuning is very large even for a small value of  $m_0$ parameter, see fig. (\ref{comptataN20}(a)). Also the BG fine-tuning is very large in a single hidden sector scenario, see fig. (\ref{fig:BGFT}). For both measures, the minimal value of fine-tuning is greater than $10^3$. In light of experiments progressively disfavoring weak scale SUSY, if we try to increase the SUSY scale in the single hidden sector, then we further increase the fine-tuning  of the SUSY parameters to give the correct Higgs mass and EWSB. While the notion of single sector SUSY breaking is convenient, it is mainly motivated from arguments of simplicity. Thus, in full generality, we must consider multiple hidden sectors. We study this and find that fine-tuning decreases drastically as we increase number of hidden sectors ($N_{HS}$), see fig. (\ref{comptataN20}(b)). We can see at a low value of $m_0$ there is negligible fine-tuning for multiple hidden sectors, while in the limit $N_{HS}=1$, we have the same order of fine-tuning ($\sim 10^3$) as conventional single sector fine-tuning. We explain below in detail the cancellation effect on fine-tuning. 
\subsection{Fine-tuning in Multiple Hidden Sectors Scenario} 
\begin{itemize}
	\item \textbf{BG Fine-tuning in multi-hidden sectors set-up:}
\end{itemize}
In our set-up, boundary conditions at GUT scale are as following (assuming universal coupling of all scalar, gaugino, and trilinear coupling):
\[m_{0}^2=\frac{d_1F_1^2}{M_{Pl}^2}+\frac{d_2F_2^2}{M_{Pl}^2}+.....+\frac{d_{N_{HS}}F_{N_{HS}}^2}{M_{Pl}^2}\]
\[M_{1/2}=\frac{s_1F_1}{M_{Pl}}+\frac{s_2F_2}{M_{Pl}}+.....+\frac{s_{N_{HS}}F_{N_{HS}}}{M_{Pl}}\]
\[A_{0}=\frac{a_1F_1}{M_{Pl}}+\frac{a_2F_2}{M_{Pl}}+.....+\frac{a_{N_{HS}}F_{N_{HS}}}{M_{Pl}}\]
where $F_{\alpha}=M_{Pl}\tilde{m}c_{\alpha}$. For coupling parameters $d_{\alpha},s_{\alpha},a_{\alpha}=1$, boundary conditions will reduce to the following form:
\be
m_{0}^2=\tilde{m}^2\sum_{\alpha=1}^{N_{HS}}c_{\alpha}^2\quad\quad
M_{1/2}=\tilde{m}\sum_{\alpha=1}^{N_{HS}}c_{\alpha}\quad\quad
A_{0}=\tilde{m}\sum_{\alpha=1}^{N_{HS}}c_{\alpha}
\label{meanvalue}
\ee
where $c_{\alpha}$ are discrete uniformly distributed charges in $\alpha^{th}$ hidden sector. Soft SUSY breaking parameters will get the contribution from all hidden sectors with mean
\be
\langle m_0^2\rangle=N_{HS}\langle c_{\alpha}^2\rangle\tilde{m}^2\quad\quad\quad\langle M_{1/2},A_0\rangle=N\langle c_{\alpha}\rangle\tilde{m}
\ee 
Renormalization group (RG) equations relate  high energy boundary conditions to those at low energy.
After RG evolution, following are simple expressions of $m_{H_u}^2(SUSY)$ and $m_{H_d}^2(SUSY)$ in terms of the fundamental parameters $c_{\alpha}$'s at SUSY scale:
\be
m_{H_u}^2=X(Q)\tilde{m}^2\sum_{\alpha=1}^{N_{HS}}c_{\alpha}^2+Y(Q)\tilde{m}^2(\sum_{\alpha=1}^{N_{HS}}c_{\alpha})^2
\ee
\be
m_{H_d}^2=\tilde{X}(Q)\tilde{m}^2\sum_{\alpha=1}^{N_{HS}}c_{\alpha}^2+\tilde{Y}(Q)\tilde{m}^2(\sum_{\alpha=1}^{N_{HS}}c_{\alpha})^2
\ee
\be
\frac{m_Z^2}{2}=-\mu^2+\frac{\tilde{m}^2[(\tilde{X}(Q)-X(Q)\tan^2\beta )\sum_{\alpha=1}^{N_{HS}}c_{\alpha}^2+(\tilde{Y}(Q)-Y(Q)\tan^2\beta )(\sum_{\alpha=1}^{N_{HS}}c_{\alpha})^2]}{\tan^2\beta-1}
\ee
\be
\frac{m_Z^2}{2}=-\mu^2+\tilde{\alpha}\tilde{m}^2\sum_{\alpha=1}^{N_{HS}}c_{\alpha}^2+\tilde{\beta}\tilde{m}^2(\sum_{\alpha=1}^{N_{HS}}c_{\alpha})^2
\ee
where $X(Q),\tilde{X}(Q),Y(Q)$ and $\tilde{Y}(Q)$ are RG coefficients and depend on the running scale Q.
\be
\tilde{\alpha}=\frac{\tilde{X}(Q)-X(Q)\tan^2\beta}{\tan^2\beta-1}\quad\quad\quad
\tilde{\beta}=\frac{\tilde{Y}(Q)-Y(Q)\tan^2\beta}{\tan^2\beta-1}
\ee
We know the definition of $\Delta_{BG}$ is the fractional change of $m_Z^2$
due to fractional variation in high scale parameters. In multiple hidden sectors scenario, charge coming from each sector are high scale parameters ($c_{\alpha}$), which are completely independent. The $\Gamma_{\alpha}$ sensitivity coefficients in this scenario are:
\[\Gamma_{\alpha}\equiv \lvert\frac{-2c_{\alpha}\partial m_{Hu}^2}{m_Z^2\partial c_{\alpha}}\lvert\]
\be
\Gamma_{\alpha}=\lvert \frac{-2c_{\alpha}\tilde{m}^2}{m_Z^2}(X(Q)(2c_{\alpha})+Y(Q)(2\sum_{\alpha=1}^{N_{HS}}c_{\alpha}))\lvert
\ee
BG fine-tuning is maxima of these sensitivity coefficients.
\[\Delta_{BG}= \max(\Gamma_{\alpha})\]
	
If we want to minimize the BG fine-tuning, we require $\Gamma_1=\Gamma_2=.....=\Gamma_{N_{HS}}$.
Suppose we have $N_{HS}=2$ then to minimize the BG fine-tuning (assuming RG coefficient X(Q) and Y(Q) are the same for both sectors):
\[\Gamma_1=\Gamma_2\]
\[\frac{-2\tilde{m}^2}{m_Z^2}(X(Q)(2c_{1}^2)+Y(Q)(2c_1^2+2c_1c_2))=\frac{-2\tilde{m}^2}{m_Z^2}(X(Q)(2c_{2}^2)+Y(Q)(2c_2^2+2c_1c_2))\]
One solution of this equation is $c_1=c_2$. There are $C^{2N_{HS}-1}_r$ solutions that satisfy $\Gamma_1=\Gamma_2=.....=\Gamma_{N_{HS}}$. r is order of polynomial. There is always a solution which is independent of RG coefficients X(Q) and Y(Q). That solution is $c_1=c_2=...=c_{N_{HS}}$. This solution motivates coherent charges coming from independent hidden sectors for the minimization of fine-tuning.\\
We have quantized charges, in which case, definition of sensitivity coefficients will be the following:
\be
\Gamma_{\alpha}\equiv \lvert\frac{-2( m_{Hu}^2(c_{\alpha}+1)-m_{Hu}^2(c_{\alpha}))}{m_Z^2}\lvert
\label{discreteBG}
\ee
\be
\label{BGeq}
\Gamma_{\alpha}=\lvert \frac{-2\tilde{m}^2}{m_Z^2}(X(Q)(1+2c_{\alpha})+Y(Q)(1+2\sum_{\alpha=1}^{N_{HS}}c_{\alpha}))\lvert
\ee
The sensitivity coefficients are very sensitive to sign of charges (due to cancellation region in $M_{1/2}$ and $A_0$, from linear sum of charges).    
BG fine-tuning is
\[\Delta_{BG}= \max(\Gamma_{\alpha})\]
\begin{figure}[h!]
	\centering
	\subfigure[]{%
		\label{fig:cohBG5}%
		\includegraphics[width=7.5cm,height=6cm]{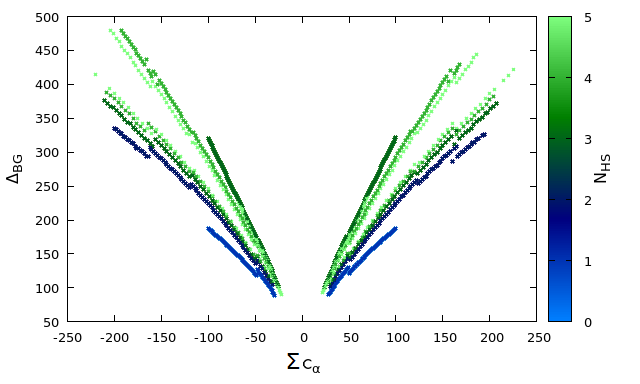}}%
	\hfill
	\subfigure[]{%
		\label{fig:noncohBG5}%
		\includegraphics[width=7.5cm,height=6cm]{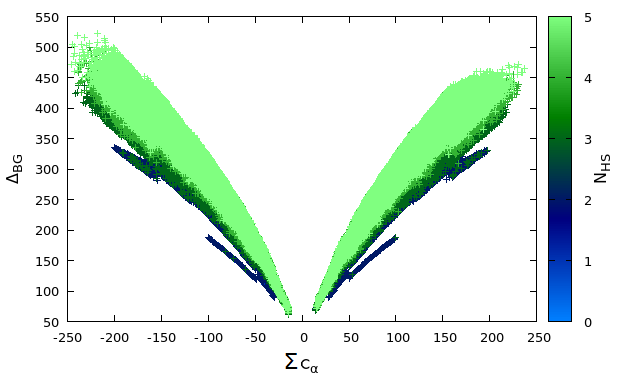}}%
	\caption{In these plots $\tilde{m}=100$ GeV with charges distributed in range of [-100:100]. fig (a) and (b) are contours of hidden sectors in plan of sum of charges and BG fine-tuning, in coherent SUSY and non-coherent SUSY scenarios, respectively. All the parameter space in all plots of this chapter are after applying the Higgs mass constraints and LHC constraints of gluino and stop masses.}
	\label{BGFT5}
 \end{figure}
 In fig. (\ref{BGFT5}(a), we have considered coherent SUSY scenario where charges coming from each sector have the same value with both positive and negative signs. In this plot, BG fine-tuning is of the order (10$^1-10^2$) for $N_{HS}=5$. It is not changing drastically with a large value of $N_{HS}$, because it strongly depends on the linear sum of charges coming from each hidden sector, eq(\ref{BGeq}). There will always be some cancellation region in summation of charges because some of them are negative, the corresponding region will then give minimum BG fine-tuning, for every value of $N_{HS}$. Even in the lower cancellation region, BG fine-tuning is not that large ($<10^3$). This is due to the discretization of charges that modifies the definition of $\Delta_{BG}$, eq. (\ref{discreteBG}) . In fig. (\ref{BGFT5}(b)), we consider non-coherent scenarios, where charge coming from each sector are random. The BG fine-tuning order of magnitude is the same as the coherent scenario: because it is not very sensitive to individual charge values coming from each sector, but is more sensitive to the sum of charges; and both coherent and non-coherent SUSY have cancellation regions. In both cases, BG fine-tuning value will be lower for larger cancellation region. In some cases occurring at very low probability, one of the RG coefficients, Y(Q), becomes zero. Then BG fine-tuning becomes more sensitive to individual charges. 
\begin{itemize}
\item \textbf{Nomura-Kitano fine-tuning in multi-hidden sectors set-up:}
\end{itemize}
The evolution of $m_{H_u}^2$ is parametrized as follows:
\be
m_{H_u}^2|_{rad}=\frac{3y_t^2}{8\pi^2}(\eta_{m_0}(Q)\tilde{m}^2\sum_{\alpha=1}^{N_{HS}}c_{\alpha}^2+\eta_{m_{1/2}}(Q)(\tilde{m}\sum_{\alpha=1}^{N_{HS}}c_{\alpha})^2)\log\left(\frac{M_{GUT}}{m_{SUSY}}\right)
\ee
According to NK definition, each term of radiative contribution coming from the soft SUSY breaking should be small. In our case, each sector independently contributes to radiative correction, and because of RG evolution, there are also interference terms that contribute to radiative correction. In the case of multiple hidden sectors:
\be
m_{H_u}^2|_{rad}=\sum_{\alpha_i}^{}(m_{H_u}^2|_{rad})_{\alpha_i}+\sum_{\alpha_i\ne \alpha_j}^{} (m_{H_u}^2|_{rad})_{\alpha_i\alpha_j}
\ee
where
\be
(m_{H_u}^2|_{rad})_{\alpha_i}=\frac{3y_t^2}{8\pi^2}(\eta_{m_0}(Q)\tilde{m}^2+\eta_{m_{1/2}}(Q)\tilde{m}^2)c_{\alpha_i}^2\log\left(\frac{M_{GUT}}{m_{SUSY}}\right)
\ee
\be
(m_{H_u}^2|_{rad})_{\alpha_i\alpha_j}=\frac{3y_t^2}{8\pi^2}\eta_{m_{1/2}}(Q)\tilde{m}^2(c_{\alpha_i}c_{\alpha_j})\log\left(\frac{M_{GUT}}{m_{SUSY}}\right)
\ee
There will be (N(N+1)/2) parameters of fine-tuning w.r.t each sector and with interaction terms defined as:
\be
\Delta^{-1}_{\alpha_i}=\frac{M_{Higgs}^2}{2(m_{H_u}^2\vert_{rad})_{\alpha_i}}
\ee
\be
\Delta^{-1}_{\alpha_i\alpha_j}=\frac{M_{Higgs}^2}{2(m_{H_u}^2\vert_{rad})_{\alpha_i\alpha_j}}
\ee
\be
\Delta^{-1}=Max[(m_{H_u}^2\vert_{rad})_{\alpha_i},(m_{H_u}^2\vert_{rad})_{\alpha_i\alpha_j}]
\label{deltaNK}
\ee

Solution for minimum fine-tuning requires $|c_1|=|c_2|=...=|c_{N_{HS}}|$ under assumption of RG coefficients being almost equal. It does not depend on sign of charges.\\
\begin{figure}[h!]
	\centering
	\subfigure[]{%
		\label{fig:NKfive}%
		\includegraphics[width=7.5cm,height=6cm]{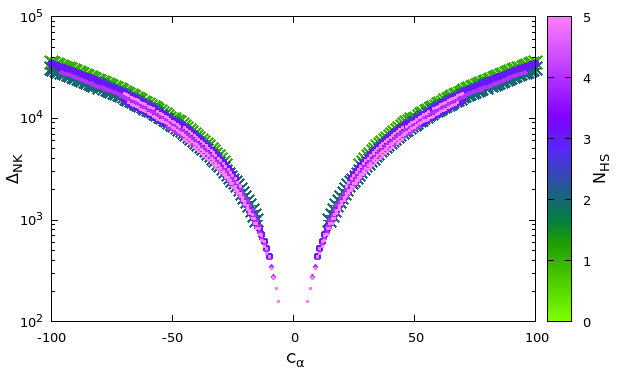}}%
	\hfill
	\subfigure[]{%
		\label{fig:NK5}%
		\includegraphics[width=7.5cm,height=6cm]{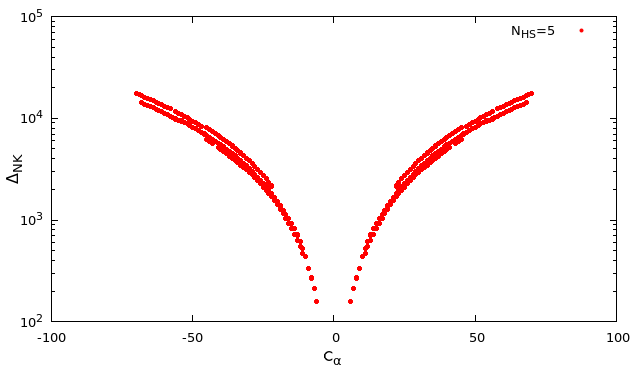}}%
	\vskip\baselineskip
	
	\subfigure[]{%
		\label{fig:NK20}%
		\includegraphics[width=7.5cm,height=6cm]{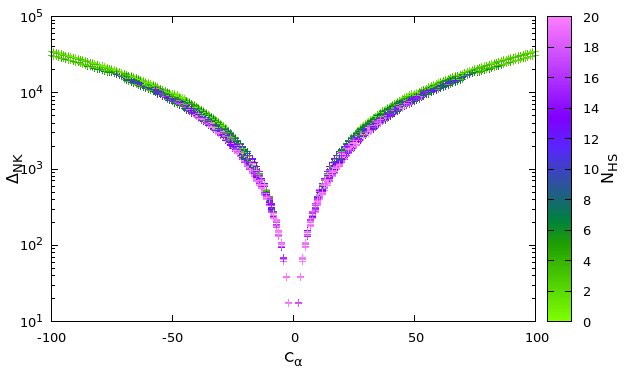}}%
	\hfill
	\subfigure[]{%
		\label{fig:msusyNK}%
		\includegraphics[width=7.5cm,height=6cm]{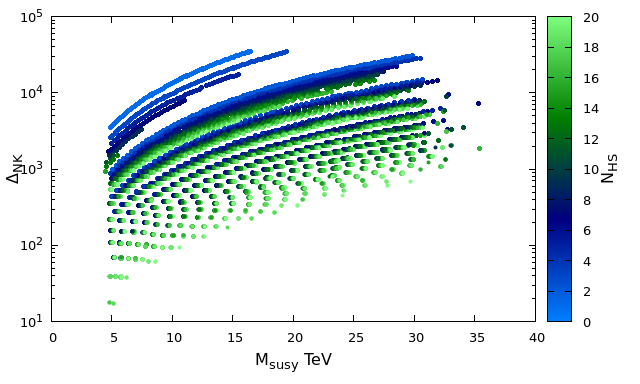}}%
	\caption{In these plots $\tilde{m}=100$ GeV and charges are distributed in range of [-100:100] in each sector. Fig. (a,b,c) are plots in plane of NK fine-tuning and charge $c_{\alpha}$, where color palette is representing $N_{HS}$. We can see effect of $N_{HS}$ on NK fine-tuning with charges and $M_{SUSY}$ scale.}
	\label{NKmulti}
\end{figure}
In fig. (\ref{NKmulti}), we have considered coherent charges.  As we discussed above, NK fine-tuning depends on individual charge coming from each sector and interaction between two-sector. If we want minimum fine-tuning, then all charges coming from each sector should be the same. We also see that the NK fine-tuning decreases as almost $1/N_{HS}$ (RG effect diverts little bit from exact $1/N_{HS}$). Fine-tuning become almost negligible even for $N_{HS}=20$, see fig. (\ref{NKmulti}(c)). It also depends on the value of charges; it will decrease as $c_{\alpha}$ coming from each sector will decrease, see fig. (\ref{NKmulti}(b)). In fig. (\ref{NKmulti}(d)), we can see the fine-tuning is almost negligible with $N_{HS}=20$ for $M_{SUSY}=5$ TeV, where for the same SUSY scale, single sector fine-tuning is around $\sim 10^3$. There are two reasons for it. First, in the single hidden sector to get 5 TeV SUSY scale, charges value should be very high, but in the multi-hidden sector, charge value coming from each sector could be low for the same scale. The second reason is that there is larger cancellation in $M_{1/2}$ parameter because of negative charges coming from some sectors, which will affect RG running of $m_{\tilde{t}}$ and reduce the SUSY scale.\\
\begin{figure}[h!]
	\centering
	\subfigure[]{%
		\label{fig:nonNKfive}%
		\includegraphics[width=7.5cm,height=6cm]{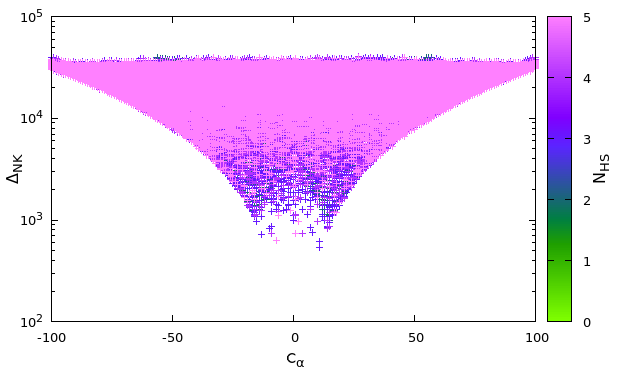}}%
	\hfill
	\subfigure[]{%
		\label{fig:nonNK5}%
		\includegraphics[width=7.5cm,height=6cm]{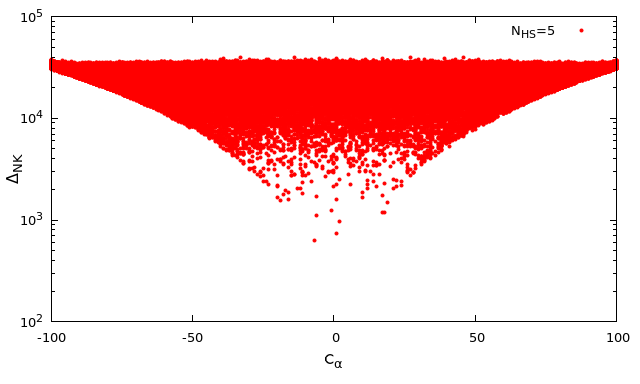}}%
	\vskip\baselineskip

	\subfigure[]{%
		\label{nonNK20}%
		\includegraphics[width=8cm,height=6cm]{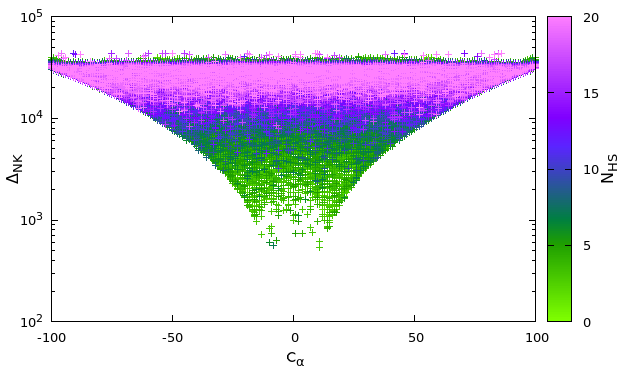}}%
	\caption{We consider non-coherent SUSY scenario with $\tilde{m}=100$ GeV and charge distributed in range of [-100:100]. We can see effect of $N_{HS}$ on NK fine-tuning with charge.}
	\label{nonNK}
\end{figure}
 In fig. (\ref{nonNK}), we have considered the non-coherent charges scenario, where charges coming from the different sectors will be completely random. $c_\alpha$ are charges coming from $5^{th}$ hidden sector in fig. (\ref{nonNK}(a)), (\ref{nonNK}(b)) and charges coming from $20^{th}$ hidden sector in fig. (\ref{nonNK}(c)); note that charges coming from other sectors are different in this non-coherent case.\\
 As we discussed in eq.(\ref{deltaNK}), NK fine-tuning is defined as maxima of fine-tuning coming from different sectors and interaction terms. In non-coherent scenario, as we start to increase the number of hidden sectors, initially we get some region where charge coming from each sector is small, see fig. (\ref{nonNK}(a),\ref{nonNK}(b)), but as we increase $N_{HS}$ further, the probability of getting small number of charges from each sector will start to decrease, i.e., there will be at least one or two hidden sectors from which charges will be large, and that will increase overall fine-tuning, see fig. (\ref{nonNK}(c)). 
 From fig. (\ref{NKmulti}) and fig. (\ref{nonNK}), we can say that coherent SUSY is the solution that gives minimum NK fine-tuning. 
 
    $\mathbf{d_{\alpha}\ne 1}$ \textbf{case}:
    
   In this case, we consider $d_{\alpha}$ (which is a complex coupling parameter of soft scalar mass effective operator) to be a random number between 1/4 to 4. Then the definition of both NK and BG fine-tuning is modified. The following are the modified equation for calculation of NK fine-tuning:
    \be
    m_{H_u}^2|_{rad}=\frac{3y_t^2}{8\pi^2}(\eta_{m_0}(Q)\tilde{m}^2\sum_{\alpha=1}^{N_{HS}}d_{\alpha}c_{\alpha}^2+\eta_{m_{1/2}}(Q)(\tilde{m}\sum_{\alpha=1}^{N_{HS}}c_{\alpha})^2)\log\left(\frac{M_{GUT}}{m_{SUSY}}\right)
    \ee   
    \be
    m_{H_u}^2|_{rad}=\sum_{\alpha_i}^{}(m_{H_u}^2|_{rad})_{\alpha_i}+\sum_{\alpha_i\ne \alpha_j}^{} (m_{H_u}^2|_{rad})_{\alpha_i\alpha_j}
    \ee   
    where
    \be
    (m_{H_u}^2|_{rad})_{\alpha_i}=\frac{3y_t^2}{8\pi^2}(\eta_{m_0}(Q)\tilde{m}^2d_{\alpha}+\eta_{m_{1/2}}(Q)\tilde{m}^2)c_{\alpha_i}^2\log\left(\frac{M_{GUT}}{m_{SUSY}}\right)
    \ee
    \be
    (m_{H_u}^2|_{rad})_{\alpha_i\alpha_j}=\frac{3y_t^2}{8\pi^2}\eta_{m_{1/2}}(Q)\tilde{m}^2(c_{\alpha_i}c_{\alpha_j})\log\left(\frac{M_{GUT}}{m_{SUSY}}\right)
    \ee
Similarly, For BG fine-tuning modified equations are:  
    \be
    m_{H_u}^2=X(Q)\tilde{m}^2\sum_{\alpha=1}^{N_{HS}}d_{\alpha}c_{\alpha}^2+Y(Q)\tilde{m}^2(\sum_{\alpha=1}^{N_{HS}}c_{\alpha})^2
    \ee		
    \be
    \Gamma_{\alpha}\equiv \lvert\frac{-2( m_{Hu}^2(c_{\alpha}+1)-m_{Hu}^2(c_{\alpha}))}{m_Z^2}\lvert
    \ee
    \be
    \Gamma_{\alpha}=\lvert \frac{-2\tilde{m}^2}{m_Z^2}(X(Q)(d_{\alpha}+2d_{\alpha}c_{\alpha})+Y(Q)(1+2\sum_{\alpha=1}^{N_{HS}}c_{\alpha}))\lvert
    \label{dalphaBG}
    \ee
\begin{figure}[h!]
	\centering
	\includegraphics[width=8cm,height=6cm]{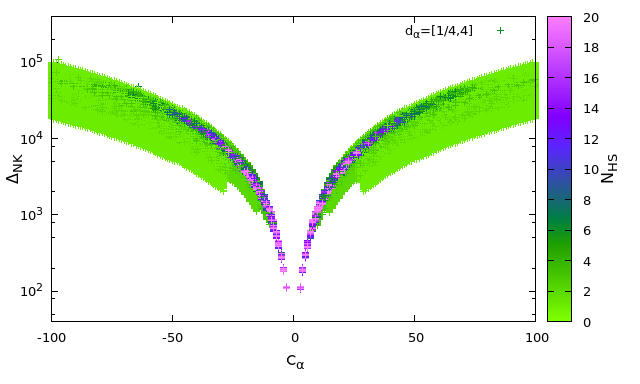}	
	\caption{We consider coherent SUSY scenario with $d_{\alpha}$ coefficient a random number in between $[1/4:4]$, $\tilde{m}=100$ GeV and charges distributed in range of [-100:100].} 
	\label{dalphaNK}
\end{figure}	
As we can see in eq. (\ref{dalphaBG}), BG fine-tuning mostly depends on sum of charges, so it will not change much from $d_{\alpha}=1$ case. And the NK fine-tuning, see fig. (\ref{dalphaNK}), will be minimum where both charge and coupling parameter of each sector is the same. Even with different value of $d_{\alpha}$, NK fine-tuning in coherent case is lower than non-coherent SUSY case as $N_{HS}$ increases.
\section{Coherent SUSY Spectrum}
	\begin{figure}[h!]
		\centering
		\subfigure[]{%
			\label{fig:muN20}%
			\includegraphics[width=7.5cm,height=6cm]{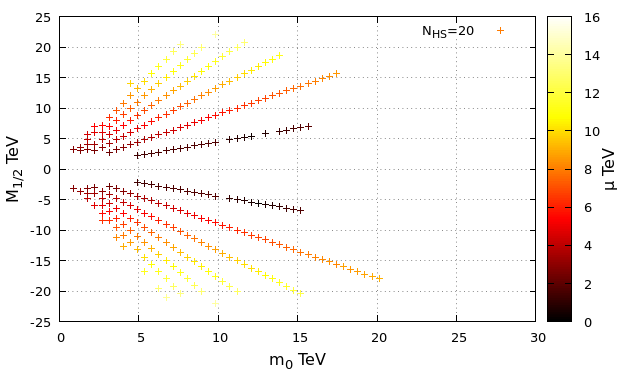}}%
		\hfill
		\subfigure[]{%
			\label{fig:stopmuN20}%
			\includegraphics[width=7.5cm,height=6cm]{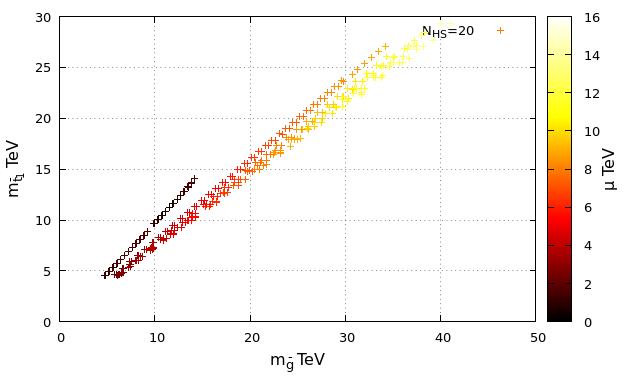}}%
		\caption{We consider coherent SUSY scenario with $\tilde{m}=100$ GeV and charges distributed in range of [-100:100] with $N_{HS}=20$.} 
		\label{spectrumN20}
	\end{figure}
\begin{figure}[h!]
	\centering
	\subfigure[]{%
		\label{fig:mhalfmu}%
		\includegraphics[width=7.5cm,height=6cm]{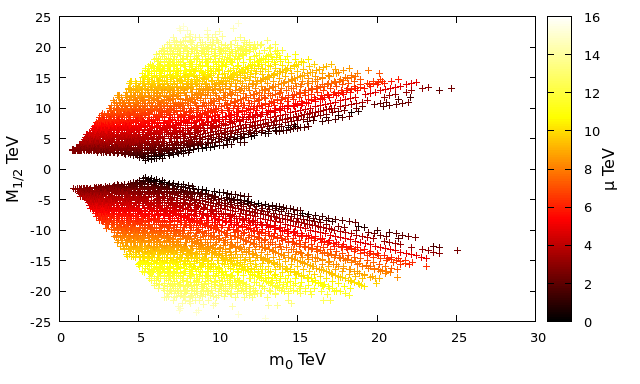}}%
	\hfill
	\subfigure[]{%
		\label{fig:mhalfgluino}%
		\includegraphics[width=7.5cm,height=6cm]{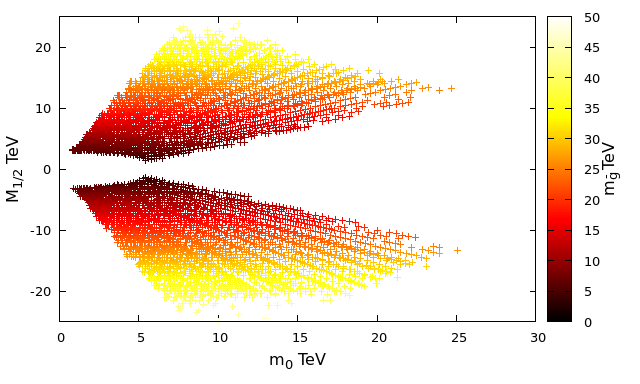}}%
	\vskip\baselineskip
	
	\subfigure[]{%
		\label{fig:mhalhstop}%
		\includegraphics[width=7.5cm,height=6cm]{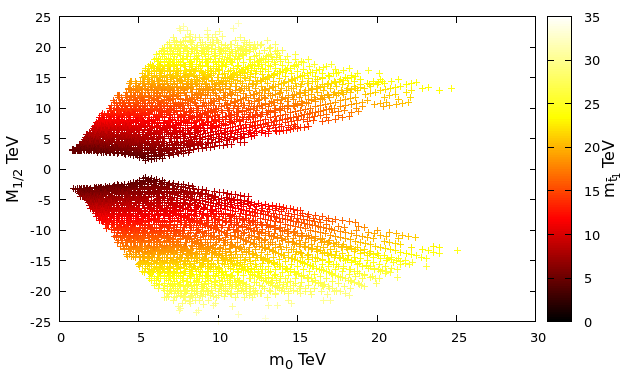}}%
	\hfill
	\subfigure[]{%
		\label{fig:mustop}%
		\includegraphics[width=7.5cm,height=6cm]{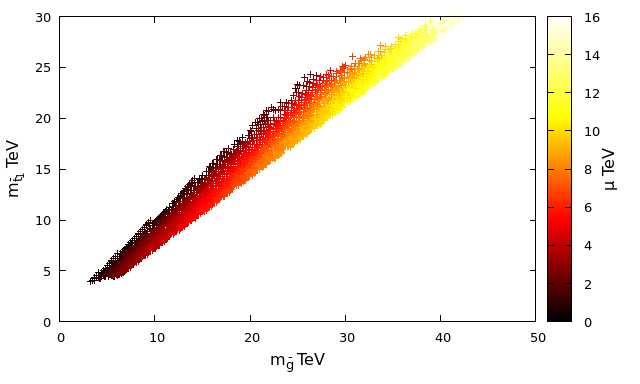}}%
	
	\caption{We consider coherent SUSY scenario with $\tilde{m}=100$ GeV and charges distributed in range of [-100:100]. Here $N_{HS}$ is varing from 1 to 20.} 			   			\label{cohspectrum}
\end{figure}

In this section, we discuss the spectrum for coherent SUSY with $d_{\alpha}=1$. A crucial point of the spectrum is the value of $\mu$.\\
As can be seen in the fig. (\ref{cohspectrum}(a)); the $\mu$ can be as small as 100 GeV. The black region is for $\mu<2$ TeV where $M_{1/2}$ is always less than $m_0$. Although, the mean of $m_0$ goes as $\sqrt{N_{HS}}$ and mean of $M_{1/2}$ goes as $N_{HS}$, see eq. (\ref{meanvalue}), still for low $\mu$ value we have $M_{1/2}<m_0$ due to large cancellation region with almost negligible fine-tuning and gluino and stop mass greater than 3 TeV, see fig.(\ref{cohspectrum}(b),\ref{cohspectrum}(c)). The gluino and stop masses are almost degenerate in large cancellation regions, see fig. (\ref{cohspectrum}(d)).\\
In fig. (\ref{spectrumN20}(b)) we show the plot for $N_{HS}=20$ where we can explicitly see that black line is diagonal in plane of mass of gluino and lightest stop with $\mu<2$ TeV.
Thus, one can conclude that the fine-tuning is not negligible just due to division into a large number of parameters; it is purely because of large cancellations which leads to low values of $\mu$ with correct EWSB. 

\section{Dark Matter in Multiple Hidden Sectors Scenario}
Dark matter is one of the nice features of R-parity conserving supersymmetry. The lightest supersymmetric particle (LSP) provides a stable, neutral, colorless, weakly-interacting cold dark matter candidate. The dark matter prediction is usually considered as one of the successes of weak-scale SUSY. However, constraints from LEP and LHC on SUSY particles and relic density measurements severely constrain the parameter space of supersymmetric models. Like, in the mSUGRA model, most of the parameter space has LSP that is almost purely bino. The bino is a gauge singlet whose annihilation in the early universe occurs through squark and slepton exchange. LHC and relic density put a severe constraint on bino like dark matter. It is almost ruled out. To satisfy relic density constraint, we need rapid annihilation of neutralinos via s-channel Higgs exchange or coannihilation of neutralinos with other sfermions to bring down the relic density to the desired range. Another possibility is that the LSP has a significant higgsino component, then the couplings of neutralinos to the Higgs or Z is enhanced. Furthermore, annihilation into gauge bosons or fermion pairs becomes very efficient. To satisfy one of these conditions, we require special relations among parameters. These are highly sensitive to small variations.
These conditions are satisfied only for narrow strips in the parameter space of the
mSUGRA model. There is great detailed literature that tells us that one of the most plausible solutions to reproducing the correct value of the dark matter density is the well-tempered neutralino, which corresponds to the boundary between a pure bino and a pure higgsino or wino. In mSUGRA, there is some focus point region where at large values of $m_0$, $\mu$ has a small value, and we get a well-tempered neutralino dark matter.\\
Below we discuss the dark matter in multiple hidden sectors with coherent SUSY and non-coherent SUSY scenarios. In multiple hidden sectors scenario, there are some regions where $M_{1/2}$ parameter is very low or $\mu$ value is very low even for high values of $m_0$ parameter, and that is purely because of high cancellation in $M_{1/2}$ parameter. That is true in both coherent and non-coherent scenario. Because of low values of $\mu$, we get a well-tempered dark matter with correct relic density or under-abundant higgsino-type dark matter.
\subsection{Non-Coherent SUSY}
\begin{figure}[h!]
		\centering
		\subfigure[]{%
			\label{fig:noncohmixing}%
			\includegraphics[width=7.5cm,height=5cm]{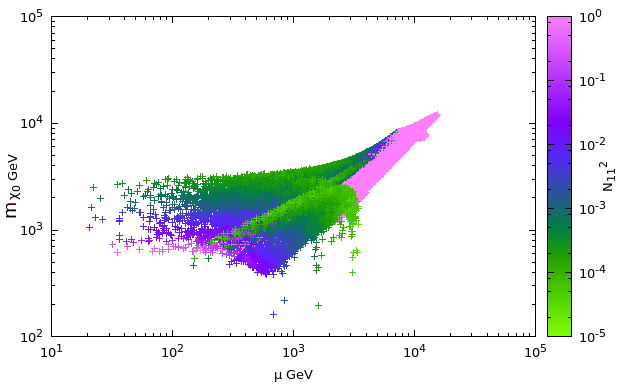}}%
		\hfill
		\subfigure[]{%
			\label{fig:nonrelic}%
			\includegraphics[width=7.5cm,height=5cm]{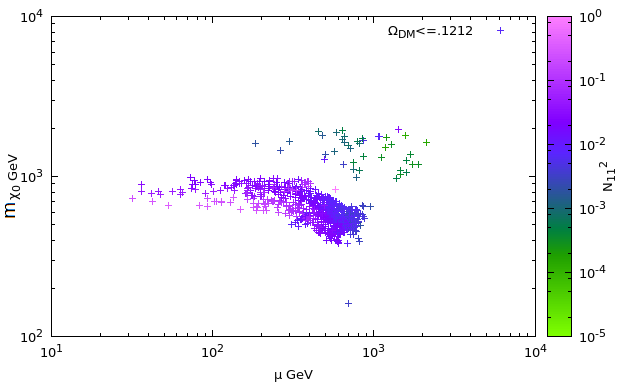}}%
		\vskip\baselineskip
		\subfigure[]{%
			\label{fig:noncohDM}%
			\includegraphics[width=7.5cm,height=5cm]{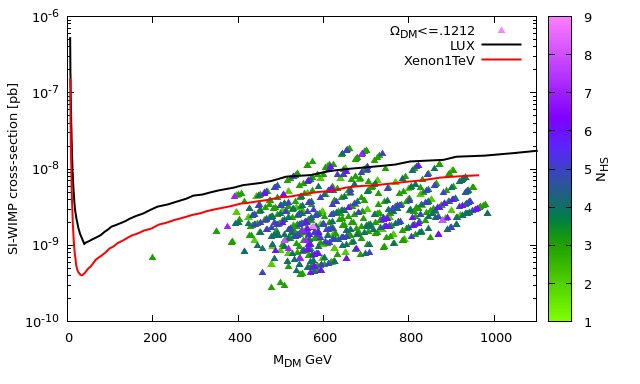}}%
		\hfill
		\subfigure[]{%
			\label{fig:noncohDMrelic}%
			\includegraphics[width=7.5cm,height=5cm]{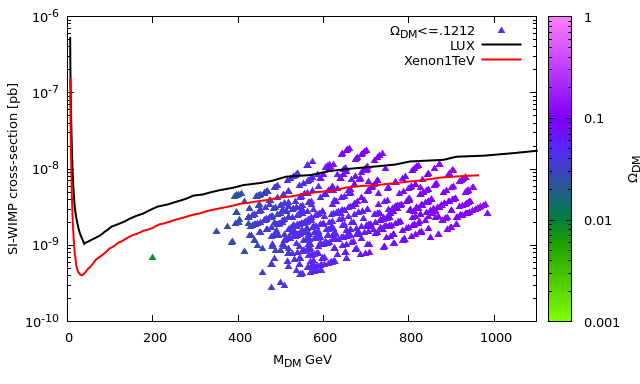}}%
	
		\caption{We consider non-coherent SUSY scenario with $\tilde{m}=100$ GeV and charges distributed in range of [-100:100]. Fig (b,c,d) are with relic density $\leq 0.1212 $ }
		\label{noncohDM}
	\end{figure}
In the non-coherent SUSY scenario, fig. (\ref{fig:noncohmixing}) is a plot of $\chi_0$ (LSP) vs. $\mu$ (higgsino parameter), where the color palette is representing the bino contribution in LSP ($N_{11}^2=1-ON(2)^2-ON(3)^2-ON(4)^2$, where ON(2), ON(3) and ON(4) are mixing parameters of wino and higgsinos respectively). $N_{11}^2=1$ means LSP is pure bino type dark matter. Pink color in fig. (\ref{noncohDM}(a)) corresponds to order 1 bino contribution in LSP. There is no relic density constraint in fig. (\ref{noncohDM}(a)). We have a large parameter region where LSP is not purely bino. The region of parameters in fig. (\ref{noncohDM}(b)) is after applying the upper bound of relic density. In fig. (\ref{noncohDM}(a)), we can see that dark green points correspond to higgsino type of dark matter with $\mu$ value in the range of 600 GeV to 1 TeV, and there is a region of well-tempered dark matter in which higgsino mixing with bino is varying from 1 to 90$\%$ with $\mu$ value from 30 to 1000 GeV. We get this small value of $\mu$ purely because of large cancellation in $M_{1/2}$ parameter with a large value of $m_0$ for these points. Fig. (\ref{noncohDM}) is a plot of a spin-independent cross-section of dark matter scattering with the proton of the nucleus vs. mass of dark matter. Here the region above the red line is ruled out by direct detection experiment Xenon1T. Similarly, the black line is for the LUX experiment. A small region of points satisfying relic abundance is ruled out by direct detection, but a large region of dark matter with a mass in the range 200-2000 GeV is allowed by all experimental constraints. Allowed parameter space has $N_{HS}$ value 2-8. For large value of $N_{HS}$, there is very little probability for large cancellation, and over all value of sum of charges is larger even with cancellation. In fig. (\ref{noncohDM}(d)),  we have allowed parameter space for dark matter with 1 to 100$\%$ relic abundance of dark matter.

\subsection{Coherent SUSY}
\begin{figure}[h!]
	\centering
	\subfigure[]{%
		\label{fig:cohDM}%
		\includegraphics[width=7.5cm,height=5cm]{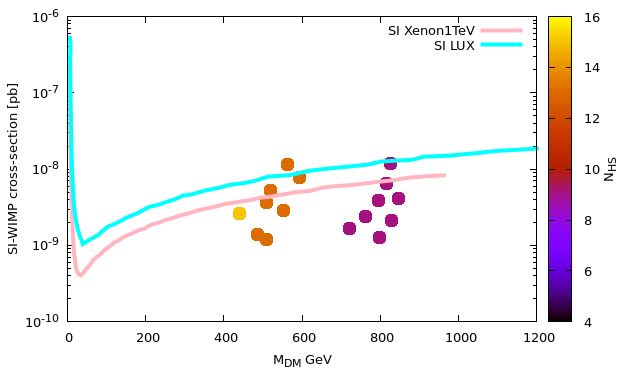}}%
	\hfill
	\subfigure[]{%
		\label{fig:cohDMrelic}%
		\includegraphics[width=7.5cm,height=5cm]{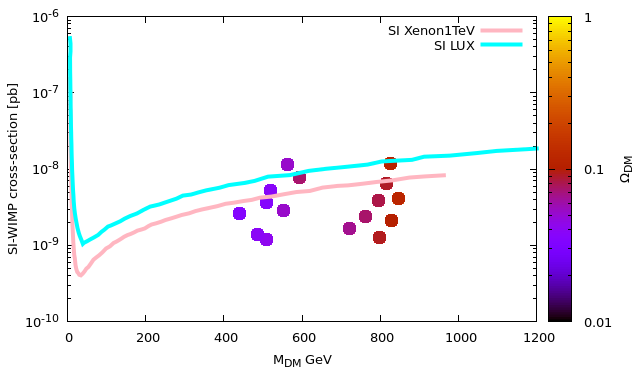}}%
	\vskip\baselineskip
	\subfigure[]{%
		\label{fig:Cohmixing}%
		\includegraphics[width=8cm,height=5cm]{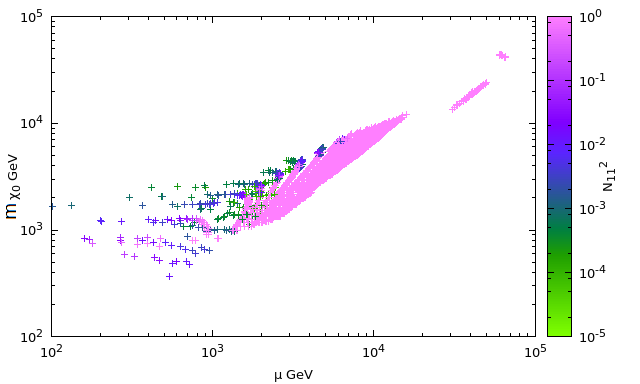}}%
	\caption{We consider coherent SUSY scenario with $\tilde{m}=100$ GeV and charges distributed in range of [-100:100]. Here $N_{HS}$ is varying from 1 to 20.}
	\label{cohDM}
\end{figure}
\begin{figure}[h!]
	\centering
	\subfigure[]{%
		\includegraphics[width=7.5cm,height=5cm]{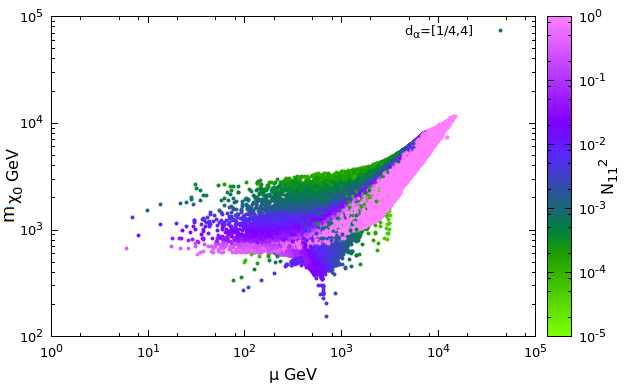}}%
	\hfill
	\subfigure[]{%
		\includegraphics[width=7.5cm,height=5cm]{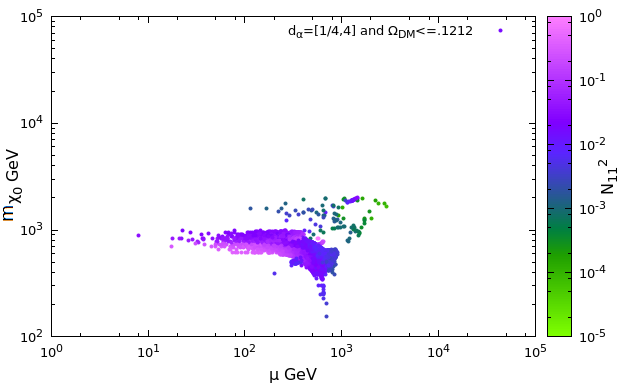}}%
	\vskip\baselineskip
	\subfigure[]{%
		\includegraphics[width=7.5cm,height=5cm]{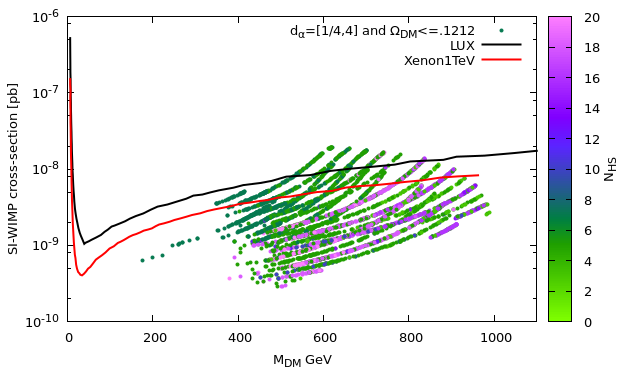}}%
	\hfill
	\subfigure[]{%
		\includegraphics[width=7.5cm,height=5cm]{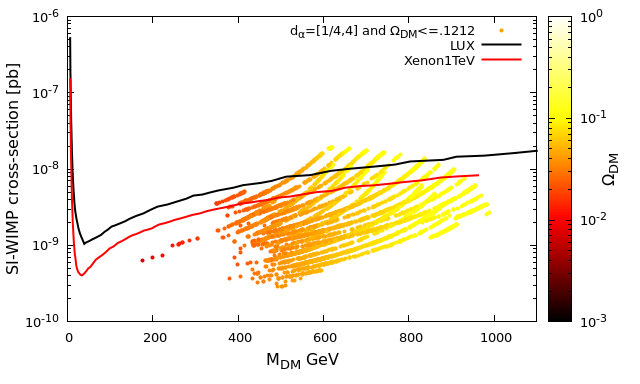}}%
	
	\caption{We consider coherent SUSY scenario with $d_{\alpha}$ coefficient a random number in between $[1/4:4]$, $\tilde{m}=100$ GeV and charges distributed in range of [-100:100].}
	\label{dalphaDM}
\end{figure}

In the coherent SUSY scenario, we study dark matter with coupling parameter $d_{\alpha}=1$ as well as $d_{\alpha}$ varying within [1/4:4]. Fig.(\ref{cohDM}(c)) is plot of $\chi_0$ (LSP) vs $\mu$ (higgsino parameter), where color is presenting bino contribution in LSP. The parameter space in fig. (\ref{cohDM}(c)) is without applying relic density constraints. Pink color in fig. (\ref{cohDM}(c)) is corresponding $\mathcal{O}(1)$ bino contribution in LSP. We have a lot of parameter regions where LSP is not purely bino. Fig. (\ref{cohDM}(a),\ref{cohDM}(c)) are with upper bound on relic density of dark matter. Fig. (\ref{cohDM}(a)) is a plot of a spin-independent cross-section of dark matter scattering with the proton of the nucleus vs. mass of dark matter. Here the region above the light-pink line is ruled out by direct detection experiment Xenon1T; the similar cyan line is for the LUX experiment. A few points satisfying relic abundance  are ruled out by direct detection, but some points with dark matter mass in the range 400-900 GeV are allowed by all experimental constraints. Allowed parameter space has $N_{HS}$ value in between 9 to 15. All points which satisfy relic density upper bound are with an odd number of hidden sectors where $M_{1/2}$ parameter is equal to a charge coming from one sector, charges coming from other sectors are canceled for these points, and $m_0$ parameter is $\sqrt{N_{HS}}$ times the charge coming from one sector. These points have maximum cancellation. Finding maximal cancellation points is less probable because of discrete charges with fixed coupling ($d_{\alpha}=1$). Fig. (\ref{cohDM}(a)), have some points with full relic density. However, if we consider $d_{\alpha}$ to be a random number in the range of 1/4 to 4, then the probability of getting maximal cancellation will increase, and we get more allowed points in parameter space with all constraints, see fig. (\ref{dalphaDM}).
\section{Conclusion}
We have presented here a model with multiple sequestered hidden sectors which are coupled at tree level to the
visible MSSM sector through effective operators. We parametrize the spurion fields in terms of integer charges within a finite range. The charges add linearly in contribution to gaugino masses and quadratically for scalar masses. The minimization of fine-tuning gives $C^{2N_{HS}-1}_r$ solutions. One of them is independent of the RG running effect and motivates coherent charges from independent hidden sectors. This leads to negligible fine-tuning with multiple hidden sectors. We study both non-coherent and coherent scenarios and see that both cases give well-tempered dark matter that satisfy all constraints, including points that reproduce exact relic density. And for both cases, fine-tuning is smaller than that in conventional single hidden sector SUSY. The cancellations in gaugino masses naturally leads to regions where the gaugino masses are suppressed, and the scalar masses are large, leading to \textit{focus point} like regions with small $\mu$. Thus one arrives at well-tempered or co-annihilation dark matter and negligible fine-tuning with a heavy spectrum, simultaneously, in a set-up with multiple hidden sectors.
\clearpage
\begin{subappendices}
\section{Example Points}
   \begin{table}[h!]
\begin{adjustbox}{max width=\textwidth}
   	$
   	\begin{array}{|c|c|c|c|c|c|c|c|c|c|c|c|}
   	\hline{\mbox{Parameter}}&{\rm{ a}}&
   	{\rm{b}} &{\rm{c}} & {\rm{d}} &
   	{\rm{e}} & {\mbox{Parameter}}&{\rm{a}}&
   	{\rm{b}} &{\rm{c}} & {\rm{d}} &
   	{\rm{e}}\\  \hline
   	C_{\alpha} &87 & 51&53 & 49 &13&\tilde{g} &-17.220 &10.153 &-10.966 &-10.179  &12.591  \\
   	\hline
   	m_0 &8.700 &5.100 &11.851 & 10.957 &2.9069&	\tilde{t_1}  &13.282 &7.970 &10.638 &9.851  &9.327  \\ \hline
   	M_{1/2} & -8.700&5.100 &-5.300 &-4.900  &6.500&\tilde{t_2}  &15.569 &9.328 &12.986 &12.025  &10.743 \\ \hline
   	M_{SUSY} &14.380 &8.622 &11.753 &10.884  &10.010  &	\tilde{c_1}  &16.350 &9.815 & 14.556 &13.481  &10.953\\ \hline
   	\mu  &5.753 &3.686 &0.659 &0.973  &4.973&	\tilde{c_2}  &17.000 &10.187 & 14.811 &13.716  &11.503 \\ \hline
   	A_{t}  &0.585 &-0.606 &0.449 &0.452  &-0.6529& \tilde{u_1}  &16.350 &9.815 &14.556 &13.481  &10.953 \\ \hline
   	\Delta_{NK} &27037.3 &10034.63 &15958.55 &13848.65&968.26  &\tilde{u_2}  &17.002 &10.188 &14.814 &13.719 &11.504  \\ \hline
   	\Delta_{BG} &172.00 &123.20&182.76 &171.04&179.05& 	\tilde{b_1}  &15.559 &9.318 &12.981 &12.021  &10.729 \\ \hline
   	h  & 0.1259& 0.1242&0.1257 & 0.1253 & 0.1247&\tilde{b_2}  &16.180 &9.714 &14.413 &13.348  &10.835 \\ \hline
   	H   &11.790 &7.064 &12.242 &11.355  &7.121& \tilde{s_1}  &16.275 & 9.769 &14.526 &13.454  &10.884 \\ \hline
   	A  &11.785 &7.062 &12.236 &11.348  &7.118 & \tilde{s_2}  &17.000 &10.187 &14.811 &13.717  &11.503\\ \hline
   	H^{\pm}  &11.790 &7.065 &12.242 &11.355  &7.121&\tilde{d_1}  &16.275 &9.769 &14.527 &13.454 &10.884  \\ \hline
   	\chi_1  & 4.053&2.365 &1.076 &1.787  &3.013 &	\tilde{d_2}  &17.002 &10.188 &14.814 &13.719  &11.504 \\ \hline
   	\chi_2  &6.046 &3.846 &1.078 &1.791  &5.124& \tilde{\tau_1}  &9.189 &5.385 &11.914 &11.015  &3.728 \\ \hline
   	\chi_3  &6.048 &3.856 &2.483 &2.296  &5.139&\tilde{\tau_2}  &10.234 &6.007 &12.233 &11.310  &5.037 \\ \hline
   	\chi_4  &7.240 &4.227 &4.420 & 4.085  &5.387 &	\tilde{\mu_1}  &9.269 &5.435 &12.007 &11.101  &3.776  \\ \hline
   	\chi_1^{\pm}  & 6.001&3.813 &1.072 &1.779  &5.078&	\tilde{\mu_2}  &10.270 &6.030 &12.278 &11.352  &5.054 \\ \hline
   	\chi_2^{\pm}  &7.083 & 4.136&4.301 &3.975  &5.284& \tilde{e_1}  &9.270 &5.436 &12.008 &11.101  &3.777\\ \hline
   	N_{HS} & 1 & 1 &5 &5 &5 &  	\tilde{e_2}  &10.271 &6.030 &12.280 &11.353  &5.055 \\ \hline
   	\end{array}
   	$
   	\end{adjustbox}
   	\caption{5 example points of spectrum with different $N_{HS}$ values. All masses are in TeV.}
   	\label{t:mod1}
   \end{table} 
   
  \begin{table}[h!]
  	\begin{tabular}{|l|l|l|l|l|l|l|l|}
  		\hline{\mbox{Parameter}}&{\rm{ a}}&
  		{\rm{b}} &{\rm{c}} & {\mbox{Parameter}}&{\rm{ a}}&
  		{\rm{b}} &{\rm{c}}\\\hline
  		m$_0$                        & 11906   & 6300    & 5408 & $m_{\tilde{t}_2}$                & 12269.0 & 6272.0  & 5122.1  \\
  		M$_{1/2}$                    & 4500    & -2100   & -1500 & $m_{\chi_0}$                & 931.1   & 846.3   & 592.45  \\
  		N$_{HS}$                     & 7       & 9       & 13    & $\Omega_{DM}$                & 0.104   & 0.113   & 0.072  \\
  		c$_{\alpha}$                 & 45      & 21      & 15    & 	$\Delta_{NK}$                & 7979.29 & 1945.46 & 1075.56 \\
  		$m_{\tilde{g}}$                  & 9495.9  & -4859.9 & -3451.7 & $\Delta_{BG}$                & 162.79  & 84.42   & 64.56  \\
  		$m_{\tilde{t}_1}$                & 9951.2  & 5011.1  & 4024.0  & $N_{11}^2$                   & 0.0015  & 0.0668  & 0.1405\\
  		$\mu$                        & 603.3   & 852.19  & 604.6  & & & &\\
  		\hline
  	\end{tabular}
  	\caption{Coherent SUSY example points for dark matter. All masses are in GeV.}
  \end{table}   	   		   
\section{Input Distributions}
\begin{figure}[h!]
	\centering
	\subfigure[]{%
		\label{fig:firstb}%
		\includegraphics[width=7.5cm,height=6cm]{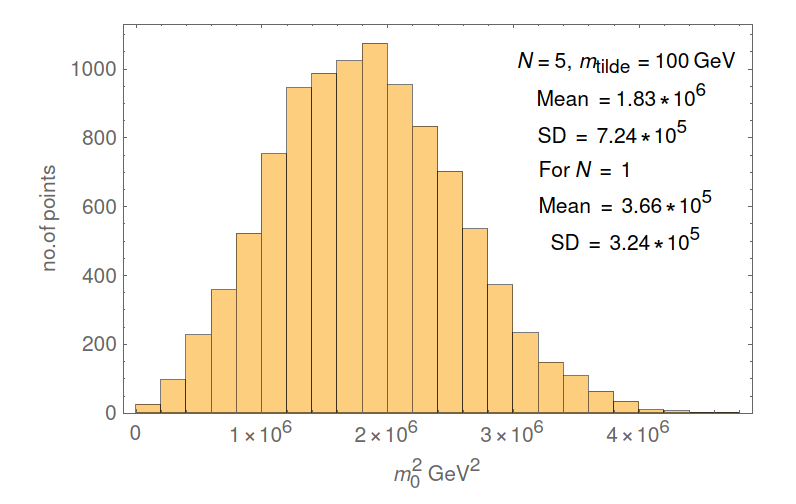}}%
	\hfill
	\subfigure[]{%
		\label{fig:secondb}%
		\includegraphics[width=7.5cm,height=6cm]{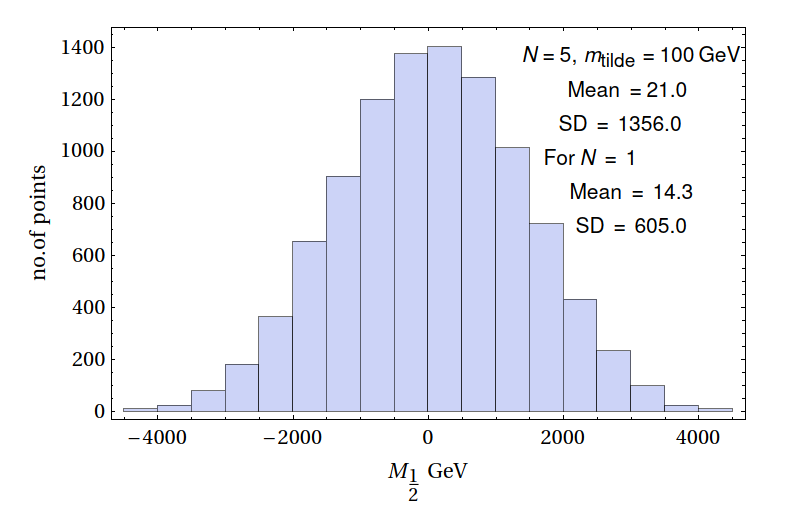}}%
	
	\caption{Fig:(a) sum of squares of $F_{\alpha}$. Its mean value is $N\mu_{uni}$ but distribution is not exactly normal, it needs high value of $N_{HS}$ to converge  to normal distribution. Fig:(b) linear sum of $F_{\alpha}$. Its mean and standard deviation goes as Normal distribution even for N=5.}
\end{figure} 

%
%


  \end{subappendices}

\chapter{Diluting SUSY Flavor Problem on the Landscape}
\label{Chapter 4}

\lhead{Chapter 4. \emph{Diluting SUSY Flavor Problem on the Landscape}}
\section{Motivation}
			We consider an explicit effective field theory example based on the Bousso-Polchinski 
			framework with a large number $N_{HS}$ of hidden sectors contributing  to  supersymmetry breaking. Each contribution comes from four form quantized fluxes, multiplied by random couplings.  
			The soft terms in the observable sector in this case become random variables, with mean values and standard deviations which are computable. We show that this setup naturally leads
			to a solution of the flavor problem in low-energy supersymmetry if $N_{HS}$ is sufficiently large. We investigate the consequences for flavor violating processes at
			low-energy and for dark matter. 
		\section{Introduction} 
		\label{introsection}

In string or supergravity based models, it has been long known that in general scenarios, it is hard to escape 
flavor violation, unless some specific conditions are chosen \cite{Brignole:1993dj,Brignole:1997dp}.	 For example, 
if the K{\" a}hler potential of the matter fields is  canonical  and independent of the moduli/hidden
sector fields, one can expect a universal, flavor-independent form for the soft terms as in minimal Supergravity.   On the other hand, 
the problem can also be avoided if supersymmetry breaking is dominantly dilaton mediated \cite{Kaplunovsky:1993rd}. Other solutions include decoupling of the first two generations \cite{Cohen:1996vb}  or imposing flavor symmetries
(see for example, \cite{Antusch:2008jf,Pomarol:1995xc,Dine:1993np,Dudas:1995eq} and references there in).  

In this chapter we address these issues from a different point of view, inspired by the landscape of string theory vacua. We consider a large number $N_{HS}$ of sectors contributing to supersymmetry breaking. Large number of sequestered
 hidden sectors have also been considered recently in \cite{Cheung:2010mc,Cheung:2010qf,Benakli:2007zza}, in models with multiple (pseudo)goldstini.  Other works which have addressed supersymmetric soft spectrum phenomenology from the landscape following \cite{Denef:2004ze} include \cite{Baer:2019tee,Baer:2019xww,Baer:2017uvn,Baer:2017cck}.
 In particular a solution to flavor and CP problems in the landscape through heavy first two generations was
 proposed in \cite{Baer:2019zfl}. 
 
Here we consider quantized four form fluxes {\`a} la Bousso-Polchinski \cite{Bousso:2000xa}.
		Each sector contributes in a quantized way, with a quanta that will be taken to be below the electroweak scale. Due to the large number of contributions, the observable soft terms become random variables 
		with Normal-type distributions around an average value. The setup also has the virtue of minimizing fine-tuning of the electroweak scale, due to the small contributions of each sector. 
		We find that our setup can address at the same time the flavor problem of low-energy SUSY by generating FCNC effects proportional to standard deviation of soft terms from their mean value, which are
		parametrically suppressed as $1/{\sqrt N_{HS}}$. By performing an RG analysis from high to low-energy, the setup also makes concrete predictions for low-energy flavor observables. 
			
The logic and setup we put forward is mainly originating from Bousso and Polchinski approach to the cosmological constant \cite{Bousso:2000xa}. We consider the implications of
		the string theory landscape for observable sector soft SUSY breaking terms. We assume a large $N_{HS}\gg1$ number 
	of SUSY-breaking sectors communicating through  gravitational couplings to the
		Supersymmetric Standard Model (SSM). 
Such models could naturally appear in string theory, where there may be several independent sources of supersymmetry breaking.

		While we impose the cancellation of the cosmological constant  \`a la Bousso-Polchinski, we do not necessarily use our framework 
		to address the cosmological constant problem. Instead we use the framework as a network of hidden 
		sectors, each contributing individually to soft supersymmetry breaking.  		
\section{Four-forms  and Fluxes}

Three form gauge potentials with (non-dynamical) four-form field strengths were considered longtime ago for addressing the cosmological constant problem \cite{Aurilia:1980xj,Witten:1983ux,Henneaux:1984ji,Brown:1987dd,Brown:1988kg,Duff:1989ah,Feng:2000if,dvali:2001sm}, the gauge hierarchy problem
\cite{Dvali:2003br,Dvali:2004tma,Herraez:2016dxn,Giudice:2019iwl}(see also \cite{Lee:2019efp}), the strong CP problem \cite{Dvali:2005an,Dvali:2005zk}, inflation \cite{Kaloper:2008fb,Kaloper:2011jz,Kaloper:2014zba,Dudas:2014pva} and supersymmetry breaking \cite{Farakos:2016hly}.  On the other hand,
it turned out to play an important role in the landscape of string theory compactifications \cite{Farakos:2017jme,Herraez:2018vae,Lanza:2019xxg} (for a recent review see e.g. \cite{Lanza:2019nfa}).  
Here we briefly review the main points of a theory containing three-forms with quantized form-forms field strengths. 

Let us start from a Lagrangian containing some scalar fields $\varphi_i$ and three-form fields $C_{mnp}^\alpha$, with the action
\begin{equation}
{\cal S}_0 = \int d^4 x \ \{ - \frac{1}{2} (\partial \varphi_i)^2 - \Lambda_0 -  \frac{1}{2 \times 4 !} F_{mnpq}^{\alpha,2} +
\frac{1}{24} \ f_\alpha (\varphi_i) \ \epsilon^{mnpq} F_{mnpq}^\alpha \} \ , \label{shift1}
\end{equation}
where 
\begin{equation}
F_{mnpq}^\alpha = \partial_m C_{npq}^\alpha + {\rm 3 \ perms.} \ . \label{shift2}
\end{equation}
For future convenience we define
\begin{equation}
F^\alpha = \frac{1}{4 !} \epsilon^{mnpq} F_{mnpq}^\alpha \ , \ F_{mnpq}^\alpha = - \epsilon_{mnpq} F^\alpha  \ . \label{shift3}
\end{equation}
 
The lagrangian (\ref{shift1}) has actually to be supplemented with a boundary term 
\begin{equation}
{\cal S}_b = \frac{1}{6} \int d^4 x \ \partial_m \left( F^{mnpq}_\alpha C_{npq}^\alpha - f_\alpha (\varphi_i) \epsilon^{mnpq}
C_{npq}^\alpha \right)  \ . \label{shift4}
\end{equation}
The total action is
\begin{eqnarray}
&& {\cal S}  = {\cal S}_0 +{\cal S}_b =   \int d^4 x \ \{ - \frac{1}{2} (\partial \varphi_i)^2 - \Lambda_0 - \frac{1}{2 \times 4 !} F_{mnpq}^{\alpha,2} - \frac{1}{6} \ \epsilon^{mnpq} \partial_m f_\alpha (\varphi_i) \  
C_{npq}^\alpha \} \nonumber \\ 
&& + \frac{1}{6} \int d^4 x \ \partial_m \left( F^{mnpq}_\alpha C_{npq}^\alpha  \right)  \ .  \label{shift5}
\end{eqnarray}
A massless three-form gauge field in four spacetime dimensions has no on-shell degrees of freedom. As such, it
can be integrated out via its field eqs. 
\begin{equation}
\partial^m F_{mnpq}^\alpha \ = \  \ \epsilon_{mnpq} \partial^m f_\alpha (\varphi_i)  \ , \label{shift6}
\end{equation}
whose solution is given by
\begin{equation}
F_\alpha = -  f_\alpha (\varphi_i)  + c_\alpha \ , \label{shift7}
\end{equation}
where $c_\alpha$ is a constant, which is to be interpreted as a flux. It was argued in \cite{Bousso:2000xa} that $c_\alpha$
are quantized in units of the fundamental membrane coupling $c_\alpha = m_\alpha e$, fact that  has important consequences for the landscape of string theory.
After doing so, the final lagrangian takes the form
\begin{equation}
{\cal S} = \int d^4 x \ \{ - \frac{1}{2} (\partial \varphi_i)^2 - \Lambda_0 - \frac{1}{2} \sum_\alpha (f_\alpha (\varphi_i)-c_\alpha)^2 \} \ . \label{shift8}
\end{equation}
The final resulting cosmological constant is therefore scanned by the flux
\begin{equation}
\Lambda =  \Lambda_0 + \frac{1}{2} \sum_\alpha (f_\alpha (\varphi_i)-c_\alpha)^2  \ . \label{shift9}
\end{equation}
Notice that the boundary term ${\cal S}_b$ is crucial in obtaining the correct action. Ignoring it leads
to the wrong sign of the last term in  (\ref{shift8}), fact that created confusion in the past. 
\subsection{Supersymmetric formulation}

The embedding of four-form fluxes in supersymmetry and supergravity proceeds by introducing three-form multiplets, defined as the real superfields
\cite{Gates:1980ay,Burgess:1995kp,Burgess:1995aa,Binetruy:1995hq,Binetruy:1996xw,Binetruy:2000zx,Groh:2012tf} 
\begin{eqnarray}
&& U_\alpha = {\bar U}_\alpha = B_\alpha + i (\theta \chi_\alpha - {\bar \theta} {\bar \chi}_\alpha) + \theta^2 {\bar M}_\alpha +
 {\bar \theta}^2 { M_\alpha} + \frac{1}{3} \theta \sigma^m {\bar \theta} \epsilon_{mnpq} C^{npq}_\alpha
+ \nonumber \\ 
&& \theta^2 {\bar \theta}  (\sqrt{2} {\bar \lambda}_\alpha + \frac{1}{2} {\bar \sigma}^m \partial_m \chi_\alpha)
+ {\bar \theta}^2 \theta  (\sqrt{2} {\lambda}_\alpha - \frac{1}{2} {\sigma}^m \partial_m {\bar \chi}_\alpha)
+ \theta^2 {\bar \theta}^2 (D_\alpha-\frac{1}{4} \Box B_\alpha) \ . \label{susy1}
\end{eqnarray}
The difference between $U_\alpha$ and a regular vector superfield $V$ is the replacement of the vector potential
$V_m$ by a three-form $C^{npq}_\alpha$. 
In order to find correct kinetic terms, the analog of the chiral field strength superfield $W_{\alpha}$
for a vector multiplet is replaced by the chiral superfield \cite{Gates:1980ay}
\begin{equation}
T_\alpha = - \frac{1}{4} {\bar D}^2 U_\alpha \quad , \quad T_\alpha (y^m,\theta) = M_\alpha + \sqrt{2} \theta \lambda_\alpha +
\theta^2 (D_\alpha + i F_\alpha) \ , \label{susy2}
\end{equation}
with $F_\alpha$ defined as in (\ref{shift3}). The definition (\ref{susy2}) is invariant under the gauge transformation $U_\alpha \to U_\alpha - L_\alpha$, where $L_\alpha$ are linear multiplets. Correspondingly, Lagrangians expressed as a function of $T_\alpha$ will have this gauge freedom.   One can therefore choose a gauge in which $B_\alpha=\chi_\alpha=0$  in  (\ref{susy1}) and the physical fields are 
complex scalars $M_\alpha$ and Weyl fermions  $\lambda_\alpha$. 

Notice that for the purpose of finding the correct on-shell Lagrangian and scalar potential, there is a simpler formulation in which $T_\alpha$ are treated as  standard chiral superfields with 
$D_\alpha+iF_\alpha$ as 
auxiliary fields, no boundary terms are included, but the superpotential of the theory is changed
according to \cite{Binetruy:1996xw,Binetruy:2000zx,Groh:2012tf}
\begin{equation}
W (\phi_i, T_\alpha) \to W' (\phi_i, T_\alpha) \ = \ W (\phi_i, S_\alpha) + i c_\alpha T_\alpha \ , \label{susy07} 
\end{equation}     
where $c_\alpha$ are the quantized fluxes. The linear terms in the superpotential shift linearly the auxiliary fields. In supergravity, the (F-term) scalar potential can be written as ($M_P=1$ in what follows)
\begin{equation}
V= K_{\alpha \bar \beta} F^{\alpha}  F^{\bar \beta} - 3 m_{3/2}^2 \quad , \quad {\rm where} \quad  F^{\alpha} = e^{\frac{K}{2}} K^{\alpha \bar \beta} \overline{D_\beta W} \ .    \label{susy08}
\end{equation}
The linear flux terms shift therefore the auxiliary fields according to
\begin{equation}
F^{\alpha} = e^{\frac{K}{2}} K^{\alpha \bar \beta} \overline{D_\beta W'} = e^{\frac{K}{2}} K^{\alpha \bar \beta} \overline{D_\beta W} - i  e^{\frac{K}{2}} K^{\alpha \bar \gamma}
 ({\bar c}_{\bar \gamma} + K_{\bar \gamma} {\bar T}_{\bar \beta} {\bar c}_{\bar \beta} )  \ ,    \label{susy09}
\end{equation}
 leading to a scanning of the cosmological constant.
 \subsection{Soft terms in supergravity}
 
 We start from a supergravity Lagrangian containing hidden sector (moduli) fields $T_\alpha$, whose auxiliary fields contain the four-form fluxes we introduced previously, coupled to matter fields
 called $Q_i$ in what follows. The Kahler potential and superpotential are defined by
 \begin{eqnarray}
 && K = {\hat K} (T_\alpha, {\bar T}_{\alpha}) + K_{i \bar j} (T_\alpha, {\bar T}_{\alpha}) Q^i {\bar Q}^{\bar j} + \frac{1}{2} \left( Z_{i j} (T_\alpha, {\bar T}_{\alpha}) Q^i {Q}^{ j} + {\rm h.c.} \right) 
 \ , \nonumber \\
 && W = {\hat W} (T_\alpha) + \frac{1}{2}  {\tilde \mu}_{i j} (T_\alpha) Q^i {Q}^{ j} +  \frac{1}{3}  {\tilde Y}_{i jk} (T_\alpha) Q^i {Q}^{ j} Q^k + \cdots    \ .   \label{soft1}
\end{eqnarray}
The low-energy softly broken supersymmetric Lagrangian is defined by the superpotential and soft scalar potential
\begin{eqnarray}    
&& W_{\rm eff} = \frac{1}{2}  {\mu}_{i j} Q^i {Q}^{ j} +  \frac{1}{3}  {Y}_{i jk}  Q^i {Q}^{ j} Q^k \ , \nonumber \\ 
&&  {\cal L}_{\rm soft} = - m_{i \bar j}^2 q^i q^{\bar j} - \left( \frac{1}{2}  {B}_{i j} q^i {q}^{ j} +  \frac{1}{3}  {A}_{i jk}  q^i {q}^{ j} q^k + \frac{1}{2} M_a \lambda_a \lambda_a + {\rm h.c.} \right) \ ,   \     \label{soft2}
\end{eqnarray} 
where $Y_{ijk} = e^{K/2}  {\tilde Y}_{i jk}$. After imposing the cancellation of the cosmological constant,  the various soft terms and the supersymmetric masses are given by 
\cite{Kaplunovsky:1993rd,Brignole:1993dj,Brignole:1997dp,Ferrara:1994kg,Dudas:2005vv}
\begin{eqnarray}
&& M_a = \frac{1}{2} g_a^2 F^{\alpha} \partial_\alpha f_a \ , \nonumber \\
&& m_{i  \bar j}^2 = m_{3/2}^2 K_{i \bar j} - F^{\alpha}  F^{\bar \beta} R_{i \bar j \alpha \bar \beta} \quad , \quad {\rm where} \quad R_{i \bar j \alpha \bar \beta} = \partial_\alpha \partial_{\bar \beta}
K_{i \bar j}  - K^{m \bar n} \partial_\alpha K_{i \bar n} \partial_{\bar \beta} K_{m \bar j} \ , \nonumber \\
&& A_{ijk} = (m_{3/2} - F^{\alpha} \partial_\alpha \log m_{3/2} ) Y_{ijk} + F^{\alpha} \partial_\alpha Y_{ijk} - 3 F^{\alpha}  \Gamma_{\alpha (i}^l Y_{ljk)} \ ,    \label{soft2}
   \end{eqnarray} 
where we have introduced also the Kahler connexion
\begin{equation}
\Gamma_{IJ}^K = K^{K \bar L} \partial_I K_{J \bar L} \ .    \label{soft3}
\end{equation}  $B_{ij}$ terms are not displayed since they will not be scanned in what follows. Similarly the
$\mu$ term is determined by the radiative electroweak symmetry breaking conditions at the weak scale.

 \subsection{Scanning soft terms and gravitino mass}
 
 Taking into account the scanning of auxiliary fields from four-forms fluxes, in what follows we use the simplified scanning
 \begin{equation}
F_\alpha = c_\alpha {\tilde m} M_{Pl} \ , \label{scan1}
 \end{equation}
where $c_\alpha$ are integers and $\tilde m M_{Pl}$ is the quantum of scanning. Taking into account the cancellation of the cosmological constant, and setting the matter fields wavefunctions in a canonical form, the formulae for the scanning will be taken to be 
\begin{eqnarray}
&& m_{3/2} = \tilde m (g_0 + \sum_{\alpha}g_\alpha c_\alpha) \quad , \quad m_{3/2}^2= \frac{1}{3} \sum_{\alpha=1}^{N_{HS}} \frac{F_\alpha^2}{M_{Pl}^2} = \frac{1}{3} {\tilde m}^2 \sum_{\alpha} {c_\alpha^2}  \ , \ \nonumber \\
&& (m_0^2)_{i \bar j} = m_{3/2}^2 \delta_{i \bar j}  +  {\tilde m}^2 \sum_{\alpha}  d_{\alpha, i \bar j}  c_\alpha^2 \ , \ \nonumber \\
&& M_{1/2}^A = {\tilde m} \sum_{\alpha}  s_{\alpha}^A  c_\alpha \ , \ \nonumber \\
 && A_{i jk} = m_{3/2} y_{i jk}  +  {\tilde m}  \sum_{\alpha}  a_{\alpha, i jk}  c_\alpha \ , \  \label{scan2}
 \end{eqnarray}
 where $\alpha = 1 \cdots N$, $-M \leq c_\alpha \leq M$ (integers), $-d_0 \leq a_\alpha, d_\alpha,s_\alpha,g_\alpha \leq d_0$ (continuous). These last couplings are taking to be continuous in order
 to take into account the couplings of the hidden sector fields with the MSSM ones, which are dependent on the hidden sector vev's and interactions. Note that the soft terms defined above are in the so-called super CKM basis which is important for the flavor discussion below. 
 
The scanning of the gravitino mass, combined with the cancellation of the cosmological constant (the Deser-Zumino relation) implies a constrained among the fluxes
\begin{equation}
g_0^2 + 2 g_0 \sum_\alpha g_\alpha c_\alpha + \sum_{\alpha,\beta}  g_\alpha g_\beta  c_\alpha  c_\beta  = \frac{1}{3} \sum_\alpha c_\alpha^2 \ . \label{scan3}
\end{equation}
Taking the average value of (\ref{scan3}) this implies in particular 
\begin{equation}
g_0^2 = \sum_\alpha \overline{(1/3- g_\alpha^2) c_\alpha^2}\simeq \frac{N_{HS}}{9}(1-d_0^2)M^2 \quad , \quad \overline{m_{3/2}} = {\tilde m}  g_0 \sim O(\sqrt{N_{HS}}) {\tilde m}  \ . \label{scan4}
\end{equation} where we have used the large flux limit  $M \gg1$. We can also compute
\begin{equation}
\overline{m_{3/2}^2} = \frac{\tilde m^2}{3} \sum_\alpha \overline{c_\alpha^2}  \sim    \frac{1}{9} N_{HS} M^2 {\tilde m^2} \ , \label{scan4}
\end{equation}
The mean values of the soft terms are therefore computed to be
\begin{equation}  
\overline{(m_0^2)_{i \bar j}} = \overline{m_{3/2}^2} \delta_{i \bar j} ,\quad \quad \overline{A_{ijk}} = \overline{m_{3/2}} y_{i jk}, \quad  \quad   \overline{M_{1/2}^A} = 0  \ .  \label{scan5}
\end{equation}
There are two type of averages : one over the flux quanta $c_\alpha$ and the other over the (continuous) couplings $d_\alpha$. Being independent variables, one can use formulae of the type
\begin{equation}
\overline{f_1 (d_\alpha) f_2 (c_\alpha)} = \overline{f_1 (d_\alpha)} \times \overline{ f_2 (c_\alpha)}  = \frac{1}{2d_0} \int_{-d_0}^{d_0} dx f_1 (x)  \ \times \
\frac{1}{2M+1} \sum_{c_\alpha=-M}^M f_2 (c_\alpha)  \ . \label{scan6}
\end{equation}
By using such formulae, one finds
\begin{eqnarray}
&& (\Delta m_0^2)^2 = (\Delta m_{3/2}^2)^2 + {\tilde m}^4 \sum_\alpha \overline{d_\alpha^2 c_\alpha^4} \simeq \frac{N_{HS}M^4}{15}\tilde{m}^4\left(\frac{4}{27} + d_0^2 \right) \ , \nonumber \\
&& {\rm where} \quad (\Delta m_{3/2}^2)^2 = \frac{{\tilde m}^4}{9} \sum_\alpha [\overline{c_\alpha^4} - (\overline{c_\alpha^2})^2 ]  \ . \label{scan7}
\end{eqnarray}
Consequently, one finds
\begin{equation}
(\delta_{ij})_{LL/RR} \equiv
\frac{\delta m_0^2}{\overline{m_0^2}} \simeq \frac{1}{\sqrt{N_{HS}}} \sqrt{ \frac{1}{5}\left(4 + 27d_0^2\right)} \ . \label{scan8}
\end{equation}
The off-diagonal entries, which have zero average values, are governed by the standard deviation $\delta m_0^2$. One concludes then that they are suppressed compared to the
diagonal entries. For a large number of hidden sector $N_{HS} \geq 10^{6} $, the flavor problem of MSSM is therefore solved. While this discussion is considering the flavor violating entries at the supergravity scale, in practice at the
weak scale, as we will see in the next section, $N_{HS}\sim 100$ would be sufficient to absolve  strong constraints from  $\Delta m_K$. For the constraint from $\mu \to e +\gamma$, however, $N_{HS}\sim 100$ is not sufficient 
and a larger value of $N_{HS}$ should be chosen.  

It should be noted however that the above discussion is pertaining to definition of $\delta_{ij}$ at the high scale. 
At the weak scale, for the leptonic sector (in the absence of right handed neutrinos), there is no significant 
change in the mean values, where as for the hadronic (squark) sector, due to the large gluino contributions
to the squark masses in RG running, the $\delta_{ij}^{q,u,d}$ would be further suppressed by a factor  from $7$ up to 
an order of magnitude. 

For the gaugino masses, one finds
\begin{equation}
\Delta M_{1/2}^2 = {\tilde m}^2 \sum_\alpha \overline{s_\alpha^2 c_\alpha^2} \simeq N_{HS} \frac{d_0^2 M^2}{9} {\tilde m}^2  \ . \label{scan9}
\end{equation}
Therefore one finds the standard deviation
\begin{equation}
\Delta M_{1/2} = d_0 \sqrt{ \overline{m_{3/2}^2}}  \ . \label{scan10}
\end{equation}
For A-terms, let us consider for definiteness
\begin{equation}
A^u = m_{3/2} y_D^u + {\tilde m} \sum_{\alpha}a_\alpha^u c_{\alpha} \ , \label{scan11}
\end{equation}
where in the mass basis for fermions and  scalar-fermion-gaugino couplings are diagonal, $y_D^u $ is diagonal in the flavor space. If $a_\alpha^u \sim y^u \tilde{a}_\alpha $, then one expects the flavor violation 
in this case to be under control. However, if this is not the case, we can use the same arguments as above. 
One then finds the standard deviation
\begin{eqnarray}
&& (\Delta A^u)^2 = \overline{m_{3/2}^2} (y_D^u)^2 + {\tilde m}^2 \sum_\alpha  \overline{(a_\alpha^u)^2 c_{\alpha}^2} - \left(\overline{m_{3/2}} y_D^u\right)^2 \nonumber \\
&& \simeq \frac{N_{HS} M^2}{9} {\tilde m}^2 d_0^2\left(1 + (y_D^u)^2 \right)  \ . \label{scan12}
\end{eqnarray}
The A-terms are such that additional flavor violation (other than from Yukawa couplings) would be from the variance
of the distribution. Thus we have 
\begin{equation}
(\delta_{ij}^u)_{LR/RL} \equiv
{\Delta A^u v_u \over \overline{m_0^2}} \sim   {3 d_0 v_u \over \sqrt{N_{HS}} \tilde{m} M} \ .  \label{scan14}
\end{equation}
From the above it is clear that, similarly  to the case of scalar masses, there is a suppression  $1/\sqrt{N_{HS}}$ coming from the large number of hidden-sector fields.

Notice that our starting expressions for soft terms (\ref{soft2}) and the scanning we performed above is different compared to one based on a naive spurion-type parametrization of soft terms (which we used in chapter \ref{Chapter 2} and \ref{Chapter 3}):
		
		\begin{equation}
		\label{sugraoperators1}
		\sum_{\alpha=1}^{N_{HS}}\frac{s_{\alpha }^a}{M_{Pl}}\int d^2\theta T_{\alpha} W^{a} W^{a}
		\quad  \to \quad \quad
		M^{A}_{1/2} =  \frac{1}{M_{Pl}}\sum_{\alpha=1}^{N}s_{\alpha }^A F_{T_{\alpha}} \ , 
		\end{equation}
		\begin{equation}
		\label{sugraoperators2}
		\sum_{\alpha=1}^{N_{HS}}\frac{d_{\alpha, i j}}{M_{Pl}^2}\int d^4\theta T^{\dagger}_{\alpha}T_{\alpha}Q^{\dagger}_iQ_j 
		\quad \to \quad \quad
		m_{\tilde{f}_{ij}}^2 = \frac{1}{M_{Pl}^2}\sum_{\alpha=1}^{N_{HS}}d_{\alpha, ij}F_{T_{\alpha}}^{\dagger}F_{T_{\alpha}} \ , 
		\end{equation}
		\begin{equation}
		\label{sugraoperators3}
		\sum_{\alpha=1}^{N_{HS}}\frac{a_{\alpha, i jk}}{M_{Pl}}\int d^2\theta T_{\alpha}Q_iQ_jQ_k
		\quad \to \quad \quad	
		A_{ijk}=\frac{1}{M_{Pl}}\sum_{\alpha=1}^{N_{HS}}a_{\alpha, ijk}F_{T_{\alpha}} \ , 
		\end{equation}
		\begin{equation}
		\label{sugraoperators4}
		\sum_{\alpha=1}^{N_{HS}}\frac{b_{\alpha }}{M_{Pl}^2}\int d^4\theta T_{\alpha}T^{\dagger}_{\alpha}H_u H_d	
		\quad \to \quad \quad
		B_{H_uH_d}=\frac{1}{M_{Pl}^2}\sum_{\alpha=1}^{N_{HS}}b_{\alpha}F_{T_{\alpha}}F_{T_{\alpha}}^{\dagger} \ , 
		\end{equation}
		\begin{equation}
		\label{sugraoperators5}
		\sum_{\alpha=1}^{N_{HS}}\frac{q_{\alpha }}{M_{Pl}}\int d^4\theta T^{\dagger}_{\alpha}H_u H_d
		\quad \to \quad \quad
		\mu=\frac{1}{M_{Pl}}\sum_{\alpha=1}^{N_{HS}}q_{\alpha}F_{T_{\alpha}}^{\dagger} \ . 	
		\end{equation}
		
		The difference is that in the SUGRA expressions  (\ref{soft2}) the flavor-blind contributions proportional to $m_{3/2}^2$ scan coherently (add up) in soft terms, whereas the other contributions, which
		are similar to the global SUSY expressions (\ref{sugraoperators5}), being multiplied by random couplings scanned around zero, average to zero. A similar scan we performed above, but
		starting from  (\ref{sugraoperators5}) would not lead a suppression of FCNC effects, unlike our scan above. 
						 
	
\section{Numerical Analysis}
Using eqs.(\ref{scan2}) as boundary conditions at the high scale, we perform a numerical analysis of the resulting soft spectrum at the weak scale and studied the phenomenology. 
For the numerical analysis, we have considered $N_{HS}$ to be 100, with $c_{\alpha}$ varying discretely and randomly from  -100 to 100.  We explored taking $\tilde{m}$ to be 20 GeV.  The maximum value of
$m_{3/2}$ is roughly about 6 TeV.  The parameters $d_\alpha,s_\alpha, a_{\alpha},g_\alpha$ are varied between \{-1/4,1/4\}. A larger
value for the $d_0$ parameters would lead to significant number of the points ruled out due to tachyonic masses
at the weak scale.  We believe that larger values  $d_0 \sim {\cal O}(1)$ would not significantly alter the
results presented here. Finally we set $\tan\beta =10$. We show that these values of $N_{HS}$ are enough to demonstrate the $1/\sqrt{N_{HS}}$ suppression on the flavor violating off-diagonal entries. 
We use Suseflav \cite{Chowdhury:2011zr} for computation of the spectrum and computing the flavor
observables.  

\begin{figure}[H]
	\centering
	\subfigure[]{%
		\label{fig:firstb}%
		\includegraphics[width=7.5cm,height=6cm]{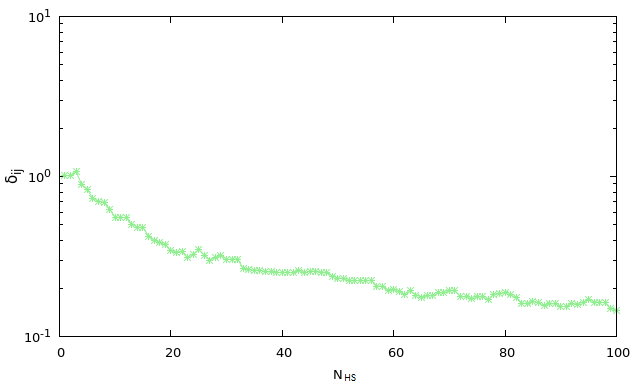}}%
	\hfill
	\subfigure[]{%
		\label{fig:secondb}%
		\includegraphics[width=7.5cm,height=6cm]{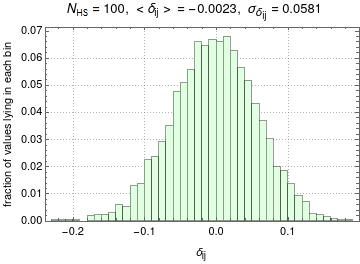}}%
	
	\caption{ A scatter plot showing the variation of maximum value of $\delta_{ij}$  of the type $LL/RR$ 
	with respect to number  of hidden sectors,  $N_{HS}$ at the high scale (left side). A histogram of the $\delta_{ij}$ is presented.  As expected the mean is very close to zero and the variance is as computed in the text.  }
	\label{deltadist.}
\end{figure}

The variation of the off-diagonal flavor entries in the sfermion mass matrices is presented in  Figs.(\ref{deltadist.}(a))
and (\ref{deltadist.}(b)).  In Fig.(\ref{deltadist.}(a)) we present the scatter plot of a typical 
$\delta_{ij}$ as defined in eq.(\ref{scan8}).  From the plot it is clear that the $\delta_{ij}$ does fall off as
 $1/\sqrt{N_{HS}}$. The second figure show the same data in terms of a histogram, where as we can see 
 the mean value is close to zero and the variance is as expected from the formulae in eqs.(\ref{scan7},\ref{scan8}). 

The high scale distributions are then evolved to the weak scale where the full soft supersymmetric 
spectrum is computed. Radiative electroweak symmetry breaking conditions are imposed.  Experimental
constraints from LHC and the Higgs mass are also taken in  consideration. As is standard practice
we consider one $\delta_{ij}$ at a time. In the present letter, we consider the two of the strongest 
constraints, \textit{i.e.} the mass difference between the neutral $K$-mesons, $\Delta M_K$ and 
the leptonic rare decay $\mu \to e+\gamma$. A more detailed analysis with rest of the flavor processes
will be presented elsewhere \cite{edplskv}. 

\begin{figure}[H]
	\centering
	\subfigure[]{%
		\label{deltamkLL}%
		\includegraphics[width=7.5cm,height=6cm]{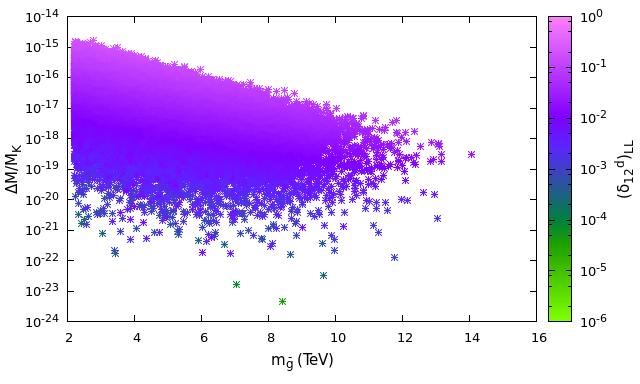}}%
	\hfill
	\subfigure[]{%
		\label{deltamkLR}%
		\includegraphics[width=7.5cm,height=6cm]{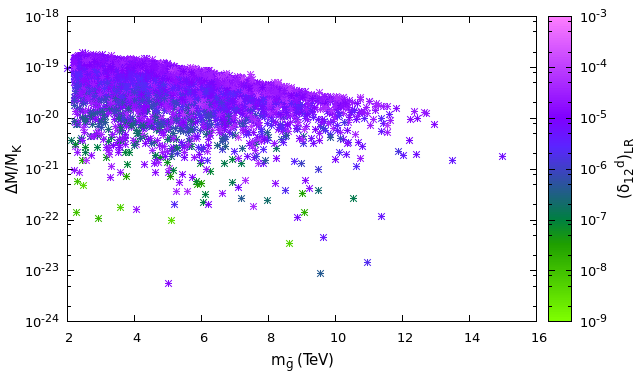}}%
	
	\caption{Regions of the parameter space which satisfy the bounds  from  LHC, Higgs mass and
	other phenomenological bounds. We have chosen $N_{HS}$ to be 100 and $\tilde{m}$ to be 20 GeV. 
	The distribution of $\delta$ values is as per the eqs. (\ref{scan8}) and (\ref{scan14}).
	  The left side plot is for $LL$ type mass insertion whereas the right hand side is for $LR$ type mass insertion.  All points satisfy the experimental constraint from $\Delta M_K$.}
	  \label{fig:dlta}
\end{figure}

At the weak scale, the diagonal entries would be enhanced due to renormalization group equation
running, while the inter-generational entries of the squark matrices  
 would only receive corrections suppressed by the product of  Yukawa couplings and CKM angles
 \cite{Ciuchini:2007ha}. Due to this the $\delta_{ij}$ would be further suppressed roughly by an additional factor 
 which is proportional to the gluino mass corrections and  roughly independent of the number of 
 hidden sector fields. In figs. (\ref{fig:dlta}(a),\ref{fig:dlta}(b)) we present the regions of the parameter space allowed by $\Delta M_k$ constraint as a function of the gluino mass. It should be noted here that  we have
 taken the weak scale values of the mass insertions of eq.(\ref{scan8}), 
  where all the parameters appearing on the RHS are computed
 at the weak scale. The  left figure is for the $LL$ mass insertion where as the right figure is for the $LR$ mass
 insertion.  As can be seen from the figure, all the points lie below the experimentally measured value of
 $\Delta M_K$ \cite{Tanabashi:2018oca}.  The spectrum at the weak scale for the first two generations
 is about 5-6 TeV and the gluino mass is shown in the figure after taking into consideration the limits from LHC. 
 For this spectrum and a diluted $\delta \lesssim 10^{-1}$ the constraint from $\Delta M_K$ is satisfied. 
 
 \begin{figure}[H]
	\centering
	\subfigure[]{%
		\label{muegLL}%
		\includegraphics[width=7.5cm,height=6cm]{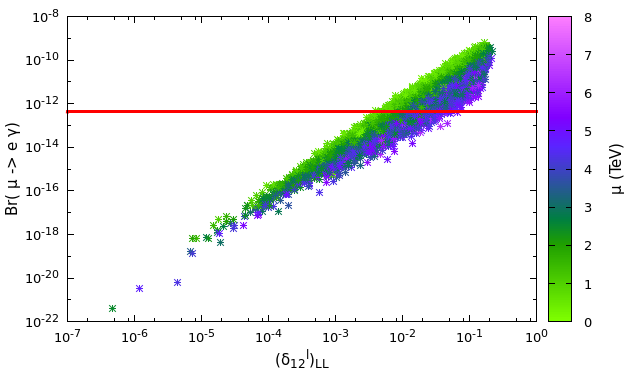}}%
			\hfill
	\subfigure[]{%
		\label{muegLR}%
		\includegraphics[width=7.5cm,height=6cm]{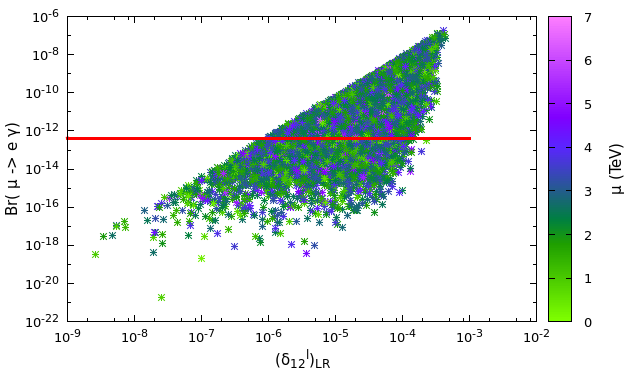}}%
		\caption{ Regions of the parameter space constrained by the leptonic rare decay $\mu \to e + \gamma$
		for $(\delta_{12})_{LL}$ (on the left) and for $(\delta_{12})_{LR}$ (on the right). The horizontal (red) line
		is the present experimental limit from MEG experiment. As can be seen, a large part of the parameter space 
		survives the experimental limit for $N_{HS}=100$.}
		\label{muggamma}
		\end{figure}
The leptonic rare process  $\mu \to e+\gamma$ is however more strongly constraining for the same set
of parameters, \textit{i.e,} $N_{HS} =100$ and $\tilde{m} = 20$ GeV. In Figs. (\ref{muggamma}(a) and \ref{muggamma}(b)
we present results of the scanning as a function of the $\delta$ parameter and $\mu$. The left figure 
is for a $LL$ type mass insertion whereas the right figure is for $RR$ type mass insertion.  As one can see
from the figures, a significantly large region of the parameter space is still compatible with the latest 
result from the MEG experiment\cite{TheMEG:2016wtm}, but the constraints from  $LR$ are significantly 
stronger, as expected.  A larger $N_{HS}$ value $\gtrsim 10^5$ would lead to  complete dilution of the $\delta$.

\begin{figure}[H]
	\centering
	\subfigure[]{%
		\label{darkmix}%
		\includegraphics[width=7.5cm,height=6cm]{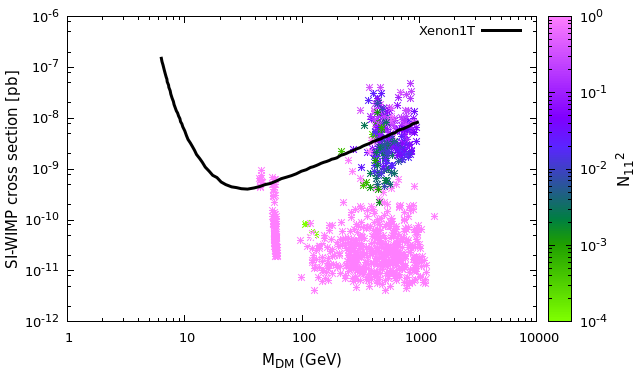}}%
	\hfill
	\subfigure[]{%
		\label{darkcoann}%
		\includegraphics[width=7.5cm,height=6cm]{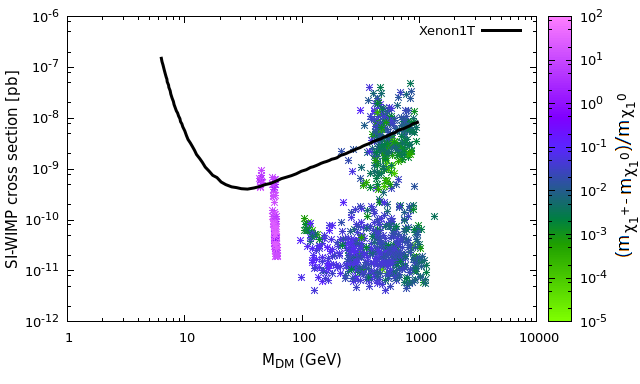}}%
	\caption{Regions of the parameter space which satisfy the relic density and the direct detection results from 
	 Xenon1T. On the left we show the spin independent cross-section with respect to the lightest 
	 neutralino mass and the bino component of the lightest neutralino. On the right, we show the same, with the
	 mass difference between chargino and bino. }
	 \label{fig:darkcoann}
\end{figure}

Finally we have also looked for regions with neutralino dark matter which could lead to correct relic density 
while satisfying the constraints from direct detection and flavor.  As can be seen from fig (\ref{fig:darkcoann}(a)), 
there  are two branches which satisfy relic density as well as the direct detection result.  The first branch has
 dark matter masses $\lesssim$ 100 GeV and the lightest neutralino is a pure bino. In the second branch 
the neutralino has a region in which it is a pure bino and another region where there is a significant admixture from
wino and  higgsino.  The regions where the neutralino are pure bino have significant co-annihilations with the
chargino as can be seen from the fig (\ref{fig:darkcoann}(b)).  These regions arise due to the non-universality in the gaugino
masses at the high scale due to the $s_\alpha$ parameters.  On the other hand, regions with bino-higgsino mixing
arise due to cancellations in  $M_{1/2}$ in contributions from various fluxes of different spurion fields. 
 As the charges/fluxes ${c}_{\alpha}$ 
take both signs, for significantly large $N_{HS}$ there is an enhanced probability of 
cancellations between the charges leading to small $M_{1/2}$ at the high scale. We numerically found that 
this probability is significantly high for $N_{HS}\gtrsim 30$.  Due to the universal 
nature of the gravitino mass, such cancellations do not occur in the 
soft scalar mass terms.  A low value for $\mu$ is very probable in these regions leading to significant 
bino-higgsino mixing. Together they lead to regions with physically viable regions of neutralino dark matter. 
More details of these regions will be presented elsewhere \cite{edplskv}. 

\section{Conclusion}

We present a novel solution to the supersymmetric flavor problem in the presence of large number of 
hidden sector (spurion) fields. Such a scenario naturally arises in the string landscape. The result does not depend on
the explicit details of the string construction, but crucially on the form of the soft terms in the supergravity
potential in the presence of a large number of hidden sector fields, eqs.(\ref{scan2}). They naturally lead to 
a suppression of the flavor violating entries as $1/\sqrt{N_{HS}}$. At the weak scale, there is further suppression
due to the renormalization group running, especially for the hadronic mass insertions. 
We have shown that numerically $N_{HS}=100$ is sufficient to remove the constraints from $\Delta M_K$, whereas  
a much larger $N_{HS}$ would be required to eliminate completely the constraints on the leptonic sector from
 $\mu \to e + \gamma$. Conversely, a discovery of such leptonic processes in forthcoming experiments could be 
 a smoking gun of such a scenario. 
 The four fluxes contribution to the soft terms presented here provides an interesting framework to further
 study the implications for low energy phenomenology. 

\begin{subappendices}	
\section{Fine-tuning Arguments}
The large number of fluxes or spurion fields can also help to alleviate the so called little gauge hierarchy problem
of MSSM in the light of heavy supersymmetric spectrum. We assume N-sequestered hidden sectors for the discussion in the
following. 
 
The evolution of $m_{H_u}^2$ is parametrized as following in our case:
\be
m_{H_u}^2|_{rad}=\frac{3y_t^2}{8\pi^2}(\eta_{m_0}(Q)\tilde{m}^2\sum_{\alpha=1}^{N_{HS}}c_{\alpha}^2+\eta_{m_{1/2}}(Q)(\tilde{m}\sum_{\alpha=1}^{N_{HS}}c_{\alpha})^2)\log\left(\frac{M_{GUT}}{m_{SUSY}}\right)
\ee
 In case of multiple hidden sectors:
\be
m_{H_u}^2|_{rad}=\sum_{\alpha_i}^{}(m_{H_u}^2|_{rad})_{\alpha_i}+\sum_{\alpha_i\ne \alpha_j}^{} (m_{H_u}^2|_{rad})_{\alpha_i\alpha_j}
\ee
where
\be
(m_{H_u}^2|_{rad})_{\alpha_i}=\frac{3y_t^2}{8\pi^2}(\eta_{m_0}(Q)\tilde{m}^2+\eta_{m_{1/2}}(Q)\tilde{m}^2)c_{\alpha_i}^2\log\left(\frac{M_{GUT}}{m_{SUSY}}\right)
\ee
\be
(m_{H_u}^2|_{rad})_{\alpha_i\alpha_j}=\frac{3y_t^2}{8\pi^2}\eta_{m_{1/2}}(Q)\tilde{m}^2(c_{\alpha_i}c_{\alpha_j})\log\left(\frac{M_{GUT}}{m_{SUSY}}\right)
\ee
The fine-tuning is defined with respect to each sector and with interaction terms. There are (N(N+1)/2) parameters of fine-tuning:
\be
\Delta^{-1}_{\alpha_i}=\frac{M_{Higgs}^2}{2(m_{H_u}^2\vert_{rad})_{\alpha_i}}
\ee
\be
\Delta^{-1}_{\alpha_i\alpha_j}=\frac{M_{Higgs}^2}{2(m_{H_u}^2\vert_{rad})_{\alpha_i\alpha_j}}
\ee
\be
\Delta^{-1}=Max[(m_{H_u}^2\vert_{rad})_{\alpha_i},(m_{H_u}^2\vert_{rad})_{\alpha_i\alpha_j}]
\ee

According to NK definition, each term of radiative contribution coming from the soft SUSY breaking should be small. In our case each sector independently will contribute in radiative correction and because of RG evolution there will be an interference term that will contribute in radiative correction. Mean of interference terms will be zero. Only independent terms will contribute in mean of NK fine-tuning. In this case mean of fine-tuning from each sector is suppressed by $1/N_{HS}$ factor as compared to single hidden sector scenario.
\be
\Delta^{-1}_{\alpha_i}=\frac{M_{Higgs}^2}{2(m_{H_u}^2\vert_{rad})_{\alpha_i}}
\ee
    BG fine-tuning: We know the definition of $\Delta_{BG}$ measures is the fractional change in $m_Z^2$
due to fractional variation in high scale parameters. In multiple hidden sectors scenario, flux coming from each sector is high scale parameters ($c_{\alpha}$), which are completely independent. The $\Gamma_{\alpha}$ sensitivity coefficients in this scenario are:
    \be
    \Gamma_{\alpha}\equiv \lvert\frac{-2( m_{Hu}^2(c_{\alpha}+1)-m_{Hu}^2(c_{\alpha}))}{m_Z^2}\lvert
    \ee
where BG fine-tuning is maxima of these parameter.
\[\Delta_{BG}= \max(\Gamma_{\alpha})\]    
Renormalization group (RG) equations do relate  high to low energy boundary condition.
After RG evolution, following is simple expression of $m_{H_u}^2(SUSY)$ and $m_{H_d}^2(SUSY)$ hold in terms of the fundamental parameters:
    \be
    m_{H_u}^2=X(Q)m_0^2 + Y(Q)M_{1/2}^2 + Z(Q) A_0^2 +W(Q)A_0M_{1/2}
    \ee	
where X(Q), Y(Q), Z(Q) and W(Q) are RG coefficients. Using mean of soft parameters given above, mean of $\Gamma_{\alpha}$ is follow:     
\[\overline{\Gamma_{\alpha}}\sim \frac{N_{HS}\tilde{m}^2}{3m_Z^2}O(1) \]
For $N_{HS}=100$ and $\tilde{m}=20$ GeV, $\overline{\Gamma}_{\alpha}$ is $\mathcal{O}(1-10)$.	 	 
  \end{subappendices}  
\chapter{A Fresh Look at Proton Decay in SUSY SU(5)}
\label{Chapter 5}

\lhead{Chapter 5. \emph{A Fresh Look at Proton Decay in SUSY SU(5)}}
\section{Introduction}
\begin{figure}[h!]
    \centering
    \includegraphics[width=12cm,height=4cm]{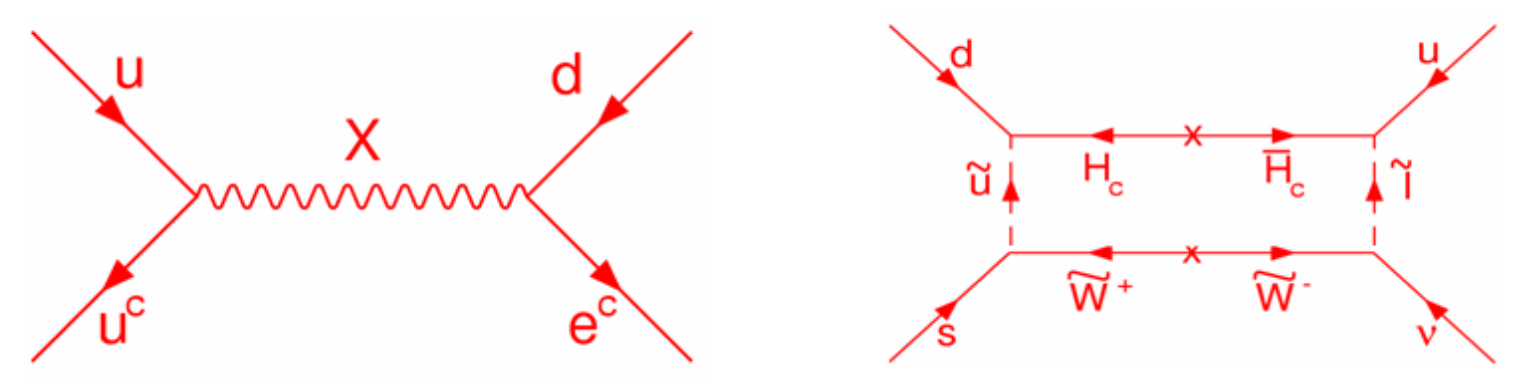}
    \caption{\textbf{left}: Diagram for $p\rightarrow e^+ \pi^0$ mediated by X gauge boson. \textbf{right}: Diagram for $p\rightarrow K^+ \bar{\nu}$ mediated by colored Higgsino with chargino dressing.}
    \label{xboson}
\end{figure}

\begin{figure}[h!]
    \centering
    \includegraphics[width=14cm,height=8cm]{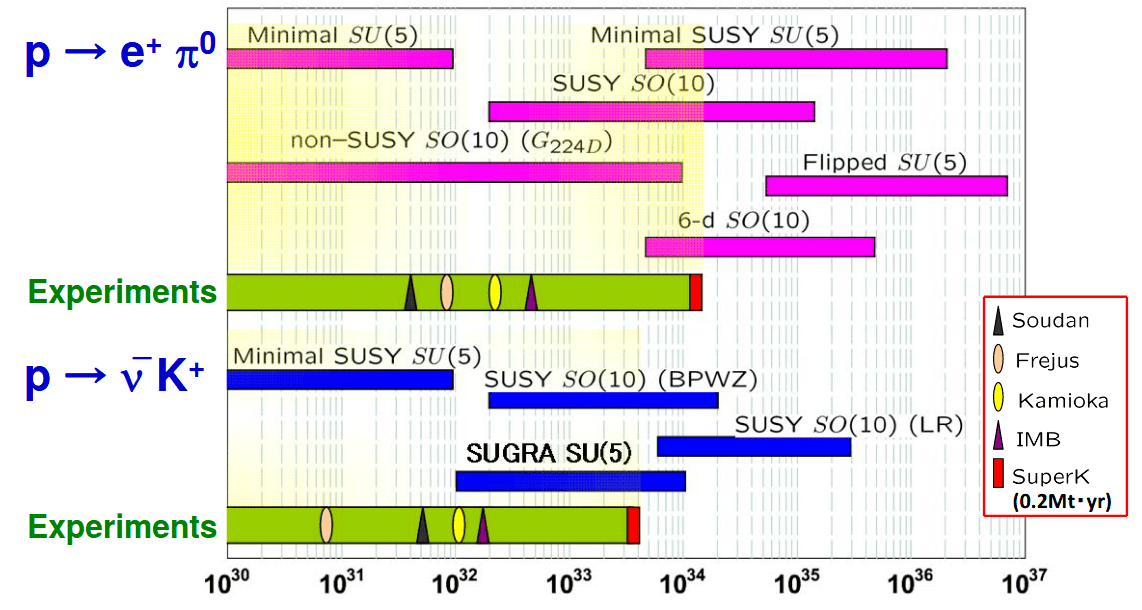}
    \caption{Summary of current status of proton decay for two major decay modes $p\rightarrow e^+ \pi^0$ and $p\rightarrow K^+ \bar{\nu}$.}
    \label{statusPD}
\end{figure}

After the success of electroweak unification, we are motivated to study a unified theory of electroweak and strong interactions, called Grand Unified Theory (GUT). SU(5) is the minimal large (simple) gauge group that can contain all the SM gauge groups as subgroups. Baryon number violation is the most impressive prediction of GUT models and it leads to proton decay. The GUT scale energy is not attainable by accelerators, but we can indirectly probe such a high scale through proton decay measurements. The minimal SU(5) GUT predicts $\tau(p\rightarrow e^+ \pi^0)\sim 10^{31}$ years as the dominant decay mode of proton decay, now completely ruled out from experimental limit, see table \ref{tab:protondecaylimits}. It also predicts $\sin^2{\theta_w}$ that is different from the value extracted from the precision measurements \cite{Langacker:1991an,Alexander:1991vi} at LEP and SLC. Thus, experimentally, minimal GUT SU(5) models have been ruled out completely. 
Here, we will study the minimal SU(5) GUT model with supersymmetry\cite{Sakai:1981pk,Hayato:1998zyn}. One of the phenomenological consequences of the supersymmetric SU(5) is the prediction of  $\sin^2{\theta_w}$, which is supported by the precision measurements.  In SUSY-SU(5), proton decay arises from D=4,5,6 baryon number violating operators
 \begin{itemize}
     \item \textbf{D=4}: In supersymmerty, there are gauge invariant dimension-4 terms in superpotential
     \begin{equation}
         W= \lambda^{ijk}L_iL_je^c_{Rk}+\lambda^{'ijk}L_iQ_jd_{Rk}^c
 + \lambda^{''ijk}u^c_{Ri}d^{c}_{Rj}d^{c}_{Rk}+\epsilon^iL_iH_2
     \end{equation}
     which violate lepton and/or baryon number and lead to very fast proton decay. The proton decay bound on $p\rightarrow e^+ \pi^0$ demands for these coupling parameters to be really small for TeV scale supersymmetry. It is more natural then to consider it to be zero which is made possible by introducing a new discrete symmetry, called R-parity, that forbids these terms.
     \item \textbf{D=5}: Dimension 5 baryon number violating operators are new in SUSY-GUT compared to the non-SUSY GUT. They are generated by the exchange of the colored Higgs multiplet, see fig. (\ref{xboson}). This is discussed below in detail.
     \item \textbf{D=6}: In a typical grand unified theory based on simple groups like SU(5) or SO(10) there are additional gauge bosons which are typically called X and Y bosons in SU(5), which carry both color and weak indices. Thus they can lead to transitions between quarks and leptons at the same vertex leading to baryon number violation.  These typically lead to six dimensional baryon number violating operators in effective field theory after integrating out the GUT gauge bosons. They are present in both SUSY-GUT and non-SUSY GUT models. I discuss these below in detail.
 \end{itemize} 
 The main decay mode of proton decay $p\rightarrow K^+ \bar{\nu}$ in SUSY-SU(5) GUT has been searched for and continues to be searched for in various underground neutrino experiments. In fact, for supersymmetry particles of about a few TeV, the proton decay bounds coming from Kamioka and IMB experiments rule out the minimal SUSY-SU(5) model, see fig. (\ref{statusPD}).  However, these models can be rescued if one concentrates on regions where there are  peculiar cancellations taking place in the proton decay matrix elements or there are  high threshold corrections which modify the GUT scale. There are a lot of detailed analyses on the proton decay in the minimal SUSY-GUT models\cite{Kim:1984df,Dimopoulos:1981dw,Chamseddine:2000nk,Goto:1998qg,Hisano:1994hb}. \\
 Now one of the famous statements during that time was that "Not even decoupling can save SUSY SU(5)". This was by Murayama and Pierce who pointed out that even if you decouple the first two generations of squarks the stability of the proton is not going to increase\cite{Murayama:2001ur}. Further Murayama and Arkani-Hamed have shown that  in the presence of decoupling, some of the masses could become tachyonic (especially the third generation) due to renormalization group running\cite{ArkaniHamed:1997ab}. Thus decoupling does not fit in naturally within the framework of Grand Unified theories.  
The proton decay computations have been refined over the years with both LLLL and RRRR operators being taken into consideration by  Goto et.al \cite{Goto:1998qg}  and also by Shirai \cite{Nagata:2013sba}. A review can be found by Pran Nath \textit{et. al}\cite{Nath:2012nh}. A recent computation on heavy supersymmetry in SUSY SU(5) can be found by Lavignac and Bajc\cite{Bajc:2015ita}. 
 All this research is twenty years old now and one might question the relevance of revisiting it in the present times. Well, in the mean time, there have several developments in the field: 1) Higgs has been discovered with a mass of 125 GeV, 2) LHC has not found supersymmetric particles and in fact  all the evidence including flavor and dark matter seems to be pointing towards a decoupling scenario 3) Super-K limits on proton decay have been pretty strong 4) The constraints on WIMP dark matter have become particularly strong. All these aspects have been  discussed in introduction in detail. Theyy seem to  point towards  supersymmetry parameter space which seems to be either partially or fully split (Dimopoulos \textit{et. al}) or a generational split in the supersymmetric parameter space\cite{Randall:2012dm}.
The other reason to look back into these results is more phenomenological. There are at least three new experiments which will be looking for proton decay in this decade. These experiments are sensitive to at least two important modes of proton decay leading to an improvement of a factor of 5 to an order of magnitude in the limits.
\section{Experimental Status and  Prospects}
Proton decay is one of the most important signatures of baryon number violation\cite{Abbott:1980zj,Ellis:1979hy,Weinberg:1979sa,Wilczek:1979hc,Machacek:1979tx,Langacker:1980js,Pati:1973uk,Georgi:1974sy}. The discovery of proton decay would indicate physics beyond Standard Model and perhaps towards a
Grand Unified Theory (GUT). The present limits on different decay modes of proton decay coming from different experiments are summarised in table (\ref{tab:PDbounds}).
\begin{table}[h]
\centering
\begin{tabular}{|c|c|}
\hline\hline
\textbf{p Decay Modes}   & \textbf{Partial mean life($10^{30}$ years)}  \\
\hline
$p\rightarrow \pi^0 e^+$ & $>$16000 \cite{Miura:2016krn}
\\
$p\rightarrow \pi^0 \mu^+$ & $>$7700 \cite{Miura:2016krn}
\\
\hline
$p\rightarrow K^0 e^+$ & $>$1000\\
$p\rightarrow K^0 \mu^+$ & $>$1600\\
\hline  
$p\rightarrow \eta^0 e^+$ & $>$10000\\
$p\rightarrow \eta^0 \mu^+$ & $>$4700\\
\hline
$p\rightarrow \pi^+ \bar{\nu}$ & $>$390 \cite{Abe:2013lua}\\
$p\rightarrow K^+ \bar{\nu}$ & $>$ 5900 \cite{Abe:2014mwa}
\\
\hline         
\end{tabular}
\caption{Partial mean life of p decay modes\cite{Tanabashi:2018oca} constrained by the experiments Super-K,IBM, etc.}
\label{tab:PDbounds}
\end{table}

\begin{figure}[h!]%
    	\centering
    	\subfigure[]{%
    		\label{fig:first1}%
    		\includegraphics[width=7cm,height=5cm]{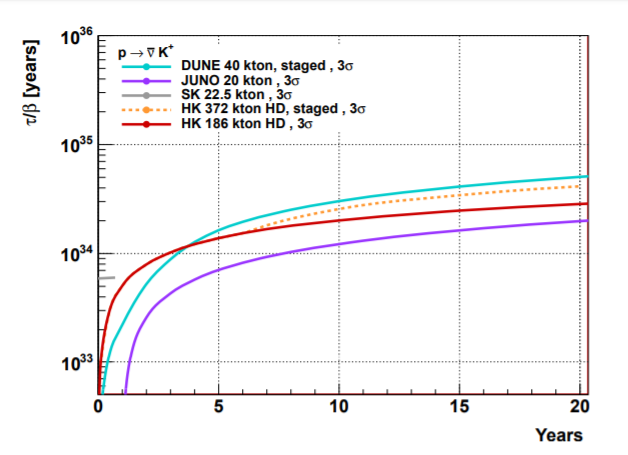}}%
    	\hfill
    	\subfigure[]{%
    		\label{fig:second1}%
    		\includegraphics[width=7cm,height=5cm]{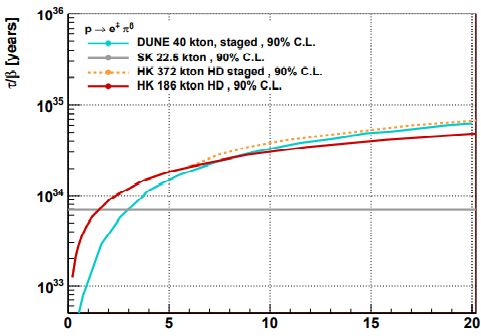}}%

    	  	\caption{The comparison of different experiments for the $p\rightarrow K^+ \bar \nu$ channel's discovery potential and 90$\%$ C.L. region as a function of the year\cite{Abe:2018uyc}.}
\label{hyperk}
    	 \end{figure}

\begin{figure}[h!]%
    	\centering    	 
    	\subfigure[]{%
    		\includegraphics[width=7cm,height=5cm]{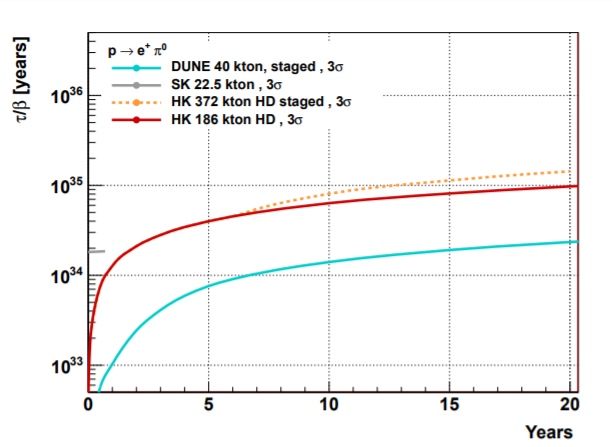}}%
    	\hfill
    	\subfigure[]{%
    		\includegraphics[width=7cm,height=5cm]{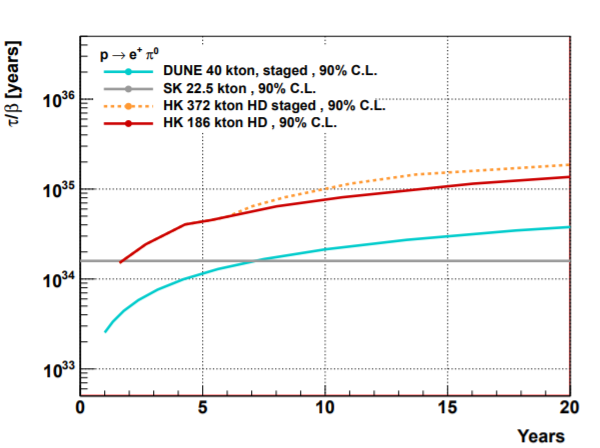}}%

    	  	\caption{The comparison of different experiments for the $p\rightarrow \pi^0 e^+$ channel's discovery potential and 90$\%$ C.L. region as a function of the year\cite{Abe:2018uyc}.}
\label{dunebound}
    	 \end{figure}

In this chapter we will study the two dominant decay modes of proton decay $p \to K^+ \bar \nu$ and $p\rightarrow e^+ \pi^0$ in full detail in split generation spectrum with and without flavor violation.
The present and upcoming limits on two main
decay modes are shown in Table. (\ref{tab:protondecaylimits}). 
\begin{table}[h]
     \centering
    \resizebox{\textwidth}{!}{\begin{tabular}{|c|ccc|}
     \hline\hline
       \textbf{p Decay Modes}   & &\textbf{Partial mean life  ($10^{33}$ years)}    & \\
       &\textbf{ Current (90$\%$ CL)} &\textbf{Future (3$\sigma$ discovery)} &\textbf{Future (90$\%$ CL)}\\
       \hline
         $p\rightarrow \pi^0 e^+$ & 16 \cite{Miura:2016krn} & DUNE: 15 (25) \cite{Abe:2018uyc} & DUNE: 20 (40) \cite{Abe:2018uyc}\\
          & & Hyper-K: 63 (100) \cite{Abe:2018uyc} & Hyper-K: 78 (130) \cite{Abe:2018uyc}\\
$p\rightarrow \pi^0 \mu^+$ & 7.7 \cite{Miura:2016krn} & Hyper-K: 69 \cite{Abe:2018uyc}& Hyper-K: 77 \cite{Abe:2018uyc}\\
         
       \hline
       
         $p\rightarrow K^+ \bar{\nu}$ & 5.9\cite{Abe:2014mwa} & JUNO: 12 (20)\cite{Abe:2018uyc}& JUNO: 19 (40)\cite{An:2015jdp}\\
         & & DUNE: 30 (50)\cite{Abe:2018uyc} & DUNE: 33 (65)\cite{Acciarri:2015uup,Abi:2018dnh}\\
         & & Hyper-K: 20 (30)\cite{Abe:2018uyc}& Hyper-K: 32 (50)\cite{Abe:2018uyc}\\
         \hline         
     \end{tabular}}
     \caption{For
future prospects, the detector operations are assumed for 10 (20) years \cite{Ellis:2019fwf}.}
     \label{tab:protondecaylimits}
 \end{table}
From table (\ref{tab:protondecaylimits}), we can see, DUNE and Hyper-K will give the highest bounds on proton decay modes $p\to K^+ \bar{\nu}$ and $p\rightarrow \pi^0 e^+$, respectively. \\
\textbf{DUNE}: The DUNE (Deep Underground Neutrino Experiment) experiment is placed in the USA, which has a very focused neutrino beam from Fermilab to a place about 1300 km away in Lead, Sanford USA. The associated lab is called LBNF (Long Baseline Neutrino facility) hosted by Fermilab. There will also be a near detector in Fermilab close to the Neutrino beam generation. Notably, the beam is highly focused and collimated. \\
\textbf{Hyper-K}: The Hyper-K is a next-generation neutrino experiment of large-scale water Cherenkov detectors with fiducial mass 187 kt in Tochibora mine, Japan. It will be situated about 295 km away from the J-PARC proton accelerator research complex in Tokai, Japan such that neutrinos generated in the J-PARC can be measured in the detector. Hyper-Kamiokande will be able to measure the leptonic CP violation with the highest precision, proton decay, the neutrino mass ordering using atmospheric neutrinos, the three-flavor neutrino oscillation paradigm precisely and solar neutrino oscillation precisely. Using the J-PARC beam, Hyper-K can extract the oscillation parameters.\\
 DUNE and Hyper-K are complementary experiments for neutrino physics. DUNE is more sensitivity to the mass hierarchy, due to large effect of matter because of long-baseline as compared to Hyper-K, which has a shorter baseline that has less matter effect, will increase sensitivity to the measurements of the CP-violation phase.
 Hyper-K is more sensitive to proton decay channel $p\rightarrow \pi^0 e^+$ and $p \rightarrow K^+ \bar{\nu}$. Hyper-K will have the highest sensitivity to  $p\rightarrow \pi^0 e^+$ till the order of $10^{35}$ years, but it will be challenging to detect the $K^+$ from nucleon decays in a water Cherenkov detector since its momentum is below the Cherenkov light threshold; thus only the decay products of $K^+$ can be used for the identification of $K^+$, so the expected limit on this channel is one order lesser than 
 $p\rightarrow \pi^0 e^+$ channel. The complementary experiment DUNE is also sensitive to nucleon decay.  Due to its smaller mass compared to Hyper-K, it is competitive mainly in modes with distinctive final state tracks such as those involving kaons. The improvement in lifetime limits expected for 10(20) years of Hyper-K and DUNE exposure given in fig. (\ref{hyperk},\ref{dunebound}).


\section{Details of 5-dimension Operator's Wilson Coefficient Calculation in SUSY-SU(5)}
Proton decay in minimal supersymmetric SU(5) model is dominated by D=5 operators, which are generated by the exchange of colored Higgs multiplet. In this section we discuss D=5 operators  relevant for the proton decay to clarify our notation and conventions.  We consider 5-plet Higgs\cite{Hisano:1992jj}. Assuming R-parity conservation, SU(5) superpotential has following Yukawa terms: 
\be  W_{Y}^{SU(5)} =  \frac{1}{4}h^{ij}\epsilon_{abcde}\chi^{ab}_i\chi^{cd}_jH^{e}+\sqrt{2}f^{ij}\chi^{ab}_i\phi_{ja}\bar{H}_b
\label{wysu5}
\ee

where i, j = 1, 2, 3 and  a,.., e = 1 to 5 are the generation and the SU(5) indices, respectively. The Yukawa couplings $h^{ij}$ and $f^{ij}$ are the complex matrices and have redundant degrees of freedom, some of which can be eliminated by the field re-definition of $\chi$ and $\phi$. Here we use $\sqrt{2}\chi^{ab} \phi_a \bar{H}_b=\frac{1}{\sqrt{2}}\chi^{ab} (\phi_a \bar{H}_b-\bar{H}_a\phi_b)$. Relative sign between these two term depends on convention of SM Yukawa. The Yukawa coupling $h^{ij}$ is a symmetric matrix. Here $\phi$ and $\chi$ are the chiral superfields of $\bar{5}$ and 10 representations of SU(5), respectively, and contain the MSSM chiral superfields as:
\be \phi =  \begin{pmatrix}   d_1^c  \cr  d_2^c \cr d_3^c  \cr e \cr-\nu  \end{pmatrix} , \chi =\frac{1}{\sqrt{2}} \begin{pmatrix}   0 & u^c_3 & -u^c_2 & u^1 & d^1 \cr -u^c_3  & 0 & u^c_1 & u^2 & d^2 \cr u^c_2 & -u^c_1 & 0 & u^3 & d^3 \cr -u^1 & -u^2 & -u^3 & 0 & e^c \cr -d^1 & -d^2 & d^3 & -e_c & 0 \end{pmatrix}  
\label{phichi}
\ee
with $L = (\nu\quad e)$, $Q^{\alpha} = (u^{\alpha} \quad d^{\alpha})$, 
where $\alpha=1,2,3$ denotes the color index. The chiral superfields $u_{\alpha}$ and $d_{\alpha}$ contain left-handed up-type and down-type quarks, $u_{\alpha}^c$ and $d_{\alpha}^c$ are the charge conjugations of right-handed up-type and down-types quarks, e and $\nu$ are left-handed leptons and neutrinos, and $e^c$ are the charge conjugations of right handed charged-leptons. $Q^{\alpha}$ and L denote chiral superfields of weak-doublet quarks and leptons, respectively. $H$ and $\bar{H}$ are a pair of the 5 and $\bar{5}$ multiplet of SU(5) which contain doublet Higgs MSSM multiplets and color-triplet partners:
\be H=\begin{pmatrix}
H_{C}^1  \cr  H_{C}^2 \cr H_{C}^3  \cr H^+_f \cr H^0_f 
\end{pmatrix},\bar{H}=
\begin{pmatrix}
	\bar{H}_{1C}  \cr  \bar{H}_{2C} \cr \bar{H}_{3C}  \cr \bar{H}^-_f \cr -\bar{H}^0_f 
\end{pmatrix}
\label{HHbar}
\ee
where 1,2,3 indices are for color. The last two components correspond to the MSSM Higgs superfields $H_u = (H^+_f \quad H_f^0)$, $H_d = (\bar{H}_f^0 \quad \bar{H}_f^-)$.
Rewriting the superpotential after redefining the fields $\phi=W^{\dagger}\phi'$ and $\chi=U^{\dagger}\chi'$ 
\be
W_{Yu}^{SU(5)}=-\frac{1}{4} U h_{ij} U^T \epsilon_{abcde}(\chi'^{ab}_i)^T\chi'^{cd}_jH^{e}-\sqrt{2}(\chi'^{ab}_i)^TUR^{\dagger}\hat{f}\phi'_{ja}\bar{H}_b
\label{wsu5yu1}
\ee
Defining $\hat{h}=U h U^T$ and $f_{ij}=R^{\dagger} \hat{f}W$,
\be 
W_{Yu}^{SU(5)}=-\frac{1}{4}\hat{h}\delta_{ij}\epsilon_{abcde}(\chi'^{ab}_i)^T\chi'^{cd}_jH^{e}-\sqrt{2}(\chi'^{ab}_i)^TUR^{\dagger}\hat{f}\phi'_{ja}\bar{H}_b
\label{wsu5yu2}
\ee
$\hat{h},\hat{f}$ are the real diagonal matrices.
Product of two unitary matrices is also a unitary matrix, so we can write $V=UR^{\dagger}$, here $V$ is unitary matrix with 3 real rotation angles and 6 phases.

We can rewrite $V=P_2\tilde{V}P_3$, here $\tilde{V}$ has 3 real rotation angles and one phase like $V_{CKM}$, P$_3$ has 3 diagonal phases and P$_2$ is diagonal matrix containing two phases. Redefining $\phi'=P_3^*\phi''$ and rewriting $(\chi_i^{ab})^T=-\chi_i^{ab}$, we get
\be 
W_{Y}^{SU(5)}=\frac{1}{4}(\delta^{ij}\hat{h})\epsilon_{abcde}\chi'^{ab}_i\chi'^{cd}_jH^{e}+\sqrt{2}\chi'^{ab}_i(P_2\tilde{V}\hat{f})^{ij}\phi''_{ja}\bar{H}_b
\label{wsu5y3}
\ee
\be
\therefore h^{ij}=\hat{h}^i\delta^{ij}, f^{ij}=(P_{2}\tilde{V}\hat{f})^{ij}
\label{hf}
\ee
 To match with MSSM Yukawa couplings, first we write SU(5) supermutiplets in terms of MSSM chiral multiplets, as following
 \begin{align}
 W_Y^{SU(5)}=-\epsilon^{\alpha\beta\gamma}(\hat{h}\delta^{ij})Q_{i\alpha}Q_{j\beta}H_{C\gamma}+ (\hat{h}\delta^{ij})e^c_iu^c_jH_C+M_{H_C}H_C\bar{H}_C-(P_{2}\tilde{V}\hat{f})^{ij}Q_iL_j\bar{H}_C+\nonumber\\
 2(P_{2}\tilde{V}\hat{f})^{ij}\epsilon^{\alpha\beta\gamma}u^c_{i\alpha}d^c_{j\beta}\bar{H}_{C\gamma}+(\hat{h}^i\delta^{ij})u^c_iQ_jH_u+(P_{2}\tilde{V}\hat{f})^{ij}Q_id^c_jH_d+(P_{2}\tilde{V}\hat{f})^{ij}e^c_iL_jH_d
\label{wsu5y4}
\end{align}
Redefining $e^c_i=(P_2\tilde{V}_{ij})^{\dagger}e^{'c}_j$ and $d^c=P_2^*d^{'c}$:
\begin{align}
W_Y^{SU(5)}=-\epsilon^{\alpha\beta\gamma}(\hat{h}\delta^{ij})Q_{i\alpha}Q_{j\beta}H_{C\gamma}+ \hat{h}_i(P_2\tilde{V}^{ij})^{\dagger}e^{'c}_ju^c_iH_C+M_{H_C}H_C\bar{H}_C-(P_{2}\tilde{V}\hat{f})^{ij}Q_iL_j\bar{H}_C+\nonumber\\
2(\tilde{V}\hat{f})^{ij}\epsilon^{\alpha\beta\gamma}u^c_{i\alpha}d^{'c}_{j\beta}\bar{H}_{C\gamma}+(\hat{h}^i\delta^{ij})u^c_i(Q_jH_u)+(\tilde{V}\hat{f})^{ij}(Q_iH_d)d^{'c}_j+\hat{f}\delta^{ij}e^c_i(L_jH_d)
\label{wsu5y5}\end{align}
Now we match the above superpotential to that in the MSSM. 
\begin{align}  
     W_Y^{MSSM}=(Q_i.H_u)\hat{Y}_u^{ii}u^{c}_i+(Q_i.H_d)(V_{CKM}\hat{Y}_d)^{ij}d^{c}_j+(L_i.H_d)\hat{Y}_e^{ii}e^c_i  
     \label{wmssm}
\end{align}
Matching with MSSM Yukawa couplings:
\begin{equation}
\hat{h}=\hat{Y_u} , \quad \hat{f}=\hat{Y}_e , \quad (\tilde{V}\hat{f})^{ij}=(V_{CKM}\hat{Y}_d)^{ij}\quad \text{If} \quad \hat{f}=\hat{Y}_d, \quad\text{then}\quad
\tilde{V}^{ij}=V_{CKM}^{ij}\quad P_2=1   
\label{yukawamatching}
\end{equation}
After matching, SU(5) Yukawa superpotential (replacing all primed fields with unprimed fields):

\begin{align}
W_Y^{SU(5)}=-\epsilon^{\alpha\beta\gamma}(\hat{Y}_u\delta^{ij})Q_{i\alpha}Q_{j\beta}H_{C\gamma}+ \hat{Y}_{ui}(V_{CKM}^{ij})^{\dagger}e^{c}_ju^c_iH_C+M_{H_C}H_C\bar{H}_C-(V_{CKM}\hat{Y}_d)^{ij}Q_iL_j\bar{H}_C+\nonumber\\
2(V_{CKM}\hat{Y}_d)^{ij}\epsilon^{\alpha\beta\gamma}u^c_{i\alpha}d^{c}_{j\beta}\bar{H}_{C\gamma}+(\hat{Y}_u^i\delta^{ij})u^c_i(Q_jH_u)+(V_{CKM}\hat{Y}_d)^{ij}(Q_iH_d)d^{c}_j+\hat{Y}_e\delta^{ij}e^c_i(L_jH_d)
\label{wsu5y6}
\end{align} 
In mass eigenstate of SM fermion $Q_i^{'}=\begin{pmatrix} u_i^{'} \cr d_i^{'} \end{pmatrix}=\begin{pmatrix} u_i \cr (V_{CKM})_{ij}d_j \end{pmatrix}$. In mass eigenstate of SM fermion $Q_i$ is replaced by $Q_1^{'}$ in Yukawa term of SU(5) superpotential. Integrating $H_C$ and $\bar{H}_C$ using their equation of motion
\be
  \Bar{H}_C=\frac{1}{M_{H_C}}(\epsilon^{\alpha\beta}(\hat{Y}_u\delta^{ij})Q_{i\alpha}^{'}Q_{j\beta}^{'}- \hat{Y}_{ui}(V_{CKM}^{ij})^{\dagger}e^{c}_ju^c_i)
  \label{HC}
\ee
\be
H_C=\frac{1}{M_{H_C}}((V_{CKM}\hat{Y}_d)^{ij}Q_i^{'}L_j-2(V_{CKM}\hat{Y}_d)^{ij}\epsilon^{\alpha\beta}u^c_{i\alpha}d^{c}_{j\beta})
\label{HCbar}
\ee  
Putting $H_C$ and $\bar{H}_C$ value in eq. (14), we get baryon number violating terms in SU(5) superpotential:
\begin{align}
 W_{BV}^{SU(5)}=- \frac{1}{M_{H_C}}\left[C_{5L}^{ijkl}\frac{1}{2}\epsilon^{\alpha\beta\gamma}(Q_{i\alpha}^{'}Q_{j\beta}^{'})(Q_{k\gamma}^{'}L_l)+C_{5R}^{ijkl}\epsilon^{\alpha\beta\gamma}(u_{i\alpha}^cd_{j\beta}^c)(u_{k\gamma}^ce_l^c)\right]  
 \label{WBVsu5}
\end{align}
using Fierz identity $((u_{k\gamma}^{c} e_l^{c})=(e_l^{c}u_{k\gamma}^{c}))$, Wilson coefficients of RRRR and LLLL baryon number violating operators can be written as
\begin{align}
  C_{5R}^{ijkl}&=\frac{2}{M_{H_C}}(V_{CKM}\hat{Y_d})^{ij}\hat{Y}_u^k(V_{CKM}^{kl})^{\dagger}\nonumber \\
  C_{5L}^{ijkl}&=\frac{2}{M_{H_C}}(\hat{Y}_u^i\delta^{ij})(V_{CKM}\hat{Y}_d)^{kl}
\end{align}

In term of mass eigenstate of fermion $Q_{i\alpha}^{'}Q_{i\beta}^{'}=2u_{i\alpha}(V_{CKM})_{in}d_{n\beta}$.
\be
(Q_{i\alpha}^{'}Q_{j\beta}^{'})(Q_{k\gamma}^{'}L_l)=2(V_{CKM})_{jn}(u_{i\alpha}d_{n\beta}u_{k\gamma}e_l)-2(V_{CKM})_{jn}(V_{CKM})_{kp}(u_{i\alpha}d_{n\beta}d{p\gamma}\nu_l)
\ee

\section{Details of Proton Decay\label{pdecaydetails}}

 In this section, we summarize the proton decay width calculations. We use the convention and formula  of \cite{Goto:1998qg}.  Since the dimension five operators consist of two fermion (quark/lepton) and two boson (squark/slepton) component fields, effective baryon number violating four-fermion operators are generated by one loop ``dressing" diagrams which involve gauginos, higgsinos and charginos. In this calculation, we use all the dressing diagrams (exchanging neutralino, chargino and gluino) given in \cite{Goto:1998qg}, taking account of various mixing effects among SUSY particles, such as flavor mixing of squarks and sleptons, left-right mixing of squarks and sleptons. To understand the results in terms of flavor violating parameters $\delta$'s, we calculate amplitudes for dominating dressing diagram, in terms of $\delta$'s, for the decay mode $p\rightarrow K^+ \bar\nu$, with and without flavor mixing (see appendix \ref{dressing}). We take SU(5) phases to be zero in our calculation and use vertices from \cite{Goto:1998qg}. The main mode of the proton decay in GUT models are those where a proton decays into a pseudoscalar meson and an anti-lepton. The operator product expansion (OPE) leads to the decay amplitude of such processes written in terms of the Wilson coefficients which contain all the details of the high energy part of a GUT, and the low energy QCD matrix elements of the proton and pseudoscalar states with three-quark operators. The partial decay width reads \cite{Aoki:2017puj}:
\begin{equation}
    \Gamma(B_i\rightarrow M_j l_k)=\frac{m_i}{32\pi}\left(1-\frac{m_j^2}{m_i^2}\right)^2\, \left|\, \sum_I C_I^{ijk}W_{0I}^{ijk} (B_i\rightarrow M_j)\, \right|^2
    \label{decaywidth}
\end{equation}
with $m_i$ and $m_j$ being the mass of the nucleon and pseudoscalar meson, and $C_I^{ijk}$ being the Wilson coefficient of the operator of chiral type I and i,j,k are flavor indices , which also enter in $W_{0I}^{ijk}$. Contribution of GUT model parameters come in the Wilson coefficients. At GUT scale, Wilson coefficient of RRRR and LLLL baryon number violating operators can be written as: 
\begin{align}
  C_{5R}^{ijkl}&=\frac{2}{M_{H_C}}(V_{CKM}\hat{Y_d})^{ij}\hat{Y}_u^k(V_{CKM}^{kl})^{\dagger}\nonumber \\
  C_{5L}^{ijkl}&=\frac{2}{M_{H_C}}(\hat{Y}_u^i\delta^{ij})(V_{CKM}\hat{Y}_d)^{kl}
  \label{C5RC5L}
\end{align}
The knowledge of bounds on partial decay width of proton from different experiments and $W_0$ from lattice calculations \cite{Aoki:2017puj} can be used to understand GUT physics and to restrict the GUT parameters. We use ref. \cite{Aoki:2017puj} for values of hadronic matrix elements in which, they use the chiral Lagrangian method \cite{Claudson:1981gh,Chadha:1983sj} to evaluate hadronic matrix elements. 

Each matrix element is proportional to low energy constants depending on chirality; $\alpha_p$ (for RL) and $\beta$ (for LL). These are defined through the nucleon to vacuum matrix elements. Writing the quark flavor explicitly, $\alpha_p$ and $\beta_p$ \cite{Aoki:2017puj} defined by 
\begin{equation}
    \langle 0| (du)_R u_L|p\rangle=\alpha_pP_R u_p,\qquad\langle 0| (du)_Lu_L|p\rangle=\beta_pP_L u_p
    \label{hadronic}
\end{equation}
with proton spinor field $u_p$.
The value  of $\alpha_p$ and $\beta_p$  at $\mu=2$ GeV in $\overline{MS}$ naive dimensional regularization scheme is\cite{Aoki:2017puj}
\begin{equation}
    \alpha_p=-0.0144(3)(21)\quad\mbox{GeV}^3 \qquad \beta_p=0.0144(3)(21)\quad\mbox{GeV}^3
    \label{LowEP}
\end{equation}

where the first error is statistical and the second is systematic. Using these low energy constants, the relevant matrix elements $W_0$ can be computed via BChPT formula as\cite{Aoki:2017puj}

 \begin{table}[h]
     \centering
     \begin{tabular}{||c||c||}
     \hline\hline
 $\langle\pi^0|(ud)_Ru_L|p\rangle = \frac{\alpha_p}{\sqrt{2}f}(1+D+F)$ & $\langle K^0|(us)_Ru_L|p\rangle = -\frac{\alpha_p}{f}\left(1+(D-F)\frac{m_N}{m_B}\right)$\\
 \hline
 $\langle\pi^0|(ud)_Lu_L|p\rangle = \frac{\beta_p}{\sqrt{2}f}(1+D+F)$
  & $\langle K^0|(us)_Lu_L|p\rangle = \frac{\beta_p}{f}\left(1-(D-F)\frac{m_N}{m_B}\right)$\\ 
  \hline
  $\langle \eta|(ud)_Ru_L|p\rangle=-\frac{\alpha_p}{\sqrt{6}f}(1+D-3F)$  & $\langle K^+|(ud)_Rs_L|p\rangle=\frac{\alpha_p}{f}\left(1+\left(\frac{D}{3}+F\right)\frac{m_N}{m_B}\right)$ \\
  \hline
 $\langle \eta|(ud)_Lu_L|p\rangle=\frac{\beta_p}{\sqrt{6}f}(3-D+3F)$ & $\langle K^+|(ud)_Ls_L|p\rangle=\frac{\beta_p}{f}\left(1+\left(\frac{D}{3}+F\right)\frac{m_N}{m_B}\right)$\\
 \hline
 $\langle K^+|(us)_Rd_L|p\rangle=\frac{\alpha_p}{f}\frac{2D}{3}\frac{m_N}{m_B}$ &  $\langle K^+|(ds)_Ru_L|p\rangle=\frac{\alpha_p}{f}\left(1+\left(\frac{D}{3}-F\right)\frac{m_N}{m_B}\right)$ \\
 \hline
 $\langle K^+|(us)_Ld_L|p\rangle=\frac{\beta_p}{f}\frac{2D}{3}\frac{m_N}{m_B}$ &
$\langle K^+|(ds)_Lu_L|p\rangle=-\frac{\beta_p}{f}\left(1-\left(\frac{D}{3}-F\right)\frac{m_N}{m_B}\right)$\\
        \hline\hline
     \end{tabular}
     \caption{BChPT formulas of $W_0$}
     \label{tab:tabel_1}
 \end{table}
with low energy parameters $D= 0.80$, $F= 0.47$.  In this paper, we use $f= 0.131$ GeV, $m_N= 0.94$ GeV and $m_B= 1.15$ GeV. 
The matching Wilson coefficients need to be calculated at scale $\mu= 2$ GeV.
The dimension five coupling constants C$_{5L}$ and C$_{5R}$ of eq. (\ref{C5RC5L}) at the GUT scale are written in terms of the Yukawa coupling constants. C$_{5L}$ and C$_{5R}$ at $M_{SUSY}$ scale evaluated by solving the RGE's given in eq. (A26) of ref. \cite{Goto:1998qg}. At the $M_{SUSY}$ scale, the coefficients C$_{5L}$ and C$_{5R}$ for dimension-5 operators are matched to the baryon-number violating four-fermion operators. The matching condition expressions are given in \cite{Goto:1998qg} from eqs. (A28-A34). Finally, we evaluate the long-distance QCD corrections to the baryon-number violating four-fermion operators below $M_{SUSY}$ scale down to the hadronic scale $\mu=2$ GeV. They are calculated at two-loop level in \cite{Nihei:1994tx,Nagata:2013sba}. The solution of two-loop RGEs is 
\begin{equation}
    \frac{C(\mu)}{C(\mu_0)}=\left[\frac{\alpha_s(\mu)}{\alpha_s(\mu_0)}\right]^{-\frac{2}{b_1}}\left[\frac{4\pi b_1+b_2\alpha_s(\mu)}{4\pi b_1+b_2\alpha_s(\mu_0)}\right]^{\left(\frac{2}{b_1}-\frac{42+4N_f+9\Delta}{18b_2}\right)}
    \label{C2GeV}
\end{equation}
with $b_1$ and $b_2$ defined by
\begin{equation}
    b_1=-\frac{11N_c-2N_f}{3}, \qquad b_2=-\frac{34}{3}N_c^2+\frac{10}{3}N_cN_f+2C_FN_f
    \label{b1b2}
\end{equation}
where $N_c=3$, $N_f$ number of quark flavors, $\Delta=0(\Delta=-10/3)$ for $C_{LL}(C_{RL})$ and $C_F$(D4) is the quadratic Casimir invariant of SU(3) color. By using above solution , we can compute the long distance factor 
\begin{equation}
    A_{LL}\equiv \frac{C_{LL}(2 \mbox{GeV})}{C_{LL}(M_{SUSY})},\qquad     A_{RL}\equiv \frac{C_{RL}(2 \mbox{GeV})}{C_{RL}(M_{SUSY})}
    \label{ALLARL}
\end{equation}

as follows:

\begin{align}
A_{LL}=\left[\frac{\alpha_s(m_t)}{\alpha_s(M_{SUSY})}\right]^{\frac{2}{7}}\left[\frac{28\pi +26\alpha_s(m_t)}{28\pi +26\alpha_s(M_{SUSY})}\right]^{-\frac{79}{546}}
\left[\frac{\alpha_s(m_b)}{\alpha_s(m_t)}\right]^{\frac{6}{23}}\left[\frac{92\pi +116\alpha_s(m_b)}{92\pi +116\alpha_s(m_t)}\right]^{-\frac{1375}{8004}}\nonumber\\
\left[\frac{\alpha_s(2\,\mbox{GeV})}{\alpha_s(m_b)}\right]^{\frac{6}{25}}\left[\frac{100\pi +154\alpha_s(2\,\mbox{GeV})}{100\pi +154\alpha_s(mb)}\right]^{-\frac{2047}{11550}}
\label{ALL}
\end{align}
and 
\begin{align}
A_{RL}=\left[\frac{\alpha_s(m_t)}{\alpha_s(M_{SUSY})}\right]^{\frac{2}{7}}\left[\frac{28\pi +26\alpha_s(m_t)}{28\pi +26\alpha_s(M_{SUSY})}\right]^{-\frac{19}{91}}
\left[\frac{\alpha_s(m_b)}{\alpha_s(m_t)}\right]^{\frac{6}{23}}\left[\frac{92\pi +116\alpha_s(m_b)}{92\pi +116\alpha_s(m_t)}\right]^{-\frac{430}{2001}}\nonumber\\
\left[\frac{\alpha_s(2\,\mbox{GeV})}{\alpha_s(m_b)}\right]^{\frac{6}{25}}\left[\frac{100\pi +154\alpha_s(2\,\mbox{GeV})}{100\pi +154\alpha_s(mb)}\right]^{-\frac{173}{825}}
\label{ARL}
\end{align}

\section{SU(5) formulae for proton decay from D=6 operators}
At the GUT scale, the Wilson coefficients of the relevant D=6 proton decay operators are:
\be
C_{RL}^{(6)ijk}(ud_id_j\nu_k)=\frac{g_{GUT}^2}{M_V^2}V_{CKM}^{1j}\delta^{ik}\, ,
\ee
\be
C_{RL}^{(6)ik}(ud_iul_k)=\frac{g_{GUT}^2}{M_V^2}\delta^{ik}, \quad \quad
C_{LR}^{(6)ik}(ud_iul_k)=\frac{g_{GUT}^2}{M_V^2}(\delta^{ik}+V_{CKM}^{1i}(V_{CKM}^{\dagger})^{k1})\, .
\ee
At $\mu = 2\, \mbox{GeV}$, we use
\be
C^{(6)} (\mu = 2\, \mbox{GeV}) = A_R\, C^{(6)} (M_{GUT})\, ,
\ee
where the renormalization factor $A_R = A^{SD}_R A^{LD}_R$ is the product of a short-distance part~\cite{Ibanez:1984ni}
\be
  A^{SD}_R \simeq \left( \frac{\alpha_3 (M_{SUSY})}{\alpha_{GUT}} \right)^{4/3}
    \left( \frac{\alpha_2 (M_{SUSY})}{\alpha_{GUT}} \right)^{-3/2}
     \left( \frac{\alpha_1 (M_{SUSY})}{\alpha_{GUT}} \right)^{-23/198}
\label{eq:A_SD_D=6}
\ee
and of a long-distance part (in which we neglect the subleading $SU(2)_L$ and $U(1)_Y$ contributions
between $M_{SUSY}$ and $m_Z$)
\be
  A^{LD}_R = \left( \frac{\alpha_s(2\,\mbox{GeV})}{\alpha_s(m_b)} \right)^{\frac{6}{25}}
    \left( \frac{\alpha_s(m_b)}{\alpha_s(m_t)} \right)^{\frac{6}{23}}
    \left( \frac{\alpha_s(m_t)}{\alpha_s(M_{SUSY})} \right)^{\frac{2}{7}}\, .
\ee
The coefficient $A^{SD}_R$ only accounts for the 1-loop renormalization of the Wilson
coefficients due to the gauge sector; the Yukawa contributions can be neglected to a good
approximation, as the proton decay operators do not include third generation fermions
(except for neutrinos). Furthermore, the small $U(1)_Y$ contribution is only approximate.
The decay rates of proton decay from D=6 operators:
\be
\Gamma(p\rightarrow K^+ \bar \nu_e) =  \frac{\pi}{2}\, m_p  \left( 1 - \frac{m_{K^+}^2}{m^2_p} \right)^2
\frac{\alpha_{GUT}^2}{M_V^4}\, A_{RL}^2 \left| V_{us} W^{\langle K^+| (ud)_R s_L |p\rangle}_0 \right|^2 
\ee
\be
\Gamma(p\rightarrow K^+ \bar \nu_{\mu}) =  \frac{\pi}{2}\, m_p  \left( 1 - \frac{m_{K^+}^2}{m^2_p} \right)^2
\frac{\alpha_{GUT}^2}{M_V^4}\, A_{RL}^2  \left| V_{ud} W^{\langle K^+| (us)_R d_L |p\rangle}_0 \right|^2
\ee

\be
\Gamma(p\rightarrow K^+ \bar \nu) =  \frac{\pi}{2}\, m_p  \left( 1 - \frac{m_{K^+}^2}{m^2_p} \right)^2
  \frac{\alpha_{GUT}^2}{M_V^4}\, A_{RL}^2 \left( \left| V_{ud} W^{\langle K^+| (us)_R d_L |p\rangle}_0 \right|^2
  + \left| V_{us} W^{\langle K^+| (ud)_R s_L |p\rangle}_0 \right|^2 \right) 
\ee
\be
\Gamma(p\rightarrow \pi^0 e^+) =  \frac{\pi}{2}\, m_p \left( 1 - \frac{m_{\pi^0}^2}{m^2_p} \right)^2
  \frac{\alpha_{GUT}^2}{M_V^4}\, A_{RL}^2 \left|W^{\langle \pi^0| (ud)_R u_L |p\rangle}_0 \right|^2
  \left( 1+(1+|V_{ud}|^2)^2 \right) 
\ee
\be
\Gamma(p\rightarrow \pi^0 \mu^+) =  \frac{\pi}{2}\, m_p \left( 1 - \frac{m_{\pi^0}^2}{m^2_p} \right)^2
  \frac{\alpha_{GUT}^2}{M_V^4}\, A_{RL}^2 \left|W^{\langle \pi^0| (ud)_R u_L |p\rangle}_0 \right|^2
  (1+V_{11}V_{21}^{\dagger})^2  
\ee
\be
\Gamma(p\rightarrow \pi^+  \bar \nu_e) =  \frac{\pi}{2}\, m_p \left( 1 - \frac{m_{\pi^+}^2}{m^2_p} \right)^2
  \frac{\alpha_{GUT}^2}{M_V^4}\, A_{RL}^2 \left|W^{\langle \pi^+| (ud)_R d_L |p\rangle}_0 \right|^2
  (1+V_{11}V_{21}^{\dagger})^2  
\ee
\be
\Gamma(p\rightarrow K^0  e^+ ) =  \frac{\pi}{2}\, m_p  \left( 1 - \frac{m_{K^0}^2}{m^2_p} \right)^2
  \frac{\alpha_{GUT}^2}{M_V^4}\, A_{RL}^2  \left|W^{\langle K^0| (us)_R u_L |p\rangle}_0 \right|^2(V_{12}V_{11}^{\dagger})^2
\ee

\be
\Gamma(p\rightarrow K^0  \mu^+ ) =  \frac{\pi}{2}\, m_p  \left( 1 - \frac{m_{K^0}^2}{m^2_p} \right)^2
  \frac{\alpha_{GUT}^2}{M_V^4}\, A_{RL}^2  \left|W^{\langle K^0| (us)_R u_L |p\rangle}_0 \right|^2(1+(1+V_{12}V_{21}^{\dagger}))^2
\ee

\section{The set up: SU(5) with sliding decoupling scale} 
\begin{figure}[h!]
\centering
\includegraphics[height=8cm]{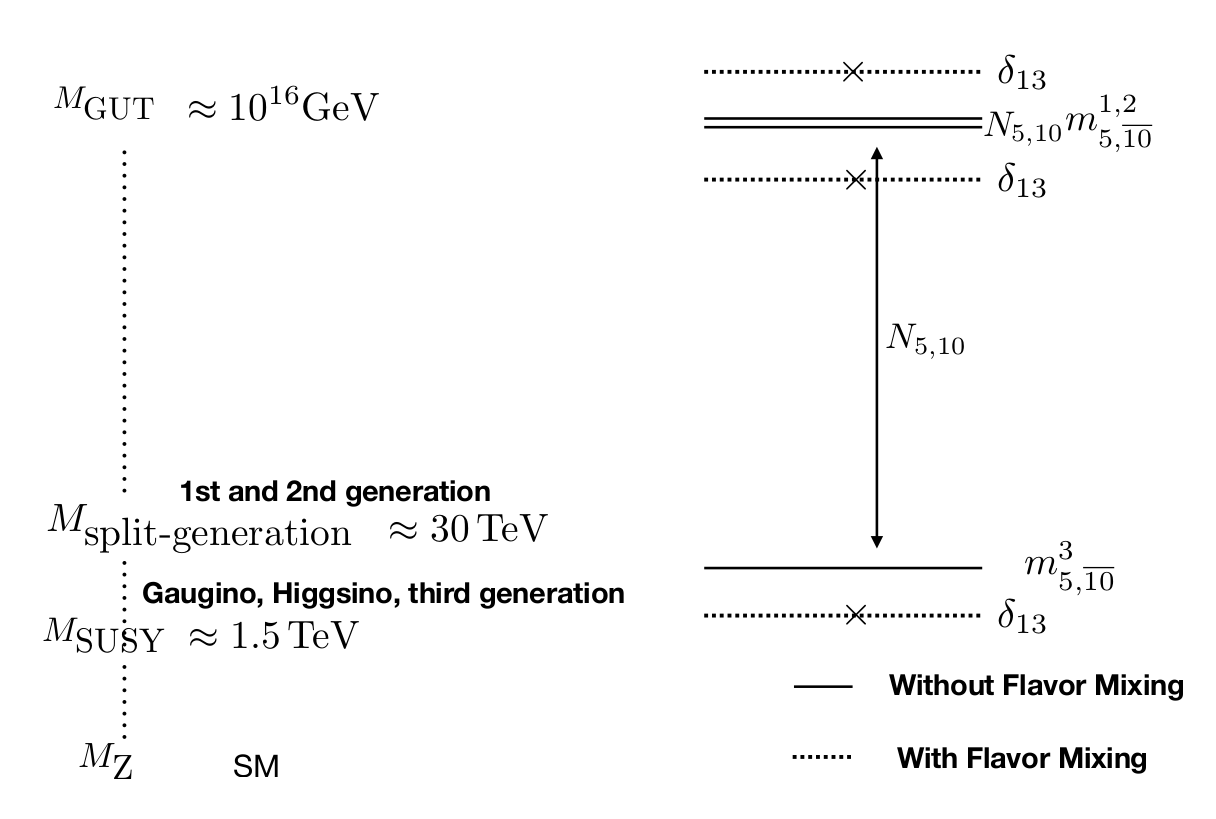} 
\caption{Illustration of scales and scalars generation separation.\label{scales} }
\end{figure} 
Weak scale supersymmetry is inconsistent with the various experimental observations of flavor changing processes like $\Delta M_K$. These constraints either need highly degenerate first two generation of squarks, or some symmetry to suppress off-diagonal entries of squark masses matrices or very heavy first two generations of squarks with masses around 100 TeV. Similarly the branching ratio for $\mu\to e \gamma$, $\mu \to e e e $ etc require first two generations of slepton masses to be around 100 TeV. From phenomenological point of view, we assume first two generations between a few TeV to 100 TeV  and all other SUSY particles to be near weak scale. The schematic energy scale diagrams of split generation and effects of flavor mixing on splitting are presented in fig. (\ref{scales}). We cannot take first two generation sfermion masses arbitrarily large within this framework as it could lead to  color and charge breaking minima (see section \ref{tachyon}). We used the term ``split-generation" for this case. In this section, we describe our notation for ``split-generation" scenario. We define splitting parameters and mass matrices of sfermions at GUT scale where we define boundary values, and at TeV scale where we calculate the SUSY spectrum. 

At $M_{GUT}$, we specify the boundary values for the soft masses (defined in the so-called super-CKM basis, in which the up-type quark mass matrices are diagonal and squarks are rotated parallel to their superpartners) as follows:
\begin{equation} 
m_{\tilde Q}^2= m_{10}^2 \begin{pmatrix}   1 & 0 & \delta^{10}_{13}/N_{10} \cr 0  & 1 & 0 \cr \delta^{10}_{13}/N_{10} & 0 & 1/(N_{10})^2  \end{pmatrix}
\label{mq}
\end{equation} 
\begin{equation}
m_{\tilde d}^2= m_{\bar 5}^2 \begin{pmatrix}   1 & 0 & \delta^{\bar 5}_{13}/N_{\bar 5} \cr 0  & 1 & 0 \cr \delta^{\bar 5}_{13}/N_{\bar 5} & 0 & 1/(N_{\bar 5})^2  \end{pmatrix}
\label{md}
\end{equation} 
$\left(Q,u,e\right)$ superfields are elements of same 10-plet supermultiplet of SU(5). Similarly $\left(d,L\right)$ superfields are also elements of same 5-plet supermultiplet of SU(5). In SUSY SU(5), mass matrices at the GUT scale have following relations: 
\begin{equation}
m_{\tilde Q}^2=  m_{\tilde u}^2  =  m_{\tilde e}^2  \quad ; \quad  m_{\tilde d}^2=  m_{\tilde L}^2 
\label{mQmd}
\end{equation} 

where, 
\be 
\delta_{13}=\frac{m_{13}^2}{\sqrt{m_1^2 m_3^2}}=\frac{m_{13}^2 N}{\sqrt{m_0^2 m_0^2}}=\frac{m_{13}^2 N}{m_0^2} \quad \implies m_{13}^2=\frac{\delta_{13}m_0^2}{N} 
\label{delta13}
\ee
where $N$ is splitting between first generation to third generation masses at GUT scale. At $M_{SUSY}$ scale, splitting between first and third generation is defined as:
\be
N_u=\frac{m_{\tilde{u}_1}}{m_{\tilde{t}_1}}, \quad
N_d=\frac{m_{\tilde{d}_1}}{m_{\tilde{b}_1}}, \quad
N_e=\frac{m_{\tilde{e}_1}}{m_{\tilde{\tau}_1}}
\label{NuNd}
\ee  
\section{Gauge coupling unification in split-generation scenario}
\begin{figure}[h]%
\centering
\subfigure[]{%
\label{fig:first1}%
\includegraphics[width=7.5cm,height=5.5cm]{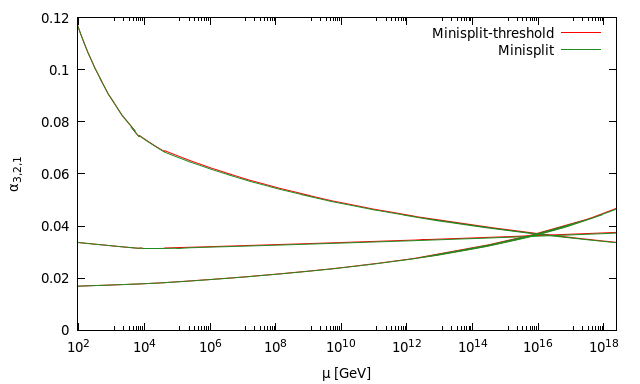}}%
\hfill
\subfigure[]{%
\label{fig:second1}%
\includegraphics[width=7.5cm,height=5.5cm]{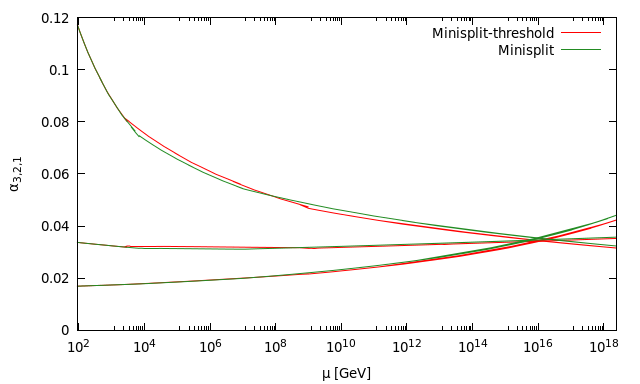}}%
\caption{(a) $M_{SUSY}=7\times10^{3}$ GeV and $M_{split-gen}=4.2\times10^{4}$ GeV, and (b) $M_{SUSY}=3\times10^{3}$ GeV and $M_{split-gen}=10^{9}$ GeV. Red lines are with SUSY threshold correction and $\beta$ coefficients at two different scale. We can see threshold corrections have negligible effect on gauge unification scale  and slope of gauge couplings.}
\label{thresholdcor}
\end{figure}

Here, we calculate the gauge coupling $\beta$-coefficients for the split-generation scenario. 
The 2-loop RGEs for the gauge couplings in MSSM\cite{Martin:1993zk}:
\be
(4\pi)^2 \frac{d}{dt}g_a=g_a^3 b_a + \frac{g_a^3}{(4\pi)^2}\left[\sum_{b=1}^{3}B_{ab}g_b^2-\sum_{x=u,d,e}^{}C_a^{x}Tr(Y_x^{ \dagger}Y_{x})\right]
\label{2loopRG}
\ee
where $t=\log(\mu)$ and $\mu$ is renormalization scale and a, b= 1, 2, 3 denote  U(1), SU(2)$_L$, SU(3)$_c$ gauge groups, respectively. We use the convention $g_1^2=(5/3)g^2$.
In general:
 \be
b_a=\left(-\frac{11}{3}C_2(G)+\frac{2}{3}\sum_{i}^{}T_i(R)+\frac{1}{3}\sum_{\alpha}^{}T_{\alpha}(R)\right)
\label{oneloopb}
\ee
for i$^{th}$ chiral fermion and $\alpha^{th}$ complex scalar. Expanding the second term:
\begin{equation}
\label{eq1}
\begin{split}
\sum_{b=1}^{3}B_{ab}g_b^2=g_a^2\left(\frac{10}{3}C_2(G_a)T(F_a)+2C_2(F_a)T(F_a)+\frac{2}{3}C_2(G_a)T(S_a)+4C_2(S_a)T(S_a)\right) \\
-g_a^2\left(\frac{34}{3}(C_2(G_a)^2)\right)+g_b^2\left(2C_2(F_b)T(F_a)+4C_2(S_b)T(S_a)\right) 
\end{split}
\end{equation}
\begin{equation}
\label{twoloopB}
\begin{split}
\sum_{b=1}^{3}B_{ab}g_b^2=g_a^2\left(\frac{10}{3}C_2(G_a)T(F_a)+2C_2(F_a)T(F_a)+\frac{2}{3}C_2(G_a)T(S_a)+4C_2(S_a)T(S_a)\right)\\
-g_a^2\left(\frac{34}{3}(C_2(G_a)^2)\right)+g_b^2\left(2C_2(F_b)T(F_a)+4C_2(S_b)T(S_a)\right)    
\end{split}
\end{equation}  
Here, $F_a, S_a$ are  representation of fermion and scalar, respectively, and T($F_a$)=$\sum T(F_a)$; similarly for S and R. These are terms which one needs to calculate for the one and two-loop $\beta$ function coefficients for any theory. However, in supersymmetry, there is one term in Kahler potential ($\phi\exp(2gV)\phi$), which looks like Yukawa term ($\bar{\varphi}(T^a_R)^i_j\lambda_a\psi^j$) with gauge coupling purely because of supersymmetry. There three more terms will contribute because of this term:
\be
-g_a^2(2C_2(R_a)T(R_a)+2C_2(G_a)T(R_a))-g_b^2(2C_2(R_b)T(R_a))
\label{newtwoloop}
\ee
where `R' is the representation of chiral supermultiplet.// For U(1)$_Y$:
\be C_2(G)=0,C_2(F)=0, T_F=\frac{3Y^2}{5}
\label{u1y}
\ee
For SU(2)$_{L}$:
\be C_2(G)=2,C_2(F)=3/4, T_F=\frac{1}{2}
\label{su2L}
\ee
For SU(3)$_{c}$:
\be C_2(G)=3,C_2(F)=4/3, T_F=\frac{1}{2}
\label{su3c}
\ee
Using eqs. (\ref{u1y}-\ref{su3c}), values of $\beta$-coefficient are as follows: \\
for SM :
\begin{equation}
b_a=\begin{pmatrix} 41/10 \\
  -19/6 \\
  -7
  	\end{pmatrix}\quad B_{ab}=
\begin{pmatrix}
		199/50 & 27/10 & 44/5 \\
		9/10 & 35/6 & 12 \\
		11/10 & 9/2 & -26
		\end{pmatrix}
	 \quad C_a^{u,d,e}=
		\begin{pmatrix}
		17/10 & 1/2 & 3/2 \\
		3/2 & 3/2 & 1/2 \\
		2 & 2 & 0
		\end{pmatrix}
		\label{SMB}
		\end{equation}
    for split-generation (low scale):
    			\begin{equation}
    			b_a=\begin{pmatrix} 79/15 \\
    			-1/3\\
    			-13/3\end{pmatrix} \quad
    			B_{ab}=
    			\begin{pmatrix}
    			407/75 & 21/5 & 176/15 \\
    			7/5 & 89/3 & 16 \\
    			22/15 & 6 & 58/3
    			\end{pmatrix}\quad
 C_a^{u,d,e}=
   	\begin{pmatrix}
   	26/5 & 14/5 & 18/5 \\
   	6 & 6 & 2 \\
   	4 & 4 & 0
   	\end{pmatrix}
   	\label{SplitB}
   	\end{equation}	
    for MSSM (split-generation high scale):
    		\begin{equation}
    		b_a=\begin{pmatrix}33/5 \\
    		1 \\
    		-3\end{pmatrix} \quad B_{ab}=
    		\begin{pmatrix}
    		199/25 & 27/5 & 88/5 \\
    		9/5 & 25 & 24 \\
    		11/5 & 9 & 14
    		\end{pmatrix} \quad
         C_a^{u,d,e}=
    		\begin{pmatrix}
    		26/5 & 14/5 & 18/5 \\
    		6 & 6 & 2 \\
    		4 & 4 & 0
    		\end{pmatrix}
    		\label{highsplitB}
    		\end{equation}	
  \textbf{Threshold corrections to gauge couplings:}
  \be
  \frac{1}{\alpha_a^+(\mu)}=\frac{1}{\alpha_a^-(\mu)}+\Delta_a^p(\mu)
  \label{alphaplus}
  \ee
  \be
  \Delta_a^p(\mu)=\frac{1}{2\pi}\sum_{p}b_a^pLog\left(\frac{M_p}{\mu}\right)
  \label{gaugethre}
  \ee
  \be
  \alpha_a^+(\mu)=\frac{\alpha_a^-(\mu)}{1+\alpha_a^-(\mu)\Delta_a^p(\mu)}
  \label{alphaplus1}
  \ee
 \begin{figure}[h]%
\centering
\subfigure[]{%
\label{fig:gaugeunifi1}%
\includegraphics[width=7.5cm,height=5.5cm]{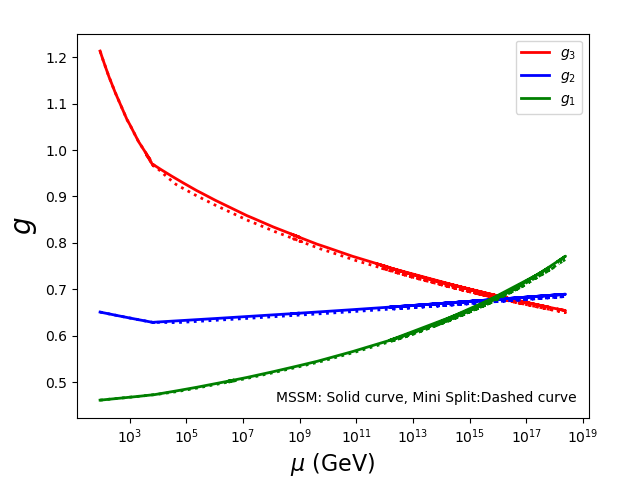}}%
\hfill
\subfigure[]{%
\label{fig:gaugeunifi2}%
\includegraphics[width=7.5cm,height=5.5cm]{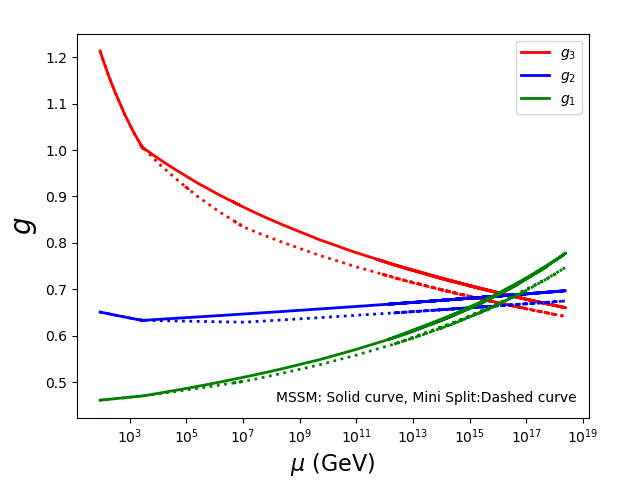}}%
\vskip\baselineskip
\subfigure[]{%
\label{fig:gaugeunifi3}%
\includegraphics[width=7.5cm,height=5.5cm]{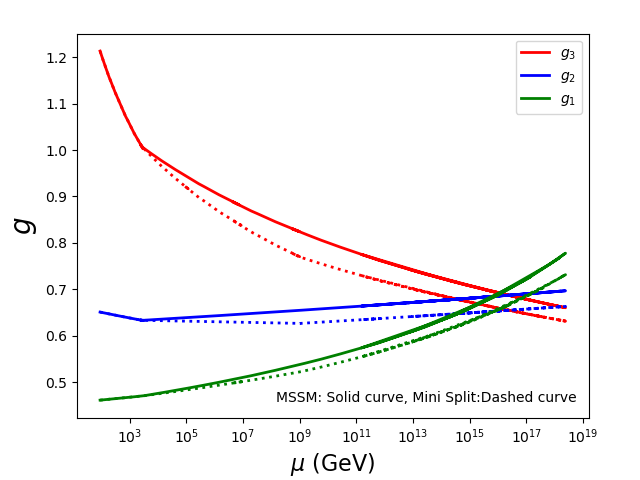}}%
\caption{ Gauge coupling running with energy scale for (a) $M_{SUSY}=7 \times 10^{3}$ GeV and $M_{split-gen}=4.2 \times 10^{4}$ GeV, (b) $M_{SUSY}=3 \times 10^{3}$ GeV and $M_{split-gen}=10^{7}$ GeV and (c) $M_{SUSY}=3 \times 10^{3}$ GeV and $M_{split-gen}=10^{9}$ GeV.\label{gaugeunifi}}
\end{figure} 
At SM:\\
   For Top quark $	b_a=\left({\begin{array}{ccc} 17/30 & 1 & 2/3\end{array}}\right)$  and Higgs $	b_a=\left({\begin{array}{ccc} 1/10 & 1/6 & 0\end{array}}\right)$ (considering full Higgs doublet.)\\
At split-generation (low scale) in mass eigenstate:
\begin{align}
       \tilde{t}::	b_a=\left({\begin{array}{ccc} 17/60 & 1/2 & 1/3\end{array}}\right) \quad  \tilde{b}::	b_a=\left({\begin{array}{ccc} 1/12 & 1/2 & 1/3\end{array}}\right) \\\nonumber
   \tilde{\tau}::	b_a=\left({\begin{array}{ccc} 1/4 & 1/6 & 0 \end{array}}\right) \quad  \tilde{\nu_{\tau}}::	b_a=\left({\begin{array}{ccc} 1/20 & 1/6 & 0\end{array}}\right) \\\nonumber   
      \tilde{B}::	b_a=\left({\begin{array}{ccc} 0 & 0 & 0\end{array}}\right) \quad  \tilde{W}::	b_a=\left({\begin{array}{ccc} 0 & 4/3 & 0\end{array}}\right) \\\nonumber
  \tilde{g}::	b_a=\left({\begin{array}{ccc} 0 & 0 & 2\end{array}}\right) \quad  \tilde{\chi_0}::	b_a=\left({\begin{array}{ccc} 2/5 & 2/3 & 0\end{array}}\right)\\ \nonumber
          A::	b_a=\left({\begin{array}{ccc} 1/20 & 1/6 & 0\end{array}}\right) \quad H^{\pm}::	b_a=\left({\begin{array}{ccc} 1/10 & 1/3 & 0\end{array}}\right)  
          \label{splitthre}
\end{align}
At split-generation (high scale) in mass eigenstate:
\begin{align}
\tilde{u}, \tilde{c}::	b_a=\left({\begin{array}{ccc} 17/60 & 1/2 & 1/3\end{array}}\right) \quad  \tilde{d},\tilde{s}::	b_a=\left({\begin{array}{ccc} 1/12 & 1/2 & 1/3\end{array}}\right)  \\ \nonumber
\tilde{e},\tilde{\mu}::	b_a=\left({\begin{array}{ccc} 1/4 & 1/6 & 0 \end{array}}\right) \quad  \tilde{\nu_e}, \tilde{\nu_{\mu}}::	b_a=\left({\begin{array}{ccc} 1/20 & 1/6 & 0\end{array}}\right)
\label{splithighthre}
\end{align}
At split-generation (low scale) in gauge state:
\begin{align}
\tilde{Q}_3::	b_a=\left({\begin{array}{ccc} 1/30 & 1/2 & 1/3\end{array}}\right) \quad  \tilde{t}::	b_a=\left({\begin{array}{ccc} 4/15 & 0 & 1/6\end{array}}\right) \\ \nonumber 
\tilde{b}::	b_a=\left({\begin{array}{ccc} 1/15 & 0 & 1/6 \end{array}}\right) \quad  \tilde{g}::	b_a=\left({\begin{array}{ccc} 0 & 0 & 2\end{array}}\right)     \\ \nonumber
\tilde{L}_3::	b_a=\left({\begin{array}{ccc} 1/10 & 1/6 & 0\end{array}}\right) \quad \tilde{\tau}::	b_a=\left({\begin{array}{ccc} 1/5 & 0 & 0\end{array}}\right)\\ \nonumber
\tilde{B}::	b_a=\left({\begin{array}{ccc} 0 & 0 & 0\end{array}}\right) \quad  \tilde{W}::	b_a=\left({\begin{array}{ccc} 0 & 4/3 & 0\end{array}}\right) \\ \nonumber  
H_d,H_u::	b_a=\left({\begin{array}{ccc} 1/10 & 1/6 & 0\end{array}}\right) \quad \mu::	b_a=\left({\begin{array}{ccc} 2/5 & 2/3 & 0\end{array}}\right)     
\label{splitthregauge}
\end{align}
At split-generation (high scale) in gauge state:
\begin{align}
\tilde{L}_1,\tilde{L}_2::	b_a=\left({\begin{array}{ccc} 1/10 & 1/6 & 0\end{array}}\right) \quad \tilde{e},\tilde{\mu}::	b_a=\left({\begin{array}{ccc} 1/5 & 0 & 0\end{array}}\right) \\ \nonumber
\tilde{Q}_1,\tilde{Q}_2::	b_a=\left({\begin{array}{ccc} 1/30 & 1/2 & 1/3\end{array}}\right) \quad  \tilde{u},\tilde{c}::	b_a=\left({\begin{array}{ccc} 4/15 & 0 & 1/6\end{array}}\right) \\ \nonumber
\tilde{d},\tilde{s}::	b_a=\left({\begin{array}{ccc} 1/15 & 0 & 1/6 \end{array}}\right)  \label{splitthregauge}
\end{align}
\textbf{Gauge coupling $\overline{MS}$ to $\overline{DR}$ conversion:}     \be
\frac{1}{\alpha_a^{\overline{DR}}(\mu)}=\frac{1}{\alpha_a^{\overline{MS}}(\mu)}+\Delta_a^{\overline{DR}} \implies
\alpha_a^{\overline{DR}}(\mu)=\frac{\alpha_a^{\overline{MS}}(\mu)}{1+\alpha_a^{\overline{MS}}(\mu)\Delta_a^{\overline{DR}}} 
\label{alphaDR}
\ee
where 
\be
 \Delta_a^{\overline{DR}}=-\frac{C_2(G_a)}{12\pi}\implies\Delta_3^{\overline{DR}}=-\frac{1}{4\pi},\quad \Delta_2^{\overline{DR}}=-\frac{1}{6\pi},\quad \Delta_1^{\overline{DR}}=0   
 \label{alphaDR1}
\ee
This scenario does not have first two generations of sparticles at low energies, unlike MSSM, but because these scalars come in complete SU(5) multiplets they do not affect the unification scale. Using two-loop RGEs of gauge couplings (SuSeFLAV code) with modified $\beta$-coefficients of one-loop and two-loop RGEs, eqs.(\ref{SMB}-\ref{highsplitB}), for split-generation spectrum, in fig. \ref{gaugeunifi} we plot gauge couplings with different split scales. We compare slope of MSSM RGEs with that of the of split-generation scenario. We can see that gauge unification scale of split-generation scenario is identical to MSSM gauge unification scale but slope of gauge coupling is changing due to RGEs modification in split-generation scenario. This slope change increases with higher split-generation scale.\\
The SUSY threshold corrections have negligible effect on slope of gauge couplings and unification scale, see fig.(\ref{thresholdcor}).

\section{How decoupled should the first two generation scalars be?}
\label{tachyon}
\begin{figure} %
\centering
\subfigure[]{%
\label{fig:weaksplitting1}%
\includegraphics[width=7.5cm,height=5cm]{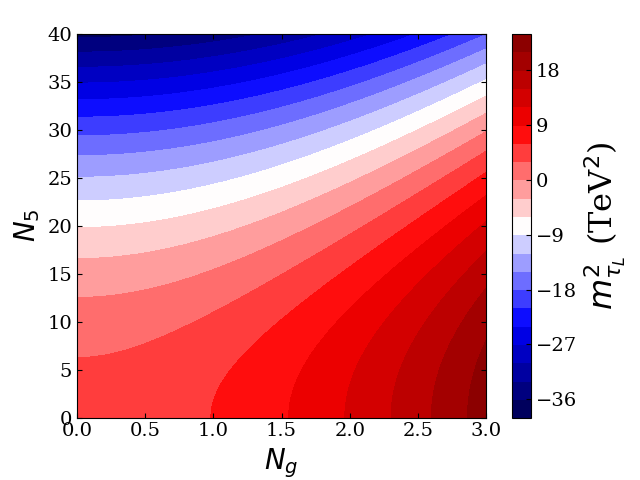}}%
\subfigure[]{%
\label{fig:weaksplitting2}%
\includegraphics[width=7.5cm,height=5cm]{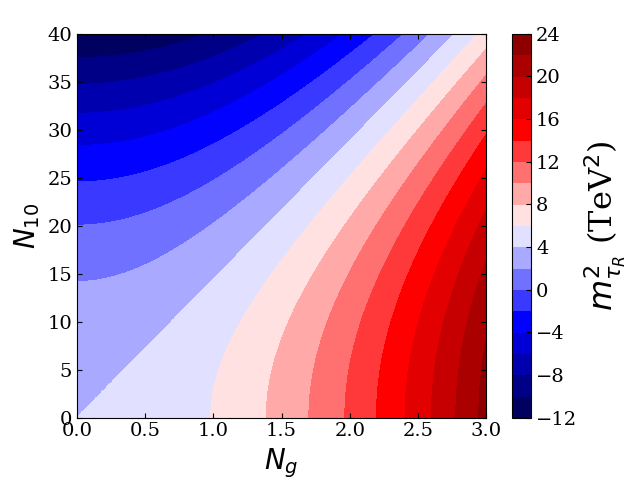}}%
\hfill
\subfigure[]{%
\label{fig:weaksplitting1}%
\includegraphics[width=7.5cm,height=5cm]{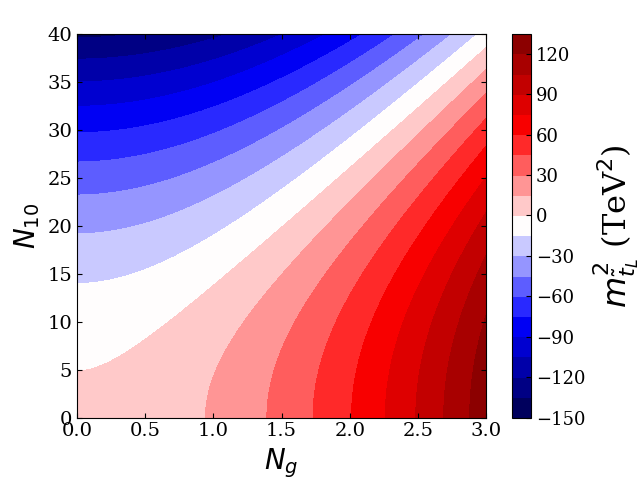}}%
\subfigure[]{%
\label{fig:weaksplitting2}%
\includegraphics[width=7.5cm,height=5cm]{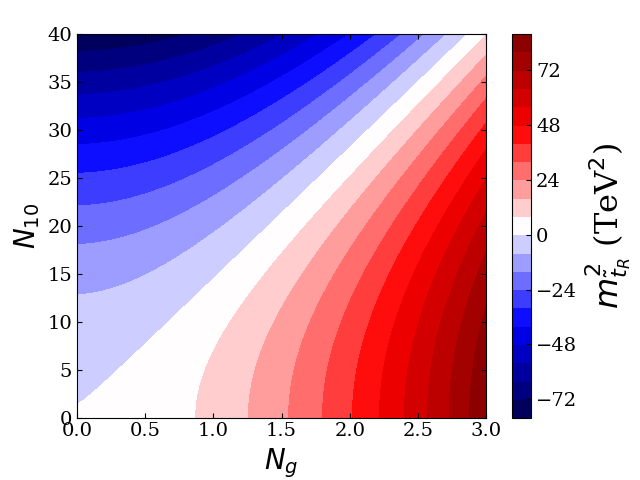}}%
\caption{Soft mass square parameters for $\tilde\tau_{L/R}$ and $\tilde t_{L/R}$ at low scale as a function of $N_g$ and $N_{5,10}$.
\label{tacyon1}} \end{figure}

\begin{figure}[h]%
\centering
\subfigure[]{%
\label{fig:tacyon5}%
\includegraphics[width=10cm,height=6.5cm]{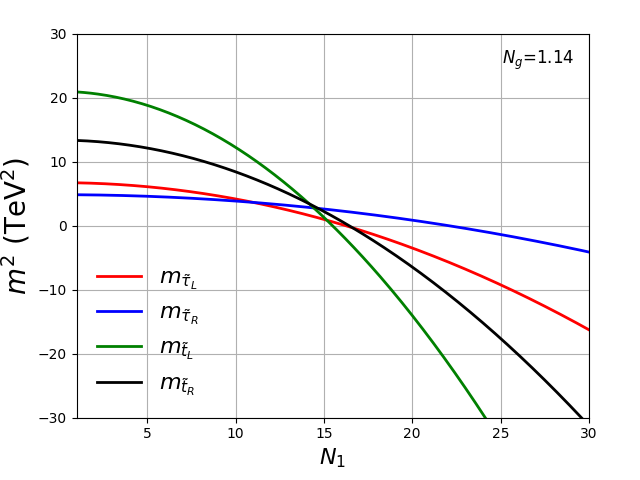}}%
\caption{Soft mass square parameters for $\tilde\tau_{L/R}$ and $\tilde t_{L/R}$ at low scale as a function $N_{5,10}$ with fixed $N_g$.\label{tacyon2}}
\end{figure}
Here, we analytically study the allowed splitting between the generations. The soft square masses at the weak scale in terms of GUT scale parameters are determined by the RGEs, including the heavy scalar contribution at the two-loop level ref.\cite{Agashe:1998zz,ArkaniHamed:1997ab}. We take $m_{\tilde{t}}(GUT)=m_{H_u}(GUT)=A_t(GUT)/\sqrt{3}$. Soft mass squares of stop and stau at the weak scale:
\be m^2 _{\tau_L} (N_g,N_{5}) = 4(1 + 0.519 N_g^2 - 0.006382 N_5^2) \ee
\be m^2_{\tau_ R}(N_g,N_{10}) = 4(1 + 0.159 N_g^2 - 0.002488 N_{10}^2).  \ee
\be m^2_{\tilde t_L} (N_g,N_{10}) = 4(0.511 + 0.1177 N_g + 3.52 N_g^2 - 0.021884 N_{10}^2) \ee
\be m^2_{\tilde t_R} (N_g,N_{10}) = 4(0.024 + 0.2372 N_g + 2.33 N_g^2 - 0.01239 N_{10}^2) \ee
These masses are plotted in terms of GUT scale splitting parameter in fig. \ref{tacyon1}. In these plots, x-axis is ratio of gluino mass to stop mass at GUT scale ($N_g=M_3 /m_{\tilde{t}}$) and y-axis is ratio of first two generation sfermion masses which is degenerate at GUT scale to stop mass at GUT scale($N_1=m_{1,2}/m_{\tilde{t}}$). Red contours have positive mass square values and blue contours have tachyonic spectrum. From fig.(\ref{tacyon1}), we can see that with very large gluino mass, maximum splitting can be 40. In fig. (\ref{tacyon2}) we choose $n_g=1.1428$ which is approximate value for our case. In this plot x-axis in $N_1$ and y-axis is square masses of third generation. From this plot we can say that max $N_1$ value is in between 15-20, for which the spectrum is not tachyonic. 



\section{Numerical analysis of proton decay rates in split-generation SUSY SU(5) GUT}
We implement split-generation scenario in SuSeFlav code\cite{Chowdhury:2011zr}  which is a supersymmetric spectrum calculator. We interface the proton decay calculations routines with the adapted SuSeFlav version. We discuss the procedure including the major modifications in the code to incorporate this scenario on top of the default set-up. We have 4 energy scales as depicted in fig. (\ref{scales}). The starting point is the experimental input for the physical parameters given in the table \ref{tab:tabel_2}. 
\begin{table}
$$
\begin{array}{cc}
\hline \hline&\vspace{-.5 cm}\\
\mbox{ Paramter }  & \mbox{Range} \\
\hline
m_{hu},m_{hd} & [-30,30] \,\,\mbox{TeV}\\
m_{10},m_{\bar 5} & [1,30] \,\,\mbox{TeV}  \\
M_{1/2} &  [0.7,3]\,\,\mbox{TeV} \\
A_0 &  [-8,8] \,\,\mbox{TeV} \\
N_{10}^1,N_{\bar 5}^1 & [1,20] \\
\delta_{13}^{10},\delta_{13}^{\bar 5} & [-1,1] \\
\tan\beta & 5 \\
\hline \hline
\end{array}
$$
\caption{Parameters range considered.}
\label{tab:tabel_range}
\end{table}

Add Higgs and top threshold correction in gauge coupling. In first iteration, we run SM Yukawas and gauge couplings from $m_Z$ to $M_{SUSY}$. Note that for the first iteration, $M_{SUSY}$ is the input parameter which we consider to be 2 TeV. Do $\overline{MS}$ to $\overline{DR}$ conversion of all gauge and Yukawa couplings.  Run all the MSSM RGEs from $M_{SUSY}$ to $M_{Pl}$ with zero soft parameters boundary values (at $M_{SUSY}$). We define $M_{GUT}$ as the scale where $g_1$ =$g_2$. Define boundary values for the soft masses as (defined in the super-CKM basis) at $M_{GUT}$  eq. (\ref{mq}-\ref{delta13})). The range considered for soft masses is given in table \ref{tab:tabel_range}. After this we run all the parameters (hard and soft) from $M_{GUT}$ to $M_{SUSY}$ and calculate the spectrum at 2 TeV scale. Program also check for $\mu$ conversion and run all the parameters to $m_Z$ scale. Now we have SUSY spectrum so we can calculate new  $M_{SUSY}$ and $M_{split-generation}$ as.
\[ M_{SUSY}= \sqrt{m_{\tilde t_1} m_{\tilde t_2}} \]
\[  M_{split-generation}=\sqrt{\frac{m^2_{\tilde s_1}+m^2_{\tilde s_2}+m^2_{\tilde c_1}+m^2_{\tilde c_2}+m^2_{\tilde u_1}+m^2_{\tilde u_2}+m^2_{\tilde d_1}+m^2_{\tilde d_2}}{8}} \]
For the second iteration, we again consider the experimental data ($\alpha_i$, fermion masses) at $m_Z$ scale. Calculate the gauge threshold corrections from top quark and Higgs boson and convert the gauge couplings and fermion masses from $\overline{MS}$ to $\overline{DR}$ scheme, Implement SUSY threshold corrections. We use modified MSSM RGEs for the gauge couplings where  we exclude the contribution of the first two generations of sfermions from $m_Z$ to $M_{split-generation}$. From $M_{split-generation}$ to $M_{GUT}$, we used full MSSM RGEs. This time we will get correct (modified) $M_{GUT}^{new}$. Now we have gauge couplings and Yukawa couplings with correct estimates of the  SUSY threshold corrections.

\section{Results}
In this section, we study both the scenarios without flavor mixing ($\delta_{ij}=0$) and with flavor mixing in first and third generation of  10-plet sfermion. We consider $m_{10}=m_{\bar{5}},   
N_{10}^1=N_{\bar{5}}^1$. We define sfermion matrix as given in eqs. (\ref{mq}-\ref{delta13}) and consider only diagonal LR mixing in sfermion mass matrices. 
\begin{figure}%
\centering
\includegraphics[height=4cm]{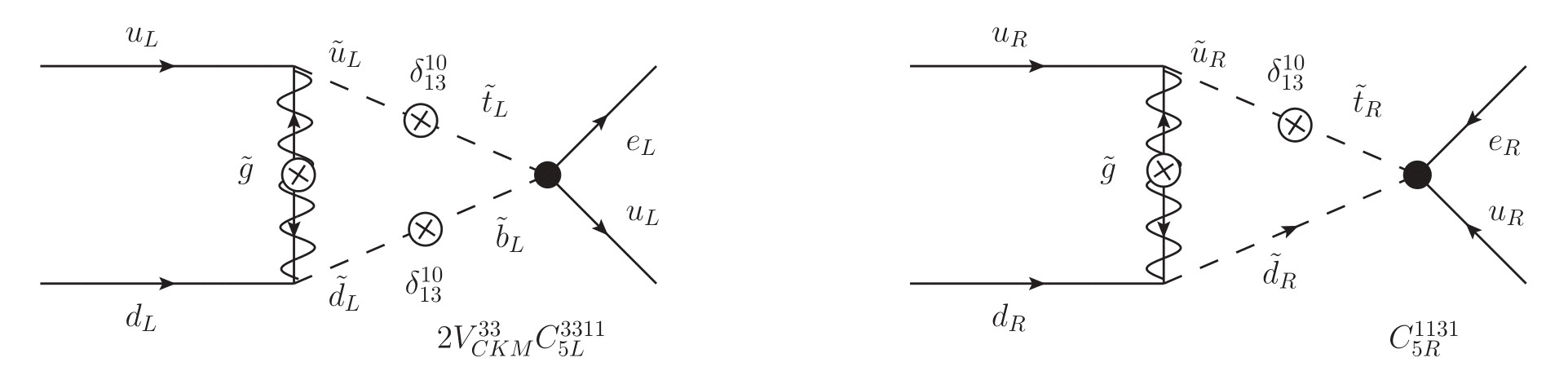} 
\caption{\label{LLRR} Gluino dressing diagram with $\delta_{13}^{10}$ insertion for $p\rightarrow \pi^0 e^+$.}
\end{figure}
\begin{figure} %
  	\centering
  	\subfigure[]{%
  		\label{fig:weaksplitting1}%
  		\includegraphics[width=7.5cm,height=5cm]{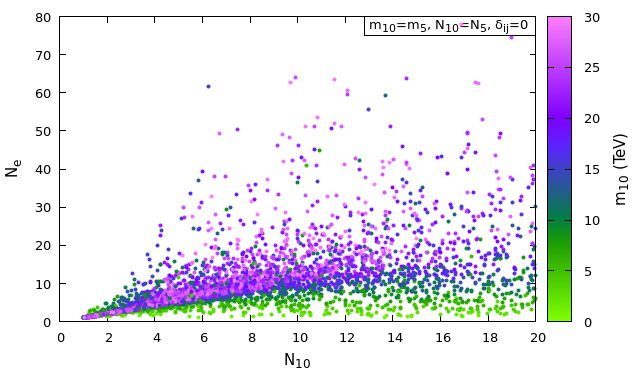}}%
  	\hfill
  	\subfigure[]{%
  		\label{fig:weaksplitting2}%
  		\includegraphics[width=7.5cm,height=5cm]{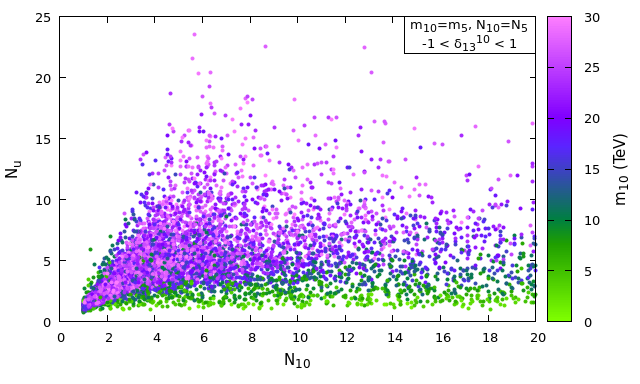}}%
  	\vskip\baselineskip
  	\subfigure[]{%
  		\label{fig:weaksplitting3}%
  		\includegraphics[width=7.5cm,height=5cm]{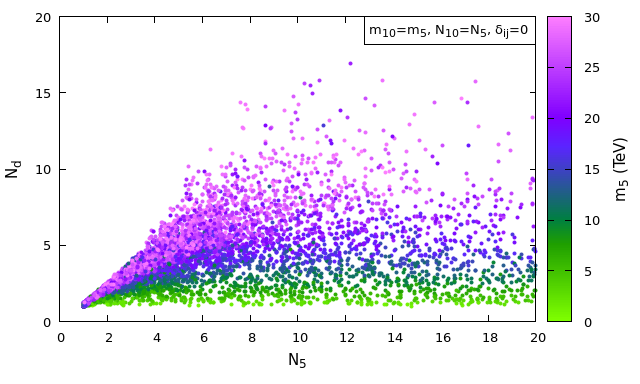}}%
  	\hfill
  	\subfigure[]{%
  		\label{fig:weaksplitting4}%
  		\includegraphics[width=7.5cm,height=5cm]{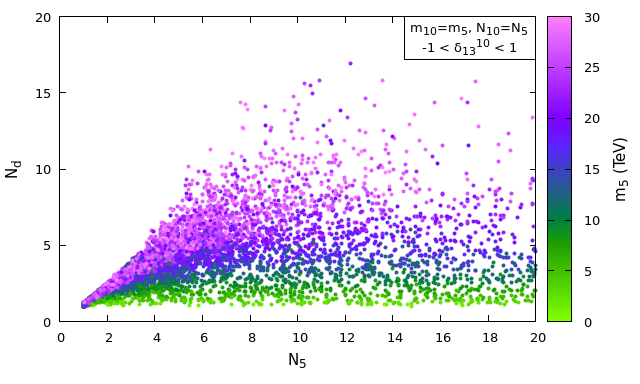}}%
  	\vskip\baselineskip
  	\subfigure[]{%
  		\label{fig:weaksplitting5}%
  		\includegraphics[width=7.5cm,height=5cm]{Pictures/protondecay/n10N2lm10.png}}%
  	\hfill
  	\subfigure[]{%
  		\label{fig:weaksplitting6}%
  		\includegraphics[width=7.5cm,height=5cm]{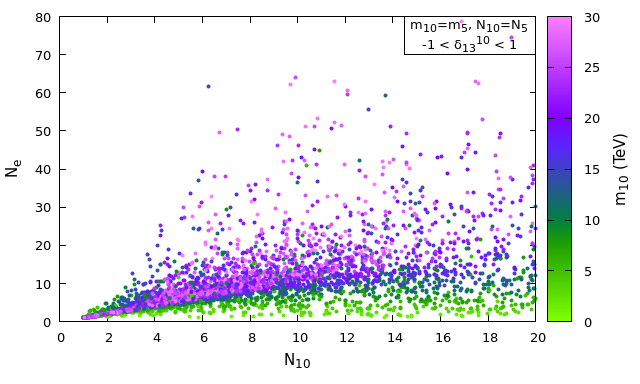}}%
  	\caption{These are plots of splitting in generation at GUT scale and weak scale. \label{weaksplitting}} 
\end{figure}
We consider generational splitting up to a factor of 20 but it could be even higher $\sim$ 30 to 40 (with large gluino mass see fig. (\ref{tacyon1},\ref{tacyon2})). The splitting factor being larger than 40 leads to negative third generation sfermion masses because of two loop RGE effect of heavy first two generations. The splitting at the weak scale could be understood from  fig. (\ref{weaksplitting}). In all the figures, projected scanned points have Higgs mass ($m_h$) in 123-128 GeV range, gluino mass ($m_{\tilde g}$) $>$ 2.2 TeV and stop mass ($m_{\tilde t}$) $>$ 1.2 TeV. In these figures, we show the GUT scale splitting (between the first two generations and third-generation) versus the same splitting at SUSY scale. Left and right panel plots are for without and with flavor mixing scenarios, respectively. Flavor mixing does not affect the weak scale splitting significantly. We can see weak splitting in up, down and lepton sfermion is quite different  because of RGE effects. Furthermore, relatively, the splitting between the slepton and squarks is also different, again because of RG effect. For example gauginos have positive contributions in RGEs, Yukawa term has negative and the first two generations at two-loop will drive the third generation more negative which increases the splitting between generations for slepton as compared to squarks. Comparing up and down sectors, the main difference arises because of large top Yukawa that drives stop mass more negative as compared to sbottom mass. Highest weak scale splitting region is  $5<N_{10}<15 $ with large $m_{10}$. The low GUT scale splitting with low $m_{10}$ is ruled out by the Higgs mass constraint while large GUT scale splitting with low $m_{10}$ drives third-generation masses very low and is ruled out by LHC bound. And if both GUT scale splitting and $m_{10}$ are large then two-loop effects become significant and spectrum becomes tachyonic.
\begin{figure}%
\centering
\subfigure[]{%
\label{fig:ppie1}%
\includegraphics[width=7.5cm,height=5.5cm]{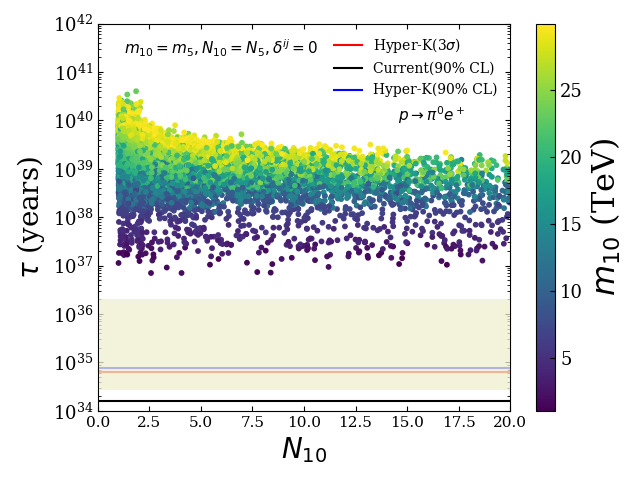}}%
\hfill
\subfigure[]{%
\label{fig:ppie2}%
\includegraphics[width=7.5cm,height=5.5cm]{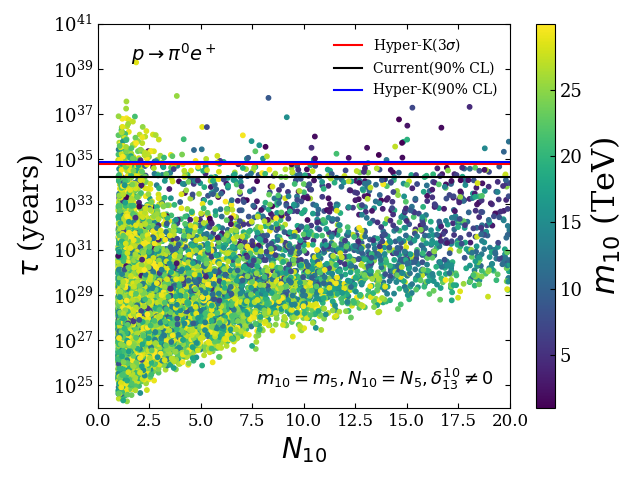}}%
\vskip\baselineskip
\subfigure[]{%
\label{fig:ppie3}%
\includegraphics[width=7.5cm,height=5.5cm]{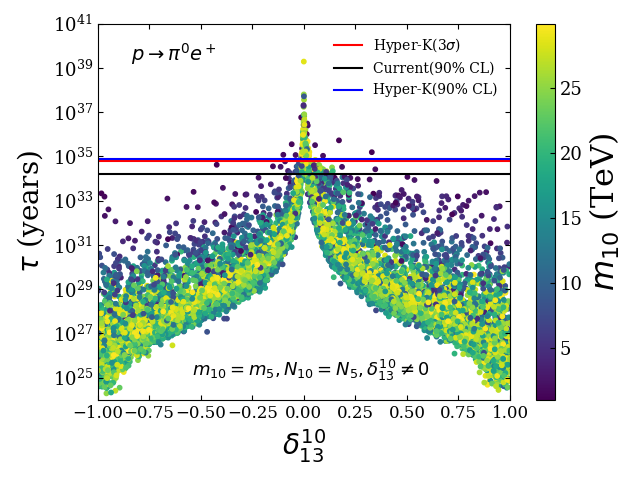}}%
\hfill
\subfigure[]{%
\label{fig:ppie4}%
\includegraphics[width=7.5cm,height=5.5cm]{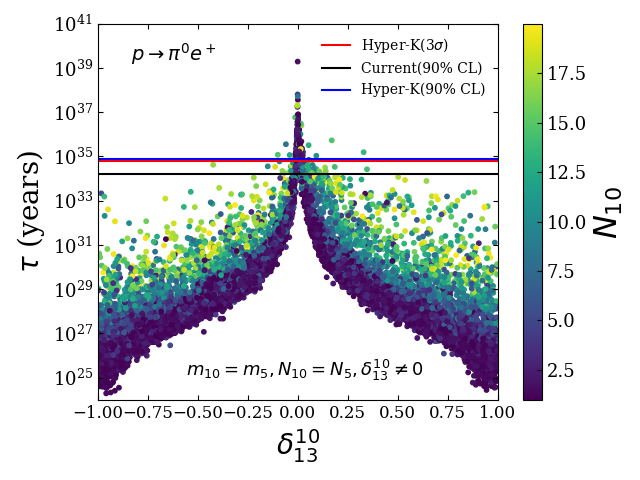}}%
\vskip\baselineskip
\subfigure[]{%
\label{fig:ppie5}%
\includegraphics[width=7.5cm,height=5.5cm]{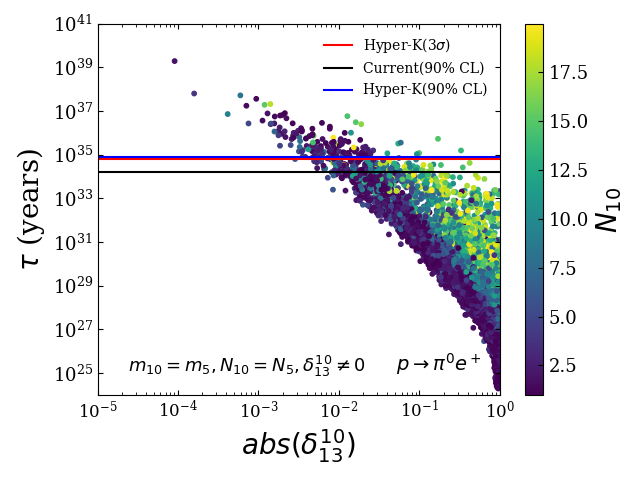}}%
\hfill
\subfigure[]{%
\label{fig:ppie6}%
\includegraphics[width=7.5cm,height=5.5cm]{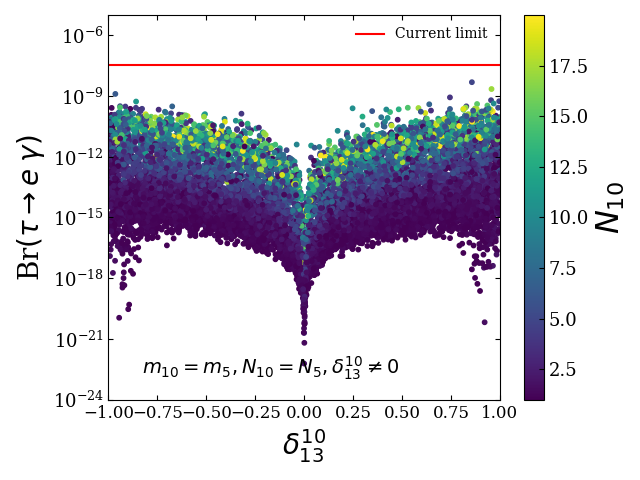}}%
\caption{\label{ppie}In these plots, we analyse proton decay mode $p\rightarrow e^+ \pi^0$ with and without flavor mixing. Fig. (f) is plot of Br$(\tau\rightarrow e \gamma)$ with flavor mixing.}
\end{figure}

We study three important modes $p\rightarrow \pi^0 e^+$, $p\rightarrow K^+ \bar{\nu_\tau}$ and $p\rightarrow K^+ \bar{\nu_e}$ of proton decay in detail with and without flavor mixing, in split-generation scenario. In section \ref{pdecaydetails}, we have discussed in detail the proton decay rate calculations.

For $p\rightarrow \pi^0 e^+$, future experiment providing highest sensitivity is Hyper-K\cite{Abe:2018uyc} and for $p\rightarrow K^+ \bar{\nu}$ DUNE experiment will be providing the highest sensitivity. 
In all figures, red line is for the highest future sensitivity and the black line is for current experimental bound.

In fig. (\ref{ppie}(a)), we show the proton life time for $p\rightarrow \pi^0 e^+$ channel. The shaded yellow region is the contribution coming from dimension six baryon violating operators. The dominant contribution in the case of no flavor mixing comes from these dimension 6 operators and will be probed in future experiments. We can see all data points are much above the current and future experimental limits. In this process without flavor mixing, only the chargino dressing diagram is most relevant which gives the highest contribution in LL and LR type of four fermion operators. With the GUT scale splitting, partial lifetime is not changing much except in the very low $N_{10}$ region; that is because of heavy degenerate spectrum which decreases the partial decay width (as also noticed by Shirai et. al). After including flavor mixing in fig. (\ref{ppie}(b)), gluino dressing diagram (see fig. (\ref{feyman})) starts contributing in LL and RR four fermion operators and  becomes the most dominating contribution that increases the partial decay width. This rules out a lot of parameter space from current and future experimental limits. In fig. (\ref{ppie}(b)) we can see the overall partial decay width is increasing with increasing $N_{10}$ but for very low value of $N_{10}$ the allowed partial lifetime varies from $10^{24}$ upto $10^{40}$ years.  This comes from the variation of $\delta_{13}^{10}$ value, see figs. (\ref{ppie}(c) and \ref{ppie}(d)). Large value of $\delta_{13}^{10}$ for low value of $N_{10}$ will destroy the degeneracy in the spectrum. Because of that, as shown in fig. (\ref{ppie}(d)), partial life time is decreasing as $|\delta_{13}^{10}|$ is increasing. If we look at figs. (\ref{ppie}(c), \ref{ppie}(d) and \ref{ppie}(e)), above the experimental limits there are both low and high $\delta_{13}^{10}$ value points. Large $\delta_{13}^{10}>.1$ have large $N_{10}$ and low $m_{10}$. These points have a highly degenerate low energy spectrum, see fig. (\ref{weaksplitting}) (around 5 TeV) at weak scale (large LR and flavor mixing causes degenerate low energy spectrum even with large GUT scale splitting). For a low $\delta_{13}^{10}$,  large $m_{10}$ with low $N_{10}$ will again have a heavy, almost degenerate, spectrum that will increase partial lifetime.  
In fig. (\ref{ppie}(f)) we calculate Br($\tau\rightarrow e \gamma$) w.r.t $\delta_{13}^{10}$. There is no bound on our parameter space from the current experiments of BR($\tau\rightarrow e \gamma$) measurements.
\begin{figure}%
\centering
\subfigure[]{%
\label{fig:pknue1}%
\includegraphics[width=7.5cm,height=5.5cm]{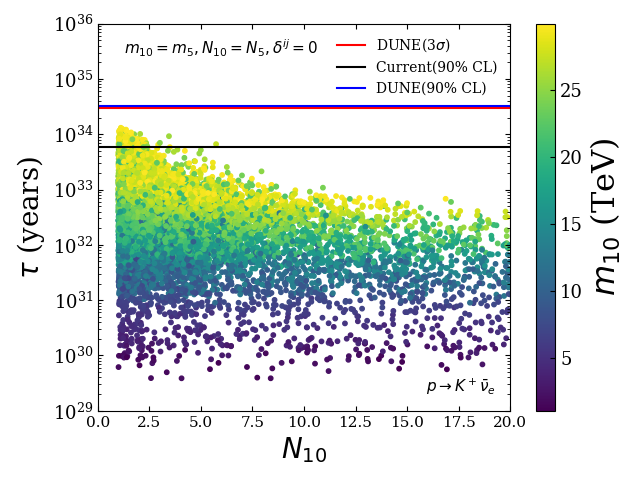}}%
\hfill
\subfigure[]{%
\label{fig:pknue2}%
\includegraphics[width=7.5cm,height=5.5cm]{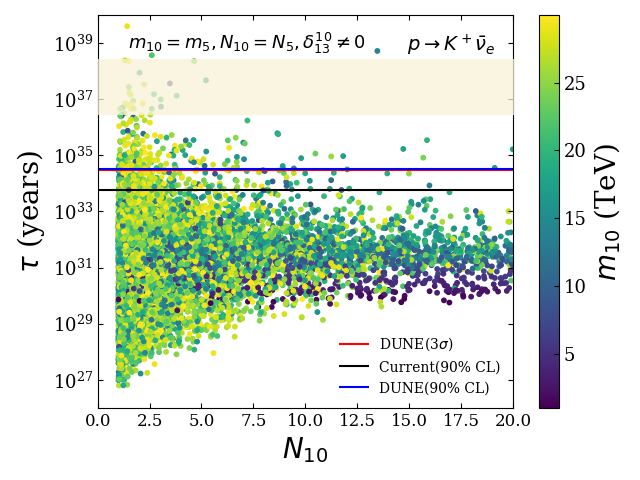}}%
\vskip\baselineskip
\subfigure[]{%
\label{fig:pknue3}%
\includegraphics[width=7.5cm,height=5.5cm]{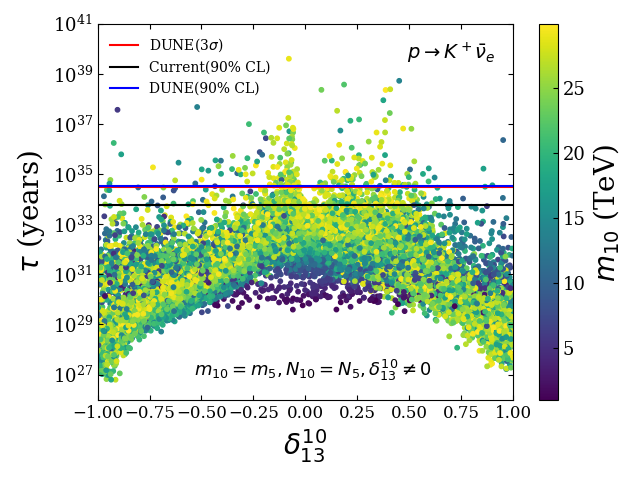}}%
\hfill
\subfigure[]{%
\label{fig:pknue4}%
\includegraphics[width=7.5cm,height=5.5cm]{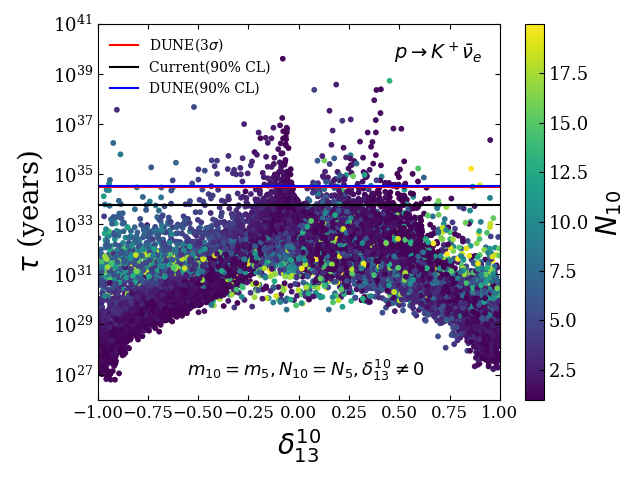}}%
\vskip\baselineskip
\subfigure[]{%
\label{fig:pknue5}%
\includegraphics[width=7.5cm,height=5.5cm]{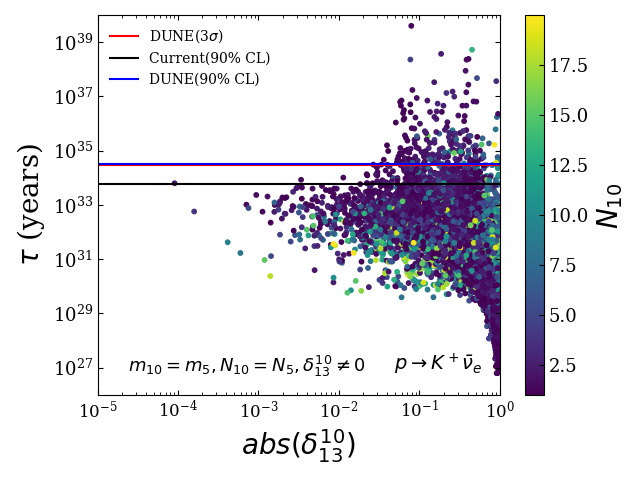}}%
\caption{\label{pknue}In these plots, we analyse proton decay mode $p\rightarrow K^+ \bar{\nu}_e$ with and without flavor mixing. }
\end{figure} 
Without flavor mixing, the decay channel $p\rightarrow K^+ \Bar{\nu}_e$ is the most constrained channel in SUSY-SU(5) model and is well known in the literature. As we can see in fig. (\ref{pknue}(a)), almost all points are ruled out by the current experimental bound, except for the low $N_{10}$ regions (which leads to the heavy, almost degenerate, spectrum). Furthermore, the allowed points will be completely covered in future DUNE experiment. So this channel will either be seen in the near future experiments or null results will rule out flavor diagonal SUSY SU(5) scenario. In this process without flavor mixing, dominant contribution comes from chargino dressing diagram in both LL type four fermion operators ($(ud)_L(s\nu)_L$) and ($(us)_L(d\nu)_L$). In figs. (\ref{pknue}(b)-\ref{pknue}(e)), when we add $\delta_{13}^{10}$, gluino and neutralino dressing contributions become relevant in addition to chargino contribution in both the above mentioned operators. These new contributions have opposite signs (depending on the sign of $\delta_{13}^{10}$ see in figs. (\ref{pknue}(c) and  \ref{pknue}(d)) w.r.t the chargino contribution which lead to both enhancement as well as cancellation in the decay rate see fig. (\ref{pknue}(e)). Because of these peculiar cancellations, even order one low $\delta_{13}^{10}$  values are allowed by future sensitivity. 
\begin{figure}%
\centering
\subfigure[]{%
\label{fig:pknutau1}%
\includegraphics[width=7.5cm,height=5.5cm]{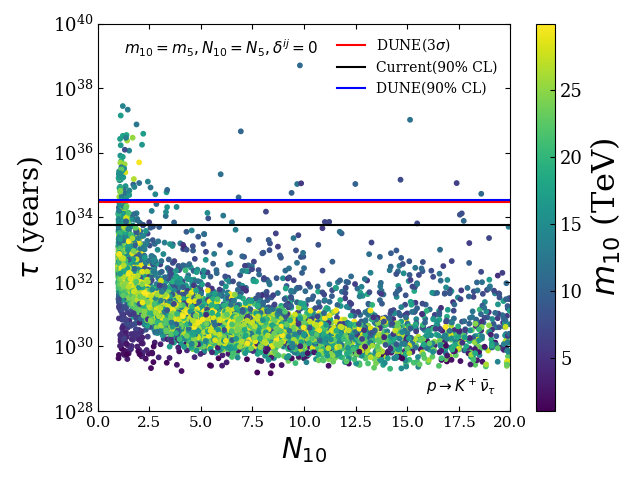}}%
\hfill
\subfigure[]{%
\label{fig:pknutau2}%
\includegraphics[width=7.5cm,height=5.5cm]{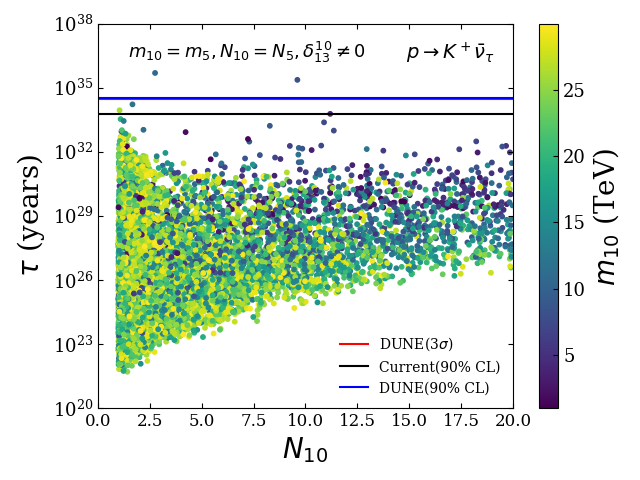}}%
\vskip\baselineskip
\subfigure[]{%
\label{fig:pknutau3}%
\includegraphics[width=7.5cm,height=5.5cm]{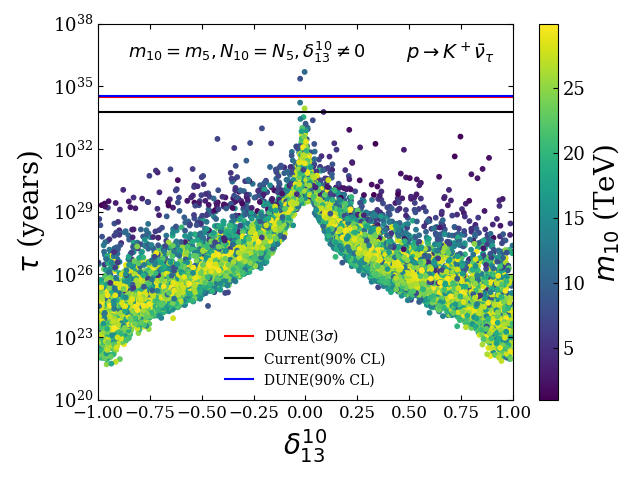}}%
\hfill
\subfigure[]{%
\label{fig:pknutau4}%
\includegraphics[width=7.5cm,height=5.5cm]{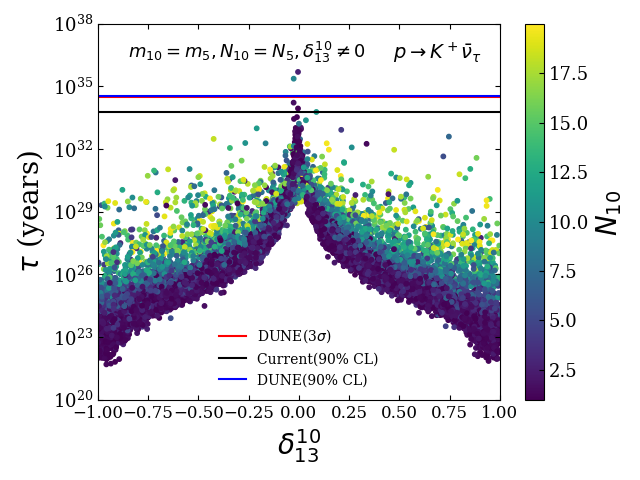}}%
\vskip\baselineskip
\subfigure[]{%
\label{fig:pknutau5}%
\includegraphics[width=7.5cm,height=5.5cm]{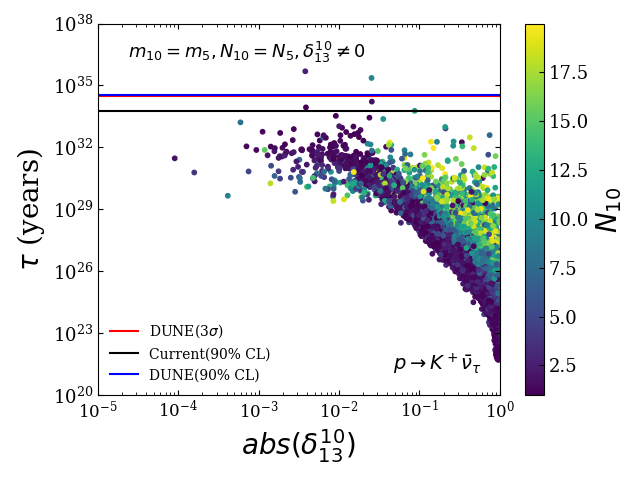}}%
\caption{\label{pknutau}In these plots, we analyse proton decay mode $p\rightarrow K^+ \bar{\nu}_{\tau}$ with and without flavor mixing. }
\end{figure}
In $p \rightarrow K^+ \nu_\tau$ channel, because of relatively larger $\tau$ Yukawa coupling, chargino dressed diagram also contributes in RL fermion operators in addition to LL  operator (present in the previous case). In fig. (\ref{pknutau}(a)), even without including flavor mixing, gluino and neutralino have relevant contribution in LL operators because tau have larger $C_{5L},C_{5R}$ contributions as compared to the electron case.  So there is already a cancellation region in fig. (\ref{pknutau}(a)) even without adding flavor. Adding flavor will only increase overall magnitude of partial decay width, reducing the life time of proton, see fig. (\ref{pknutau}(b)).


\section{Conclusion}
In this chapter, we have considered a novel decoupling scenario named `split-generation'. In this scenario first and second generation of sfermions are assumed heavy (order of 10s of  TeV) and remaining SUSY spectrum lies around couple of TeV.  We demonstrate the following phenomenological aspects of this scenario in case of minimal SUSY SU(5):
\begin{enumerate}
    \item Gauge coupling unification: We calculate full two-loop $\beta$-functions and SUSY thresholds for this scenario. We find that for large splitting the $\beta$-coefficients change the slope of gauge couplings significantly, but not the unification scale. Also, the exact unification always holds. While SUSY thresholds have negligible effect on slope of gauge couplings and unification scale.
    \item D=6 and D=5 baryon number violating operators contributing to proton decay: We study proton decay rates for two dominant channels ($p \rightarrow e^+ \pi^0$ and $p \rightarrow K^+ \bar\nu$) in minimal `split-generation' SUSY SU(5), both with and without flavor mixing in heavy and light (third) generation. The contribution of dimension 5 baryon number violating operators to the decay mode $p \rightarrow e^+ \pi^0$ is usually beyond the reach of future experiments, but here it is brought within the reach of Hyper-K, DUNE and JUNO due to the flavor mixing that opens up the gluino contribution to the amplitudes. The most dominating decay mode $p \rightarrow K^+ \bar\nu_e$ which essentially rules this model out for the range of masses we consider, is now able to survive and further interestingly can be explored at DUNE and Hyper-K due to peculiar cancellations in different dressing diagrams by introduction of flavor mixing. The split-generation in scalars has interesting consequences for the mode $p\rightarrow K^+ \bar\nu_\tau$
\end{enumerate}
We are still making progress in this work to account for all other decay modes and with other flavor mixings ($\delta$'s).

\begin{subappendices}
\section{Amplitude of dressing diagram of decay mode $p \rightarrow K^+ \bar{\nu}$ in term of $\delta$'s using the Mass Insertion Method. '}
\label{dressing}
In this section we calculate all dominating diagram with and without $\delta$'s for $p \rightarrow K^+ \bar{\nu}$. To our knowledge these amplitudes are presented for the first time in literature in Mass insertion format.
\subsection{Without delta}
\textit{Chargino Dressing} 
\begin{figure}[h!]
\centering
\includegraphics[width=15cm,height=10cm]{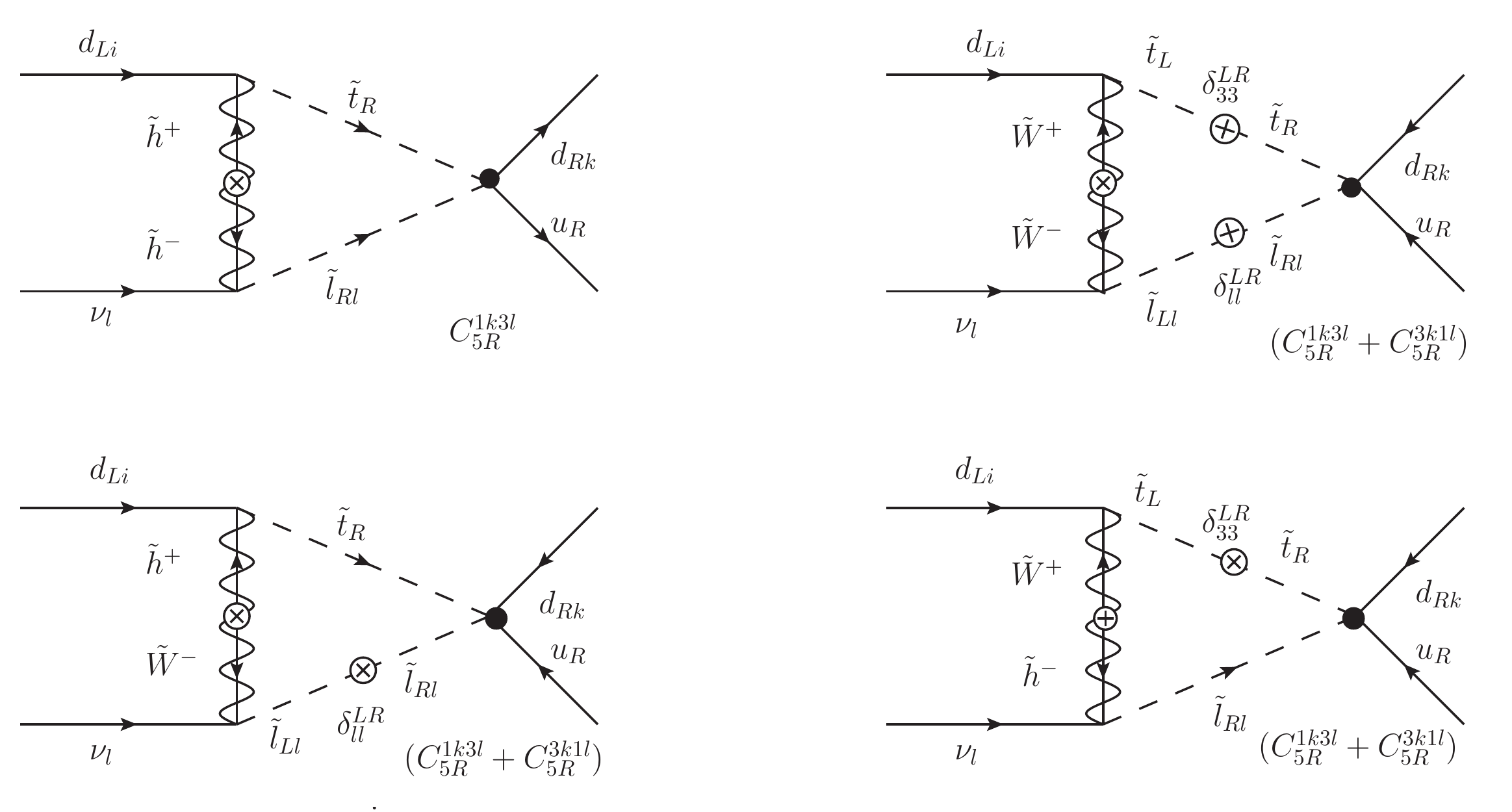} 
\caption{\label{withoutdeltaLR}Dominant chargino dressing diagrams of LR operators for $p\rightarrow K^+ \bar{\nu}$ without $\delta_{LL,RR}$ and with A-term. }
\end{figure}

\begin{figure}[h!]
\centering
\includegraphics[width=15cm,height=6cm]{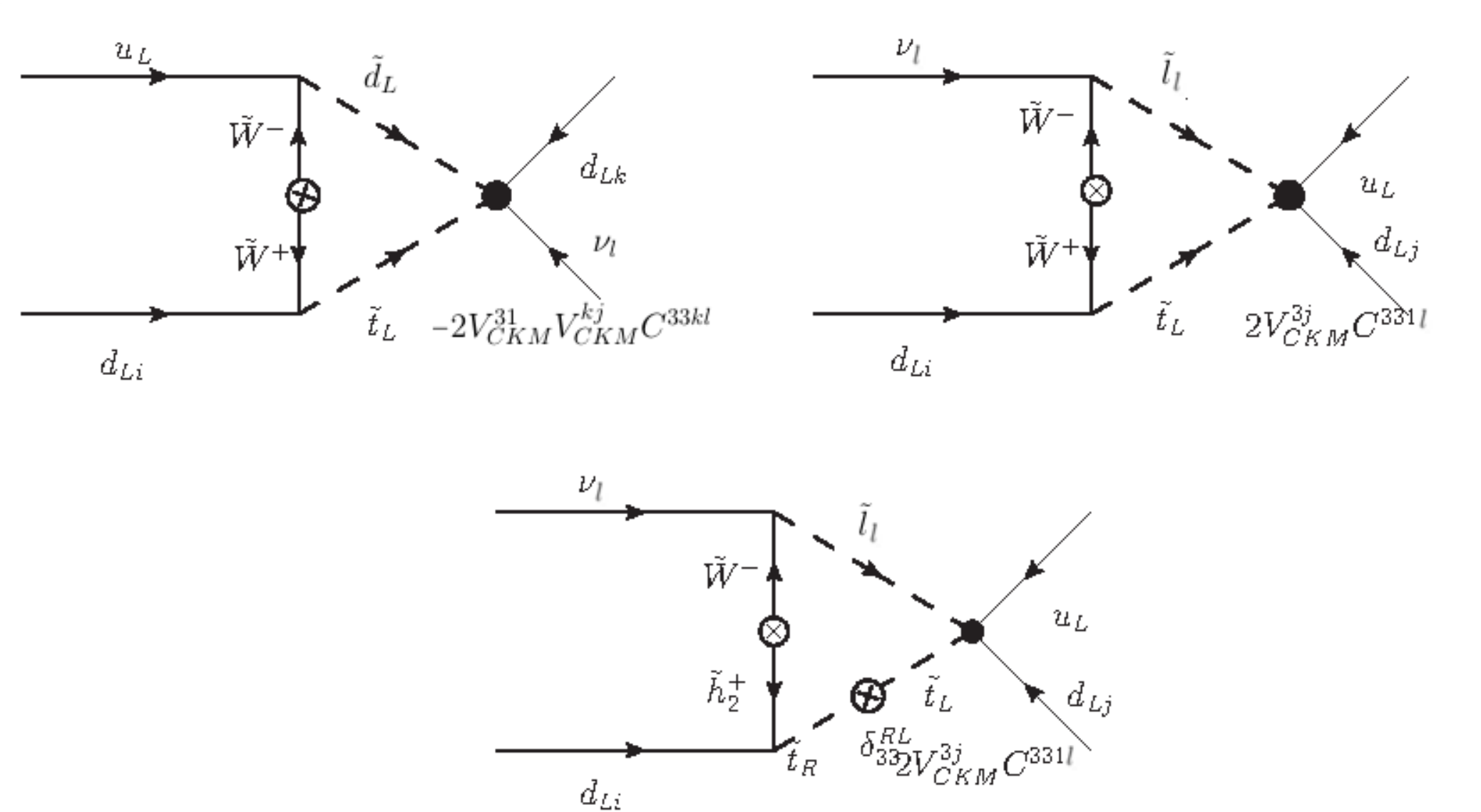} 
\caption{ Dominant chargino dressing diagrams of LL operators for $p\rightarrow K^+ \bar{\nu}$ without $\delta_{LL,RR}$ and with A-term. As $\tilde{W}^-\tilde{h_2}^+$ dressing there will be one with $\delta^{RL}$ insertion with $\tilde{W}^+\tilde{h_2}^-$ dressing.}
\label{feyman}
\end{figure}

The chargino dressing contribute in RL operators where diagram in fig. \ref{withoutdeltaLR} are the dominant contribution:
\begin{align}
C_{\tilde{h}^\pm}&=-[(C_{5R}^{113l}+C^{311l})V_{ts}+(C_{5R}^{123l}+C_{5R}^{321l})V_{td}]y_ty_{e_l}\mu P_L 2I(0,m_{\tilde{t}_R}^2,m_{\tilde{e}_{Rl}}^2)\nonumber \\
C_{\tilde{W}^\pm}&=-[(C_{5R}^{321l}+C^{123l})V_{td}+(C_{5R}^{311l}+C_{5R}^{311l})V_{ts}]g_2^2 M_2 m_t m_{e_l}(A_0-\mu \cot{\beta})(A_0-\mu\tan{\beta})P_L
 \nonumber\\
&\quad\quad 2I(0,m_{\tilde{t}_L^2},m_{\tilde{t}_R^2},m_{\tilde{e}_{Rl}^2},m_{\tilde{e}_{Ll}^2}) \nonumber \\
C_{\tilde{W}^\pm\tilde{h}^{\mp}}&=-[(C_{5R}^{321l}+C^{123l})V_{td}+(C_{5R}^{311l}+C_{5R}^{311l})V_{ts}]g_2\sqrt{2}M_WP_L
[y_t\sin{\beta}m_{e_l}(A_0-\mu\tan\beta)
\nonumber \\
&\quad\quad 2I(0,m_{\tilde{t}_{R}^2},m_{\tilde{e}_{Rl}^2},m_{\tilde{e}_{Ll}^2})+y_{e_l}\cot\beta m_t(A_0-\mu \cot\beta)2I(0,m_{\tilde{t}_{L}^2},m_{\tilde{t}_{R}^2},m_{\tilde{e}_{Rl}^2})]
\end{align}
The chargino dressing contribute in LL operators where diagram in fig. \ref{feyman} are the dominant contribution:\\
\begin{align}
   C_{\tilde{W}^\pm}&=-2C_{5L}^{33kl}(V_{CKM})_{31}((V_{CKM})_{k1}V_{ts}+(V_{CKM})_{k2}V_{td})g_2^2M_2 P_L2I(0,m_{\tilde{d}_L}^2,m_{\tilde{t}_L}^2)\nonumber \\
  &+2(C_{5L}^{113l}((V_{CKM})_{31}(V_{CKM})_{12}+(V_{CKM})_{32}(V_{CKM})_{11})+C_{5L}^{331l}((V_{CKM})_{31}(V_{CKM})_{32}\nonumber\\
  &+(V_{CKM})_{32}(V_{CKM})_{31}))g^2M_2P_L2I(0,m_{\tilde{e}_{Ll}}^2,m_{\tilde{t}_L}^2)\nonumber\\
  C_{\tilde{W}^\pm\tilde{h}^{\mp}}&=g_2y_{e_l}\sqrt{2}M_W\cos\beta m_{e_l}(A_0-\mu\tan\beta))P_L(2C_{5L}^{331l}((V_{CKM})_{31}(V_{CKM})_{31}+(V_{CKM})_{32}(V_{CKM})_{32})\nonumber\\
  &+2C_{5L}^{113l}((V_{CKM})_{31}(V_{CKM})_{11}+(V_{CKM})_{32}(V_{CKM})_{12}))2I(0,m_{\tilde{e}_{Rl}}^2,m_{\tilde{e}_{Ll}}^2,m_{\tilde{t}_{L}}^2)
\end{align}
\textit{Gluino Dressing}

\begin{figure}[h!]
\centering
\includegraphics[width=15cm,height=15cm]{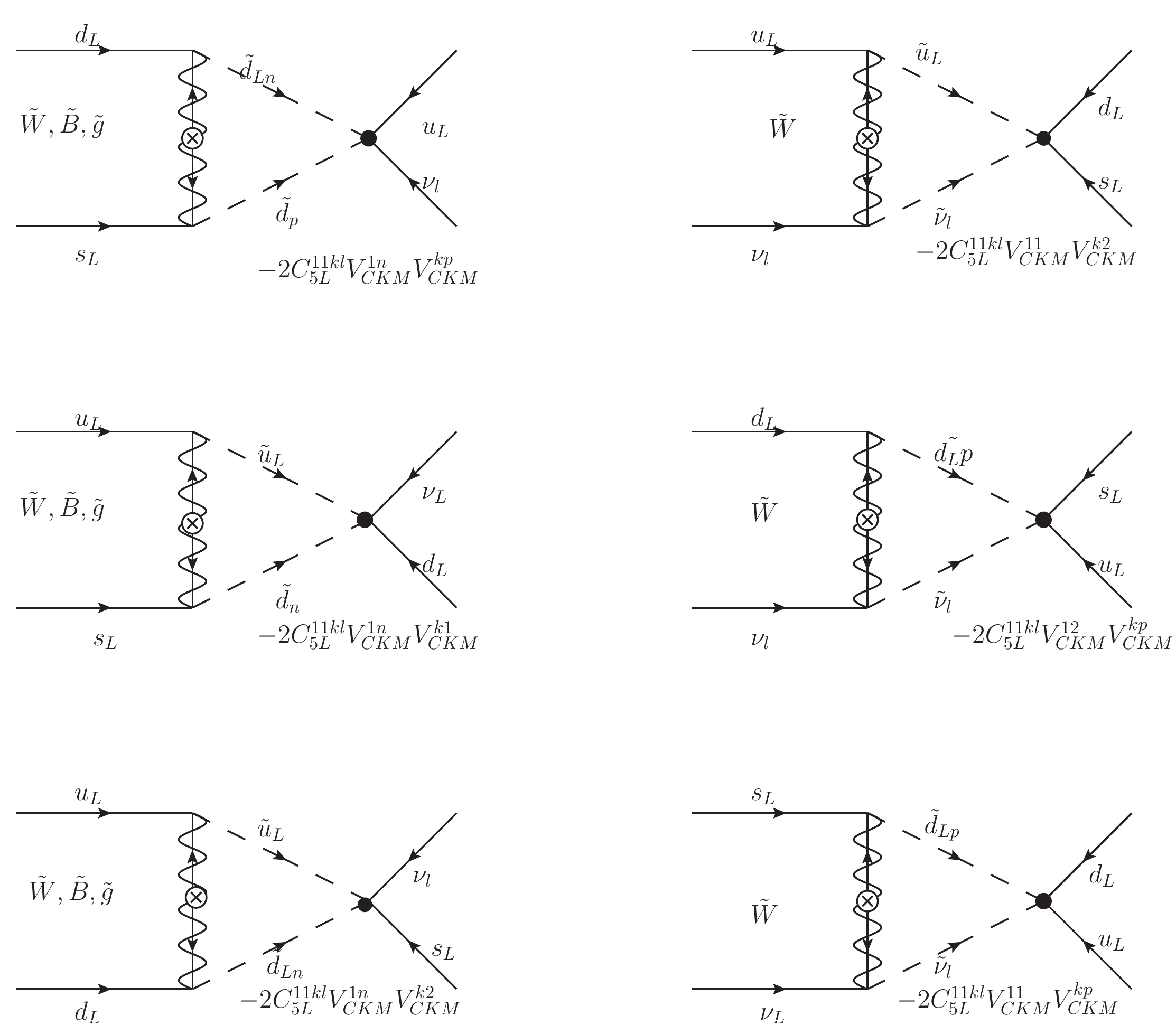} 
\caption{\label{withoutdeltaLL} Dominant gaugino dressing diagrams of LL operators for $p\rightarrow K^+ \bar{\nu}$ without $\delta_{LL,RR}$. }
\end{figure}

The gluino dressing contribute in LL operators where diagram in fig. \ref{withoutdeltaLL} are the dominant contribution:\\
\begin{align}
  C_{\tilde{g}}&=6g_3^2M_3(V_{CKM})_{P2}(V_{CKM})_{n1}(-2C_{5L}^{11kl}((V_{CKM})_{1n}(V_{CKM})_{kp}+(V_{CKM})_{1p}(V_{CKM})_{kn}))P_L\nonumber\\
  &*2I(0,m_{\tilde{d}_{Ln}}^2,m_{\tilde{d}_{Lp}}^2) 
  +6g_3^2M_3(V_{CKM})_{n2}(-2C_{5L}^{11kl}((V_{CKM})_{1n}(V_{CKM})_{k1})P_L2I(0,m_{\tilde{u}_{L}}^2,m_{\tilde{d}_{Ln}}^2)\nonumber\\
  &+  6g_3^2M_3(V_{CKM})_{n1}(-2C_{5L}^{11kl}((V_{CKM})_{1n}(V_{CKM})_{k2})P_L2I(0,m_{\tilde{u}_{L}}^2,m_{\tilde{d}_{Ln}}^2)
\end{align}
\textit{Bino Dressing}

The bino dressing contribute in LL operators where diagram in fig. \ref{withoutdeltaLL} are the dominant contribution:\\
\begin{align}
  C_{\tilde{B}}&=\frac{1}{18}g_2^2\tan^2\theta_w M_1(V_{CKM})_{P2}(V_{CKM})_{n1}(-2C_{5L}^{11kl}((V_{CKM})_{1n}(V_{CKM})_{kp}+(V_{CKM})_{1p}(V_{CKM})_{kn}))P_L\nonumber\\
  &*2I(0,m_{\tilde{d}_{Ln}}^2,m_{\tilde{d}_{Lp}}^2) 
  +\frac{1}{18}g_2^2\tan^2\theta_w M_1(V_{CKM})_{n2}(-2C_{5L}^{11kl}((V_{CKM})_{1n}(V_{CKM})_{k1})P_L2I(0,m_{\tilde{u}_{L}}^2,m_{\tilde{d}_{Ln}}^2)\nonumber\\
  &+  \frac{1}{18}g_2^2\tan^2\theta_w M_1(V_{CKM})_{n1}(-2C_{5L}^{11kl}((V_{CKM})_{1n}(V_{CKM})_{k2})P_L2I(0,m_{\tilde{u}_{L}}^2,m_{\tilde{d}_{Ln}}^2)
\end{align}
\textit{Wino Dressing} 

The wino dressing contribute in LL operators where diagram in fig. \ref{withoutdeltaLL} are the dominant contribution:\\
\begin{align}
  C_{\tilde{W}}&=\frac{1}{2}g_2^2M_2(V_{CKM})_{P2}(V_{CKM})_{n1}(-2C_{5L}^{11kl}((V_{CKM})_{1n}(V_{CKM})_{kp}+(V_{CKM})_{1p}(V_{CKM})_{kn}))P_L\nonumber\\
  &*2I(0,m_{\tilde{d}_{Ln}}^2,m_{\tilde{d}_{Lp}}^2) 
  +\frac{1}{2}g_2^2M_2(V_{CKM})_{n2}(-2C_{5L}^{11kl}((V_{CKM})_{1n}(V_{CKM})_{k1})P_L2I(0,m_{\tilde{u}_{L}}^2,m_{\tilde{d}_{Ln}}^2)\nonumber\\
  &+  \frac{1}{2}g_2^2M_2(V_{CKM})_{n1}(-2C_{5L}^{11kl}((V_{CKM})_{1n}(V_{CKM})_{k2})P_L2I(0,m_{\tilde{u}_{L}}^2,m_{\tilde{d}_{Ln}}^2)\nonumber\\
  &+\frac{1}{2}g_2^2M_2(-2C_{5L}^{11kl}((V_{CKM})_{11}(V_{CKM})_{k2}+(V_{CKM})_{12}(V_{CKM})_{k1}))P_L2I(0,m_{\tilde{u}_{L}}^2,m_{\tilde{\nu}_{Ll}}^2)\nonumber\\
  &+\frac{1}{2}g_2^2M_2(V_{CKM})_{p1}(-2C_{5L}^{11kl}((V_{CKM})_{12}(V_{CKM})_{kp})P_L2I(0,m_{\tilde{d}_{Lp}}^2,m_{\tilde{\nu}_{Ll}}^2)\nonumber\\
  &+  \frac{1}{2}g_2^2M_2(V_{CKM})_{p2}(-2C_{5L}^{11kl}((V_{CKM})_{11}(V_{CKM})_{kp})P_L2I(0,m_{\tilde{d}_{Lp}}^2,m_{\tilde{\nu}_{Ll}}^2)
\end{align}
\subsection{With delta}
\textit{Chargino Dressing} 

\begin{figure}[h!]
\centering
\includegraphics[width=15cm,height=15cm]{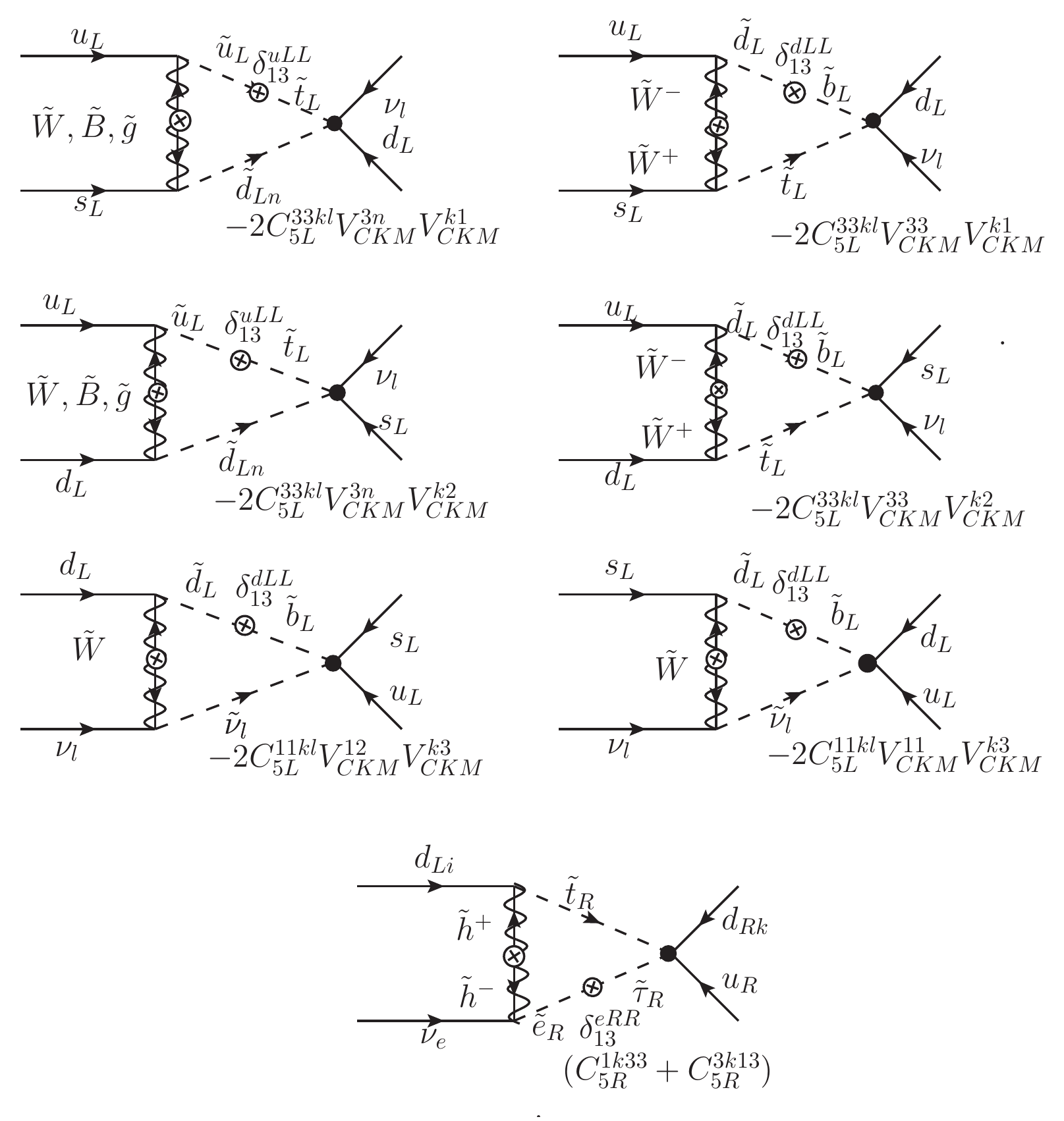} 
\caption{\label{withdelta} Dominant chargino and gaugino dressing diagrams of LL operators for $p\rightarrow K^+ \bar{\nu}$ with $\delta_{13}^{10}$. }
\end{figure}

The chargino dressing contribute in RL operators where diagram in fig. \ref{withdelta} are the dominant contribution:
\begin{align}
C_{\tilde{h}^\pm}&=-[(C_{5R}^{1133}+C^{3113})V_{ts}+(C_{5R}^{1233}+C_{5R}^{3213})V_{td}]y_ty_{e_1}\mu(\delta_{13}^{10}m_{\tilde{e}_R}m_{\tilde{\tau}_R})P_L2I(0,m_{\tilde{t}_R}^2,m_{\tilde{e}_{R1}}^2,m_{\tilde{e}_{R3}}^2)
\end{align}
The chargino dressing contribute in LL operators where diagram in fig. \ref{feyman} are the dominant contribution:\\
\begin{align}
   C_{\tilde{W}^\pm}&=-2C_{5L}^{33kl}(V_{CKM})_{33}((V_{CKM})_{k1}V_{ts}+(V_{CKM})_{k2}V_{td})g_2^2M_2(\delta_{13}^{10}m_{\tilde{d}_L}m_{\tilde{b}_L}) P_L2I(0,m_{\tilde{d}_L}^2,m_{\tilde{b}_L}^2,m_{\tilde{t}_L}^2)
\end{align}
\textit{Gluino Dressing} 

\begin{align}
  C_{\tilde{g}}&=6g_3^2M_3(V_{CKM})_{P2}(V_{CKM})_{11}(-2C_{5L}^{11kl}((V_{CKM})_{13}(V_{CKM})_{kp}+(V_{CKM})_{1p}(V_{CKM})_{k3}))P_L\nonumber\\
  &*2I(0,m_{\tilde{d}_{Ln}}^2,m_{\tilde{d}_{Lp}}^2)(\delta_{13}^{10}m_{\tilde{d}_L}m_{\tilde{b}_L})
  +6g_3^2M_3(V_{CKM})_{n2}(-2C_{5L}^{33kl}((V_{CKM})_{3n}(V_{CKM})_{k1})P_L(\delta_{13}^{10}m_{\tilde{u}_L}m_{\tilde{t}_L})\nonumber\\
  &*2I(0,m_{\tilde{u}_{L}}^2,m_{\tilde{t}_{L}}^2,m_{\tilde{d}_{Ln}}^2)
  + 6g_3^2M_3(V_{CKM})_{n1}(-2C_{5L}^{33kl}((V_{CKM})_{3n}(V_{CKM})_{k2})P_L\nonumber\\
  &*2I(0,m_{\tilde{u}_{L}}^2,m_{\tilde{t}_{L}}^2,m_{\tilde{d}_{Ln}}^2)(\delta_{13}^{10}m_{\tilde{u}_L}m_{\tilde{t}_L})
\end{align}
\textit{Bino Dressing} 

\begin{align}
  C_{\tilde{B}}&=\frac{1}{18}g_2^2\tan^2\theta_w M_1(V_{CKM})_{P2}(V_{CKM})_{11}(-2C_{5L}^{11kl}((V_{CKM})_{13}(V_{CKM})_{kp}+(V_{CKM})_{1p}(V_{CKM})_{k3}))P_L\nonumber\\
  &*2I(0,m_{\tilde{d}_{Ln}}^2,m_{\tilde{d}_{Lp}}^2)(\delta_{13}^{10}m_{\tilde{d}_L}m_{\tilde{b}_L})
  +\frac{1}{18}g_2^2\tan^2\theta_w M_1(V_{CKM})_{n2}(-2C_{5L}^{33kl}((V_{CKM})_{3n}(V_{CKM})_{k1})P_L\nonumber\\
  &*(\delta_{13}^{10}m_{\tilde{u}_L}m_{\tilde{t}_L})2I(0,m_{\tilde{u}_{L}}^2,m_{\tilde{t}_{L}}^2,m_{\tilde{d}_{Ln}}^2)
  + \frac{1}{18}g_2^2\tan^2\theta_w M_1(V_{CKM})_{n1}(-2C_{5L}^{33kl}((V_{CKM})_{3n}(V_{CKM})_{k2})P_L\nonumber\\
  &*2I(0,m_{\tilde{u}_{L}}^2,m_{\tilde{t}_{L}}^2,m_{\tilde{d}_{Ln}}^2)(\delta_{13}^{10}m_{\tilde{u}_L}m_{\tilde{t}_L})
\end{align}
\textit{Wino Dressing} 

\begin{align}
  C_{\tilde{W}}&=\frac{1}{2}g_2^2M_2(V_{CKM})_{P2}(V_{CKM})_{11}(-2C_{5L}^{11kl}((V_{CKM})_{13}(V_{CKM})_{kp}+(V_{CKM})_{1p}(V_{CKM})_{k3}))P_L\nonumber\\
  &*2I(0,m_{\tilde{d}_{Ln}}^2,m_{\tilde{d}_{Lp}}^2)(\delta_{13}^{10}m_{\tilde{d}_L}m_{\tilde{b}_L})
  +\frac{1}{2}g_2^2M_2(V_{CKM})_{n2}(-2C_{5L}^{33kl}((V_{CKM})_{3n}(V_{CKM})_{k1})P_L\nonumber\\
  &*(\delta_{13}^{10}m_{\tilde{u}_L}m_{\tilde{t}_L})2I(0,m_{\tilde{u}_{L}}^2,m_{\tilde{t}_{L}}^2,m_{\tilde{d}_{Ln}}^2)
  + \frac{1}{2}g_2^2M_2(V_{CKM})_{n1}(-2C_{5L}^{33kl}((V_{CKM})_{3n}(V_{CKM})_{k2})P_L\nonumber\\
  &*2I(0,m_{\tilde{u}_{L}}^2,m_{\tilde{t}_{L}}^2,m_{\tilde{d}_{Ln}}^2)(\delta_{13}^{10}m_{\tilde{u}_L}m_{\tilde{t}_L})
  +\frac{1}{2}g_2^2M_2(-2C_{5L}^{33kl}((V_{CKM})_{31}(V_{CKM})_{k2}+(V_{CKM})_{32}(V_{CKM})_{k1}))\nonumber\\
  &*P_L2I(0,m_{\tilde{u}_{L}}^2,m_{\tilde{t}_{L}}^2,m_{\tilde{\nu}_{Ll}}^2)(\delta_{13}^{10}m_{\tilde{u}_L}m_{\tilde{t}_L})
  +\frac{1}{2}g_2^2M_2(V_{CKM})_{11}(-2C_{5L}^{11kl}((V_{CKM})_{12}(V_{CKM})_{k3})P_L\nonumber\\
  &*2I(0,m_{\tilde{d}_{L}}^2,m_{\tilde{b}_{L}}^2,m_{\tilde{\nu}_{Ll}}^2)(\delta_{13}^{10}m_{\tilde{d}_L}m_{\tilde{b}_L})
  +  \frac{1}{2}g_2^2M_2(V_{CKM})_{21}(-2C_{5L}^{11kl}((V_{CKM})_{11}(V_{CKM})_{k3})P_L\nonumber\\
  &*2I(0,m_{\tilde{d}_{L}}^2,m_{\tilde{b}_{L}}^2,m_{\tilde{\nu}_{Ll}}^2)(\delta_{13}^{10}m_{\tilde{d}_L}m_{\tilde{b}_L})
\end{align}

\section{Physical Input Parameters}
 \begin{table}[h]
\centering
\begin{tabular}{|c|c||c|c|}
\hline\hline
$M_{t}$(pole) & 173.1\,\, \mbox{GeV} \quad \mbox{(PDG 2018)} & $m_u$ & 1.27 $\times 10^{-3}$\,\,\mbox{GeV} \quad \cite{Bajc:2015ita}\\
$M_H(pole)$ & 125.18\,\, \mbox{GeV}\quad \mbox{(PDG 2018)} & $m_d$ & $2.9 \times  10^{-3}$\,\, \mbox{GeV} \quad \cite{Bajc:2015ita}\\
$M_{W}$(pole) & 80.379\,\, \mbox{GeV}\quad \mbox{(PDG 2018)} & $m_e$ & $0.5 \times 10^{-3}$\,\,\mbox{GeV} \quad \cite{Bajc:2015ita}\\
$m_{\tau}$(pole) &   1.77686\,\,  \mbox{GeV}\quad \mbox{(PDG 2018)} & $m_c$ & 0.619\,\, \mbox{GeV} \quad \cite{Bajc:2015ita} \\
$m_b$(pole) & 4.78\,\,\mbox{GeV} \quad \mbox{(PDG 2018)} & $m_s$ & $55.0 \times 10^{-3}$\,\, \mbox{GeV}\quad \cite{Bajc:2015ita}  \\
$m_Z$(pole) & 91.1876\,\, \mbox{GeV}\quad \mbox{(PDG 2018)} &  $m_{\mu}$ & $103.0 \times 10^{-3}$\,\,  \mbox{GeV}  \quad \cite{Bajc:2015ita} \\
$m_{\pi^0}$ & 0.1349\,\, \mbox{GeV} \quad \mbox{(PDG 2018)} & $M_{Pl}$ & $2.4 \times 10^{18}$\,\, \mbox{GeV} \\
$m_{\pi^+}$ & 0.1395\,\, \mbox{GeV} \quad \mbox{(PDG 2018)} & $\alpha_s$ & 0.1181\quad \mbox{at $m_Z$ in $\overline{MS}$ scheme} \\
$m_{K^0}$ & 0.4976 \,\,\mbox{GeV} \quad \mbox{(PDG 2018)} & $\alpha_{em}^{-1}$ & 127.918 (\mbox{at $m_Z$ in $\overline{MS}$ scheme}) \\
$m_{K^+}$ & 0.4936\,\, \mbox{GeV} \quad \mbox{(PDG 2018)} & $\sin^2{\theta_w}$ & 0.231 \\
$m_{\eta^0}$ & 0.5478\,\, \mbox{GeV} \quad \mbox{(PDG 2018)} & \mbox{sgn}($\mu$) & 1 \\
$A_{ckm}$ & 0.836  & $l_{ckm}$ & 0.22453 \\
$\bar\rho_{ckm}$ & 0.122 & $\quad \bar\eta_{ckm}$ & 0.355\\
\hline
\end{tabular}
\caption{Physical parameter inputs}
\label{tab:tabel_2}
\end{table} 
  \end{subappendices}
\chapter{Outlook}
\label{Chapter 6}

\lhead{Chapter 6. \emph{Outlook}}

In this thesis, I worked on different aspects of heavy scale supersymmetry. This thesis is broadly divided into two parts. In the first part, we considered multiple hidden sectors of SUSY breaking and studied different low energy implications. We found exciting features like well-tempered or co-annihilating dark matter, almost negligible fine-tuning, novel solution for SUSY flavor problem and TeV scale spectrum with high probabilities of getting correct EWSB and Higgs mass.\\
In multiple hidden sectors scenario, we used gravity as mediation. This can further be applied for other mediation mechanisms like gauge mediation, whose phenomenology would be more varied than the one presented here. The key to low fine tuning with heavy supersymmetric spectrum seem to be having N sequestered hidden sectors; the theoretical and phenomenological aspects of this idea should be further explored. For example, it would be interesting to study the implementation of the N-sequesteredness in the full supergravity Lagrangian.  We can also explore how the multiple hidden sectors dynamics affects RG running.\\
Phenomenologically, it would be interesting to study in detail the Goldstini which are naturally present in this setting. 
In the limit of completely sequestered $N_{HS}$ supersymmetric sectors, we have an N-fold enhanced Poincare symmetry\cite{Cheung:2010mc}. When spontaneous SUSY-breaking occurs in each sector through F-term breaking, there are massless Majorana spin 1/2 fermion in each sector, known as goldstino. Then, SUSY breaking in $N_{HS}$ hidden sectors will provide us with $N_{HS}$ goldstino.  One  linear combination of these $N_{HS}$ goldstino (linear combination of goldstino is known as goldstini) is eaten by gravitino via the super-Higgs mechanism, the remaining $(N_{HS}-1)$ goldstini are in the spectrum. Due to the gravitation effect, they receive mass at tree-level that is twice of gravitino mass. They also have a direct coupling to Supersymmetric Standard Model fields.
Each of the Goldstini  couplings and mass are  dependent on the  individual dynamics of all the sectors. And if it is lighter than the lightest supersymmetric particle, it will affect cosmology and be stable. In any case, it would be interesting to see the cosmological implications of $N$ sequestering. 
Multiple goldstini would also be interesting in gauge mediation models. Formally it would also be interesting to study the R-symmetry breaking mechanisms in these models. \\

In the second part of the thesis, we studied  proton decay in the minimal SUSY-SU(5) model with flavor violation in the split-generation scenario. The conclusions of this work are presented at the end of Chapter 5. The results of this work entail further detailed phenomenological exploration in terms of emitted neutrino flavors at specific experiments like DUNE and Hyper-K.


\addtocontents{toc}{\vspace{2em}} 

\appendix 


\chapter{Minimal Supersymmetric Standard Model} 

\label{Appendix A} 

\lhead{Appendix A. \emph{MSSM}} 

The Standard Model is the most successful theory in particle physics. It describes the strong and electroweak interactions. It is based on the gauge symmetry group $SU(2)_L\times U(1)_Y$ of weak left-handed isospin and hypercharge, respectively, and the strong interaction between colored quarks based on the symmetry group $SU(3)_c$.
High-precision measurements that have been carried out in the previous century at LEP have provided a vital test of the Standard Model and ensure a comprehensive description of the strong and electroweak interactions, at present energies.  These high-precision tests give us information at the quantum correction level. Even though many observed processes test its accuracy, several experimental hints suggest the need to extend the SM for attaining a full picture. The phenomenological and theoretical needs to go beyond the Standard Model include:
\begin{itemize}
	\item The 20th-century studies evidenced the existence of dark matter. 25$\%$ of the present energy density of the Universe is known to be Dark Matter. However, we do not know anything about its fundamental nature. A stable, reasonably massive, electrically neutral particle with very weak interactions is a good candidate for dark matter. The SM does not include any such candidate particle and thus calls for New physics to explain dark matter in the Universe. 
	\item The discovery of a neutral scalar boson with mass 125.09 GeV completes the SM, but from the theoretical point of view, "light" scalars are unnatural in SM due to huge radiative corrections from quadratic divergences. Even if no new physics is introduced, then radiative corrections will become very large because of the cut-off scale $M_{Pl}$. In that case, for correct Higgs mass, we have to tune our bare tree parameter at the level of $10^{-32}$, and that looks very unnatural from a theoretical perspective. It is known as the "Hierarchy problem". For a natural theory, either these divergences should be canceled using some symmetry of new physics or there must exist a new TeV-scale physics that will make the Higgs insensitive to any higher scales. 
	\item One of the big problems in particle physics is matter dominance over antimatter in the Universe. The BBN and CMB measurements in different experiments also give evidence of Baryon asymmetry while there is no baryon asymmetry in SM processes. Thus, we need new physics to explain it.
	\item The existence of instanton solutions in nonabelian gauge theories needs $\theta$ vacuum\cite{Callan:1976je,Jackiw:1976pf} which introduces an effective CP-violating interaction term in the QCD Lagrangian ($\frac{\theta}{32\pi^2}G_{\mu\nu}^a\bar{G}^{a\mu\nu}$). This directly enhances the electric dipole moment of the neutron. The observed neutron EDM\cite{Burghoff:2011xk} puts very tight bound on the $\theta$ value, requiring it to be a very small parameter. If we set it to zero by hand, it should enhance SM's symmetry, but it does not, like Higgs mass. Thus it is natural to expect for it to have an $\mathcal{O}(1)$ value and the problem is ``why it should be small''? Unlike Higgs mass, if we set $\theta$ to be small at some scale, it stays small by SM RG evolution.
 \item  The analysis of vacuum stability is sensitive to quantum corrections. An NNLO analysis of vacuum stability in the SM tells us that the requirement of the vacuum to be stable up to the Planck scale gives \cite{Degrassi:2012ry}:
 \be
 m_h > 129.4 + 1.4\left(\frac{m_t-173.1}{0.7}\right)-0.5\left(\frac{\alpha_s(m_Z)-0.1184}{0.0007}\right) \pm 1.0
 \ee
 where masses are in GeV. ``$\pm1.0$'' is the theoretical error in the evaluation of $m_h$. The experimental errors on the top mass and the analysis of ref. \cite{Degrassi:2012ry}, gives 
 \be
 m_h > 124.4 \pm 1.8 GeV
\ee
This excludes the vacuum stability for $m_h < 126$ GeV in SM at the 2$\sigma$ level. The discovery of Higgs mass $125.01$ GeV renders EW vacuum into meta-stability around $10^{10}$ GeV. Thus, the stability of the electroweak vacuum up to the Planck scale would require new physics.
\item Neutrinos are elementary particles with zero electric charge and three flavors (the electron neutrino, the muon neutrino, and the tau neutrino). They interact via weak fundamental force. The SM has only left-handed neutrinos. Therefore they are massless in the SM. However, the discovery of `neutrino oscillations' confirmed, neutrinos are massive\cite{Fukuda:1998mi}. The neutrino oscillation is the change of one flavor of neutrino into another, and depends on the difference in the squares of neutrino masses. We need new physics to explain the mixing and masses of neutrinos. 
\item The SM is a unified electroweak theory that motivates us to study the unification of the electroweak and strong interactions theoretically. Nevertheless, SM does not have an exact unification of electroweak and strong interactions. Thus, a new fundamental Grand Unified Theory (GUT) is required for an exact unification.
\end{itemize}
In addition to these, there is a lot more at QCD non-perturbative scale that we do not understand completely. As a consequence, many models have been constructed as attempts to answer these questions. For example, to solve the hierarchy problem, many solutions have been proposed in literature such as SUSY, composite Higgs, Warped extra dimensions models, Technicolor, Large extra dimensions, anthropic explanations in the Multiverse relaxion type mechanisms, etc. Similarly, the dark matter problem has been investigated from various directions such as the effective operator approach, extra U(1) and kinetic mixing model, Higgs portal, Z' portal, anomalies as dark matter, axionic dark matter, SUSY, etc. 
One of the solutions for these problems, "Supersymmetry", is discussed here in detail. It is one of the most liked and attractive extensions of the Standard Model in the literature.

Supersymmetry is both a theoretically and phenomenologically rich theory. It is a non-trivial extension of the Poincare group in QFT, which predicts the existence of a partner to all particles, which differs by spin 1/2. It incorporates gravity if it is local. These features make it is a theory of all known fundamental interactions.  However, the most attractive parts of SUSY are the phenomenological ones. The softly-broken SUSY theories can solve many problems mentioned above. The main reason for introducing low energy SUSY theories was to solve the hierarchy problems. This new symmetry prevents the Higgs boson mass from getting large radiative corrections. If symmetry is exact, there would be no radiative corrections to Higgs boson mass due to cancellation between radiative correction coming from the SM particle and its partner particle (called superpartner), which differs by spin 1/2. This cancellation stabilizes the GUT and electroweak scale hierarchy, so we do not need to fine-tune the bare Higgs parameter. Superpartners have not been observed experimentally, which tells us SUSY is not an exact symmetry; they should have masses much larger than their SM partners. Thus SUSY must be broken. There are several problems with SUSY breaking that are discussed in \ref{Appendix B}. If SUSY is broken softly, then one need not reintroduce the hierarchy problem. It helps to understand the origin of the hierarchy problem and the source of electroweak symmetry breaking in radiative gauge symmetry breaking. Thus, the low energy softly-broken SUSY provides possible solutions to the hierarchy, dark matter, and gauge unification problems from the eighties. Extensive efforts have been made to understand the pattern of soft SUSY-breaking mechanism, Lagrangian, and towards a determination of the predicted superpartners' properties. 
Following is a summarised overview of the Minimal Supersymmetric Standard Model (MSSM) that establishes the notations and terms used in the thesis (The discussion closely follows the lectures by Prof. Vempati presented in ref .\cite{Vempati:2018pph}).
\section{ The Basics of MSSM}
The MSSM is a minimal set of particles required for the Supersymmetric Standard model. As we know, supersymmetry transforms a fermion into a boson and vice-versa. So the quantum fields of particles are upgraded to superfields, which are multiples of fermion-boson pairs that transform into each other. Superfields are functions of ordinary space and two fermionic directions ($x_{\mu},\theta,\bar{\theta}$). In MSSM, we use two kinds of superfields, chiral superfields, and vector superfields.
A chiral superfield has an expansion :

\be
\Phi = \phi + \sqrt{2} \theta \psi + \theta \theta F,
\ee 
where $\phi$ is the scalar, $\psi$ is the Weyl spin 1/2 fermion
and $F$ is the auxiliary scalar field. A vector superfield in Wess-Zumino gauge has an expansion :
\be
V = -\theta \sigma^\mu \bar{\theta} A_\mu + i \theta \theta \bar{\theta} \bar{\lambda} - i \bar{\theta} \bar{\theta} \theta \lambda + {1 \over 2} \theta \theta \bar{\theta} \bar{\theta} D 
\ee 
where $A_{\mu}$ is the vector spin one boson, $\lambda,\bar{\lambda}$ are the Majorana spin 1/2 fermion, and $D$ is the auxiliary scalar field. $\theta ,\bar{\theta}$ are Grassman variables. The MSSM Lagrangian is written in terms of these superfields. Like the Yukawa couplings, the interaction part of the Lagrangian is written in superpotential, which is a holomorphic function of chiral superfields. That means the superpotential can either be purely a function of superfields or purely a function of its conjugate superfields. The kinetic part is written using vector superfields such that the combination remains gauge invariant. In the MSSM, every SM matter field is replaced by a chiral superfield, and a vector superfield replaces every vector field. In doing so we use the conjugates of right-handed particles as superfields instead of the right-handed particles themselves, and we need two Higgs superfields to give masses to all fermions. One Higgs doublet superfield gives mass to up-type quarks, and the other gives mass to down-quarks and leptons. The particle spectrum of the MSSM and their 
transformation properties under the SM gauge group are given by,

\bea
\label{mssmspectrum}
 H_1 \equiv \left(\begin{array}{cc} H^{0}_1 & \tilde{H}_1^0 \\ H^{-}_1 & \tilde{H}_1^{-} 
\end{array}\right) 
\sim \left( 1,~2,~-{1 \over 2} \right) &\;\;& H_2 \equiv  \left(\begin{array}{cc} H^+_2 & \tilde{H}_2^+  \\ H^0_2 & \tilde{H}_2^0 \end{array}\right)\sim \left( 1,~2,~{1 \over 2} \right) \nonumber\\
Q_i \equiv \left(\begin{array}{cc}
 u_{L_i} & {\tilde u}_{L_i} \\
 d_{L_i} & \tilde{d}_{L_i}
\end{array}\right) \sim 
\left( 3,~2,~{1 \over 6} \right)
&\;\;& U_i^c \equiv \left(\begin{array}{cc} u_i^c  & \tilde{u}^{c}_i \end{array}\right)
 \sim \left( \bar{3},~ 1,~ -{2 \over 3} \right) \nonumber \\
&\;\;& D_i \equiv \left(\begin{array}{cc} d_i^c & \tilde{d}^{c}_i  \end{array}\right)
\sim \left( \bar{3},~ 1,~ {1 \over 3} \right) \nonumber \\
L_i \equiv \left(\begin{array}{cc} \nu_{L_i} & \tilde{\nu}_{L_i}  \\ e_{L_i}& 
\tilde{e}_{L_i} \end{array}\right) 
\sim \left( 1,~ 2,~ -{1 \over 2} \right)
&\;\;& E_i \equiv \left(\begin{array}{cc} e_i^c & \tilde{e}^{c}_i \end{array}\right) \sim 
 \left( 1,~ 1,~ 1 \right)
\eea
where  `i' is the generation index, and the `\~{}' fields are superpartners of SM fields. The scalar Higgs has its fermionic superpartner called `Higgsinos'. Similarly, fermions have scalar superpartners called `sfermions'. For example, the top quark superpartner is known as `stop'. The c in superscript indicates that this chiral superfield is made up of conjugates of SM right-handed fields. There are also gauge bosons, and their superpartners called gauginos:
\bea
V_s^A & : & \left(\begin{array}{cc} G^{\mu A} & \tilde{G}^A \end{array}\right) ~~ \sim~~ (8,1,0) \nonumber \\
V_W^I & : & \left(\begin{array}{cc} W^{\mu I}& \tilde{W}^I \end{array}\right) ~~~\sim~~ (1,3,0)  \nonumber \\
\label{vsf}
V_Y & :  &\left(\begin{array}{cc} B^{\mu} & \tilde{B} \end{array}\right)~~\sim~~(1,1,0)
\eea
where index `A' and `I' runs from 1 to 8 and 1 to 3, respectively. The $G$'s,$W$'s and $B$'s 
are the $SU(3)_c$, $SU(2)$ and $U(1)$ gauge bosons, respectively. And the `\~{}' fields are their superpartners `gluinos', `winos' and `bino' respectively.
\section{The MSSM Lagrangian}
In general, the supersymmetric invariant Lagrangian is constructed from three functions of superfields.
\begin{itemize}
    \item The K{\"a}hler potential, $K$, which is a real function of the superfields. 
 \item The superpotential $W$, which is a
holomorphic (analytic) function of the superfields.
\item The gauge kinetic function $f_{\alpha \beta}$ which appears in supersymmetric gauge theories as the coefficient of the product of field strength superfields,  $\mathcal{W}_\alpha \mathcal{W}^\beta$. $f_{\alpha \beta}$ determines the normalisation for the gauge kinetic terms. In MSSM, $f_{\alpha \beta}=\delta_{\alpha \beta}$
\end{itemize}
 For the MSSM, the K{\"a}hler potential is:
 \be
 \label{kahlar}
 \mathcal{L}_{kin} = \int d\theta^2 d\bar{\theta}^2 \sum_{\scriptstyle{\text{SU(3)},\text{SU(2)},\text{U(1)}}} 
\Phi_\beta^\dagger~ e^{ gV} \Phi_\beta 
 \ee
where the index $\beta$ runs over all the matter fields in eq.(\ref{mssmspectrum}), this part of MSSM Lagrangian gives us the kinetic terms for the SM fermions, sfermions, Higgs fields, and Higgsinos. It also gives gauge boson interaction with fermion, sfermion, and Higgs fields. Also, they give Yukawa type terms fermion-sfermion-gaugino and Higgs-Higgsino-gaugino vertices.

 The superpotential W is the holomorphic(analytic) function of the superfields. If we demand renormalisability, then dimension of $[W]\leq3$, which allows only products of three or less number of chiral superfields. Then the SM gauge invariant form of superpotential for the MSSM is:
 \be
\mathcal{L}_{yuk}  = \int d\theta^2~~ W (\Phi) + 
\int d\bar{\theta}^2~~ \bar{W}(\bar{\Phi})
\ee
 where $W=W_1+W_2$ and
 \bea
\label{superyuk}
W_1 &=& h^u_{ij} Q_i U_j^c H_2 + h^d_{ij} Q_i D_j^c H_1 + 
h_{ij}^e L_i E_j^c H_1 + \mu H_1 H_2 \\
\label{superrpv}
W_2&=& \epsilon_i L_i H_2 + \lambda_{ijk} L_i L_j E_k^c + \lambda'_{ijk} L_i Q_j D_k^c 
+ \lambda''_{ijk} U_i^c D_j^c D_k^c
\eea
The $W_1$ part of Lagrangian gives us the interaction terms independent of the gauge coupling (Yukawa terms), Higgsino mass, the fermion-sfermion-higgsino couplings, and scalar terms. The $W_2$ part of Lagrangian violates baryon ($\mathrm{B}$) and family
lepton numbers ($\mathrm{L}_{e,\mu,\tau}$). These terms arise because the fermions and the bosons have the same quantum number under the SM gauge group. For example, the Higgs superfield, $H_1$, and the lepton superfields $L_i$ carry the same quantum numbers under the SM gauge group.
They both are chiral superfields. There is no way of distinguishing between them. Therefore, the $W_2$ part of the superpotential appears in the supersymmetric version of the SM but not in SM. These interactions can lead to large proton decay widths. As we know experimentally, a proton is quite stable, which tells us the couplings in $W_2$ terms must to be taken to be very small. A more natural way of dealing with such small numbers for these couplings is to set them to zero. This is achieved by imposing a discrete symmetry on the Lagrangian called R-parity. R-parity is defined as:
\be
R_p = (-1)^{3 (B-L) + 2s},
\ee
where  B and L represent the Baryon and Lepton number, respectively, and s represents the spin of the particle. The R-parity of any particle is  the opposite from its superpartner. It makes sure that every interaction vertex has at least two supersymmetric partners, when R-parity is conserved. The lightest supersymmetric particle (LSP) cannot decay into a pair of SM particles and remains stable, becoming a naturally good candidate for dark matter.

The Lagrangian, which gives the kinetic terms for the field strength tensors for the gauge fields and kinetic terms for gauginos, is:
 \be
\mathcal{L}  = \int d\theta^2~~ \mathcal{W}^\alpha\mathcal{W}_\alpha + 
\int d\bar{\theta}^2~~ \mathcal{W}^{\dot{\alpha}}\mathcal{W}_{\dot{\alpha}}
\ee
where $\mathcal{W}^\alpha$ is field strength superfield. Every vector superfield has an associated field strength superfield. Putting together all these three Lagrangian terms, we have the Lagrangian for MSSM with SUSY unbroken.

As we know, F and D are auxiliary fields, and they are not physical. We have to remove these fields using equations of motions (on-shell), which have solutions as:
\be
F_i=\frac{\partial W}{\partial \phi_i} ~~~~\;;\;~~~~~D_A=-g_A  ~\phi^\star_i~T^A_{ij}~\phi_j
\ee
where index A runs over all gauge groups. The F and D terms together form the scalar, which is given as
\be
V = \sum_i ~|F_i~|^2 +~\frac{1}{2}~D^A D_A
\ee
There is also an additional contribution to scalar potential, which comes from the supersymmetry breaking. Similarly, in the Lagrangian, there is a contribution coming from supersymmetry breaking. As we discuss in Appendix \ref{Appendix B}, SUSY is broken in the hidden sector and communicated to Supersymmetric SM through messenger fields. The communication of SUSY breaking in hidden sector to visible sector leads to SUSY breaking terms in the Lagrangian. Since SUSY is not broken spontaneously within MSSM but can be broken by adding explicitly supersymmetry breaking terms (source will be hidden sector) in the Lagrangian. However, we can only add those terms which don't re-introduce quadratic divergences back into theory (usually in spontaneous SUSY breaking only these kinds of terms are generated). These terms are known as "soft SUSY breaking terms". There are four types of soft breaking terms in Lagrangian, and all these terms are invariant under the SM gauge group.
\begin{itemize}
\item Scalar mass terms: mass terms for scalar partners of SM fermions and two Higgs doublets in chiral superfields of the MSSM.
    \bea
 \mathcal{L}_{soft} &\supset &m_{Q_{ij}}^2  \tilde{Q}_i^\dagger \tilde{Q}_j +
m_{u_{ij}}^2  \tilde{u^c}_i^\star \tilde{u^c}_j  + 
m_{d_{ij}}^2  \tilde{d^c}_i^\star \tilde{d^c}_j + 
m_{L_{ij}}^2  \tilde{L}_i^\dagger \tilde{L}_j \\ \nonumber  
&+& m_{e_{ij}}^2  \tilde{e^c}_i^\star \tilde{e^c}_j +
m_{H_1}^2H_1^{\dagger}~H_1 + m_{H_2}^2H_2^{\dagger}~H_2
\eea 
here mass parameters can be complex. These terms can violate both CP and flavor.
\item Gaugino mass terms: mass terms for fermionic partners (gauginos) of gauge boson in vector superfields of the MSSM.
\be
 \mathcal{L}_{soft} \supset M_1 \tilde{B}\tilde{B}~+~M_2 \tilde{W}_I \tilde{W}_I~+~\tilde{G}_A\tilde{G}_A
 \ee
 \item Trilinear scalar couplings: They correspond to the cubic scalar terms in the superpotential. All these couplings could be complex numbers and become sources of flavor and CP violation.
 \be
 \mathcal{L}_{soft} \supset A_{ij}^u\tilde{Q}_i\tilde{u}^c_jH_2~+~A_{ij}^d\tilde{Q}_i\tilde{d}^c_jH_1~+~A_{ij}^e\tilde{L}_i\tilde{e}^c_jH_1
 \ee
 \item Bilinear scalar couplings: SUSY breaking terms like bilinear terms in the superpotential.
  \be
 \mathcal{L}_{soft} \supset B~H_1H_2
 \ee
\end{itemize}
Adding above SUSY breaking terms to the supersymmetric Lagrangian terms completes the MSSM Lagrangian. In the next sections, we consider $SU(2)_L\times U(1)_Y$ symmetry breaking and write the spectrum on a physical basis.

\section{Electroweak Symmetry Breaking in MSSM}
AS we know, MSSM has two Higgs doublets. The total scalar potential for the neutral Higgs part in the MSSM is given as: 
\bea
\label{vhneutral}
V_{neutral-higgs}&=& (m_{H_1}^2 + \mu^2) |H_1^0|^2 + (m_{H_2}^2 + \mu^2 ) |H_2^0|^2 -
(B_\mu \mu H_1^0 H_2^0 + H.c) \nonumber \\
&+&  {1 \over 8} ( g^2 + g^{\prime 2}) ({H_2^0}^2 - {H_1^0}^2)^2
\eea
where we parametrise $B\equiv B_{\mu}\mu$, and $H_1^0,H_2^0$ are neutral Higgs scalars. The first three terms of the potential come from superpotential, soft breaking terms (F-terms), and the last term comes from the D-term of the Lagrangian. We choose a field configuration where quartic terms (D-terms) vanish. The first thing we require of a potential is for it to be bounded from below. It gives a condition on the potential:
\be
\label{belowbounded}
2 B_\mu < 2 |\mu|^2~ +~ m_{H_2}^2~ + m_{H_1}^2 
\ee 
 Secondly, for spontaneous symmetry breaking (nonzero vev), at least one of the Higgs mass squared need to be negative. It gives us another condition on potential:
  \be
 B_\mu^2 >  ( |\mu|^2~ +~ m_{H_2}^2~)~(|\mu|^2 + m_{H_1}^2)
\ee 
This equation is valid only at and below the energy scale where the spontaneous breaking of electroweak symmetry becomes operative. Together with nonzero vev, we have to ensure it can reproduce the SM relations like correct gauge boson masses. In MSSM, we can insist that both neutral Higgs get vevs.
\be
\label{vevs}
<H^0_1> = {v_1 \over \sqrt{2}} \;\;\;\;; \;\;\; <H_2^0> = {v_2 \over \sqrt{2}} 
\ee 
where we can write, gauge boson masses as:
\be
M_W=\frac{g_2}{2}(v_1^2+v_2^2)^{1/2}, \;\;\;M_Z=\frac{(g_Y^2+g_2^2)^{1/2}}{2}(v_1^2+v_2^2)^{1/2}
\ee
where we define $v_1^2+v_2^2=v^2=(246\; \texttt{GeV})^2$. ``$v$" is SM Higgs vev. The potential at the minimum eq. \ref{vevs} is given by eq. (\ref{vhneutral}):
\bea
\label{minpot}
V_{neutral-higgs}^{min}&=& \frac{1}{2}(m_{H_1}^2 + \mu^2)v_1^2 + \frac{1}{2}(m_{H_2}^2 + \mu^2 ) v_2^2 - B_\mu \mu~v_1v_2  \nonumber\\ 
&+&  {1 \over 32} ( g^2 + g^{\prime 2}) (v_2^2 - v_1^2)^2
\eea
The consistency conditions for the minimum in eq. (\ref{minpot}) is the vanishing of $\partial V_{neutral-higgs}^{min}/\partial v_1$ and $\partial V_{neutral-higgs}^{min}/\partial v_2$. This gives the following minimization conditions:
\bea
\label{minime}
{1 \over 2} M_Z^2 &=& {m_{H_1}^2 - \tan^2 \beta~ m_{H_2}^2 \over
\tan^2 \beta  - 1} - \mu^2  \nonumber \\
\mbox{Sin} 2 \beta &=& { 2 B_\mu ~\mu~ \over m_{H_2}^2 + m_{H_1}^2 + 2 \mu^2 }
\eea
where  $~tan{\beta}~$  is the ratin of the vevs $v_2/v_1$. These minimization conditions relate the SUSY conserving ($\mu$) and the SM parameters ($M_Z$) with SUSY breaking soft parameters ($m_{H_1}^2,m_{H_2}^2,B_{\mu}$). These minimization conditions in eq.(\ref{minime}) are given for the "tree-level" potential. These conditions are modified for one-loop corrections of potential. In the above discussion, only neutral Higgs scalar potential is considered. The full scalar potential will add further conditions coming from color and charge breaking.
 \section{Mass Spectrum}
After the EWSB, both SM particles and their superpartners will get the mass, and superpartners also will gat mass from soft SUSY breaking terms. Here we discuss the mass spectrum on a physical basis.
\subsection{The Higgs sector}
Let's start with the Higgs fields. We have two Higgs doublets in MSSM. Each Higgs doublet has 4 degrees of freedom. Thus for two Higgs doublets there is a $8\times 8$ Higgs mass matrix. This mass matrix is divided into charged Higgs sector, CP even and CP odd Higgs neutral sector. We will discuss scalar Higgs fields masses here. From the total scalar potential, we can derive the mass matrix using the following definition:
\be
\label{stdscalarmass}
m_{ij}^2 = \left( \begin{array}{cc}  {\partial^2 V \over \partial \phi_i \partial \phi_j^\star} &
  {\partial^2 V \over \partial \phi_i \partial \phi_j }\\
 {\partial^2 V \over \partial \phi_i^\star \partial \phi_j^\star} &
  {\partial^2 V \over \partial \phi_i^\star \partial \phi_j } \end{array} \right)
\ee
\textbf{Charged Goldstones and Higgs}: \\
The total charged Higgs mass matrix
\be
\resizebox{\linewidth}{!}{%
$\displaystyle
\left( \begin{array}{cc} H_1^+ & H_2^+ \end{array} \right) 
\left( \begin{array}{cc} 
(m_{H_1}^2+\mu^2) + {1 \over 8} (g_1^2 + g_2^2) (v_1^2 - v_2^2) + {1 \over 4} g_2^2 v_2^2 & 
(B_{\mu}\mu)^2 + {1 \over 4} g_2^2 v_1 v_2 \\ 
(B_{\mu}\mu)^2 + {1 \over 4} g_2^2 v_1 v_2  &
(m_{H_2}^2+\mu^2) - {1 \over 8} (g_1^2 + g_2^2) (v_1^2 - v_2^2) + {1 \over 4} g_2^2 v_2^2 
\end{array} \right) 
\left( \begin{array}{c} H_1^- \\ H_2^- \end{array} \right) 
$}
\ee
Using the minimisation conditions in eq. (\ref{minime}), this matrix becomes,
\be
\left( \begin{array}{cc} H_1^+ & H_2^+ \end{array} \right) 
({m_3^2 \over v_1 v_2} + {1 \over 4} g_2^2) 
\left( \begin{array}{cc} 
v_2^2 & v_1 v_2 \\ 
v_1 v_2 & v_1^2  \\ 
\end{array} \right) 
\left( \begin{array}{c} H_1^- \\ H_2^- \end{array} \right) 
\ee
which has determinant zero. The following are eigenvalues of the above matrix:
\begin{eqnarray}
m_{G^\pm}^2& =& 0 \nonumber \\
m_{H^\pm}^2& =& \left({m_3^2 \over v_1 v_2} + {1 \over 4} g_2^2\right) (v_1^2 + v_2^2)
\end{eqnarray}
where $G^\pm$ are the Goldstone modes and $H^\pm$ are charged scalar Higgs. The corresponding mass diagonal fields in the unitary gauge are: 
\be
\left( \begin{array}{c} H^\pm \\ G^\pm \end{array} \right) 
= \left( \begin{array}{cc} 
sin\beta & cos\beta \\
-cos\beta & sin\beta
\end{array} \right) 
\left( \begin{array}{c} H_1^\pm \\ H_2^\pm \end{array} \right) 
\ee
\noindent
\textbf{Neutral Goldstone and CP odd Higgs}: \\[2pt]
Choosing $\phi_{i,j}$ in eq.(\ref{stdscalarmass}) to be $\Im(H_{1,2}^0)$, we have corresponding mass matrix: 
\be
\left( \begin{array}{cc} Im H_1^0 & Im H_2^0 \end{array} \right) 
\left( \begin{array}{cc} 
m_1^2 + {1 \over 8} (g_1^2 + g_2^2) (v_1^2 - v_2^2)  & 
m_3^2 \\ 
m_3^2  &
m_2^2 - {1 \over 8} (g_1^2 + g_2^2) (v_1^2 - v_2^2)  
\end{array} \right) 
\left( \begin{array}{c} Im H_1^0 \\ Im H_2^0 \end{array} \right) 
\ee
using the minimisation conditions, this matrix becomes,
\be
\left( \begin{array}{cc} Im H_1^0 & Im H_2^0 \end{array} \right) 
m_3^2
\left( \begin{array}{cc} 
v_2/ v_1  & 1 \\ 
1 & v_1/v_2 \\ 
\end{array} \right) 
\left( \begin{array}{c} Im H_1^0 \\ Im H_2^0 \end{array} \right) 
\ee
The vanishing determinant and non vanishing trace give us a massless neutral Goldstone mode G and a neutral CP odd scalar with mass:
\begin{eqnarray}
m_{G^0}^2& =& 0 \nonumber \\
\label{pseudohiggs}
m_{A^0}^2& =& \left({m_3^2 \over v_1 v_2} \right) (v_1^2 + v_2^2)~~ =~~ {2 m_3^2 \over sin 2 \beta}
\end{eqnarray}
The mixing angle between these two states 
in the unitary gauge is again just tan$\beta$. 
\be
{1 \over \sqrt{2}} 
\left( \begin{array}{c} A^0 \\ G^0 \end{array} \right) 
= 
\left( \begin{array}{cc} 
sin\beta & cos\beta \\
-cos\beta & sin\beta
\end{array} \right) 
\left( \begin{array}{c} \Im(H^0_1) \\ \Im(H_2^0)\end{array} \right) 
\ee
\noindent
\textbf{Neutral CP even Higgs}: \\
Choosing $\phi_{i,j}$ in eq.(\ref{stdscalarmass}) to be $\Re(H_{1,2}^0)$, we have corresponding mass matrix: 
\be
\left( \begin{array}{cc}\Re(H_1^0) & \Re(H_2^0) \end{array} \right) 
~{1 \over 2}~ 
\left( \begin{array}{cc} 
2 m_1^2 + {1 \over 4} (g_1^2 + g_2^2) (3 v_1^2 - v_2^2)  & 
-2 m_3^2 - {1 \over 4} v_1 v_2 (g_1^2 + g_2^2)  \\ 
-2 m_3^2 - {1 \over 4} v_1 v_2 (g_1^2 + g_2^2) & 
2 m_2^2 + {1 \over 4} (g_1^2 + g_2^2) (3 v_2^2 - v_1^2)  
\end{array} \right) 
\left( \begin{array}{c} \Re(H_1^0) \\ \Re(H_2^0) \end{array} \right) 
\ee
Using the minimisation conditions in eq. (\ref{minime}) and the definition of $m_A^2$ from eq.(\ref{pseudohiggs}), this matrix becomes:
\be
\label{cpevenhiggs}
\left( \begin{array}{cc} Re H_1^0 & Re H_2^0 \end{array} \right) 
\left( \begin{array}{cc} 
m_A^2 sin^2 \beta + M_z^2 cos \beta  & 
- (m_A^2 + m_Z^2) sin\beta cos\beta  \\ 
- (m_A^2 + m_Z^2) sin\beta cos\beta  &
m_A^2 cos^2 \beta + M_z^2 sin \beta  
\end{array} \right) 
\left( \begin{array}{c} Re H_1^0 \\ Re H_2^0 \end{array} \right) 
\ee
The eigenvalues of the matrix are: 
\be
m^2_{H,h} = {1 \over 2} \left[ m_A^2+ m_Z^2 \pm \{ (m_A^2 + m_Z^2)^2 - 4 m_Z^2 m_A^2 cos^2 2 \beta\}^{1/2} \right]
\ee 
where $m_h\leq m_H$. In the decoupling limit, where $m_{A^0}>>m_z$, the tree level light Higgs
mass is $m_h\leq m_Z.|\cos{2\beta}|$. The mixing angle
between these two states is: 
\be
tan~ 2\alpha = {m_A^2 + m_Z^2  \over m_A^2 - m_Z^2} ~tan ~ 2 \beta 
\ee

\subsection{The sfermion sector}
We have different scalar fields for the right and left handed fermions. There is mixing term for right and left handed sfermions along with mixing due to EWSB. Furthermore, we can have
mass terms which mix different generations. Thus, in  general
the sfermion mass matrix is a $6 \times 6$ mass  matrix given
as : $$ \tilde{\mathbf{f}}^\dagger ~M_{\tilde f}^2 \tilde{\mathbf{f}}~~ ; ~~~ \tilde{\mathbf{f}} =\{{\tilde f_{L_i}},{\tilde f_{R_i}}\} $$ 
where i stands for generation indices. From the total  scalar potential, the mass matrix for these sfermions can be derived
using eq.(\ref{stdscalarmass}). Thus $\mathbf{\tilde{f}}$ can be $\tilde{u}, \tilde{d}, \tilde{e}, \tilde{\nu}$ except that
$\tilde{\nu}_R=0$ . The sfermion mass matrix can be written as a $2\times 2$ Hermitian matrix of $3\times 3$ blocks in the generation space as following:
\be
M_{\tilde f}^2 \; = \;
\left(\begin{array}{cc}
 m_{\tilde{f}_{L_iL_j}}^2    &  m_{\tilde{f}_{L_iR_j}}^2  \\
  m_{\tilde{f}_{R_iL_j}}^2    &  m_{\tilde{f}_{R_iR_j}}^2
\end{array}\right)
\ee 
where 
\bea
m^2_{\tilde f_{L_i L_j}} &= &  M^2_{\tilde f_{L_i L_j}} + m^2_f \delta_{ij} 
+ M_Z^2 \cos 2 \beta (T_3 + \sin^2 \theta_W Q_{em}) \delta_{ij} \nonumber \\
m_{{\tilde f}_{L_i R_j}}^2& =& \left (\big(  Y^A_f\cdot ^{v_2}_{v_1} - m_f \mu ^{\tan \beta}_{\cot \beta}\big) \;\; \mbox{for}\; f=^{e,d}_u \right) \delta_{ij}  \nonumber \\[.2cm] 
m_{\tilde{f}_{R_iR_j}}^2 &=&  M^2_{\tilde f_{R_iR_j}} + \left( m^2_f + M_Z^2 \cos 2 \beta
\sin^2 \theta_W Q_{em} \right) \delta_{ij}
\eea
In the above, $M^2_{\tilde f_{L}}$ represents the soft mass term for the corresponding fermion ($L$ for left, $R$ for right).
The  sfermion mass matrices are Hermitian  and are thus diagonalised by a unitary rotation:
\be
R_{\tilde f} \cdot M_{\tilde f} \cdot R_{\tilde f}^\dagger = 
\mbox{Diag.}(m_{\tilde f_1},m_{\tilde f_2},\dots,m_{\tilde f_6})
\ee 
\subsection{The neutralino sector}
In this section we will study mass eigenstate for neutral non matter fermions. The neutralinos are mixtures of neutral gauginos and neutral higgsinos. The neutralino
mass matrix in the basis 
$$ \mathcal{L}~ \supset~ {1 \over 2}~ \Psi_N \mathcal{M}_N \Psi_N^T~ + H.c$$ 
where $\Psi_N = \{\tilde B,~\tilde W^0,\tilde H_1^0,\tilde H_2^0\}$
is given as :

\be
\mathcal{M}_N ~=~
\left(\begin{array}{cccc}M_1&0 & - M_Z \cos{\beta}~ \sin{\theta_W}& M_Z \sin{\beta}~ \sin{\theta_W}
\cr0& M_2&  M_Z \cos{\beta}~ \cos{\theta_W}& M_Z \sin{\beta}~ \cos{\theta_W} \cr
- M_Z \cos{\beta}~ \sin{\theta_W}& M_Z \cos{\beta}~ \cos{\theta_W} &0& -\mu\cr
M_Z \sin{\beta}~ \sin{\theta_W}& -M_Z \sin{\beta} ~\cos{\theta_W} & -\mu&0\end{array} \right),
\ee
where $\theta_w$ is the Weinberg angle. The neutralino mass matrix is Majorana mass matrix, hence it is complex symmetric in nature and diagonalised by a unitary matrix N,
\bea
N^* \cdot M_{\tilde N}
\cdot N^\dagger = \mbox{Diag.}
(m_{\chi_1^0}, m_{\chi_2^0}, m_{\chi_3^0}, m_{\chi_4^0})
\eea 
The states are rotated by $\chi_i^0 = N^\star \Psi$ to go the physical basis.
 \subsection{The chargino sector}
In a similar manner to the neutralino sector, all the 
fermionic partners of the charged gauge bosons and the 
charged  Higgs bosons mix after electroweak symmetry breaking.
However, they combine in a such a way that a wino-Higgsino
Weyl fermion pair forms a Dirac fermion. This mass matrix
is given as 
\be
\mathcal{L} \supset -\frac{1}{2} \left(\begin{array}{cc}
\tilde W^- & \tilde H_1^- 
\end{array}\right) \;
\left(\begin{array}{cc}M_2  & \sqrt{2} M_W \sin \beta \cr \sqrt{2} M_W \cos
\beta & \mu\end{array} \right) \left(\begin{array}{cc}\tilde W^+ & \tilde H_2^+
\end{array} \right),
\ee
Given the  non-symmetric (non-Hermitian) matrix  nature of this
matrix, it is diagonalised by a bi-unitary transformation, $U^* \cdot M_C \cdot 
V^\dagger = \mbox{Diag.}(m_{\chi_1^+},m_{\chi_2^+})$. The chargino
eigenstates are typically represented by $\chi^\pm$ with mass eigenvalues
$m_{\chi^\pm}$. The explicit forms for $U$ and $V$ can be found by the 
eigenvectors of $M_C M_C^\dagger$ and $M_C^\dagger M_C$ respectively. 
\chapter{Supersymmetry breaking} 

\label{Appendix B} 

\lhead{Appendix B. \emph{Supersymmetry Breaking}} 

In supersymmetry theories, the biggest issue is the breaking mechanism. Below is a brief discussion of the most commonly used SUSY breaking mechanisms in literature and their problems. We need more elaborate work on the supersymmetry breaking mechanism to understand nature with supersymmetry.

\section{Spontaneous SUSY Breaking}

It is natural to assume models in which SUSY is spontaneously broken (for detail, see review\cite{Luty:2005sn}). There are two ways of breaking: 
\begin{itemize}
\item F-type breaking of SUSY
\item D-type breaking of SUSY
\end{itemize}
 
Suppose we consider tree-level SUSY Lagrangian and spontaneous SUSY breaking, then by using potential minimization condition, we get a tree-level condition which demands the `supertrace' (the `supertrace' is a sum of the squared masses of all particles with spin multiplicities and opposite sign for fermions and bosons) of the mass matrix should vanish. This condition puts strong constraints on how SUSY should break spontaneously. In most general spontaneous SUSY breaking models, vanishing supertrace, electroweak symmetry breaking, and non-observation of superpartners together rule out spontaneous SUSY breaking. In these models, either a few particle masses become tachyonic, or a few of superpartners get masses lighter than the most massive observed fermion mass. Both are disastrous phenomenologically.
 We need either non-renormalizable terms in the Kahler potential or large loop corrections to make a viable SUSY breaking model. These sources are suppressed by high mass scales and by loop factors, respectively. That is why we need a sector (called ``hidden sector") at a larger scale than the electroweak scale where SUSY is broken. SUSY breaking is then communicated to the supersymmetric SM fields (called ``observable sector") through 'messenger' interactions, which break SUSY softly, and this mechanism of breaking is called `Hidden sector SUSY breaking.'
 \section{Hidden Sector SUSY breaking}
The hidden sector fields are singlets under the SM gauge group to prevent tree-level interactions. Therefore we need messenger fields to propagate the breaking effect to the visible sector. 
\subsection{Gravity as a messenger field}
One of the most obvious candidates for a messenger is gravity because it couples to all forms of energy. The gravity couples to the stress-energy tensor also implies that general relativity (an effective theory of gravity) is flavor-blind. But since the different flavors have different masses, and it would still couple differently according to their masses. However, UV completion of gravity theories happen above the Planck scale. It is not clear that gravity's fundamental theory can have flavor symmetries that guarantee the UV couplings of gravity are flavor-blind. There is some hint from the black hole and string theory (a fundamental theory of gravity), which says it is unlikely in the fundamental theory of gravity to allows exact global symmetries such as flavor.  At low-energy, we can parametrize gravity's effect at the Planck scale by higher dimension operators suppressed by powers of the Planck scale. Let us assume there is F-type breaking taking place in the hidden sector where SUSY breaking vev $\langle F_X\rangle\neq0, \langle X\rangle = 0$, X is hidden sector superfield. The detail of these effective operators is given in Chapter \ref{Chapter 2}. The effective operators which will be used for flavor discussion, the Kahler term for masses of scalar fields, and the superpotential term for trilinear coupling:
\be
\frac{z_{ij}}{M_{pl}^2}X^{\dagger}XQ^{i\dagger}Q^j, \frac{a_{ij}}{M_{pl}}XQ^iH_u(U^c)^j+…..
\label{flavorterm}
\ee
These terms would violate flavor symmetry. Even if we use some symmetries to prevent flavor violation through A-term, still $z_{ij}$, which is invariant under all symmetries, will violate flavor. There is no reason for this term to only have flavor-diagonal contribution unless we have flavor symmetry at the Planck scale. It is known as the \textbf{flavor problem of gravity mediated SUSY breaking models}. There are few ways to get rid of this problem, like, gauge flavor symmetry at Planck scale, minimal SUGRA where one assumes scalar masses as universal and diagonal (canonical wavefunction of matter fields). Still, this ansatz breaks electroweak symmetry radiatively (positive point of these ansatzes) and four-form fluxes (will discuss in detail in chapter 2). Another way is to use some other fields as messenger fields instead of gravity. However, in the above discussion, we only talk about integrating out heavy physics at the scale $M_{pl}$. We did not include the effects of the supergravity fields (supersymmetric generalization of Einstein gravity, SUGRA). We consider their effects in Chapter \ref{Chapter 4}.

\subsection{Gauge fields as messenger fields}

To deal with the flavor, we can take natural flavor-blind messengers. There is a problem with tree level breaking, so we try to break SUSY through the loop effect. In this framework, we can assume that SUSY breaking is communicated to the SM via heavy chiral super multiplets, and these superfields are charged under the SM gauge symmetries\cite{Giudice:1998bp,Giudice:1997ni}. The masses of these messenger fields will not remain exact supersymmetric due to SUSY breaking in the hidden sector, and integrating out these fields in the visible sector will also raise SUSY breaking in the visible sector. 
If there is no direct coupling of the hidden sector and visible sector, the leading effect of SUSY breaking on the visible sector will come from the loop diagram involving the messenger fields in loops. In this kind of model, whole observable physics in the visible sector depends only on the messenger mass parameters. That reduces the number of free parameters compared to gravity mediation. In this model, gauginos receive mass at one loop, and scalars receive mass at two-loop as:
\be
M_{i} \sim \frac{g_i^2}{16\pi^2}\frac{F}{\Lambda_{messenger}}, \quad 
m_{scalar}^2 \sim (\frac{g^2}{16\pi^2})^2|\frac{F}{\Lambda_{messenger}}|^2+…
\ee
where F is the SUSY breaking scale, A-terms are negligible in this model. The superpartners' masses are controlled by gauge couplings in this model, so colored states will always be much heavier than uncolored states, by at least a factor of 10. Including experimental lower bound on right-handed selectron, which is around 100 GeV, will imply stops around 1 TeV, and that implies a sizeable fine-tuning for EWSB. After the discovery of the Higgs mass, fine-tuning become severe in these models. The correct Higgs mass needs heavy scalars around 8-10 TeV scale or large A-term that almost vanishes in this model. For large A-terms, the messenger scale has to be very high so that RG running will generate A-terms with 1-2 TeV stop mass, but that requires heavy gluinos or tachyonic stops at a large scale\cite{Draper:2011aa}.
The other constraint comes from gravitino dark matter ($m_{3/2}=F/\sqrt(3)M_{pl}$) which is LSP in this model. Decreasing upper bounds on gravitino mass from cosmological data decreases the upper bound on the SUSY breaking scale ($\sqrt{F}$). The increasing bounds on the gauginos masses from LHC and the Higgs mass ($\sqrt(F)\gsim10^3$ TeV, which give gravitino mass heavier than 100 eV) would increase the lower bound on the SUSY breaking scale. The cosmological bound on gravitino mass($m_{3/2}<4.7$ eV, together with Higgs discovery and direct searches in LHC, excludes minimal gauge mediation with high reheating temperatures (for more details see \cite{Hook:2018sai,Moroi:1993mb}). There is another breaking mechanism in literature such as the Anomaly mediated SUSY breaking\cite{Pomarol:1999ie,Luty:2002hj}, Gaugino mediated SUSY breaking\cite{Kaplan:1999ac,Chacko:1999mi}, Gravity-Gauge mediated SUSY breaking\cite{Iyer:2014iha} and SUSY-breaking using extra dimensions. These both have a fine-tuning problem.

\addtocontents{toc}{} 

\backmatter

\label{Bibliography}

\lhead{\emph{Bibliography}} 

	\bibliographystyle{ieeetr}
	\bibliography{NSUSY.bib}
  \end{document}